\documentclass[nolinenumber]{aastex631}

\usepackage{multirow}
\usepackage{amsmath} 
\usepackage{amsthm} 
\usepackage{amssymb} 

\begin{document}
\pagenumbering{arabic}

\title{Gravitational-Wave and Gravitational-Wave Memory Signatures of Core-Collapse Supernovae}

\author[0000-0001-5386-0133]{Lyla Choi}
\affiliation{Department of Physics, Princeton University, Princeton, NJ 08544}

\correspondingauthor{Lyla Choi}
\email{lc3535@princeton.edu}

\author[0000-0002-3099-5024]{Adam Burrows}
\affiliation{Department of Astrophysical Sciences, Princeton University, Princeton, NJ 08544; Institute for Advanced Study, 1 Einstein Drive, Princeton, NJ 08540}

\author[0000-0003-1938-9282]{David Vartanyan}
\affiliation{Carnegie Observatories, 813 Santa Barbara St., Pasadena, CA 91101}

\date{\today}

\begin{abstract}
In this paper, we calculate the energy, signal-to-noise ratio, detection range, and angular anisotropy of the matter, matter memory, and neutrino memory gravitational wave (GW) signatures of 21 three-dimensional initially non-rotating core-collapse supernova (CCSN) models carried to late times. We find that inferred energy, signal-to-noise ratio, and detection range are angle-dependent quantities, and that the spread of possible energy, signal-to-noise, and detection ranges across all viewing angles generally increases with progenitor mass. When examining the low-frequency matter memory and neutrino memory components of the signal, we find that the neutrino memory is the most detectable component of a CCSN GW signal, and that DECIGO is best-equipped to detect both matter memory and neutrino memory. Moreover, we find that the polarization angle between the $h_+$ and $h_{\times}$ strains serves as a unique identifier of matter and neutrino memory. Finally, we develop a galactic density- and stellar mass-weighted formalism to calculate the rate at which we can expect to detect CCSN GW signals with Advanced LIGO. When considering only the matter component of the signal, the aLIGO detection rate is around 24$\%$ of the total galactic supernova rate, but increases to 92$\%$ when incorporating the neutrino memory component. We find that all future detectors (ET, CE, DECIGO) will be able to detect CCSN GW signals from the entire galaxy, and for the higher-mass progenitors even into the local group of galaxies. 
\end{abstract}

\section{Introduction}
\label{intro}

The routine detection by the Laser Interferometer Gravitational-Wave Observatory (LIGO) \citep{abbott2016}, Virgo Interferometer \citep{Virgo_design}, and Kamioka Gravitational Wave Detector (KAGRA) \citep{Kagra_design} of gravitational waves (GWs) from the merger of black hole and neutron star binaries \citep{LVK_O3} has made the possibility of detecting gravitational waves from core-collapse supernovae (CCSNe) more tangible. There is a long tradition for estimating the GW signature of core-collapse supernovae \citep{wheeler1966, finn1990, muller1982, Moenchmeyer1991, burrows_hayes_1996, muller1997, muller2004, muller2012, ott2009, murphy2009, yakunin:10, couch2013, Kotake2013, muller2013} and the specific features of the theory have recently come into sharper focus \citep{andresen2017, morozova2018, Radice_2019, Mezzacappa2020, vartanyan2020, Shibagaki_2021, jakobus,vartanyan2023, Mezzacappa2024}. The gravitational waves generated in the core of a massive star experiencing a supernova explosion offer insights in real time into the core-collapse supernova process, since each stage of the internal dynamics (core bounce, neutrino-driven convection, explosion, episodic mass accretion, neutron star or black hole formation) has a characteristic GW signature \citep{Marek_2009, Hayama_2018, morozova2018, Radice_2019, vartanyan2023, burrows2023_bh, Mezzacappa2024}. While to date no GWs from supernovae have been observed \citep{Aartsen2014, abbott2016_ccsn, Abbott_2020, Abbott2021, szczepanczyk2023}, both current GW detectors, such as Advanced LIGO \citep{AdvLIGO}, Virgo, and KAGRA, as well as future detectors such as the Cosmic Explorer (CE) \citep{CosmicExplorer}, the Einstein Telescope (ET) \citep{EinsteinTelescope}, and DECIGO \citep{Decigo} are expected to have the sensitivities necessary to observe and characterize a galactic CCSN GW event. 

Hence, at high frequencies ($\sim$100 $-$ 2000 Hz), we expect to be able to capture the GW signals due to violent matter motions and proto-neutron star (PNS) modal oscillations \citep{murphy2009, morozova2018, torres2019, bizouard2021, bruel2023, vartanyan2023}. At lower frequencies (approximately $<$ 10 Hz), there is a gravitational ``memory" component due to anisotropic matter motion and aspherical emission of neutrinos \citep{Epstein1978, Turner1978, burrows_hayes_1996}. Interestingly, the latter two sources may comprise strains one to two orders of magnitude greater than those of the high-frequency matter component \citep{vartanyan2020, vartanyan2023, Takiwaki_2018, Motizuki_2004, muller2012, muller1997, powell2024}. This low-frequency ``memory” can result in, among other things, a permanent shift to the surrounding spacetime after the CCSN explosion subsides \citep{Christodoulou1991, Thorne1992, burrows_hayes_1996}. \citet{Richardson_2022}, \citet{richardson2024}, and \citet{gill2024} have begun to explore the detectability of this memory effect, but there is still no consensus. \citet{richardson2024} use a match-filtering technique to argue for detectability in Earth-based detectors, whereas \citet{gill2024} claims that such low-frequency signals can be detected only with proposed lunar detectors. 

An additional complication arises due to the chaotic and multidimensional nature of CCSN explosions, which causes the GW signals from the matter and neutrino memory emissions to be highly anisotropic and to depend on the angle at which we happen to observe the source, otherwise known as the \textit{viewing angle}. This anisotropy leads to the question posed by \citet{Pajkos2023}: are there viewing angles that correspond to stronger GW signals and how do they change over time? In addition, one can ask the question: how can viewing angle bias our estimates of the total GW energy radiated by a supernova, as well as our estimates of the signals-to-noise and detection ranges derived from three-dimensional simulations?

In this paper, we study in detail the gravitational-wave signature from a suite of twenty-one sophisticated, three-dimensional radiation/hydrodynamic CCSN models with progenitor masses ranging from 9 to 60 $M_{\odot}$ carried out to late times \citep{burrows2024}. These models are the longest-run 3D supernova simulations generated to date, and, for initially non-rotating progenitors, capture the entire gravitational-wave signal of the CCSN phenomenon above $\sim$1 Hz. We include a detailed discussion of 1) the low-frequency memory component of the matter and neutrino components; 2) the anisotropy (viewing angle) dependence of the matter and neutrino strains \citep{Pajkos2023}; 3) the total radiated energies; 4) the signals-to-noise; and 5) the detection ranges \citep{srivastava2019, afle2023}. Understanding the angle dependence of quantities such as total radiated energy, signal-to-noise, and the range out to which we can detect CCSN GWs enables one to understand the error associated with signal analyses that assume spherical symmetry. The full angle dependence of the signal-to-noise and detection range also provides a more nuanced understanding of the detection prospects of GWs from CCSNe compared to many existing CCSN simulation studies in the literature, since any true GW detection from a CCSN event will come only from a particular angular direction, without knowledge of the full, three-dimensional solid-angle profile of the GW strain from the supernova. 

While the anisotropy of CCSN emission has been explored in works such as \citet{muller2012}, \citet{vartanyan2023}, and \citet{Pajkos2023}, those studies were largely restricted to the anisotropy of the matter and/or neutrino strains and not related quantities such as the radiated energy and signal-to-noise. In particular, studies such as \citet{powell2024} highlighted the potential importance of neutrino memory and calculated the anisotropy for a rotating and non-rotating 15-$M_{\odot}$ progenitor \citep{Sukhbold2018}, but for too short a time to witness memory's full strength and frequency content.  Moreover, they focused on only a handful of viewing angles. Here, we analyze the full angular distribution of the GW strain, inferred energy, signal-to-noise, and detection range for different frequency (low and high) regimes across most of the progenitor continuum and for the entire effective duration of the GW emission for most models. Among other factors, what emerges is a new appreciation for the role of neutrino memory at the lower frequencies as an important and detectable feature of the core-collapse supernova gravitational wave signature.

Finally, we use our detection range calculations to develop a stellar-mass distribution weighted mean rate for  galactic CCSN GW detection using Advanced LIGO. Such a calculation has not been performed before and is intended to provide a realistic estimate for the mean time between detections via gravitational radiation of galactic supernovae by constantly-running GW detectors. This estimate takes into consideration the different detection ranges for different progenitor masses that reside along the (Salpeter) mass function of massive stars.

The paper is organized as follows. In Section \ref{summary} we introduce the F{\sc{ornax}} simulation models and explain the numerical simulation setup. We then provide an overview of the strain, total radiated energy, signal-to-noise, and detection ranges due to GWs from matter motions in Section \ref{matter_motion}. In Section \ref{matter_angle} we discuss the viewing angular anisotropy of the GW strain, inferred radiated energy, SNR, and detection range. In Section \ref{matter_memory} we explore the low-frequency matter memory component of the GW signal. We further investigate the effects of memory on the polarization angle between the emitted $h_+$ and $h_{\times}$ strains in Section \ref{polar}. In addition to the low-frequency matter memory, in Section \ref{neutrino_memory} we study the low-frequency strain contribution from neutrino emission and explore its angular anisotropy, energy, SNR, and detection range contributions. Finally, in Section \ref{rate}, using the combined detection range calculated from the matter and neutrino strain, we develop a stellar density-weighted rate for the detection of GWs from CCSN using a galactic density models developed by \citet{McMillan2016}. We summarize our general conclusions in Section \ref{conclusions}.

\section{Simulation Summary}
\label{summary}

In this paper, we analyze the GW emission of twenty-one initially non-rotating progenitor mass models, ranging from 9 to 60 $M_{\odot}$ \citep{Sukhbold2016, Sukhbold2018} that were evolved in three dimensions using the radiation/hydrodynamic code F{\sc{ornax}} \citep{Skinner2019}\footnote{For the GW signatures of models with initial spin, the reader is referred to \citet{ott2009}; \citet{kuroda:14}; \citet{fuller2015}; \citet{richers2017}; \citet{tk18}; \citet{Shibagaki_2021}; and \citet{powell2024}}. Many of these models have already been published in other contexts \citep{vartanyan2023,burrows2023_bh,burrows2024}.
In particular, \citet{burrows2024} focus on the correlation of CCSN observables with progenitor structure and with one another. That paper contains a detailed discussion of the simulation parameters and specifications, to which we refer.

These initially non-rotating models span much of the entire supernova progenitor mass continuum and are also the longest 3D core-collapse simulations to date, with the simulations lasting from one-and-a-half to over four seconds, whereas other simulations in the literature run to around only $\sim$0.5$-$1.0 seconds \citep{richardson2024, Mezzacappa2024, powell2024}. This is significant because simulations running to only $\sim$0.5$-$1.0 seconds do not capture the entire time evolution of the GW signal and cannot probe sufficiently low frequencies. Additionally, while most simulations dump data at a rate of one millisecond, a F{\sc{ornax}} simulation  dumps data at a sufficiently high cadence to avoid Nyquist sampling limitations \citep{Radice_2019,vartanyan2023}. Depending upon the model, the Nyquist frequency for the matter signal ranges from 5000-8000 Hz, enabling a much higher frequency resolution of the GW signal. However, we sample the neutrino data only every millisecond.

In Table \ref{model_summary} we summarize the duration of the simulation, the total gravitational-wave energy radiated, the compactness $\xi_{1.75}$ \citep{O'Connor_2011}\footnote{Defined as $\frac{M/M_\odot}{R(M)/1000\text{km}}$, we have set $M$ equal to 1.75 $M_{\odot}$.}, and whether a neutron star (NS) or black hole (BH) is formed for each of our progenitor models. While seventeen of our models leave behind a neutron star, four of our models (models 12.25, 14, 19.56, 40 $M_{\odot}$) result in black holes. Two of the latter (12.25 and 14) collapse into black holes after minutes to hours without exploding and two (19.56 and 40) explode vigorously, but birth black holes within seconds of that explosion \citep{burrows2023_bh}. The difference between models 9a and 9b is that the former has imposed perturbations due to convection in the initial model, while the latter does not. \citet{burrows2024} and \citet{wang2023} have suggested that it is only for the lowest mass progenitors that explode quickly that the imposition of initial physical perturbations makes a demonstrable difference in model aspects of the outcomes.

\begin{deluxetable*}{ccccc}
\tablecolumns{5}
\tablewidth{0pt}
\label{model_summary}

\begin{minipage}{\textwidth}
  \centering
  \textbf{Model Summary} \\  
\end{minipage}

\tablehead{\colhead{Progenitor} & \colhead{Duration (s)} & \colhead{Total GW Energy Radiated ($M_{\odot}c^2$)} & \colhead{Compactness $\xi_{1.75}$} & \colhead{Final Object}}
\startdata
    9a & 1.72 & 2.69 $\times 10^{-11}$ & 6.7 $\times 10^{-5}$ & NS\\
    9b & 2.01 & 3.59 $\times 10^{-11}$ & 6.7 $\times 10^{-5}$ & NS\\
    9.25 & 2.75 & 2.93 $\times 10^{-10}$ & 2.5 $\times 10^{-3}$ & NS\\
    9.5 & 2.14 & 5.56 $\times 10^{-10}$ & 8.5 $\times 10^{-3}$ & NS\\
    11 & 3.08 & 5.31 $\times 10^{-9}$ & 0.12 & NS\\
    12.25 & 2.01 & 3.67 $\times 10^{-9}$ & 0.34 & BH*\\
    14 & 2.49 & 6.97 $\times 10^{-9}$ & 0.48 & BH*\\
    15.01 & 3.80 & 1.28 $\times 10^{-8}$ & 0.29 & NS\\
    16 & 4.16 & 3.57 $\times 10^{-9}$ & 0.35 & NS\\
    17 & 1.95 & 4.83 $\times 10^{-8}$ & 0.74 & NS\\
    18 & 4.23 & 1.02 $\times 10^{-8}$ & 0.37 & NS\\
    18.5 & 3.85 & 2.70 $\times 10^{-8}$ & 0.80 & NS\\
    19 & 4.05 & 1.73 $\times 10^{-8}$ & 0.48 & NS\\
    19.56 & 3.86 & 4.33 $\times 10^{-8}$ & 0.85 & BH\\
    20 & 3.84 & 2.50 $\times 10^{-8}$ & 0.79 & NS\\
    21.68 & 1.57 & 2.89 $\times 10^{-8}$ & 0.84 & NS\\
    23 & 4.20 & 1.98 $\times 10^{-8}$ & 0.74 & NS\\
    24 & 3.82 & 2.65 $\times 10^{-8}$ & 0.77 & NS\\
    25 & 3.83 & 2.77 $\times 10^{-8}$ & 0.80 & NS\\
    40 & 1.62 & 9.36 $\times 10^{-8}$ & 0.87 & BH\\
    60 & 4.45 & 2.16 $\times 10^{-8}$ & 0.44 & NS\\
\enddata
\caption{Summary of CCSN models from recent 3D F{\sc{ornax}} simulations \citep{burrows2024}, indexed by their mass in solar mass ($M_{\odot}$) units. Here ``Total Energy Radiated" refers to energy radiated in GWs due to matter motions. There are four BH-forming models, and the asterisk * denotes the BH-forming models which do not explode (12.25 and 14 $M_{\odot}$). The higher-mass progenitors run for longer on average and radiate more energy, but this relationship is not strictly monotonic, though it is monotonic with compactness \citep{vartanyan2023}.}
\end{deluxetable*}

\section{Gravitational Waves from Matter Motion}
\label{matter_motion}
\subsection{Strain due to Matter Motion}
\label{matter_strain}
The plus and cross polarizations of the GW strain are calculated from the second time derivative of the quadrupole tensor as \citep{oohara1997, muller2012}:
    \begin{equation}
    \begin{aligned}
    \label{h_strain_pre}
    h_+^{TT}(\theta, \phi) &= \frac{G}{c^4D}\left(\frac{d^2Q_{\theta\theta}}{dt^2}-\frac{d^2Q_{\phi\phi}}{dt^2}\right) \\
    h_\times^{TT}(\theta, \phi) &= \frac{2G}{c^4D}\frac{d^2Q_{\theta\phi}}{dt^2} \, ,
    \end{aligned}
    \end{equation}
    where $D$ is the distance to the source and $ 0 \leq \theta \leq \pi$ (measured from the z-axis) and $ 0 \leq \phi \leq 2\pi$ (measured from the x-axis) are the \textit{viewing angles}. The quadrupole tensor in spherical coordinates is given in terms of the Cartesian components as
    \begin{equation}
    \begin{aligned}
    \label{spherical_quads}
    Q_{\theta\theta} &=(Q_{xx}\cos^2\phi+Q_{yy}\sin^2\phi+2Q_{xx}\sin\phi\cos\phi)\cos^2\theta +Q_{zz}\sin^2\theta-2(Q_{xz}\cos\phi+Q_{yz}\sin\phi)\sin\theta\cos\theta \\
    Q_{\phi\phi} &= Q_{xx}\sin^2\phi + Q_{yy}\cos^2\phi - 2Q_{xy}\sin\phi\cos\phi \\
    Q_{\theta\phi} &= (Q_{yy} - Q_{xx})\cos\theta\sin\phi\cos\phi + Q_{xy}\cos\theta(\cos^2\phi-\sin^2\phi)+Q_{xz}\sin\theta\sin\phi -Q_{yz}\sin\theta\cos\phi \,\, .
    \end{aligned}
    \end{equation}

    \begin{figure*}[htbp!]
    \centering
    \includegraphics[width=0.42\linewidth]{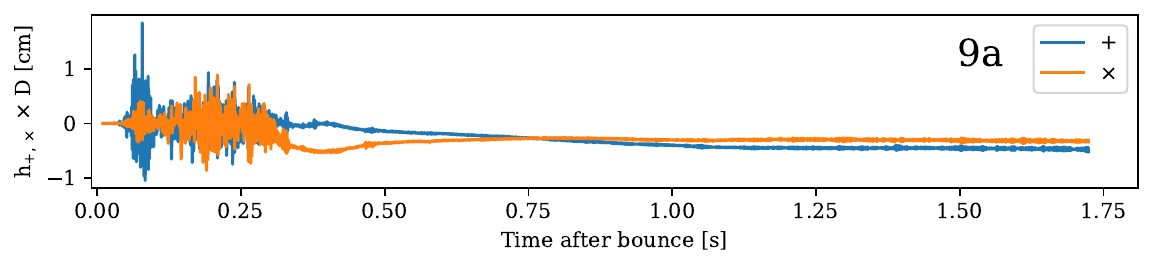}
    \includegraphics[width=0.42\linewidth]{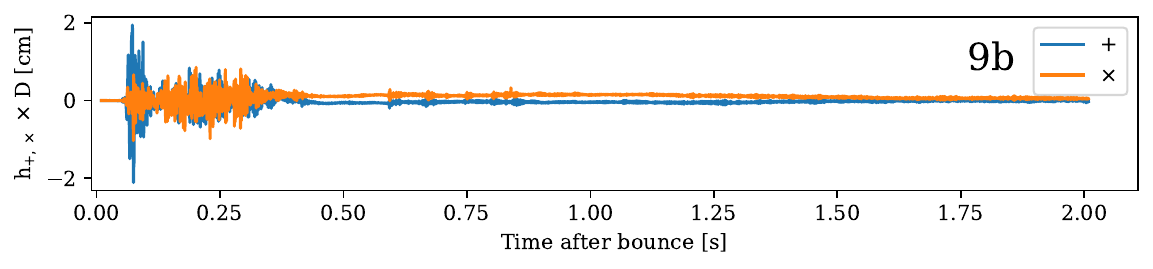}
    \includegraphics[width=0.42\linewidth]{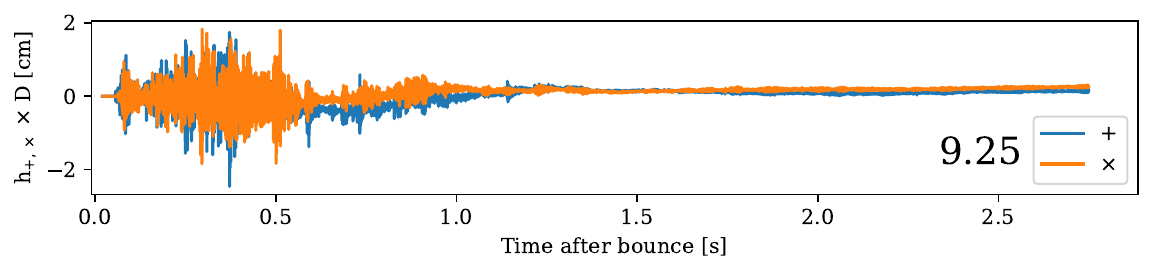}
    \includegraphics[width=0.42\linewidth]{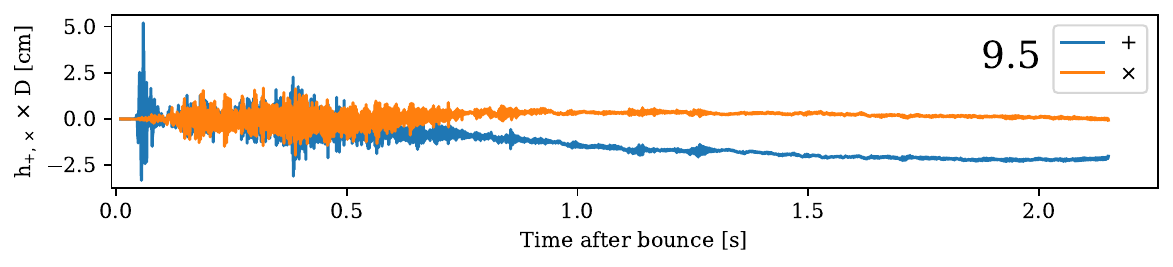}
    \includegraphics[width=0.42\linewidth]{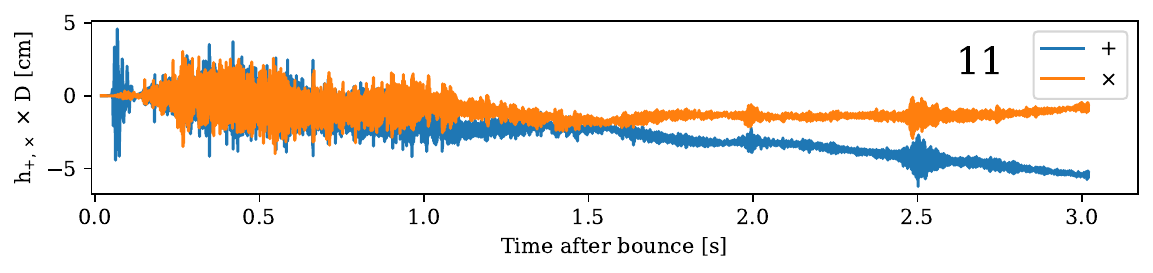}
    \includegraphics[width=0.42\linewidth]{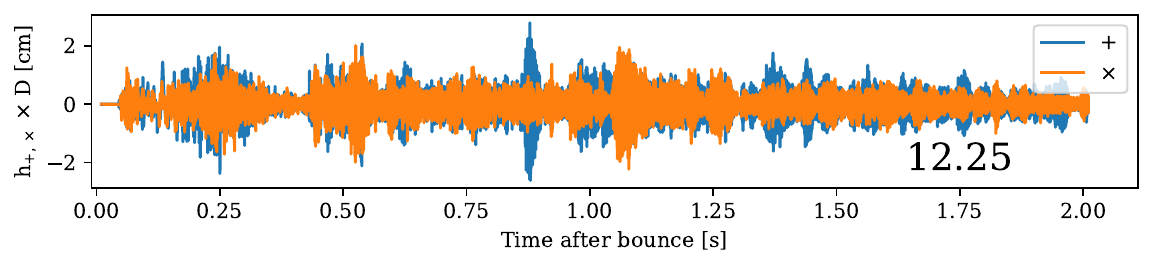}
    \includegraphics[width=0.42\linewidth]{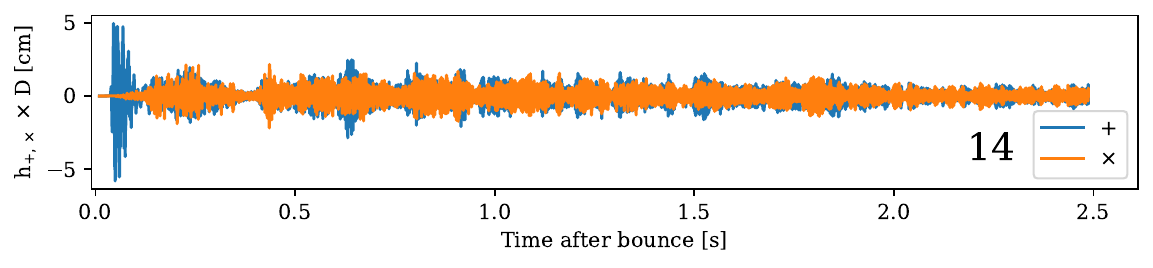}
    \includegraphics[width=0.42\linewidth]{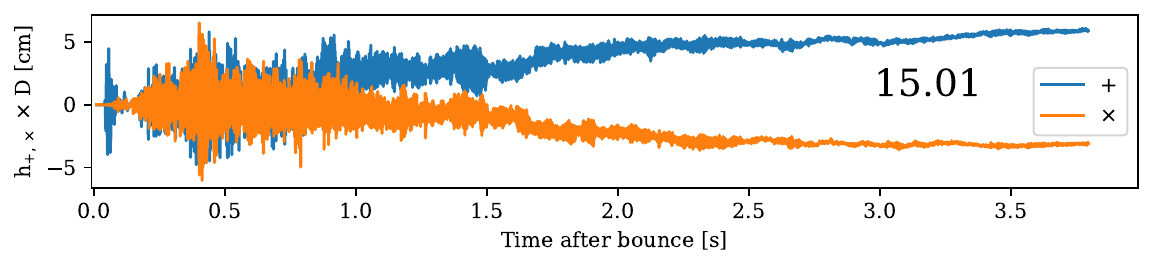}
    \includegraphics[width=0.42\linewidth]{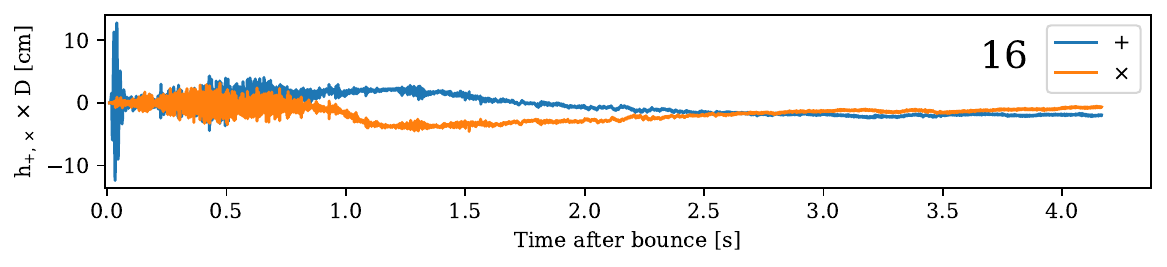}
    \includegraphics[width=0.42\linewidth]{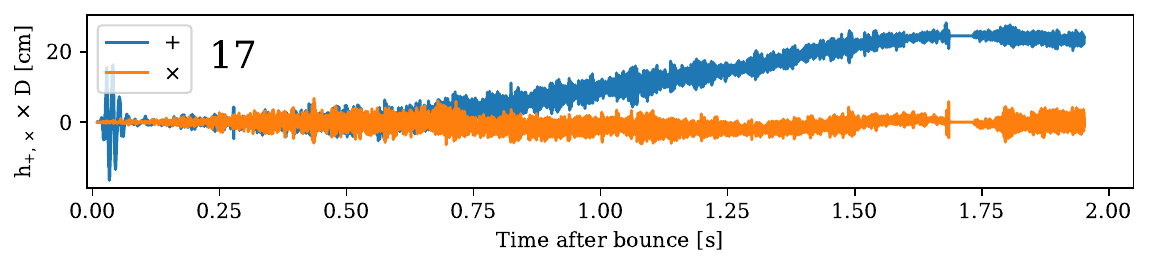}
    \includegraphics[width=0.42\linewidth]{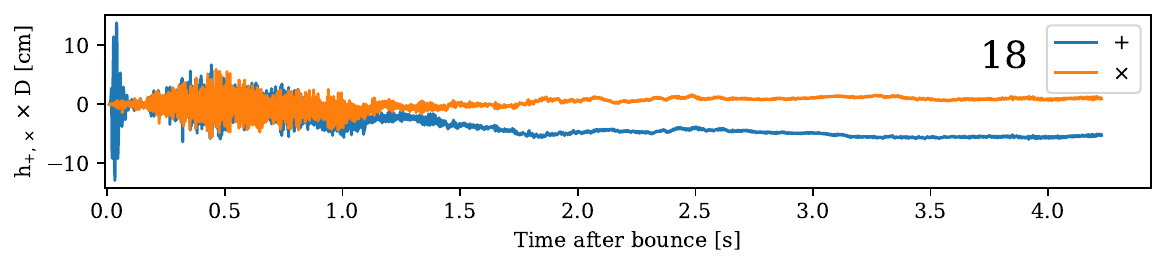}
    \includegraphics[width=0.42\linewidth]{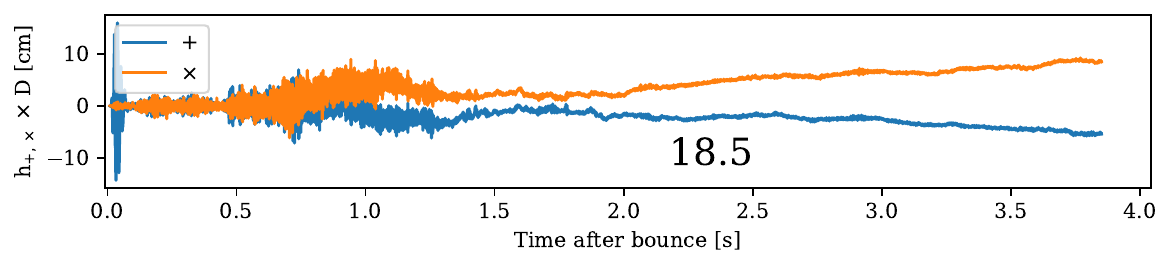}
    \includegraphics[width=0.42\linewidth]{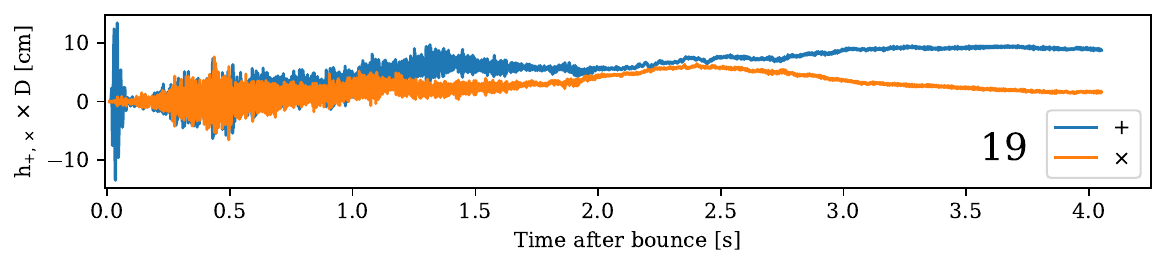}
    \includegraphics[width=0.42\linewidth]{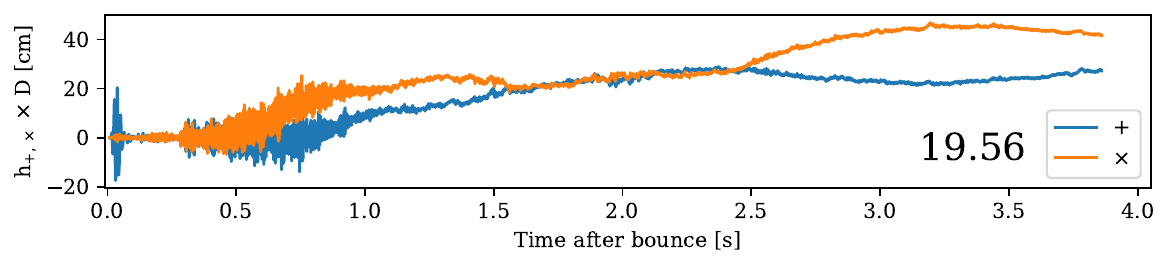}
    \includegraphics[width=0.42\linewidth]{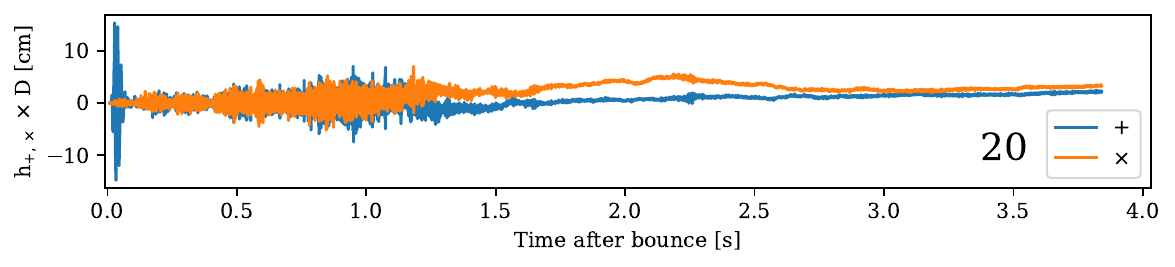}
    \includegraphics[width=0.42\linewidth]{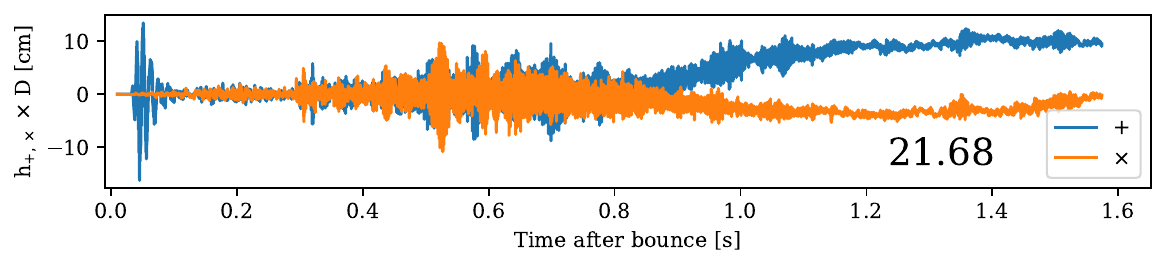}
    \includegraphics[width=0.42\linewidth]{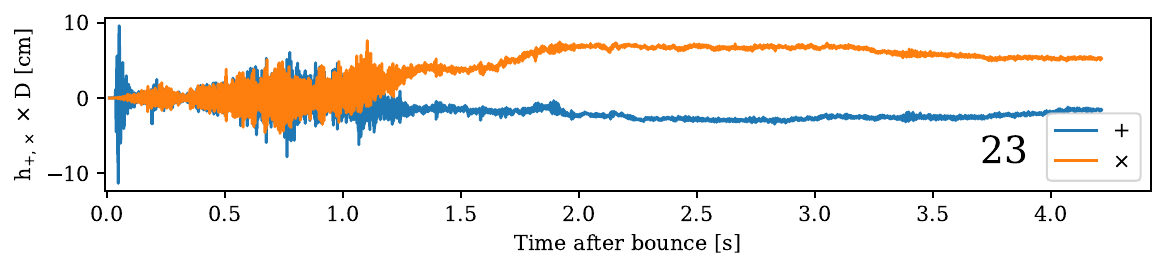}
    \includegraphics[width=0.42\linewidth]{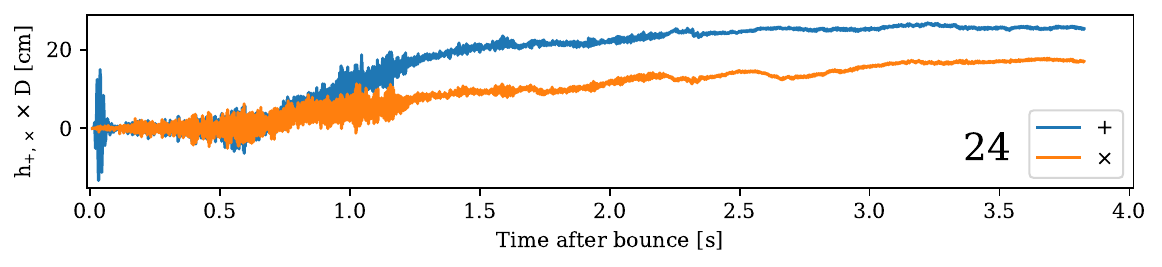}
    \includegraphics[width=0.42\linewidth]{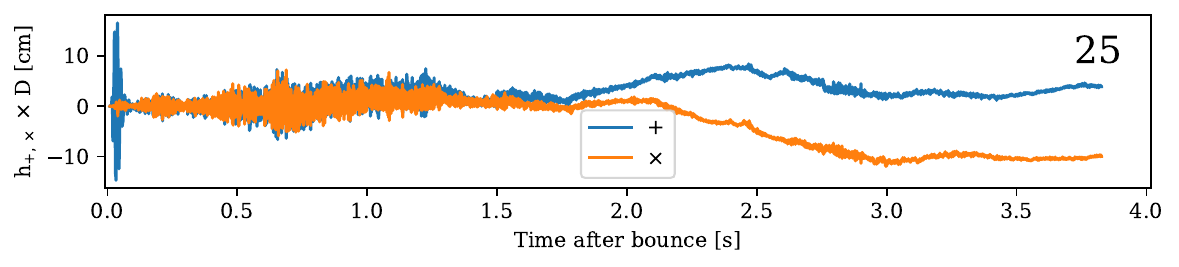}
    \includegraphics[width=0.42\linewidth]{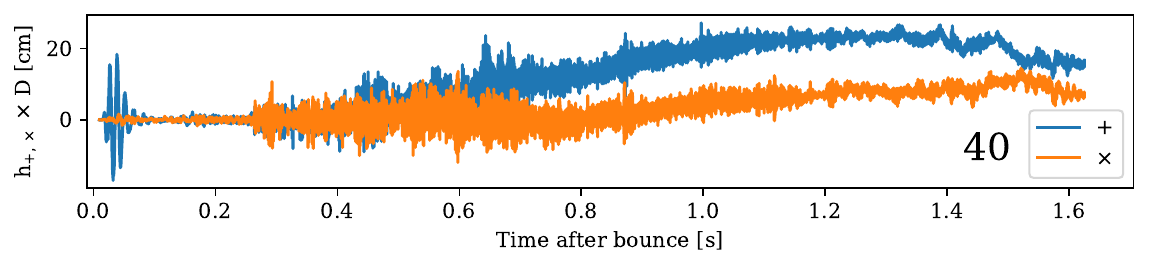}
    \includegraphics[width=0.42\linewidth]{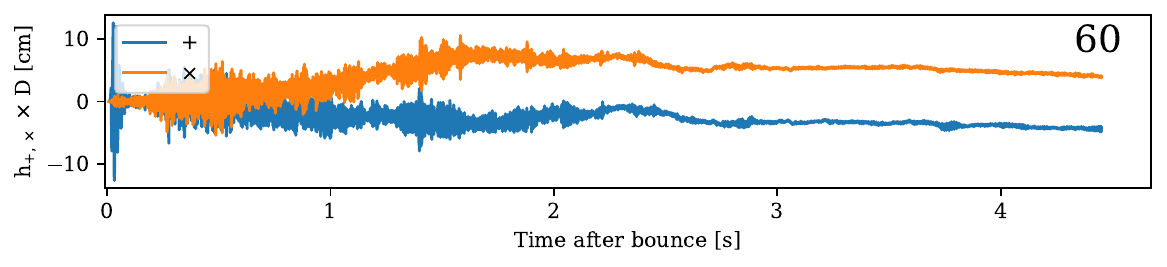}
    \caption{The $h_{+, \times}$ GW strains times distance due to matter motions for all progenitor models viewed from the x-direction. The exploding models have a high-frequency burst within the first $\approx$ 0.5-1.5 seconds. In general, the higher progenitor masses achieve greater strain values than lower-mass progenitors, but this relationship is not strictly monotonic. The 12.24 and 14 progenitor models do not explode and are steadily accreting matter at the end of the simulation. Hence, they will eventually collapse into black holes, but at a time beyond the duration of the simulations. Thus, unlike the other models, their GW signals do not taper after a few seconds after bounce. We note that the strains of almost all of the exploding models do not return to zero by the end of the simulation, indicating the permanent GW memory effect. \label{all_strains}}
    \end{figure*}

    \begin{figure*}[ht]
    \centering
    \includegraphics[width=0.75\linewidth]{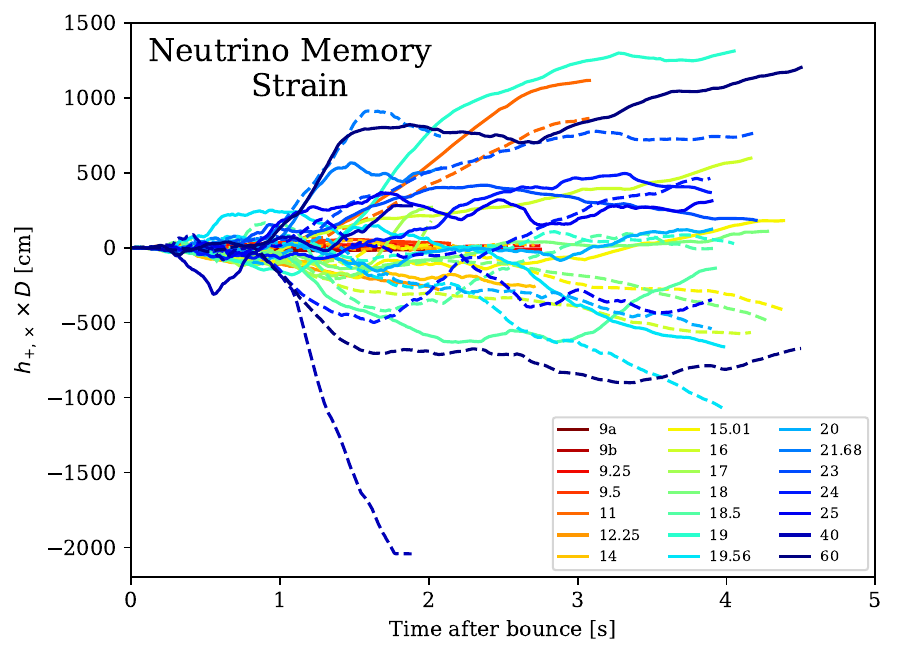}
    \caption{The $h_+$ (solid lines) and $h_{\times}$ (dashed lines) GW memory strain due to neutrino emission viewed from the x-direction for all three neutrino species combined. Unlike the matter strains in Figure \ref{all_strains}, this strain has only a low-frequency component and does not have the high-frequency burst associated with matter strains. The magnitude of the strain increases roughly with progenitor mass. Most notably, the magnitude of the strain is almost two orders of magnitude larger than the strain due to matter motions, reaching almost 2000 cm (divided by distance) for the 40 progenitor model. The neutrino memory strain also continues to grow for all progenitor models until the end of the simulation, indicating the need to carry out CCSN simulations to late times.  \label{neu_strain_plot}}
    \end{figure*}

In Figure \ref{all_strains}, we plot the plus and cross polarizations of the matter component of the GW strain in the x-direction $(\theta=\pi/2, \phi=0)$ multiplied by the distance to the source for all core-collapse models in this study. We also plot the corresponding neutrino memory strain in the x-direction in Figure \ref{neu_strain_plot}, which will be further discussed in Section \ref{direction_memory}. In the matter strain for all of the exploding NS-forming models, we can see a burst of emission in the first $\sim$50-150 milliseconds, followed by signals of significant strength lasting $\sim$0.5$-$1.5 seconds, after which the signal settles down into a lower frequency regime until the end of the simulation. The strain, however, does not return to zero for the majority of the models, an indication of the GW memory effect. These overall phases of the matter strain are described in more detail in \citet{vartanyan2023} and we refer to this paper for a more comprehensive discussion of the associated astrophysics. More massive progenitors generally experience larger strains for a longer period of time, and for the NS-forming models there is a good correlation between the strain and model compactness \citep{vartanyan2023}. Moreover, though the strains seem wildly stochastic, there is an underlying order in the frequency-time domain, with a mixed g/f-mode whose frequency increases inexorably with time carrying most of the GW power. For the accreting, non-exploding BH-formers (12.25 and 14), the high-frequency, stochastic, accretion plume excited signal is maintained throughout the entire simulation time and will likely continue for some time before collapsing into a BH after the end of the simulation \citep{burrows2023_bh}. Note that these non-exploding models manifest no appreciable memory component.
Overall and as stated earlier, the supernova models for the lower-mass progenitors have smaller strains which last for a shorter amount of time. The earlier explosion times for the lower-mass models are due to the steeper mass density profiles in their cores \citep{burrows2021, burrows2024}.
    
\subsection{Total Radiated Energy}
\label{total_energy}
The total energy radiated in GWs is calculated from the third time derivative of the quadrupole tensor \citep{vartanyan2023, oohara1997}: 
    \begin{equation}
    \begin{aligned}
    \label{E_GW}
    E_{GW} &= \frac{G}{5c^5}\int_0^{t}dt'\left[\left(\frac{d^3Q_{xx}}{dt^3}\right)^2+\left(\frac{d^3Q_{yy}}{dt^3}\right)^2+\left(\frac{d^3Q_{zz}}{dt^3}\right)^2 +2\left(\left(\frac{d^3Q_{xy}}{dt^3}\right)^2+\left(\frac{d^3Q_{xz}}{dt^3}\right)^2+\left(\frac{d^3Q_{yz}}{dt^3}\right)^2\right)\right] \, .
    \end{aligned}
    \end{equation}
Understanding the energy emitted by the supernova provides us with valuable insights into the compactness of the progenitor core and thus the dynamics of the high-pressure matter involved in the supernova process (\cite{vartanyan2023}). 

   \begin{figure*}[ht]
    \centering
    \includegraphics[width=0.5\textwidth]{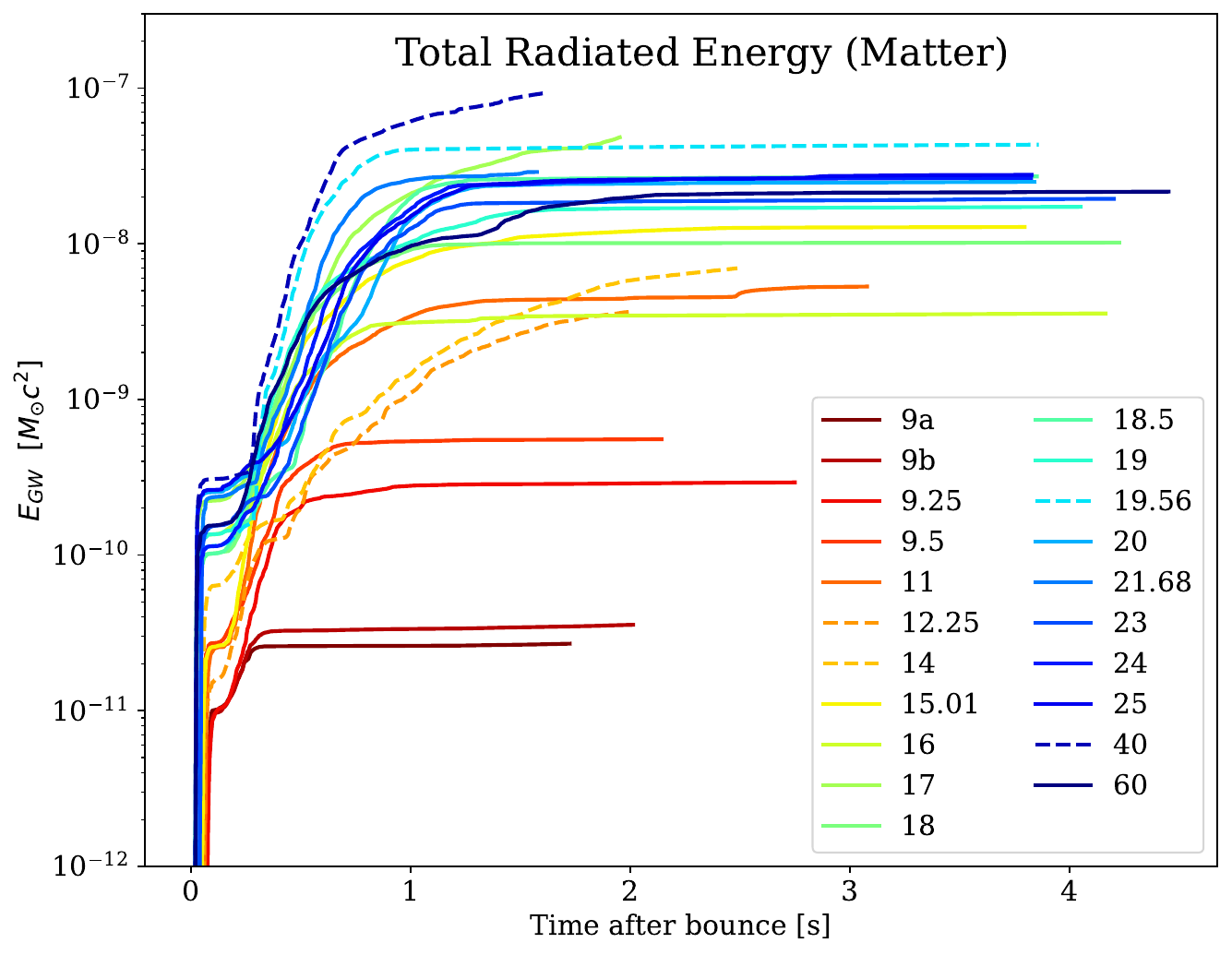}\hfill
    \includegraphics[width=0.5\textwidth]{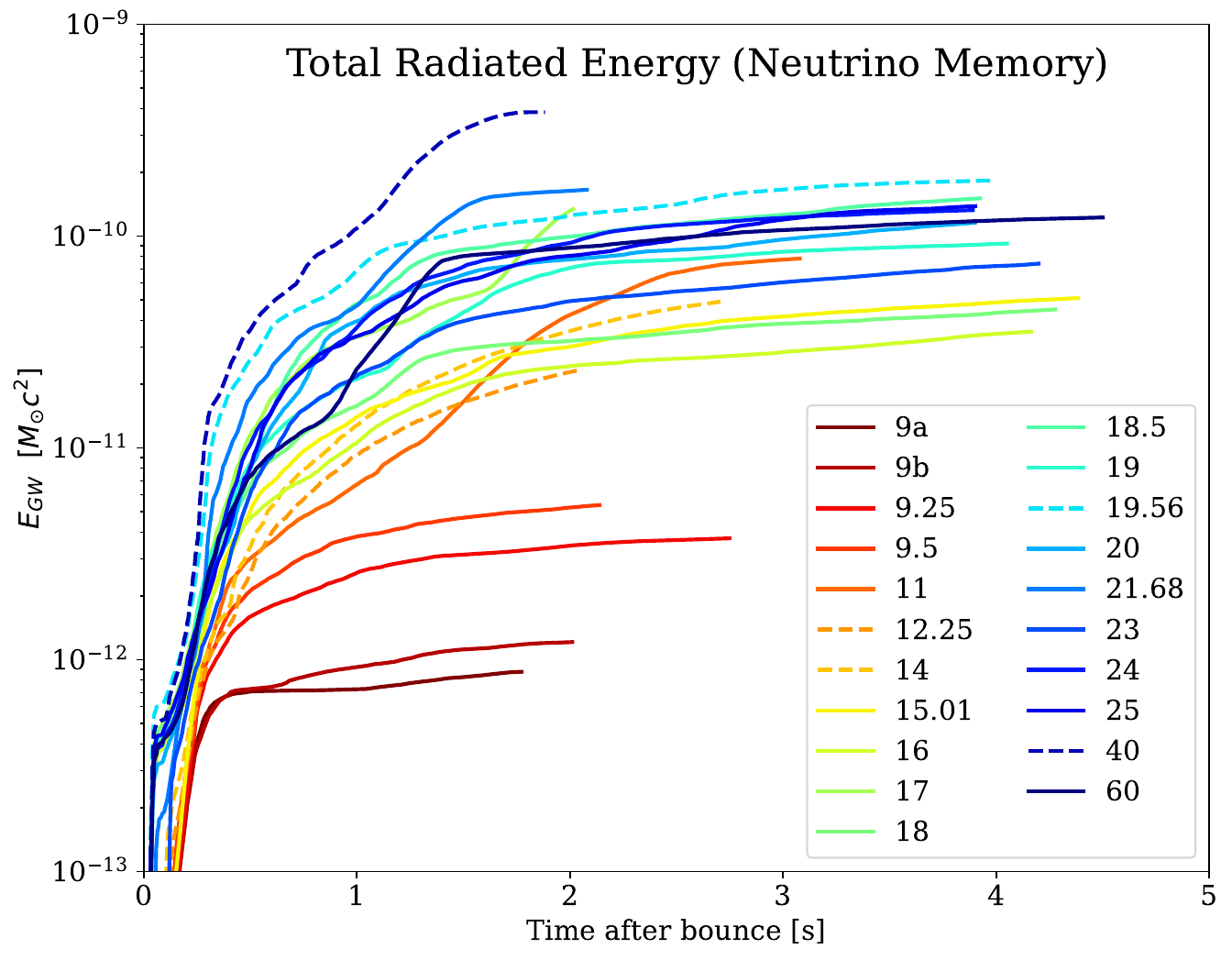}
    \caption{(Left) the total energy radiated in GWs due to matter motion for each progenitor model as a function of time after bounce. (Right) The total energy radiated in GWs due to neutrino emission for each progenitor model as a function of time after bounce. The BH-forming models are plotted with dotted lines. For the radiated GW energy due to matter motions, the total GW energy differs by around three orders of magnitude from the least massive 9a and 9b models to the BH-forming 40 and 19.56 progenitor models. Additionally, the total energy radiated from GWs due to matter motions is around two orders of magnitude greater than the energy radiated in GWs due to neutrino memory. For the matter plot, all models' radiated energy grows several orders of magnitude within the first second after bounce and then asymptotes to a constant value, whereas the energy radiated due to GWs from neutrino memory continues to grow until the end of the simulation for all progenitor models, demonstrating the importance of carrying out CCSN simulations to late times. \label{EGW_tot}}
    \end{figure*}

On the left side of Figure \ref{EGW_tot} we plot the integrated radiated GW energy due to matter motions using equation \eqref{E_GW}. The final integrated energies by the end of the simulation are also reported in Table \ref{model_summary}. On the right side of Figure \ref{EGW_tot}, we plot the integrated radiated GW energy due to neutrino memory for comparison, which will be further discussed in Section \ref{nu_memory_energy}. For the energy radiated due to matter motions, the total energy radiated roughly increases with progenitor mass; as the 40 $M_{\odot}$ model radiates the most GW energy, while the 9 $M_{\odot}$ models radiate the least, but this relation is not strictly monotonic. From Table \ref{model_summary}, we can see instead that total radiated energy is roughly monotonic with compactness. The increasing blast explosion energy with progenitor mass is related to the trend of heavier progenitors having shallower density profiles, since shallower density profiles result in a higher accretion-powered neutrino luminosities and greater neutrino optical depths in the ``gain" region behind the stalled shock, which together are implicated in higher CCSN explosion energies \citep{burrows2021,burrows2024}. These explosion energies vary by about a factor of ten. However, the corresponding radiated GW energy differs by three orders of magnitude from the least massive to the most massive progenitor. For all of the models, the radiated energy steeply increases for the first 0.5 seconds of the simulations, and while the lower-mass models plateau earlier, all of the models have radiated almost all of their GW energy after the first 1.5 seconds, with the exception of the non-exploding BH-forming models whose energy curves continue to increase even by the end of the simulation.

\subsection{Signal-to-Noise Ratio}
\label{snr}
To determine the ``detectability" of the GW signals from the various progenitors, we calculate the signal-to-noise ratio (SNR) which is given by \citet{flanagan_hughes_1998} and \citet{andresen2017} as
    \begin{equation}
    \label{SNR}
    \rho^2 = 4\int_0^{\infty} df \frac{|\tilde{h}(f)|^2}{S_n(f)} \,\, ,
    \end{equation}
    where $\tilde{h}(f)$ is the Fourier transform of the strain and $S_n(f)$ is the detector noise/sensitivity curve. Here, we define the Fourier transform as
    \begin{equation}
    \label{fourier}
    \tilde{h}(f) = \int_{-\infty}^{\infty}e^{2i\pi ft}h(t)dt \,\, .
    \end{equation}
    Note that eq. \eqref{fourier} has no normalization factor outside of the integral and that the exponent is positive. This Fourier transform follows the convention of \citet{flanagan_hughes_1998}. 
    Here, $h$ is defined as a linear combination of the $h_+$ and $h_{\times}$ polarizations, given by
    \citet{moore2014} and \citet{afle2023} as
    \begin{equation}
    \label{h_antenna}
    h(t) = F_+(l, b, \psi)h_+ + F_{\times}(l, b, \psi)h_{\times} \,\, ,
    \end{equation}
    where $F_+(l, b, \psi)$ and $F_{\times}(l, b, \psi)$ are known as the \textit{antenna pattern functions} defined as \citep{schutz2011, flanagan_hughes_1998, Mezzacappa2024}: \footnote{Note that in the GW literature $l, b,$ are typically labeled as $\theta, \phi$, but here we relabel them to avoid confusion with the viewing angles $\theta, \phi$.}
    \begin{equation}
    \label{antenna_pattern}
    \begin{aligned}
    F_+(l, b, \psi) &=  \frac{1}{2}[1+\cos^2l]\cos(2l)\cos(2\psi) -\cos l\sin(2b)\sin(2\psi) \\
    F_{\times}(l, b, \psi) &= \frac{1}{2}[1+\cos^2l]\cos(2l)\sin(2\psi) + \cos l\sin(2b)\cos(2\psi) \,\, .
    \end{aligned}
    \end{equation}
    These antenna pattern functions weight the signal according to its location and orientation in the sky, as the strength of the signal depends on its orientation relative to the orientation of the detector. 
    
    If we assume the source at a given location in the sky to have a randomly-oriented polarization and that the ensemble of such sources contains systems with all possible polarization angles \citep{schutz2011}, the average power SNR over that ensemble is given by the average over the polarization angle $\psi$ where we take $\langle ... \rangle_{\psi} = \frac{1}{\pi}\int_0^{\pi}d\psi$ \citep{flanagan_hughes_1998}. Using the fact that $\langle F_+F_{\times} \rangle_{\psi} = 0$ and $\langle F_+^2 \rangle_{\psi} = \langle F_{\times}^2 \rangle_{\psi}$, the SNR becomes \citep{schutz2011}
    \begin{equation}
    \label{rho_avg}
        \langle\rho^2\rangle_{\psi} = 2[F_+^2(l, b) + F_{\times}^2(l, b)]\int_0^{\infty}\frac{|\tilde{h}_+(f)|^2+|\tilde{h}_{\times}(f)|^2}{S_n(f)}df \,\, .
    \end{equation}    

\citet{afle2023} and \citet{srivastava2019} calculate quantities such as signal-to-noise and detection range assuming an optimally-oriented source, which is when $(l, b) = (0, 0)$ so that $F_+^2(0, 0) + F_{\times}^2(0, 0) = 1$. This provides the most optimistic value for detection, as averaging over all angles in the antenna pattern function results in a lower factor of $\frac{1}{5}$ \citep{flanagan_hughes_1998}. We follow the more optimistic convention of $(l, b) = (0, 0)$ instead of averaging over the angles, so that our final expression is:
    \begin{equation}
        \label{SNR_final}
        \rho = \sqrt{2\int_0^{\infty}\frac{|\tilde{h}_+(f)|^2+|\tilde{h}_{\times}(f)|^2}{S_n(f)}df} \,\, ,
    \end{equation}
which matches equation (6) in \citet{schutz2011} for $l=b=0$.

\begin{figure*}[ht]
    \centering
    \includegraphics[width=0.49\linewidth]{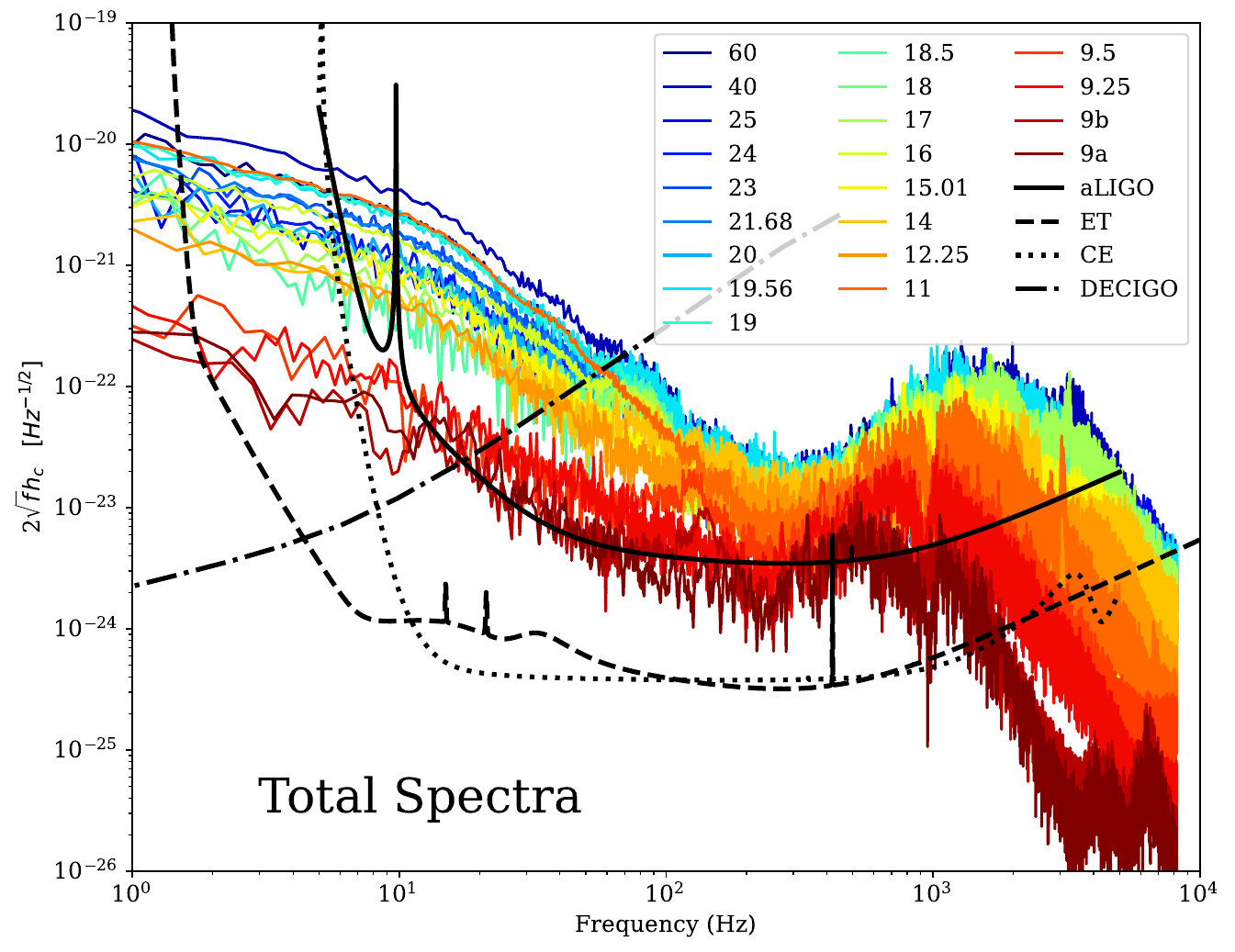}
    \includegraphics[width=0.49\linewidth]{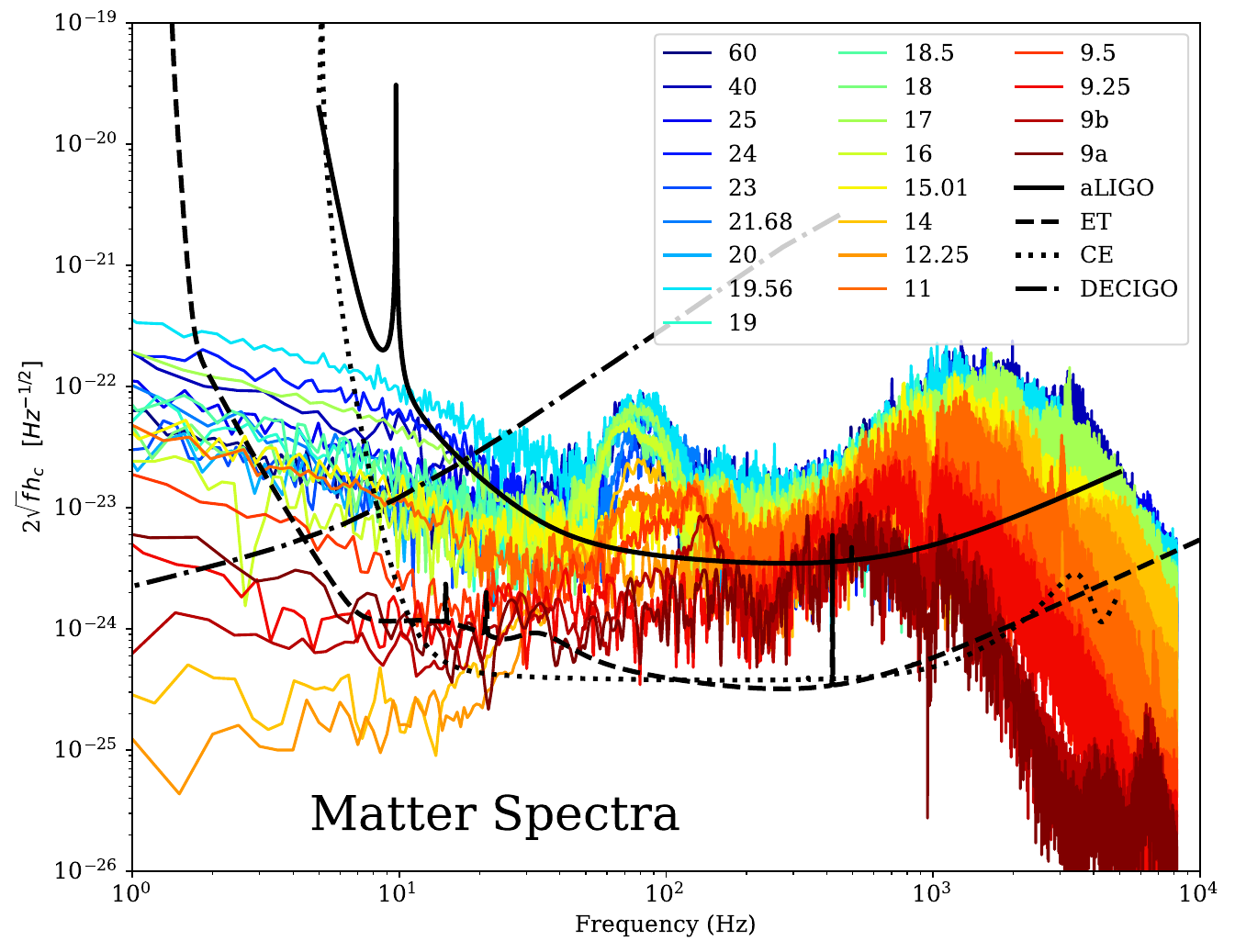}
    \includegraphics[width=0.49\linewidth]{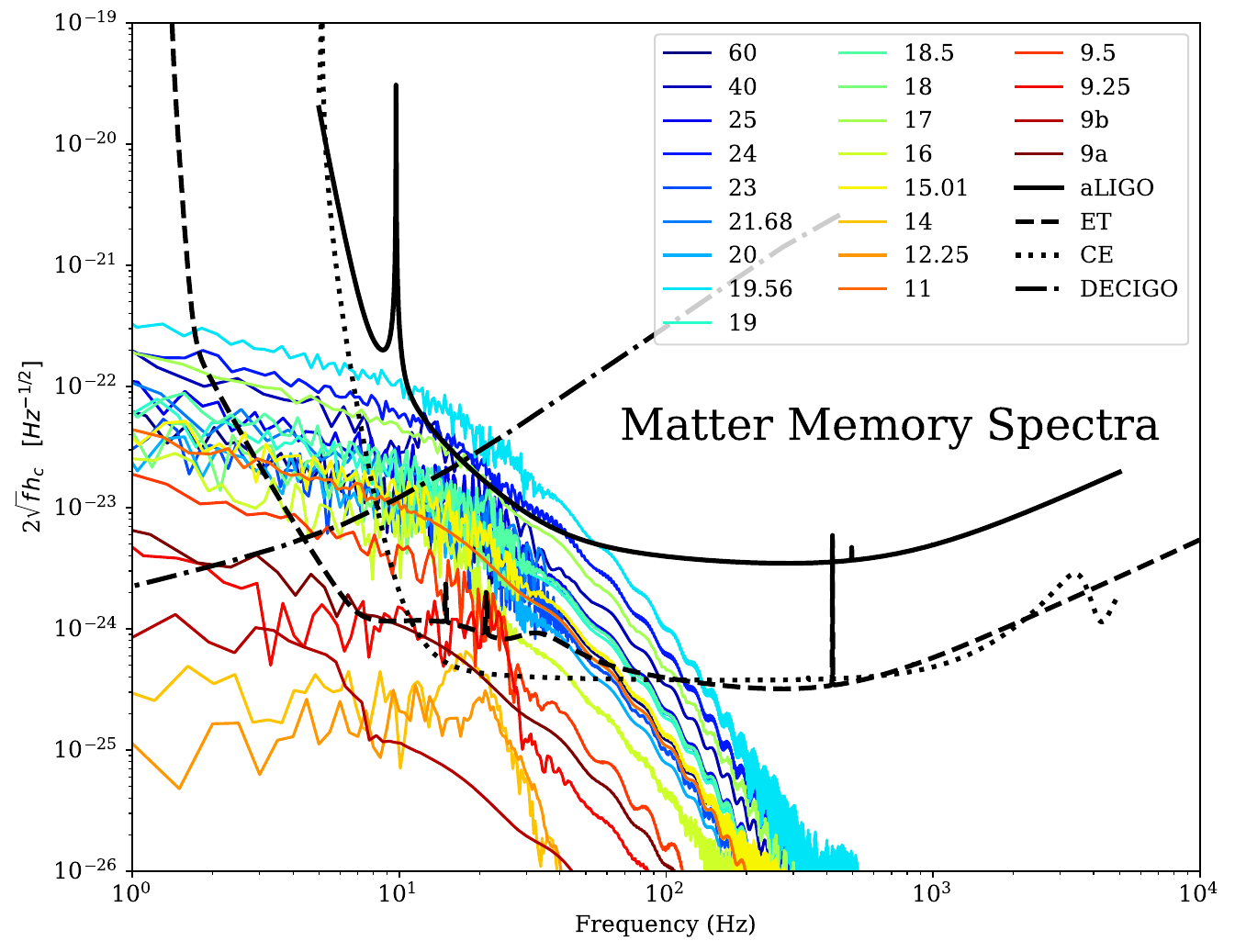}
    \includegraphics[width=0.49\linewidth]{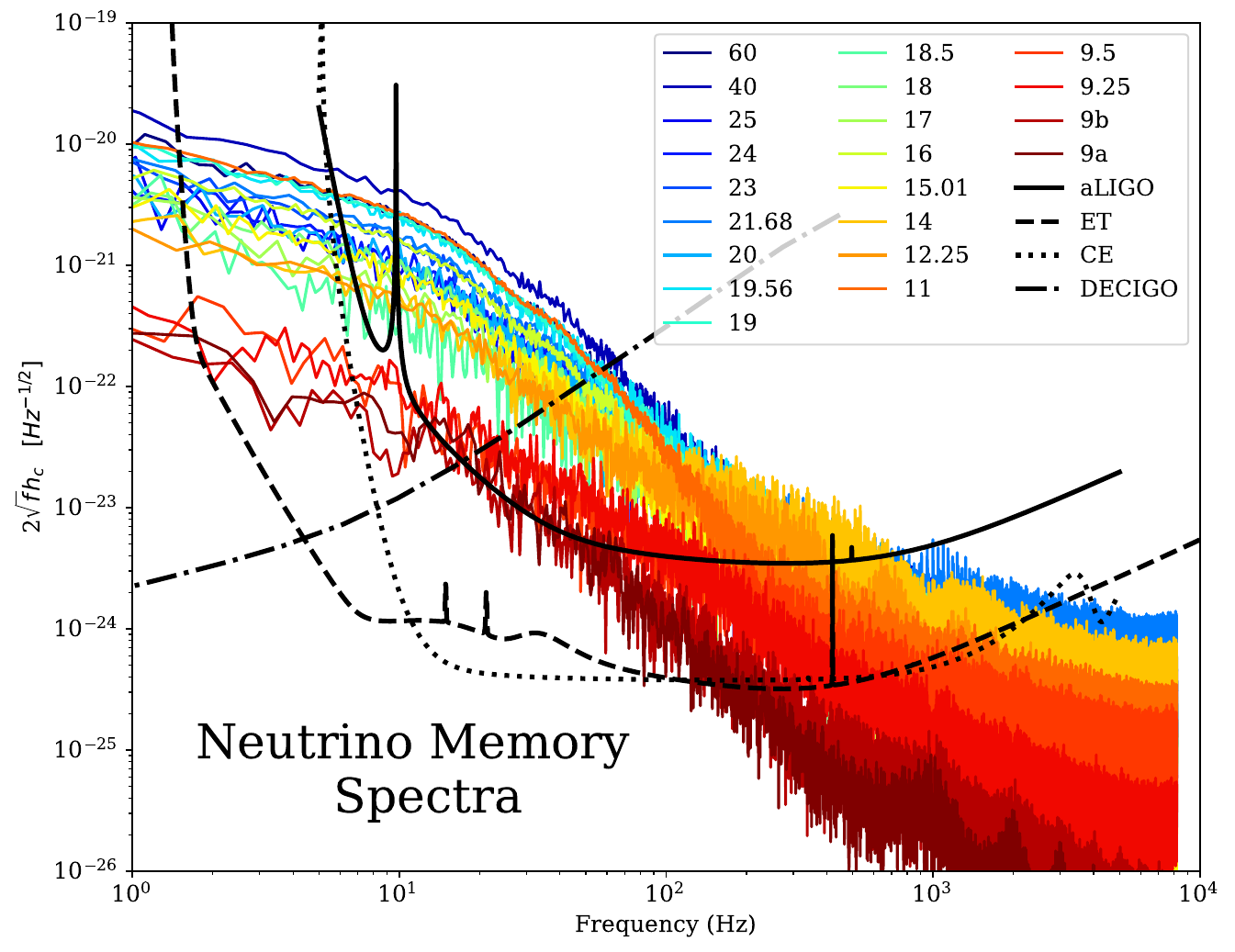}
    \caption{Integrated spectra for the total matter and neutrino memory signal (top left), the full strain due to matter motions (top right), the strain due to the low-frequency matter memory (bottom left), and the strain due to neutrino memory (bottom right). All plots are calculated from the strain viewed from the x- direction ($\theta =\pi/2, \phi$=0) and assume a distance of 10 kpc. We also plot the noise curves of the Advanced LIGO, ET, CE, and DECIGO detectors obtained from \citet{LIGO_curves} and \citet{DECIGO_curves}. Here, we use the characteristic strain $h_c(f) = \sqrt{0.5(|\tilde{h}_+(f)|^2+|\tilde{h}_{\times}(f)|^2)}$ as defined in \cite{Takami2014}. Note that the spectra for the neutrino component of the signal are not reliable in the $> 500$ Hz range due to the Nyquist sampling of the neutrino data. Overall, the heavier progenitor models have stronger matter and neutrino memory signals. The neutrino signal dominates the lower-frequency ($<$10 Hz) component of the spectrum, reaching even higher signals than the higher-frequency matter component. The ratio of the matter signal component to the noise curve for all models are greatest at around 1000 Hz.  While at 10 kpc the Advanced LIGO detector (solid black line) will be able to capture higher-frequency features of the matter component of the signal for higher-mass progenitor models, future detectors such as ET and CE will capture almost the full matter and neutrino memory signal for all progenitor models, and DECIGO will be able to capture the full, matter memory and neutrino memory component of the signal for the majority of progenitor models. 
    \label{sensitivities}}
    \end{figure*}

In Figure \ref{sensitivities} we plot the spectra for all progenitor models for the strain from the full matter and neutrino memory signals combined (top left), the strain due to matter motions (top right), the strain due to matter memory (bottom right), and the strain due to neutrino memory (bottom left). We assume a distance of 10 kpc, and plot the spectra alongside the noise curves of the aLIGO, ET, CE, and DECIGO detectors. The aLIGO, ET, and CE noise curves were obtained from  publicly-available data released by the LIGO Collaboration in \citet{LIGO_curves}. The DECIGO noise curve is from \citet{DECIGO_curves}. In this section, we focus on the isolated matter strain (top right) and discuss the remaining plots in subsequent sections. We observe that the more massive progenitors have a stronger signal, with the 40 model almost two orders of magnitude stronger than the signals from the 9a and 9b models. Overall, at 10 kpc the lower-mass progenitors are not well-captured by the aLIGO sensitivity curve. However, all models are well within the sensitivity band at this distance for the future ET and CE detectors between around 10 and 2000 Hz. Across all models, the greatest sensitivity occurs at around 1000 Hz. Additionally, all models contain a narrow gap in their spectra near 1000$-$1200 Hz due to the avoided crossing between an f- and g-mode, as discussed in \citet{vartanyan2023}.

In Table \ref{SNR_reg_table}, we calculate the SNR for each progenitor model based on the sensitivities of the LIGO, ET, CE, and DECIGO detectors averaged over the viewing solid angle assuming a distance of 10 kiloparsecs. Using the SNR $\geq 8$ detection threshold used by GW detection studies such as \citet{ligo_sensitivity2012}, \citet{abadie2010}, and \citet{Fairhurst2010}, we determine that while the GW signals from all progenitors are detectable for the next-generation ET and CE detectors, both Advanced LIGO and DECIGO will be able to see signals from only higher-mass ($> 15.01 \: M_{\odot}$) progenitors. For almost all models, the SNR values for ET and CE are around ten times those of aLIGO, and can be as high as 225. Note that SNR=8 is the standard threshold for detectability, but not necessarily for a clear resolution of a GW signal. Therefore, the SNR values in the hundreds attained by the ET and CE detectors open the possibility of not just a more confident detection, but a clearer resolution of the evolution of the strain. 
\begin{deluxetable}{ccccc}
\tablecolumns{5}
\tablewidth{0pt}
\label{SNR_reg_table}

\begin{minipage}{\textwidth}
  \centering
  \textbf{Matter SNR} \\  
\end{minipage}

\tablehead{\colhead{Progenitor} & \colhead{aLIGO} & \colhead{ET} & \colhead{CE} & \colhead{DECIGO}}
\startdata
$9a$ & 1.49 & 15.02 & 14.62 & 2.44 \\
$9b$ & 1.72 & 17.32 & 16.71 & 1.70 \\
$9.25$ & 2.97 & 28.30 & 29.13 & 2.75 \\
$9.5$ & 3.91 & 37.39 & 39.03 & 7.49\\
$11$ & 8.07 & 73.43 & 81.06 & 16.49 \\
$12.25$ & 4.84 & 44.17 & 46.87 & 0.06\\
$14$ &5.88 & 54.63 & 58.19 & 0.16\\
$15.01$ & 10.21 & 91.19 & 101.88 & 10.41 \\
$16$ & 8.82 &82.67 & 92.52 & 9.71 \\
$17$ & 15.69 & 144.21 & 163.25 & 46.88 \\
$18$ & 12.29 & 112.62 &127.34 & 18.52 \\
$18.5$ & 11.59 & 109.88 & 125.48 & 38.48\\
$19$ & 13.17 & 119.91 & 134.83 & 15.92\\
$19.56$ & 19.93 & 196.53 & 225.74&94.11 \\
$20$ & 11.32 & 102.94 & 116.14 & 10.24 \\
$21.68$ & 15.38 & 138.41 & 157.93 & 22.21\\
$23$ & 9.13 & 81.35 & 92.20 & 12.39 \\
$24$ & 13.35 & 126.92 & 144.54& 49.70 \\
$25$ & 12.09 & 111.72 & 126.79 & 29.82 \\
$40$ & 21.07 & 194.39 & 225.49 & 62.90 \\
$60$ & 13.02 & 118.47 & 132.72 & 22.38 \\
\enddata
\caption{SNR calculated for the strain due to matter motions assuming a distance of D=10 kpc for all models with the Advanced LIGO, ET, CE, and DECIGO detectors averaged over all viewing angles for the total frequency range ($< 2000$ Hz) range. For a detection threshold of SNR $\geq$ 8, both the Advanced LIGO and DECIGO detectors are only able to detect GW signals due to the matter motion component out to distances of 10 kpc from the higher-mass progenitor models, since the SNRs for the models under 14 are less than eight. However, both ET and CE detectors will be able to detect this CCSN GW component from all progenitor masses, with SNRs around ten times those of aLIGO, reaching as high as 225.}
\end{deluxetable}

While we report the results for a single aLIGO detector to allow for easier comparison with previous works, we note that for a network of detectors, such as LIGO-Virgo-KAGRA or even the two LIGO Livingston and Hanford detectors, the antenna pattern functions of each individual detector are transformed to a common coordinate system and the total SNR squared is the sum of the squared SNRs calculated for each detector \citep{schutz2011}:
    \begin{equation}
        \rho_N^2 = \sum_{k=1}^{N_D}\rho_k^2 \,\, ,
    \end{equation}
    where $N$ refers to the network of detectors and $N_D$ is the number of detectors in the network. Therefore, using the expression for the SNR in a single detector averaged over $\psi$ from eq. \eqref{rho_avg}, the SNR of a network of $N_D$ detectors is given by \citep{schutz2011}
    \begin{equation}
    \label{rho_avg_det}
    \langle \rho^2_N \rangle_{\psi} = 2\sum_{k=1}^{N_D}(F_{+, k}^2+F_{\times, k}^2)\int_0^{\infty}\frac{|\tilde{h}_+(f)|^2+|\tilde{h}_{\times}(f)|^2}{S_n(f)}df \,\, .
    \end{equation}

When considering a network of detectors, it is impossible for the source to be seen overhead at the optimal sky location $(l, b) = (0, 0)$ in all detectors. Therefore, the contribution of subsequent detectors to the SNR will be smaller than the SNR given in eq. \ref{SNR_final}. For example, for a network of the LIGO Livingston and Hanford detectors, if we assume the source to be located overhead at $(l, b) = (0, 0)$ for one detector, then the other detector is rotated by angles $\Delta\theta = 27.224$ and $\Delta\phi = 71.72$ \citep{Rakhmanov2003}. Substituting these angles into the antenna pattern function coefficient in \eqref{rho_avg_det}, we have that $F_+(\theta, \phi)^2+F_{\times}(\theta, \phi)^2 = 0.569$. Therefore, adding an additional aLIGO detector to the SNR calculated in eq. \ref{SNR_final} only increases the SNR by a factor of $\sqrt{1+0.569} = 1.252$. Similarly, \citet{schutz2011} calculates the ``mean horizon distance" (a quantity related to SNR by a multiplicative factor) of a network of LIGO Hanford, LIGO Livingston, and VIRGO detectors to be 1.43 times that of a single LIGO detector.

To conclude this section, we emphasize that the results presented in Table \ref{SNR_reg_table} and subsequent results related to SNR are for \textit{individual} aLIGO, ET, CE, and DECIGO detectors assuming optimal orientation of the supernova relative to the detector to provide a sense of the detectors' relative CCSN detection capabilities. In practice, including a network of LIGO Hanford and Livingston detectors or Hanford-Livingston-Virgo-KAGRA detectors would result in larger SNR values in the second column of Table \ref{SNR_reg_table}, but the exact factor of increase would depend on the source orientation relative to all detectors in the network. A more in-depth discussion of the effects of different detectors and source orientations on SNR can be found in works such as \citet{schutz2011} and \citet{szczepanczyk2021}.

\section{Angle Dependence of Gravitational Waves from Matter Motion}
\label{matter_angle}
While a CCSN is a three-dimensional, highly-anisotropic event \citep{vartanyan2020}, in practice, Earth-based GW detectors will detect the strain from only a single angle ($\theta, \phi$) of emission, making it difficult to determine quantities such as total radiated energy, which depend on the three-dimensional quadrupolar tensor. Additionally, quantities such as the SNR of the signal and the distance in kiloparsecs out to which we could observe a CCSN GW signal will also depend on our angle of observation. In this section, we calculate the variation of quantities including strain, total radiated energy, SNR, and detection range over all possible viewing angles of the supernova, determine how the variation depends on progenitor mass, and estimate the errors associated with the assumption that the signal at the observed angle is characteristic of the signals in all directions. 

\subsection{Angle Dependence of Radiated Gravitational-Wave Energy}
\label{matter_energy_strain}
While the total energy radiated in GWs by the supernova depends on the supernova's three-dimensional quadrupole tensor (see eq. \eqref{E_GW}), in practice, when observing GWs from CCSNe, the full quadrupole tensor of the event will be unknown to us and we will need to extrapolate calculated quantities such as total energy based on the GW strain received from a particular viewing angle. Therefore, the total radiated energy which we infer from the supernova will be an angle-dependent quantity. One method of estimating this quantity from a given direction is to use the first time derivative of the observed strain and assume the strain is the same in all directions by multiplying by a factor of $4\pi$ instead of integrating the strain over the solid angle $d\Omega$ \citep{afle2023}:
    \begin{equation}
    \label{E_GW_angle}
    E_{GW}(\theta, \phi) = \frac{c^3}{4G}\int_0^{t}dt'\left[\frac{dh_+(\theta, \phi)^2}{dt}+\frac{dh_{\times}(\theta, \phi)^2}{dt}\right] \,\, .
    \end{equation}

\begin{figure*}[ht]
    \centering
    \includegraphics[width=0.35\linewidth]{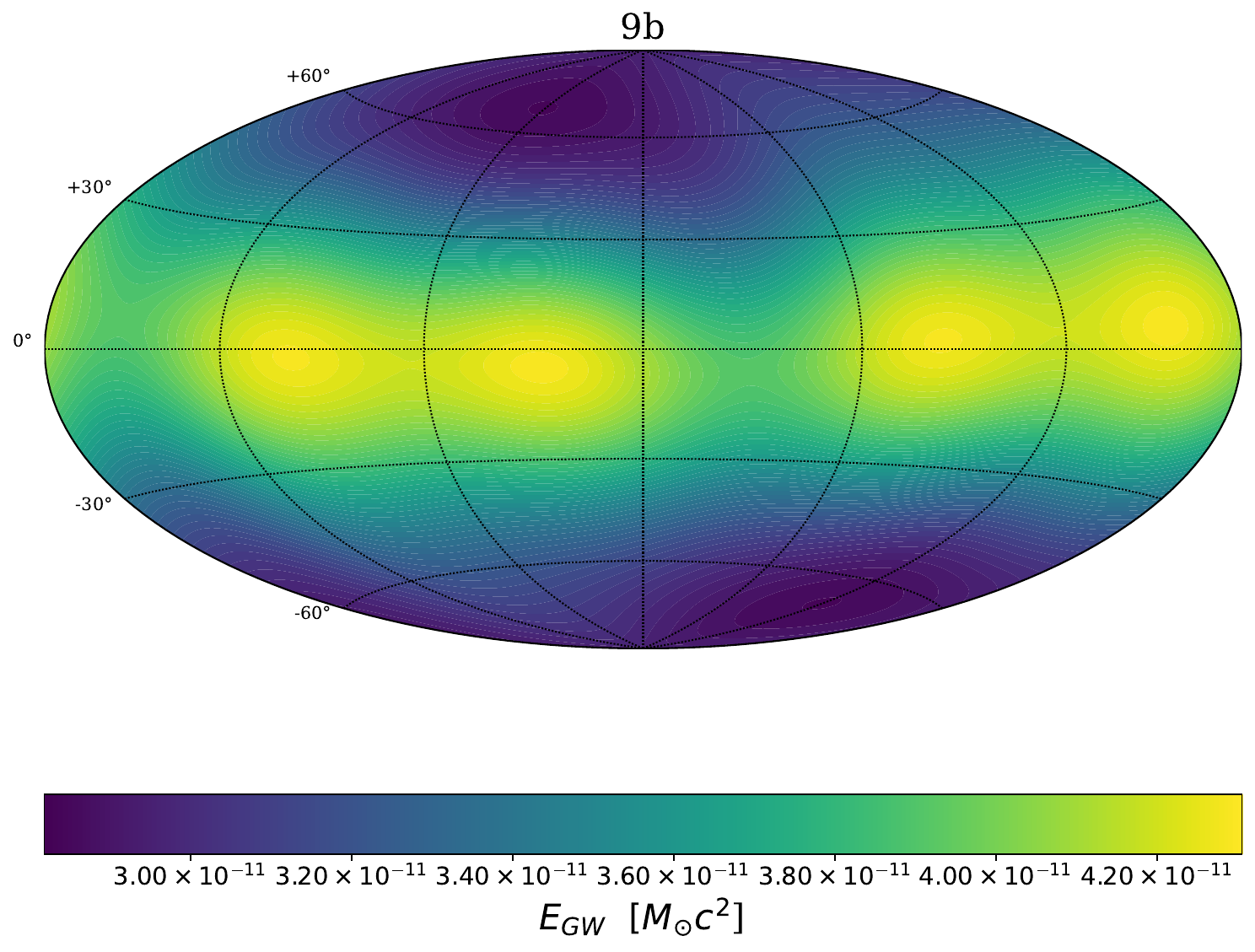}
    \includegraphics[width=0.35\linewidth]{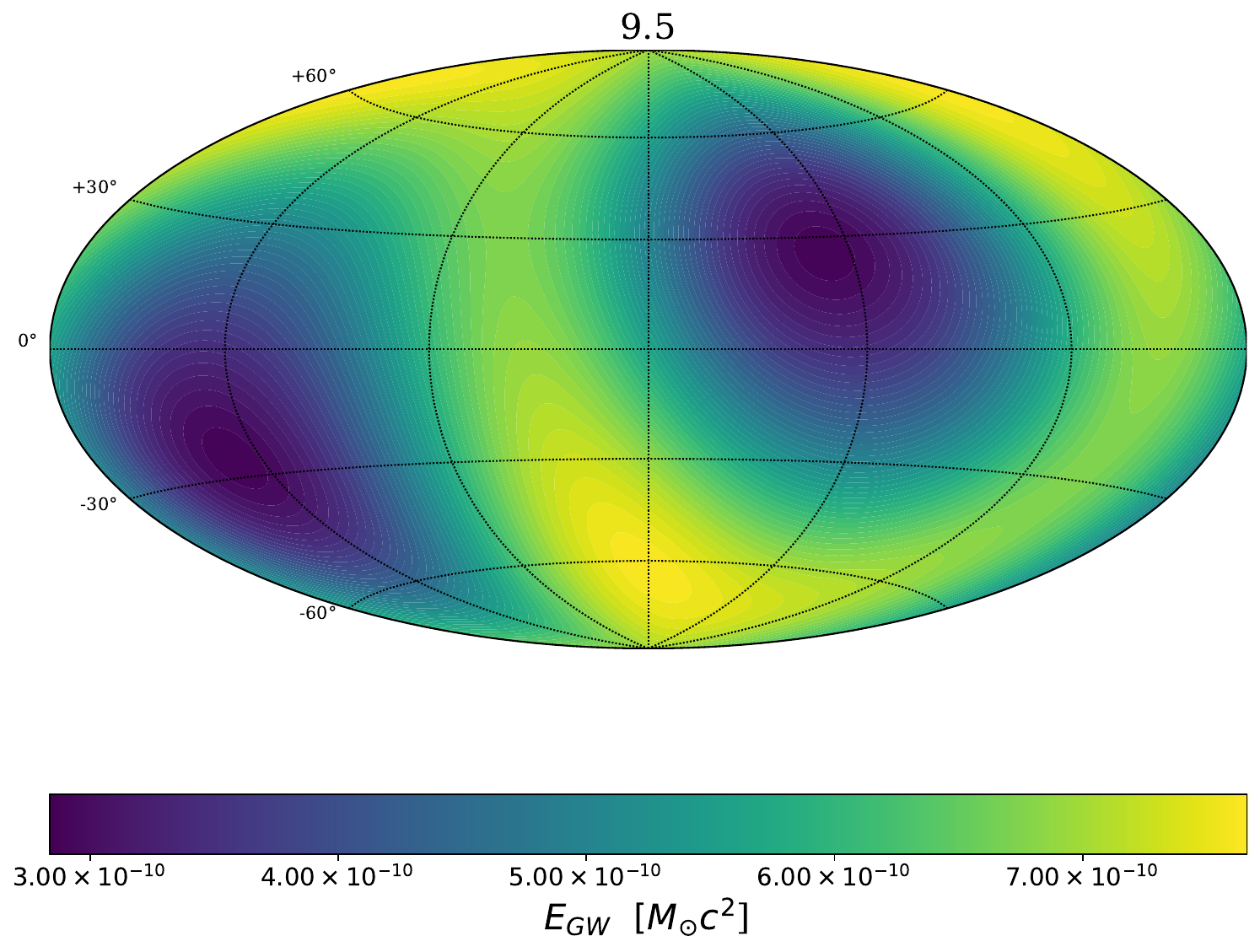}
    \includegraphics[width=0.35\linewidth]{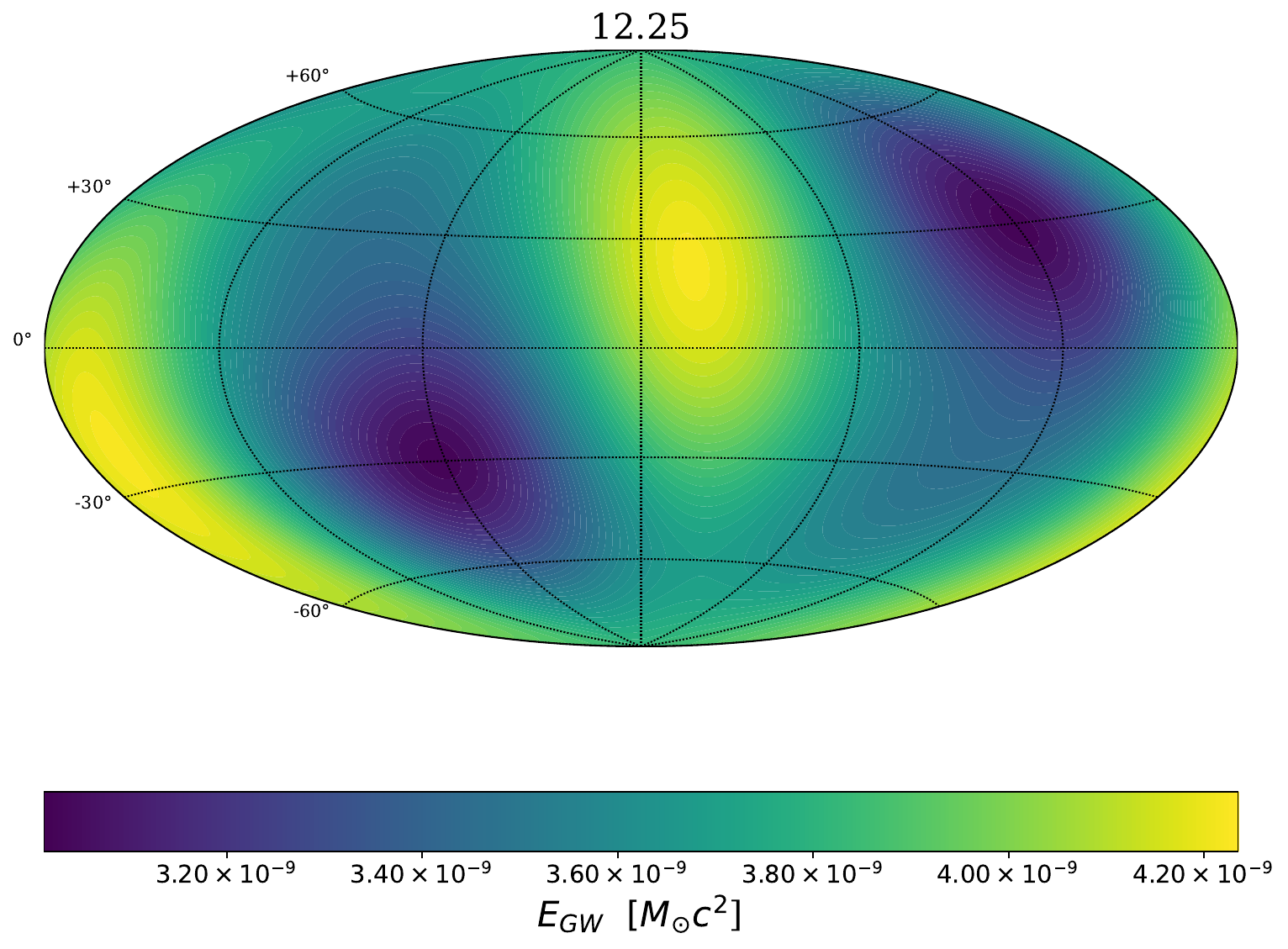}
    \includegraphics[width=0.35\linewidth]{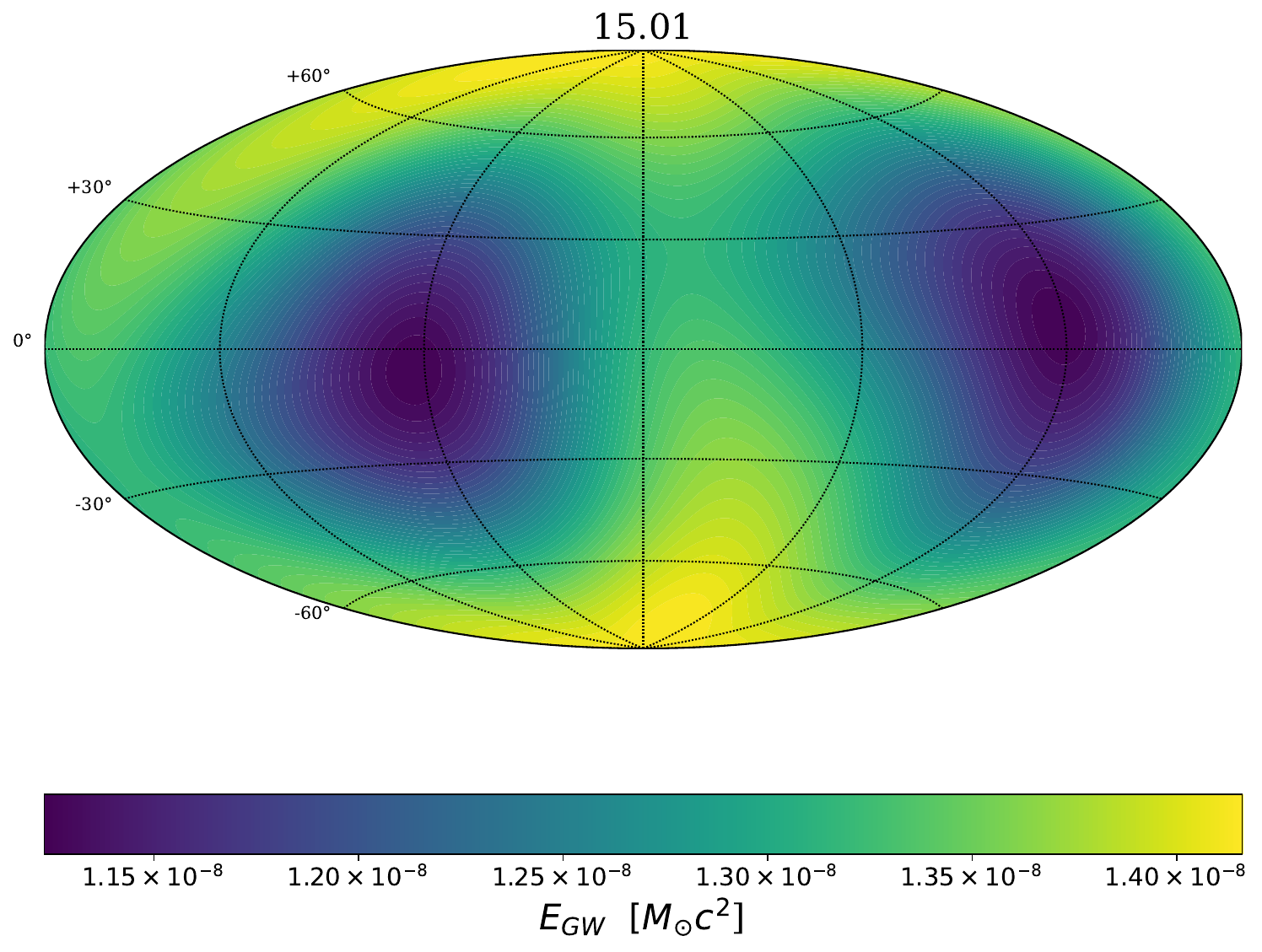}
    \includegraphics[width=0.35\linewidth]{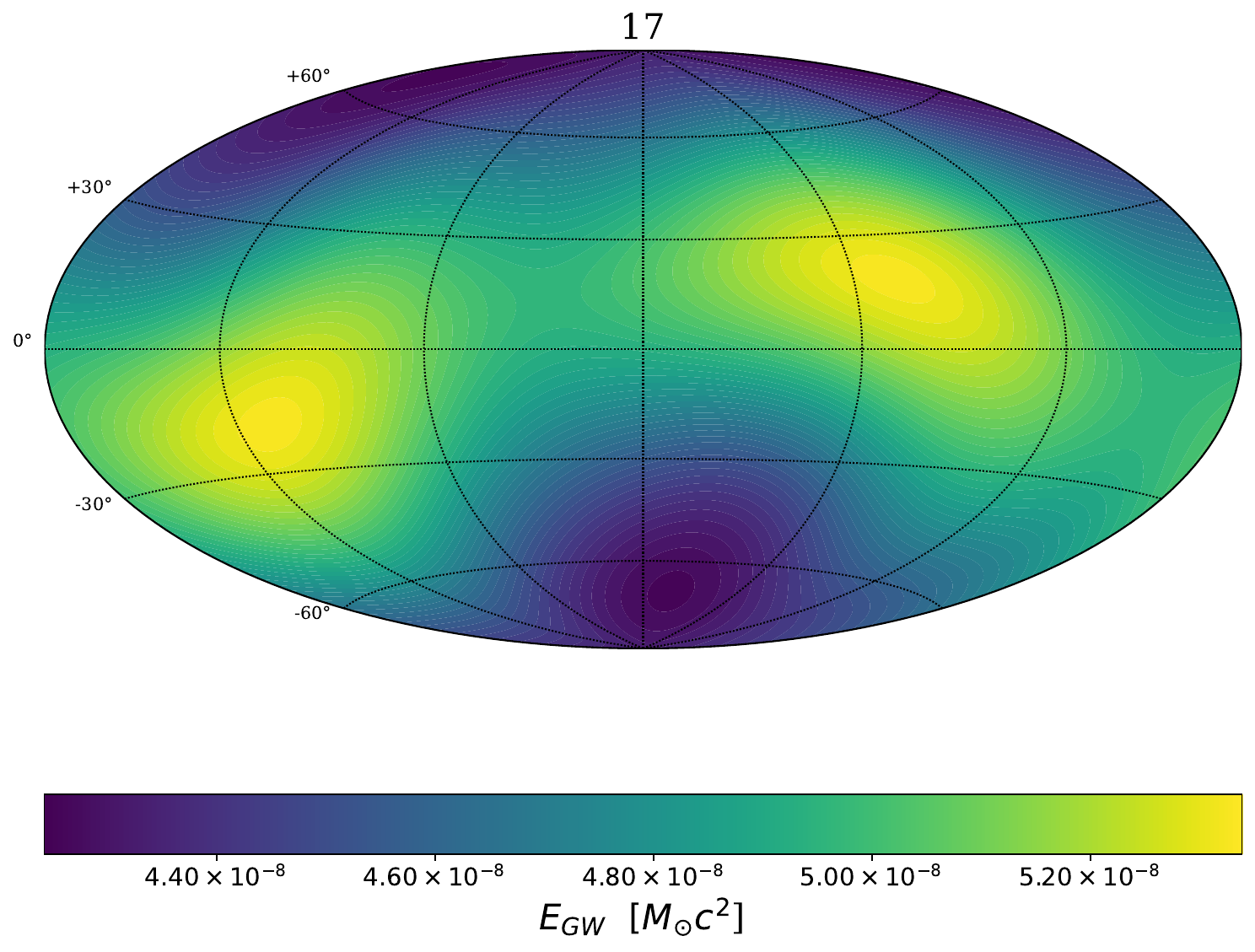}
    \includegraphics[width=0.35\linewidth]{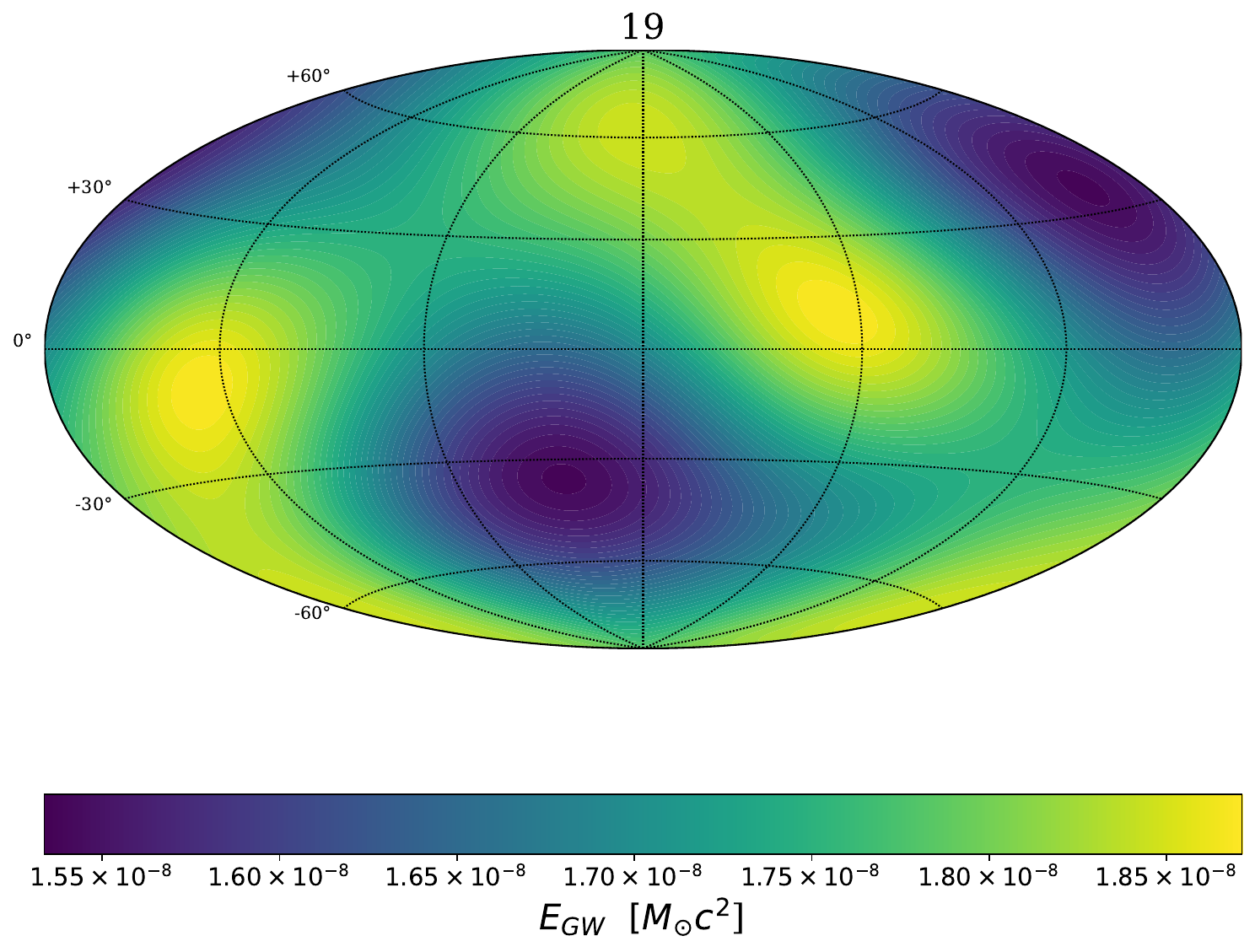}
    \includegraphics[width=0.35\linewidth]{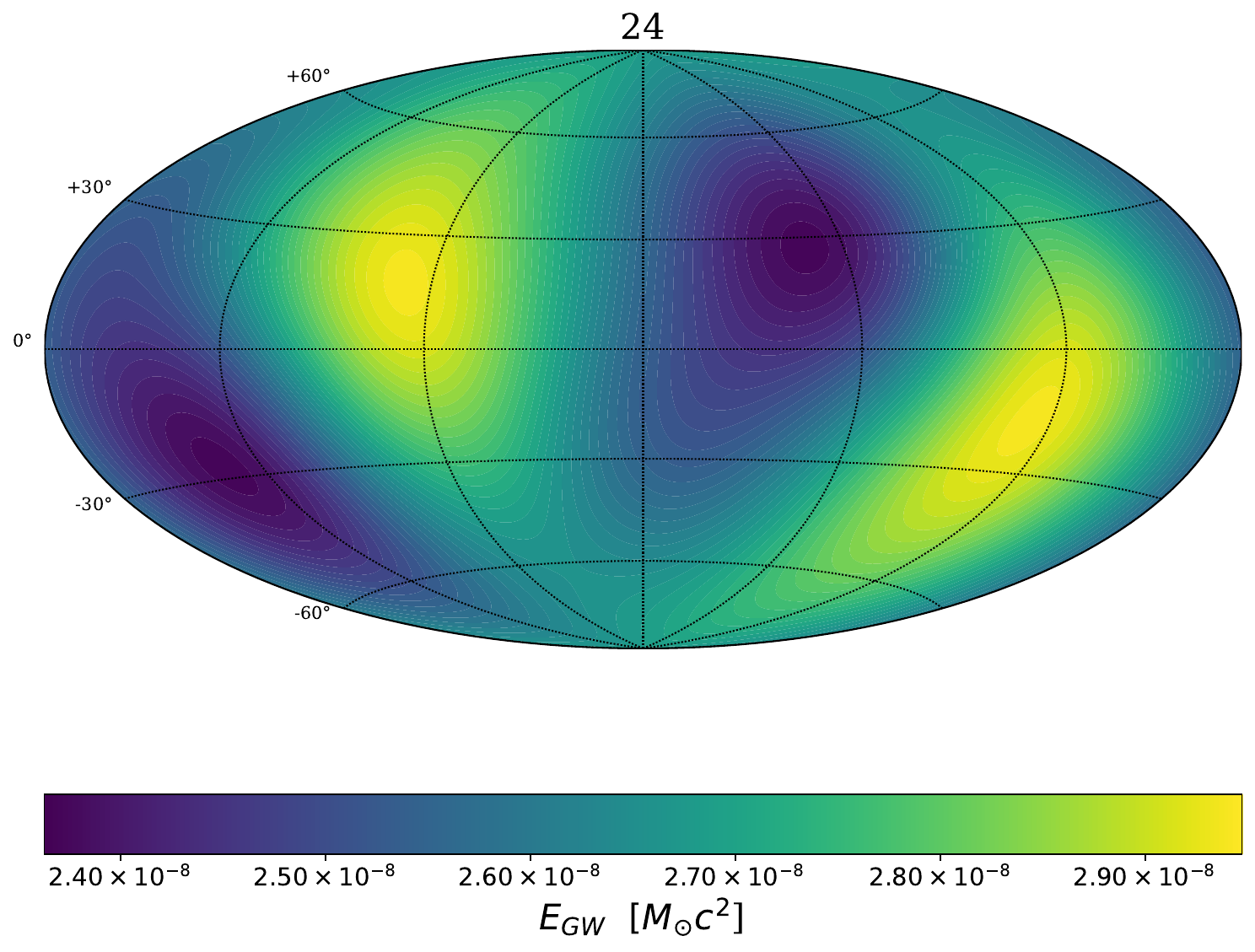}
    \includegraphics[width=0.35\linewidth]{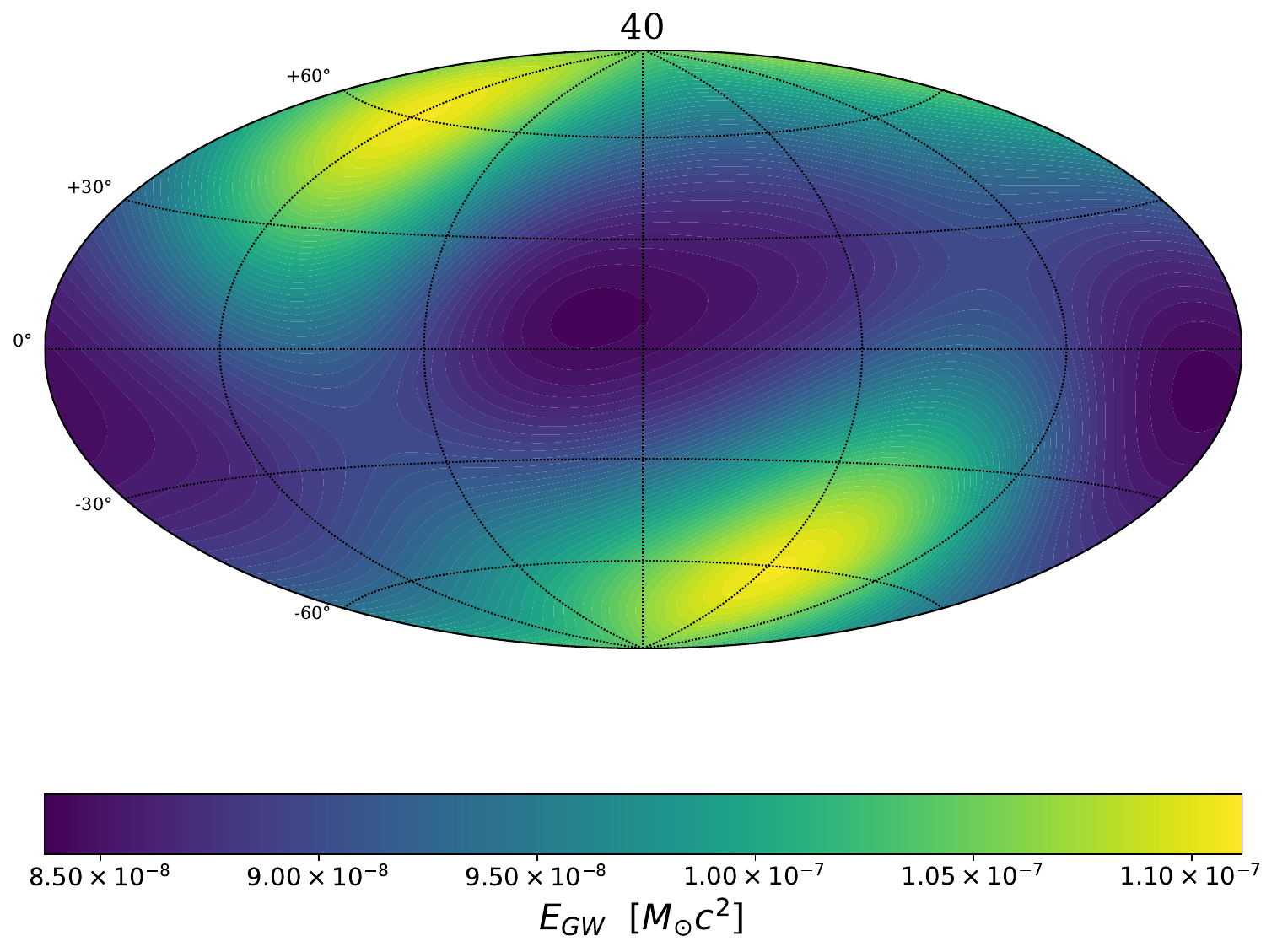}
    \caption{Total inferred energy radiated in GWs due to matter motion as a function of viewing angle ($\theta, \phi$) for the 9b, 9.5, 12.25, 15.01, 17, 19, 24, and 40 progenitor models. Observing the CCSN event from an angle in the dark blue region results in a lower estimate of the event's total radiated energy compared to an observation in the yellow regions. Between these eight examples, there are no consistent regions of higher and lower energy estimates, as the direction of greatest inferred energy differs significantly between each model. In addition to having larger total radiated energy values, the higher-mass progenitor models also have a larger spread in possible inferred energy values. \label{energy_globes}}
    \end{figure*}
    
In Figure \ref{energy_globes}, we plot the total inferred energy as a function of viewing angle $(\theta, \phi)$ for the 9b, 9.5, 12.25, 15.01, 17, 19, 24, and 40 progenitor models. Observing the CCSN event from an angle in the dark blue region results in a lower estimate of the event's total radiated energy compared to an observation in the yellow regions. Since the yellow and blue regions are distributed across different viewing angles for each model, there is not one direction or region that consistently results in greater radiated energy estimates for all CCSNe. While in Figure \ref{E_GW} we observed that higher-mass progenitors have larger total radiated energy values, in Figure \ref{energy_globes} we also observe that larger-mass progenitors also have a greater range in possible inferred energy values compared to lower-mass progenitors, and thus more room for error in total radiated energy calculations. 

\begin{figure*}[htpb!]
    \centering
    \includegraphics[width=0.49\linewidth]{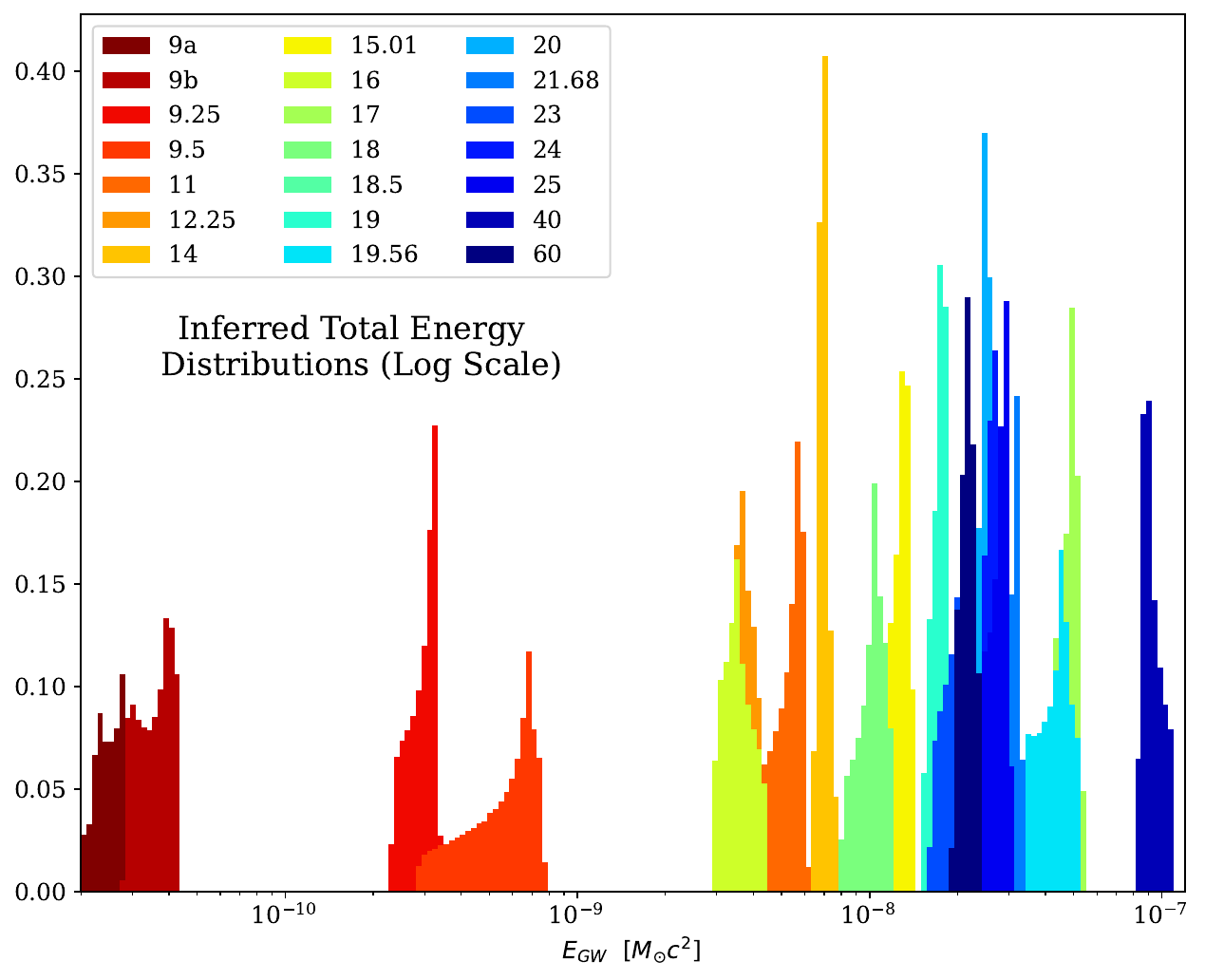}
    \includegraphics[width=0.49\linewidth]{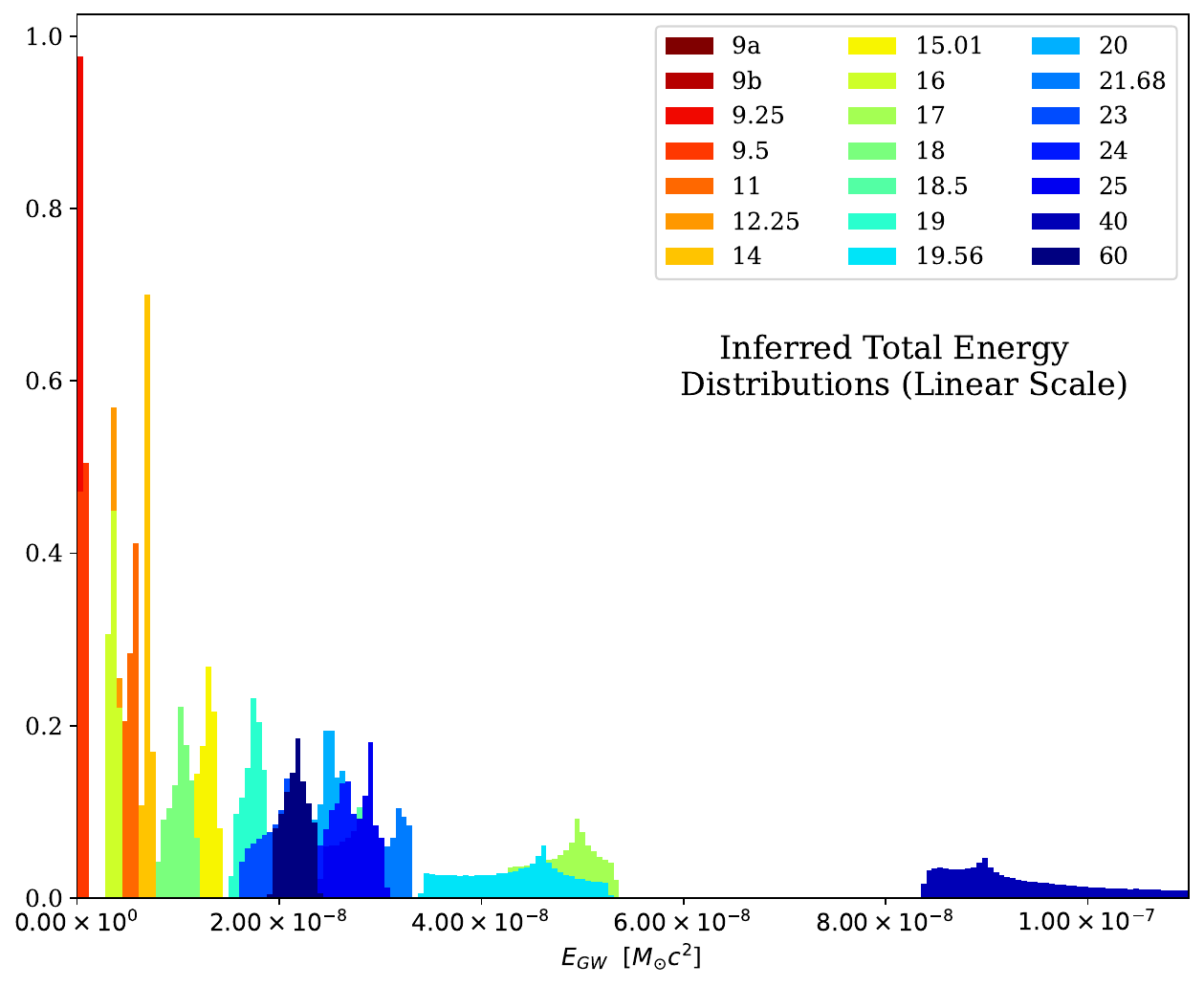}
    \includegraphics[width=0.49\linewidth]{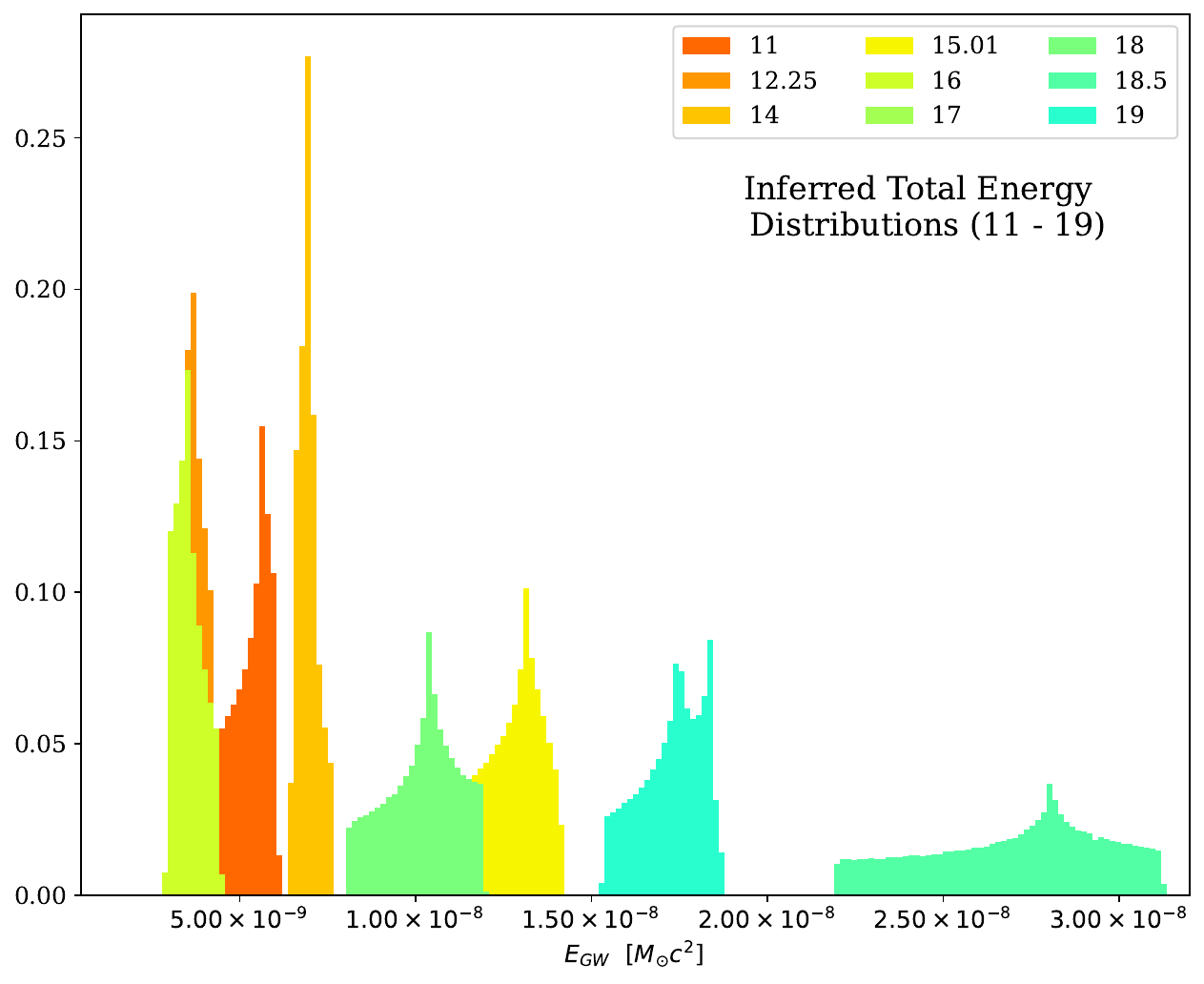}
    \includegraphics[width=0.49\linewidth]{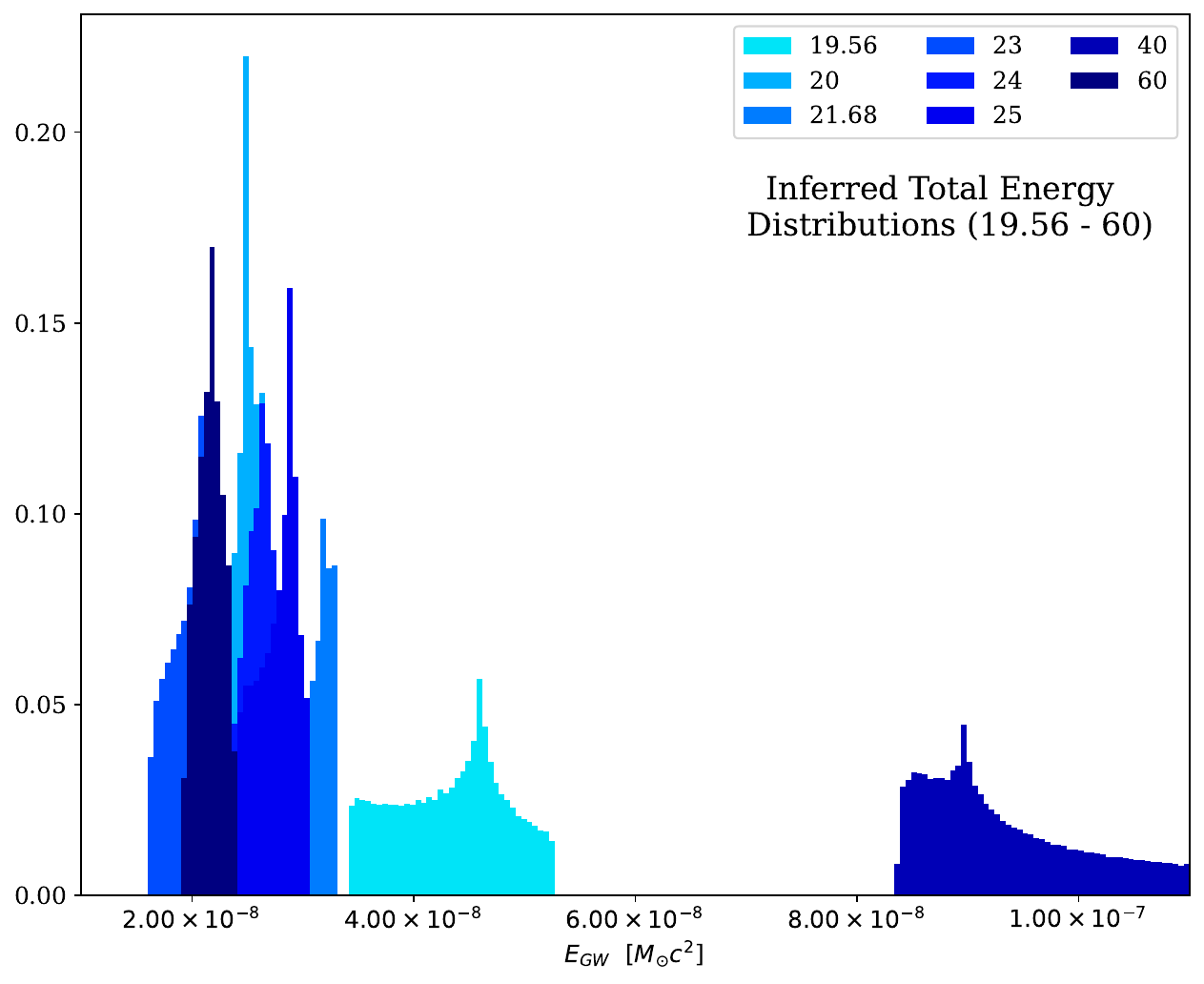}
    \caption{Inferred total radiated energy distributions due to matter motion across all viewing angles for all progenitor mass models in log scale (top left), linear scale (top right), focusing on intermediate mass models 11-19  $M_{\odot}$ (bottom left), and on high mass models 19.56-60  $M_{\odot}$ (bottom right). The area under each histogram has been normalized to one. Looking at the linearly-scaled histograms, we observe that the greater progenitor masses have a larger spread of possible inferred energy values and thus has the potential for larger energy calculation biases compared to lower mass progenitors. Almost all histograms demonstrate a defined peak, indicating a ``preferred" inferred energy value. We also note overlap in inferred energy values between different progenitor models (e.g., 12.25 and 16) depending on the viewing angle. \label{energy_hists_combined}}
    \end{figure*}

In Figure \ref{energy_hists_combined} we plot the total inferred energy values across all viewing angles decomposed into histograms for all progenitor models. We plot all models on the same axes in log (top left) and linear (top right) scale to compare their spread, and focus on the intermediate mass (11 - 23 $M_{\odot}$) and high mass (19.56 - 60 $M_{\odot}$) models in the lower plots. As demonstrated most clearly in the top right plot, the higher-mass progenitor models have a wider spread in possible energy values compared to the lower-mass progenitor models, and therefore have a higher possibility of a biased total energy calculation. Almost all histograms demonstrate a clear peak, indicating a preference towards a most likely or ``preferred" inferred energy value across all viewing angles. However, this ``preferred" value is not necessarily the same as the actual total energy value in Table \ref{model_summary}. Finally, we observe significant overlap between the models' inferred energy values, despite the single-valued nature of the true total energy in Figure \ref{E_GW}. 

\subsection{Angle Dependence of the Signal-to-Noise Ratio}
\label{snr_angle}
\begin{figure*}[ht]
    \centering
    \includegraphics[width=0.49\linewidth]{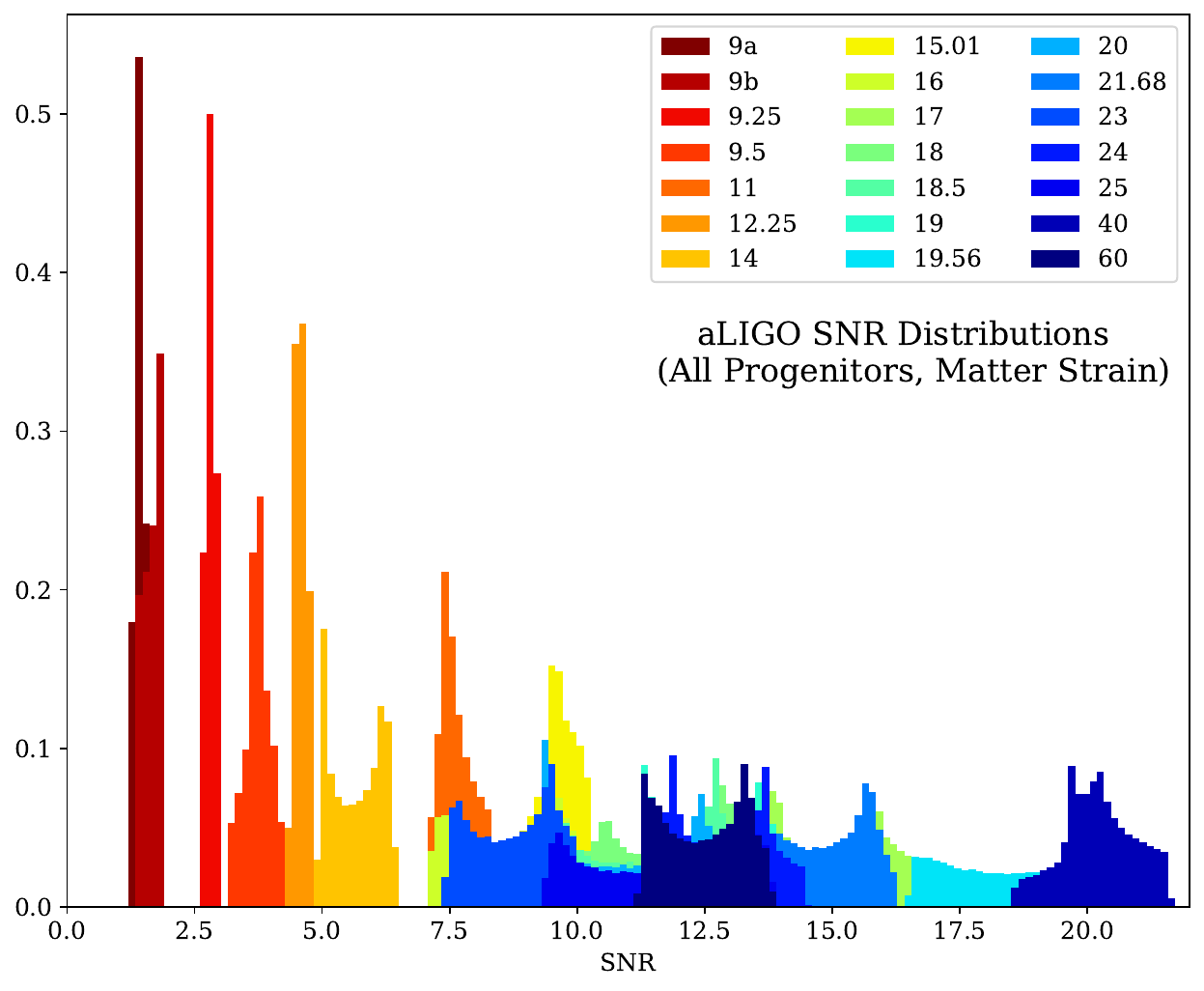}
    \includegraphics[width=0.49\linewidth]{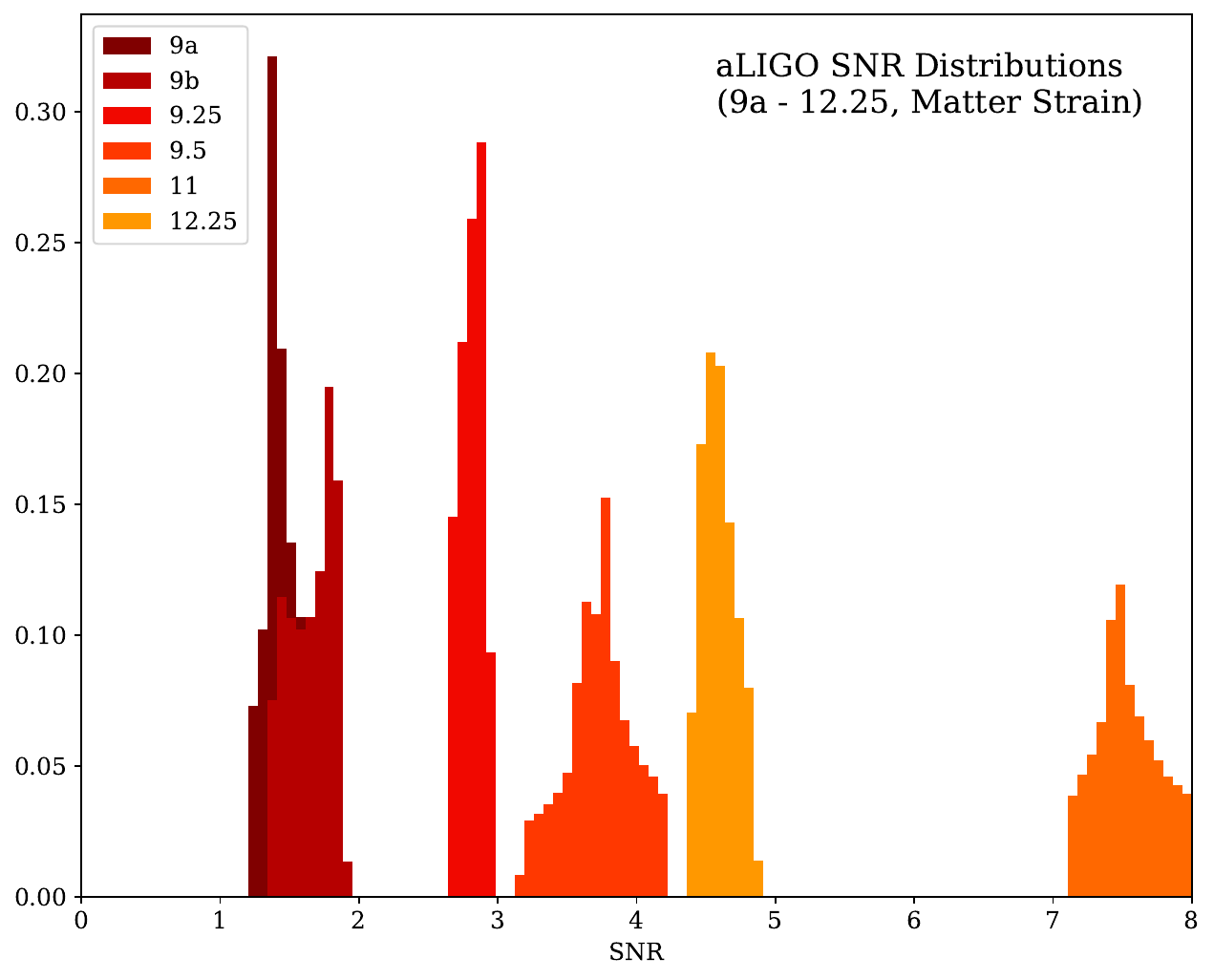}
    \includegraphics[width=0.49\linewidth]{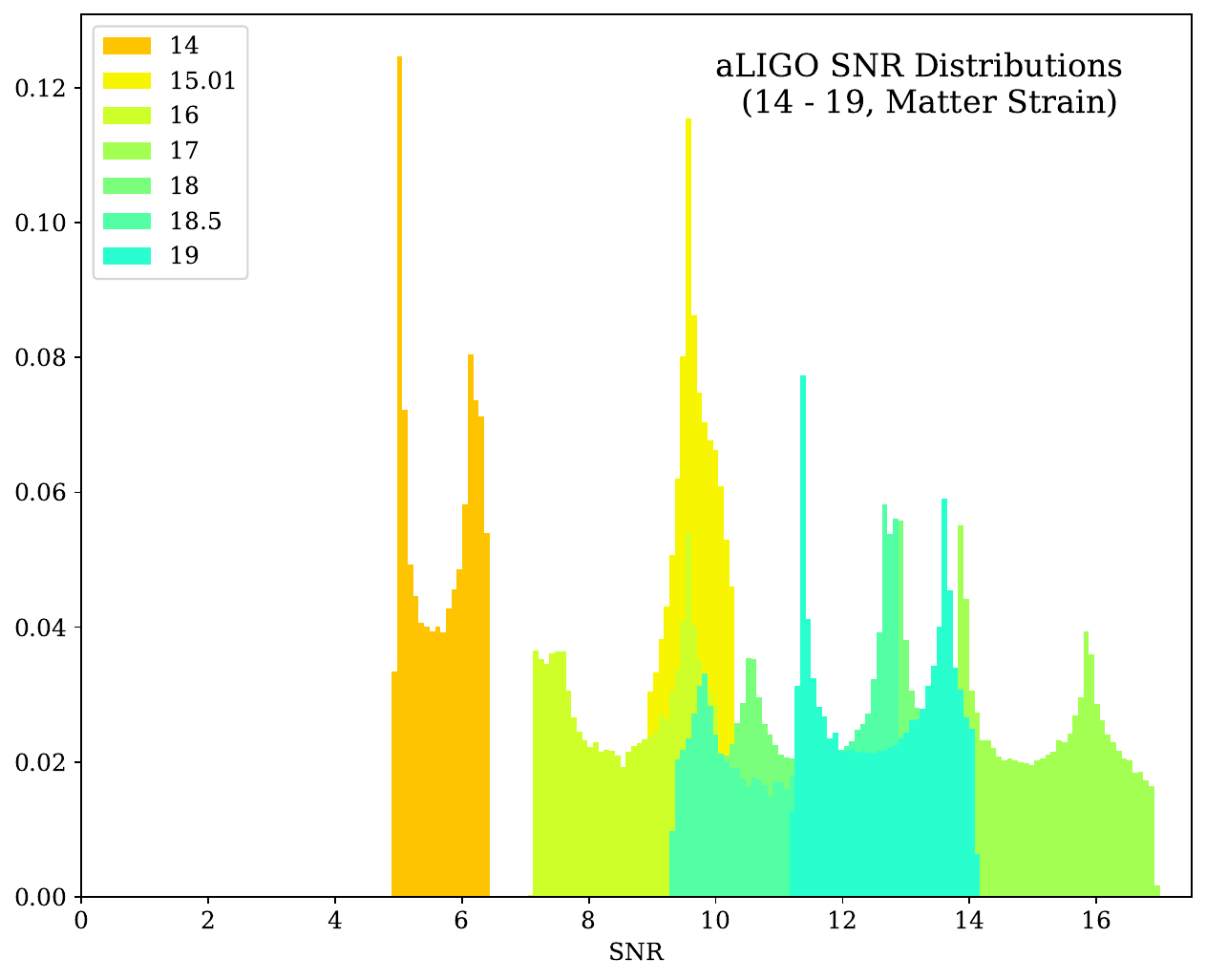}
    \includegraphics[width=0.49\linewidth]{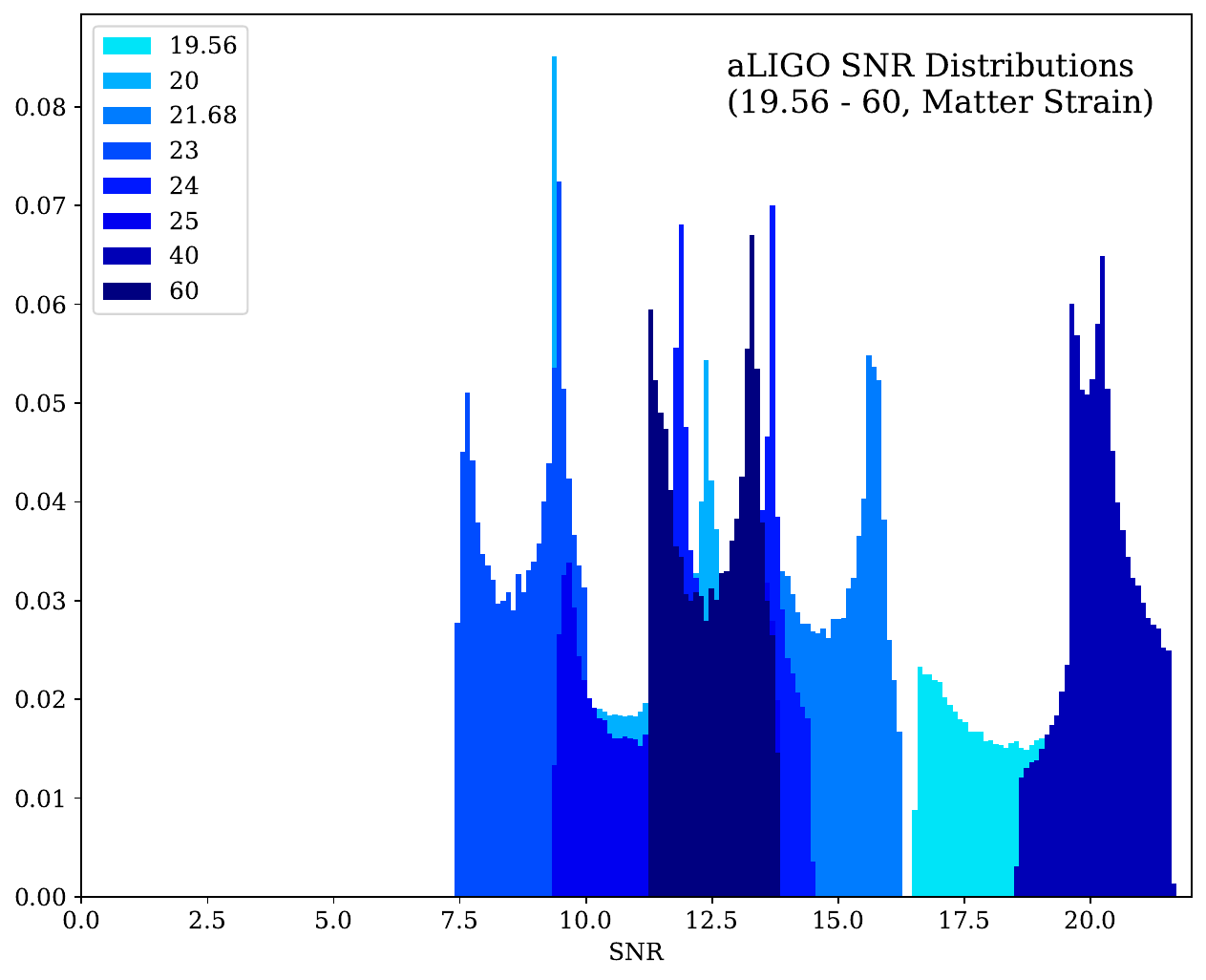}
    \caption{Distribution of possible SNR values calculated from the matter strain obtained by a single aLIGO detector over all viewing angles for all models (top left), lower-mass models (top right), intermediate-mass models (bottom left) and higher-mass models (bottom right), all at a distance 10 kpc from Earth. The area of all histograms has been normalized to one. Unlike the inferred energy histograms, these histograms do not show a defined peak or a ``most likely" SNR value. We find that larger progenitor masses not only have larger average SNRs, but also a wider spread in possible SNR values obtained depending on the viewing angle, differing by up to $\Delta\rho = 3$. There is substantial overlap in the angle-depending SNR values between progenitor models. For the 16 model, some viewing angles result in an SNR less than eight which is not considered a detection, whereas other viewing angles result in SNR values close to 10 for the same progenitor. Different detectors or network configurations will result in different SNR values on the horizontal axis, but the overall trends and shapes of the histograms will remain the same. } \label{SNR_hists}
    \end{figure*}

As with inferred radiated energy due to matter motion, the SNR also depends on the GW strain observed from a particular direction. In Figure \ref{SNR_hists}, we plot the histograms of SNR for all progenitor models across all viewing angles, assuming all CCSNe occur at a distance of 10 kpc from Earth. These SNR values are calculated using a single Advanced LIGO detector. We plot the histograms of all progenitor models together (top left), then isolate the lower-mass models 9a-12.15 (top right), intermediate mass models 14 - 19 (bottom left), and high mass models 19.56-60 (bottom right). Both the SNR values themselves along with the range of possible SNR values \textit{increases} with progenitor mass, with the possible SNR values differing by up to three for the higher-mass progenitors. The increasing range with progenitor mass mirrors the trend observed with the angular-dependence of inferred radiated energy values in the previous section. However, the SNR histograms in Figure \ref{SNR_hists} are not as cleanly peaked as the histograms of radiated energy values and thus do not demonstrate a ``preferred" or most likely value. For some models such as 16 $M_{\odot}$, this spread in angle-dependent SNR values straddles the detectability threshold of $\rho = 8$, indicating that the viewing angle may impact whether or not we consider a CCSN GW signal to be a detection. While incorporating future detectors or a network of detectors may shift the SNR values on the horizontal axis of these plots, the overall trends and shapes of the histograms will remain the same.

While \citet{Pajkos2023} found that there is no single optimal orientation of observation for a given CCSN for its entire duration, we find there are certain orientations which result in a greater SNR accumulated over the duration of the CCSN. In other words, the histograms in Figure \ref{SNR_hists} demonstrate that for every progenitor model, there are certain viewing angles that correspond to stronger and more detectable CCSN GW signals. 


\subsection{Angle Dependence of Matter Signal Detection Range}
\label{detection_range}
Another angle-dependent quantity that can be calculated from the SNR is the \textit{detection range}, or the distance (in kpc) out to which we can detect GWs from a CCSN event with a minimum SNR of eight \citep{Fairhurst2010,abadie2010}. While other ``detection range" conventions involve averaging over all sky orientations and viewing angles, here we keep the dependence on viewing angle and choose a particular sky orientation, as explained below.

Using the same sky and polarization angle assumptions as outlined in Section \ref{snr} (averaging over the polarization angle $\psi$ and choosing the optimal sky orientation $l = b =0$), we can factor out the distance to the source, invert the expression for SNR given in eq. \eqref{SNR_final} and set $\rho = 8$ to calculate the \textit{range} out to which we can detect GWs from each supernova with an SNR of eight. This results in the following expression \citep{powell2024, afle2023, schutz2011}:
    \begin{equation}
    \label{d_opt}
    d =\frac{1}{8}\sqrt{2\int_{0}^{\infty}df\frac{|\tilde{h}_+(f)|^2+|\tilde{h}_{\times}(f)|^2}{S_n(f)}} \,\, .
    \end{equation}

\begin{figure*}[ht]
    \centering
    \includegraphics[width=0.4\linewidth]{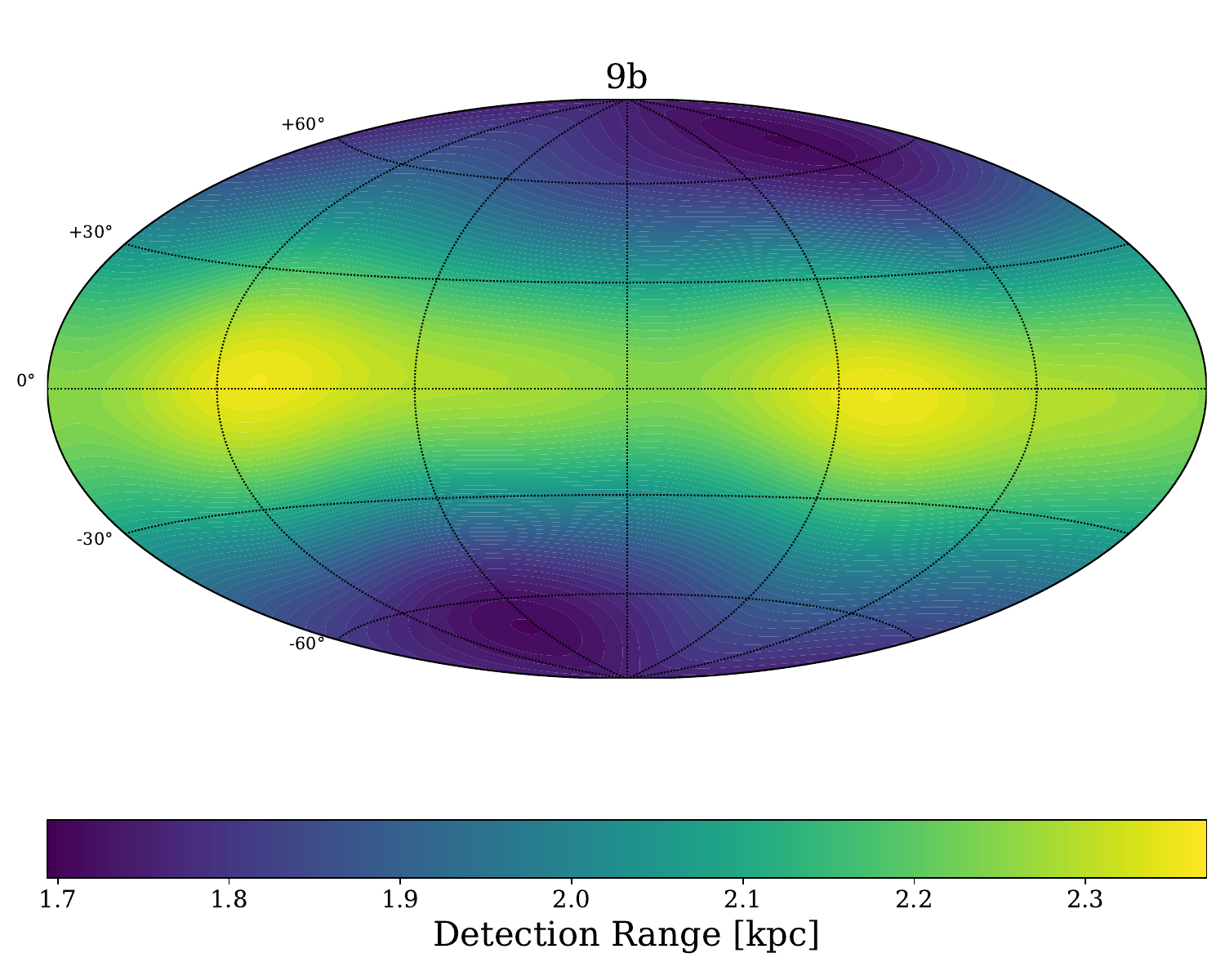}
    \includegraphics[width=0.4\linewidth]{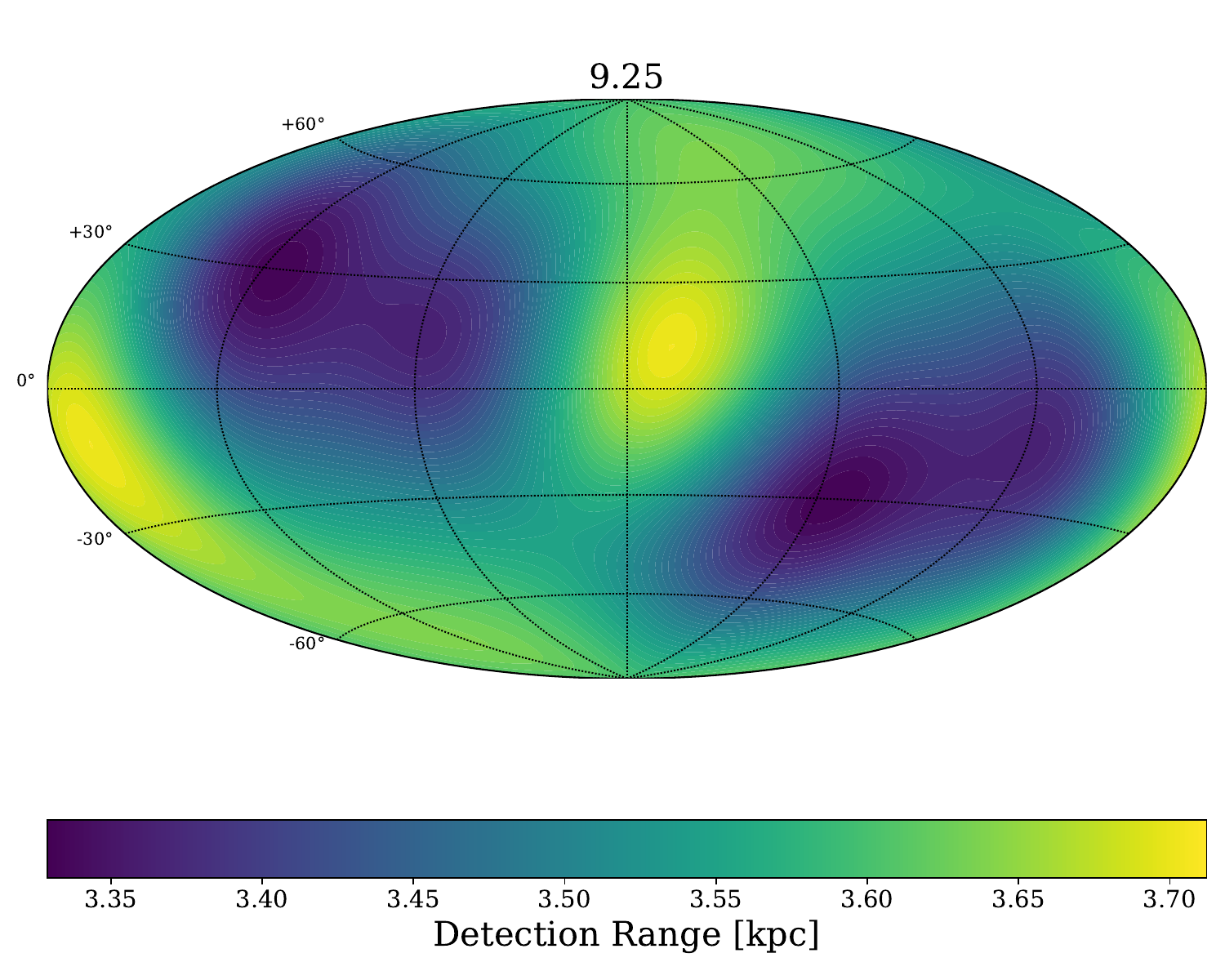}
    \includegraphics[width=0.4\linewidth]{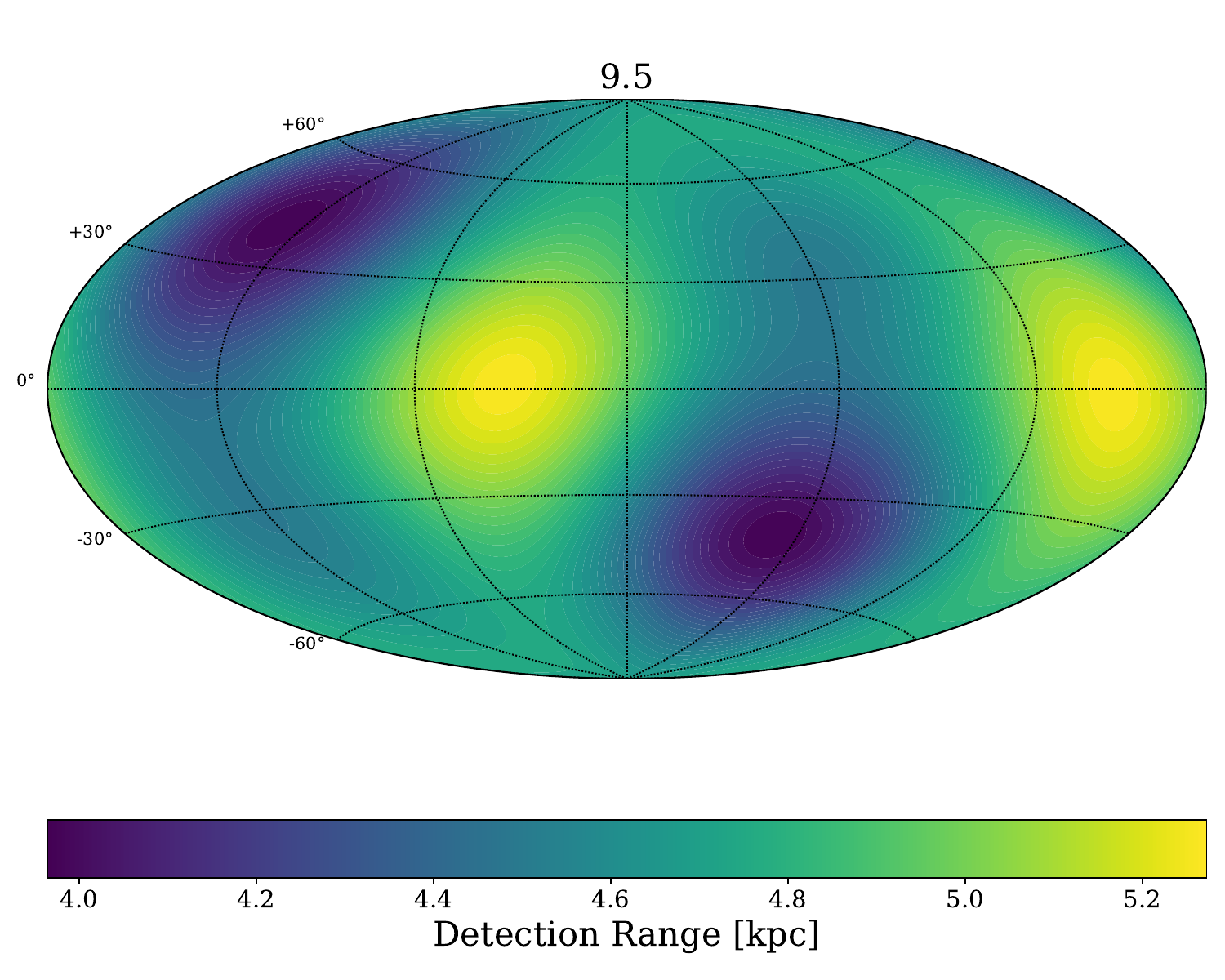}
    \includegraphics[width=0.4\linewidth]{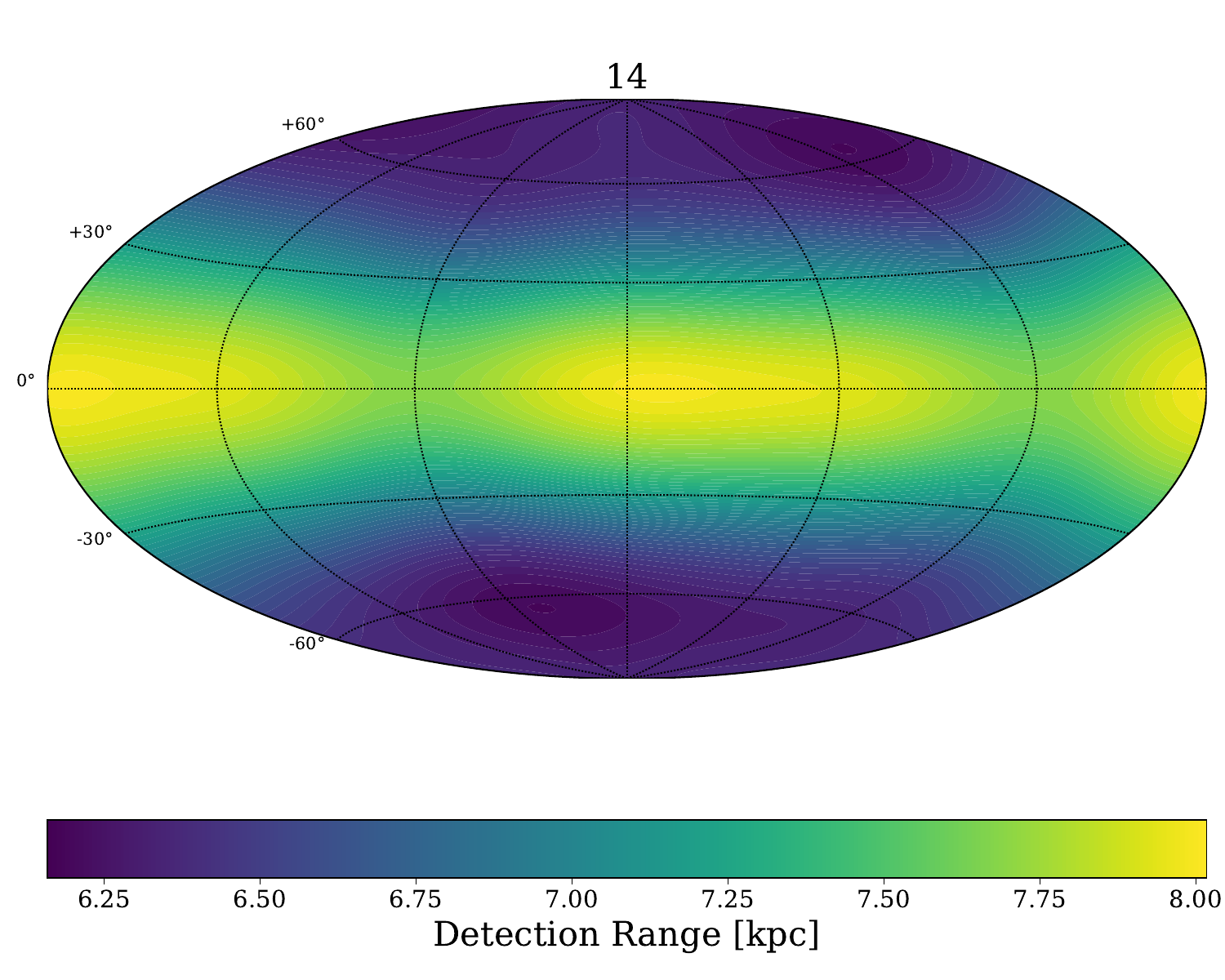}
    \includegraphics[width=0.4\linewidth]{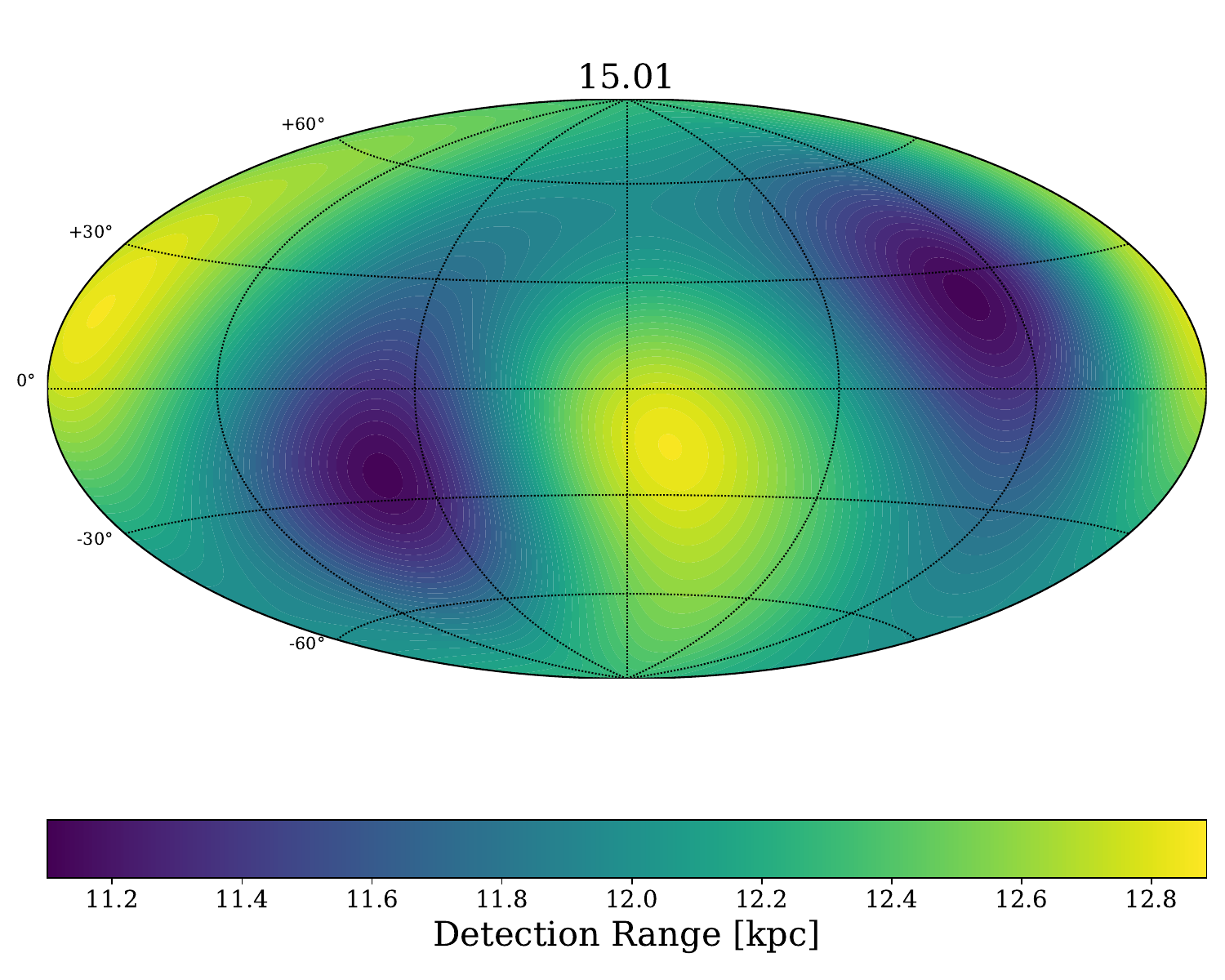}
    \includegraphics[width=0.4\linewidth]{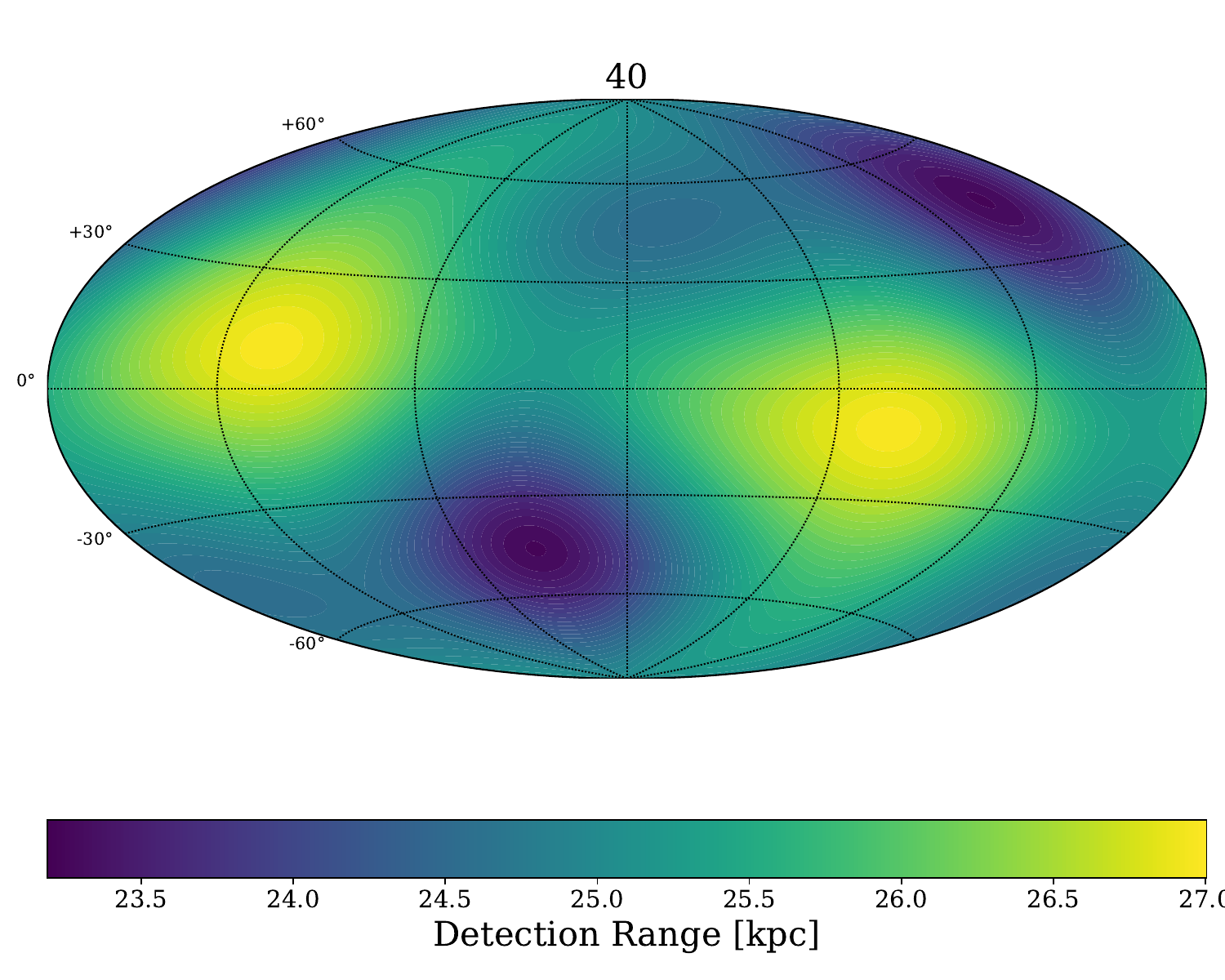}
    \caption{Detection range due to the matter strain in kiloparsecs for the 9b, 9.25, 9.5, 14, 15.01, and 40 progenitor models as a function of viewing angle for an aLIGO detector. The dark blue regions indicate shorter detection ranges whereas the yellow regions indicate longer ranges. As with the inferred energy distributions, the regions of greater and lower detection ranges varies between each progenitor model, with the greatest detection range concentrated towards the equator for the 9b and 14 plots, but are more irregularly distributed for the other models. The spread of attainable detection ranges as a function of angle also increases with progenitor mass, as the possible detection ranges differ by around 0.7 kpc for the 9b model, but around 4 kpc for the 40 model. Therefore, the ability to detect GWs from CCSNe depends on the right combination of progenitor mass, distance, and viewing angle. Using a different detector or a network of detectors will result in different detection ranges on the horizontal axis, but the histogram shapes will remain the same. \label{d_opt_globes}}
    \end{figure*}

In Figure \ref{d_opt_globes}, we plot the detection range in kiloparsecs for the 9b, 9.25, 9.5, 14, 15.01, and 40 progenitor models as a function of viewing angle and using an Advanced LIGO detector. As with the total radiated energy in Figure \ref{energy_globes}, there is no common viewing angle or region that gives the largest detection range.  The greatest detection ranges occur when observed from the equator for the 9b and 14 progenitors, but at other directions for the other progenitors. We once again observe a greater spread in detection ranges for higher mass progenitors compared to those at lower masses; the detection ranges vary by at most only 0.7 kiloparsecs for the 9b progenitor, but by around 4 kiloparsecs for the 40 mass progenitor. By examining the angle-dependence of various progenitors' detection ranges, we can see that the SNR=8 detectability of GWs from a CCSN event depends on the right combination of distance, progenitor mass, and viewing angle. 
 
\begin{figure*}[ht]
    \centering
    \includegraphics[width=0.48\linewidth]{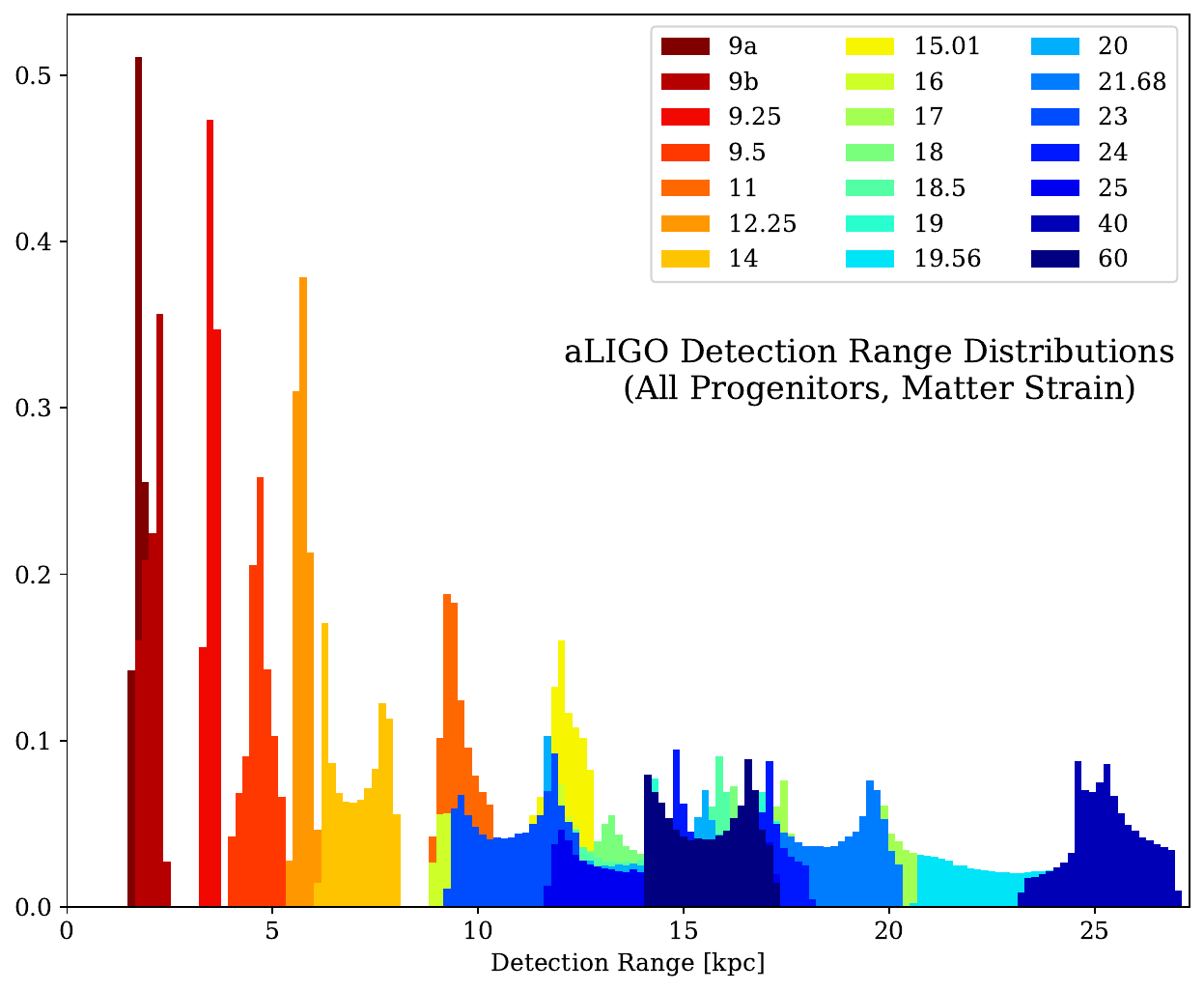}
    \includegraphics[width=0.48\linewidth]{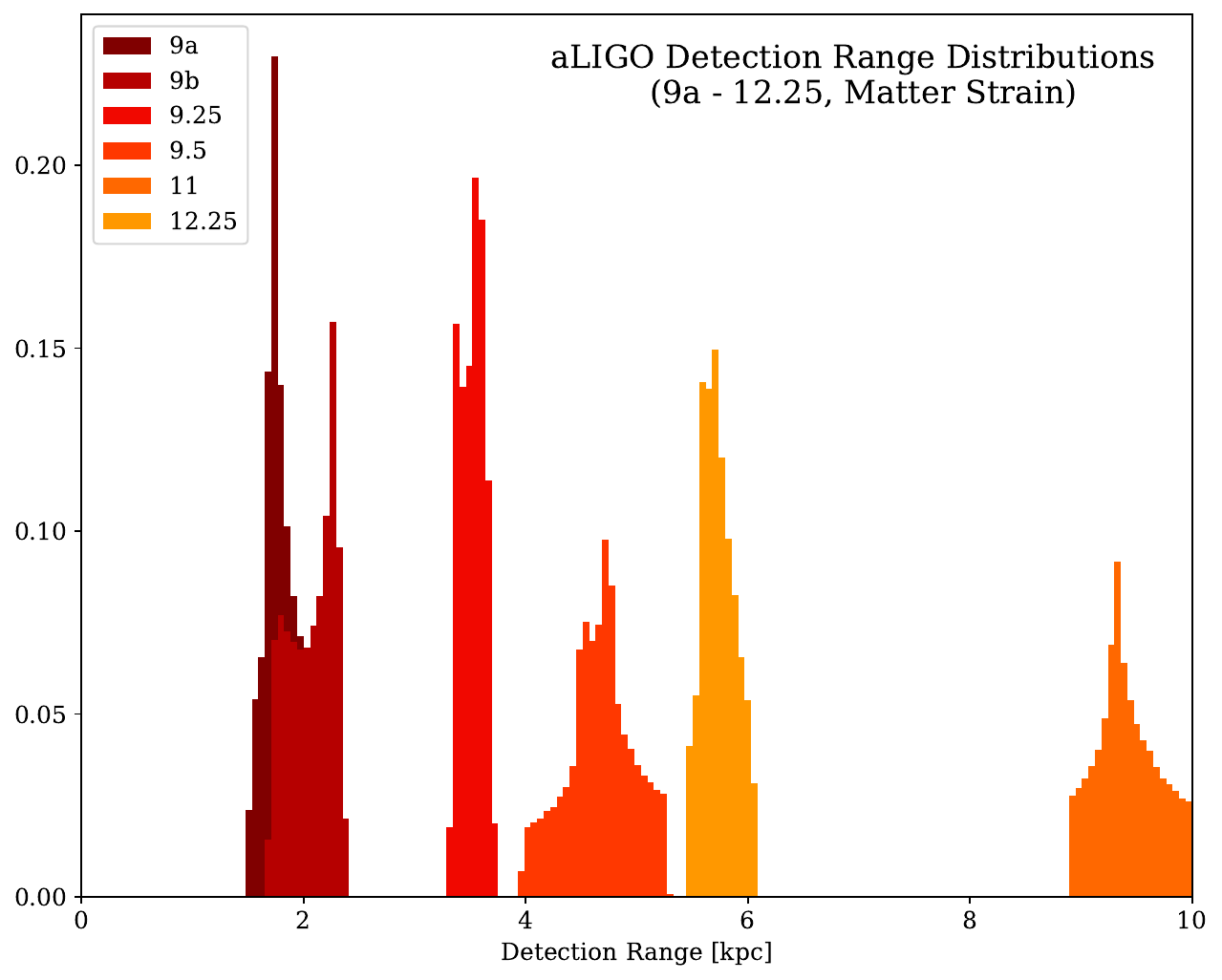}
    \includegraphics[width=0.48\linewidth]{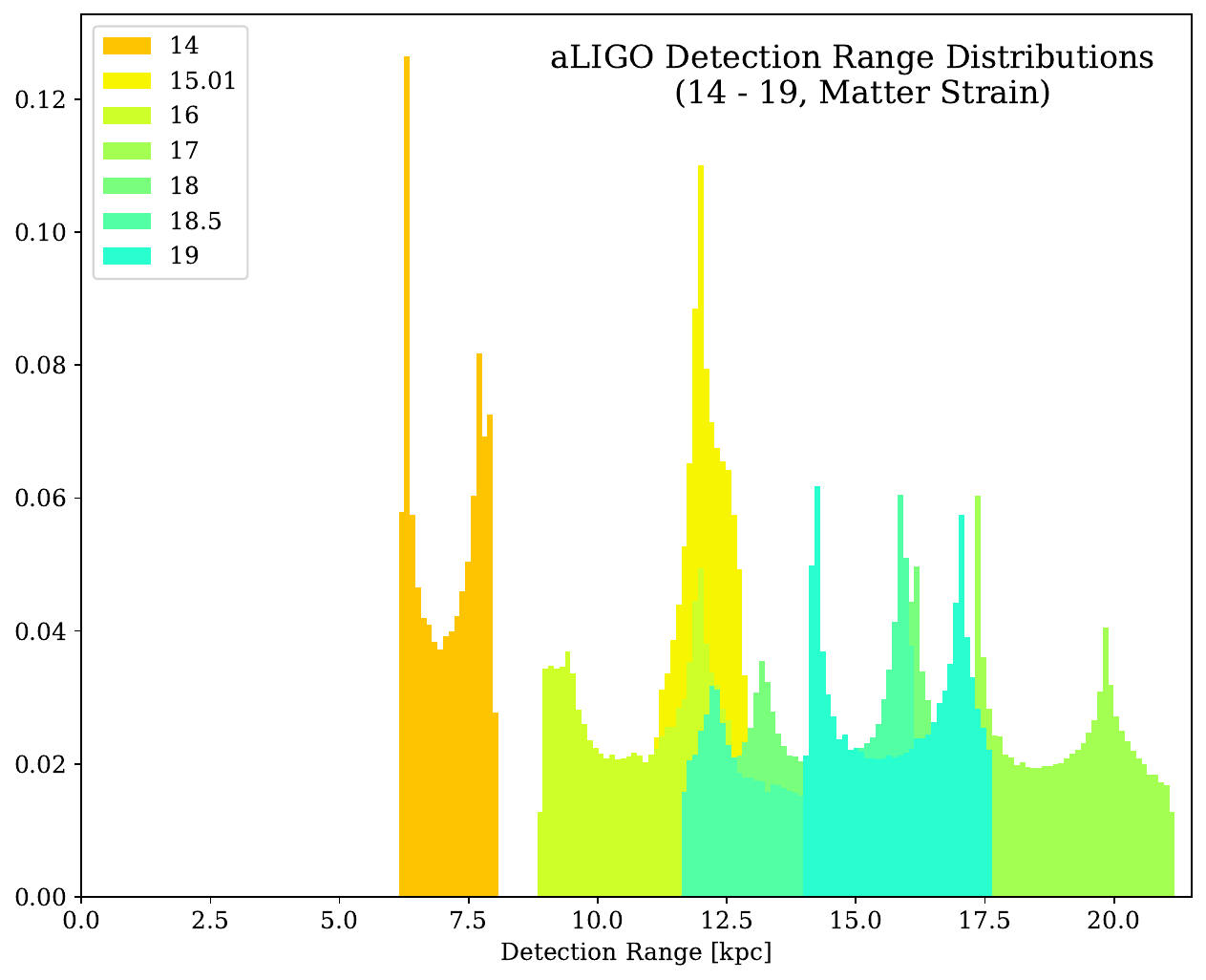}
    \includegraphics[width=0.48\linewidth]{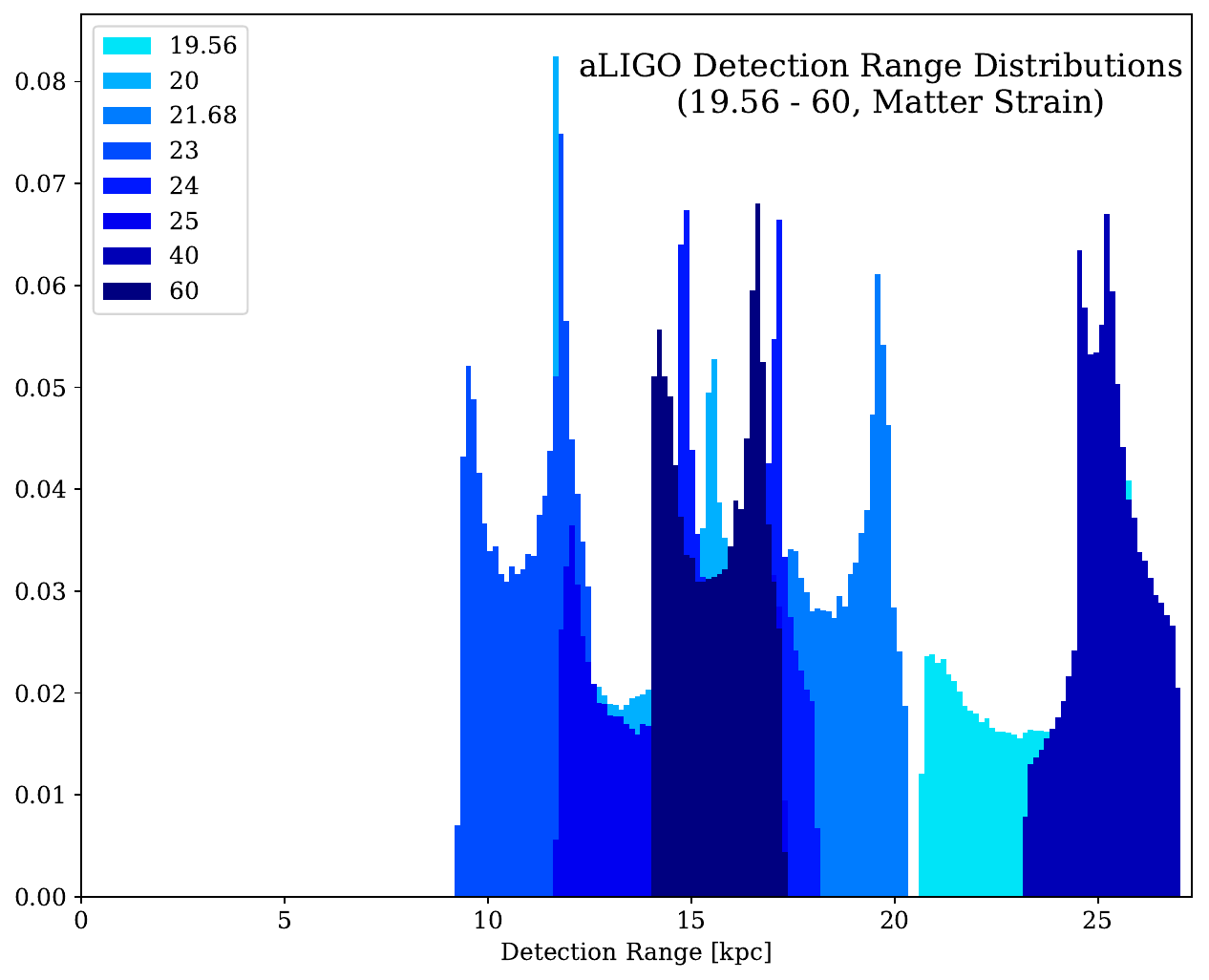}
    \caption{Detection range distributions due to matter motion across all viewing angles for all progenitor mass models (top left), low mass models $9-12.25 \: M_{\odot}$(top right), intermediate mass models $14-19 \: M_{\odot}$(bottom left), and high mass models $19.56-60 \: M_{\odot}$ (bottom right) for an aLIGO detector. The area under each histogram has been normalized to one. Note that these histograms take the same shape as those in Figure \ref{SNR_hists}, since the SNR and detection range expressions in eqs. \eqref{SNR} and \eqref{d_opt}, respectively, contain the same integrand. We observe that higher-mass progenitor models have both greater average detection ranges and also a wider spread in possible detection ranges depending on the angle of observation compared to less-massive progenitors. 
    \label{d_opt_hist_combined}}
    \end{figure*}

On the top left of Figure \ref{d_opt_hist_combined} we plot the histograms of detection ranges across all viewing angles for all progenitors on the same axes to compare their distributions. We also plot the histograms of the low mass models 9-12.25 (top right), intermediate mass models 14-19 (bottom left), and the high mass models 19.56-60 (bottom left) separately to more clearly resolve the features of the histograms. Note that since the expressions in eqs. \eqref{SNR} and \eqref{d_opt} contain the same integral, the histograms in Figure \ref{d_opt_hist_combined} take the same shape as those of Figure \ref{SNR_hists}, but with values of distance measured in kiloparsecs instead of dimensionless SNR. 

By examining the distributions in Figure \ref{d_opt_hist_combined}, we observe that the detection range increases with progenitor mass, a result consistent with previous work on matter signal detection ranges, such as \citet{srivastava2019} and \citet{afle2023}. As with the previous SNR and inferred energy plots, we also find that the spread of possible detection range values is larger with larger progenitor masses, with the possible detection ranges with a spread up to 4 kiloparsecs for higher-mass progenitors. Therefore, the distance out to which we can detect GWs from CCSNe is a highly angle-dependent quantity. 

Finally, in Tables \ref{d_opt_table}, \ref{d_opt_table_ET}, \ref{d_opt_table_CE}, and \ref{d_opt_table_decigo} we calculate the maximum, minimum, and solid-angle-averaged detection range in kiloparsecs for each progenitor model for the Advanced LIGO, ET, CE, and DECIGO detectors, respectively. From the first four columns of the Advanced LIGO in Table \ref{d_opt_table}, we find that since detection range roughly increases with progenitor mass, to detect GWs from a lower-mass CCSN progenitor, the event would have to occur extremely close to Earth (i.e., within around 5 kiloparsecs) whereas higher mass progenitors such as the 40 $M_{\odot}$ model can be observed from the other side of the Milky Way, over 20 kiloparsecs away. While the general trend of larger detection ranges from matter for more massive progenitors also applies to the ET and CE detectors, the matter ranges for both detectors are large enough to see the entire galaxy, even with the lower-mass progenitors. This mass-detection range trend is less consistent with DECIGO, as it largely captures the memory component of the strain which we discuss in the following section. Our results in Table \ref{d_opt_table} are roughly consistent with those in Table I of \citet{afle2023} and are lower than the detection ranges calculated from the non-rotating models of \citet{powell2024}. 

As with the SNR results, the detection ranges in this section are calculated for an individual aLIGO, ET, CE, or DECIGO detector to facilitate the comparison between the CCSN GW detection capabilities of current and future detectors. In practice, considering a network of detectors will cause the detection range results discussed in this section to increase by a factor dependent on the sensitivity of the detectors in the network and the orientation of the supernova relative to the detectors.

\section{Gravitational Wave Memory due to Matter Motions}
\label{matter_memory}
While the previous sections examined features of the full GW signal due to matter motions from each progenitor, in this section, we focus on the low-frequency, matter component of the signal, otherwise known as the matter \textit{memory}. This memory component is associated with the large-scale motions of the matter ejecta and can cause permanent alterations to the surrounding spacetime, well after the explosion occurs \citep{vartanyan2023, gill2024, richardson2024, Mezzacappa2024}. 

To isolate this low-frequency component of the signal, we employ a Butterworth low-pass filter, as done in \citet{richardson2024} and \citet{Pajkos2023}. The gain of a Butterworth filter is as follows:
    \begin{equation}
    \label{butterworth}
    G(f)^2 = \frac{G_0^2}{1+(\frac{f}{f_0})^{2n}} \,\, ,
    \end{equation}
where $f_0$ is the cutoff frequency and $n$ is the order which governs the steepness of the cutoff. To maximize steepness, we use the order $n=10$. We use a cutoff frequency of $f_0 = 10$ Hz as done in \citet{gill2024} and \citet{richardson2024}. Despite the high $n$ order, there is still some $> 10$ Hz leakage into the memory signal, as seen in the bottom left plot of Figure \ref{sensitivities}, which plots the spectrum of the low-frequency memory component of the matter signal in isolation. Therefore, while the results in this section focus on the low-frequency matter memory component of the signal, there is still some vestige of the $> 10$ Hz component of the signal. We also note that the 10 Hz memory definition is ambiguous, as works such as \citet{vartanyan2023} use a 15 Hz cutoff for their Butterworth filter and \citet{Richardson_2022} examine memory features under 50 Hz. Since the matter memory curves in the bottom left of Figure \ref{sensitivities} are peaked at lower frequencies, this signal component is outside of the sensitivity band of Advanced LIGO for almost all the progenitor models, and are only within the sensitivity bands for future ET, CE, and DECIGO detectors for higher-mass progenitors.

\begin{figure*}[ht]
    \centering
    \includegraphics[width=0.9\linewidth]{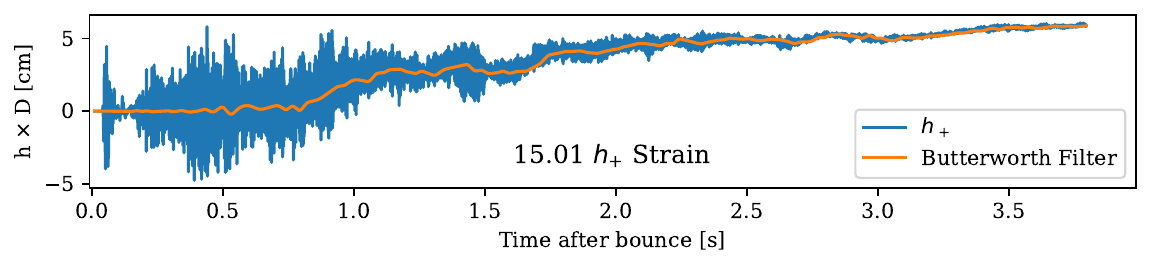}
    \includegraphics[width=0.9\linewidth]{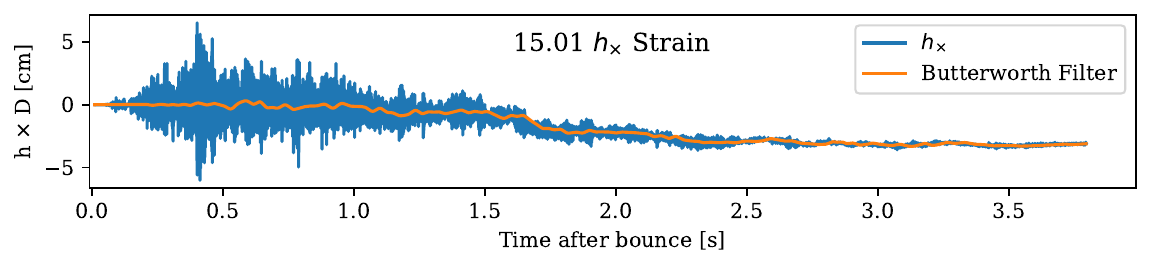}
    \caption{Butterworth low-pass filter applied to the 15.01 progenitor model strain in the x-direction. The orange, filtered curve isolates the low-frequency memory component of the strain due to large-scale matter motion which leads to a permanent alteration to the surrounding space. \label{butter_examples}}
    \end{figure*}

\begin{figure*}[ht]
    \centering
    \includegraphics[width=0.9\linewidth]{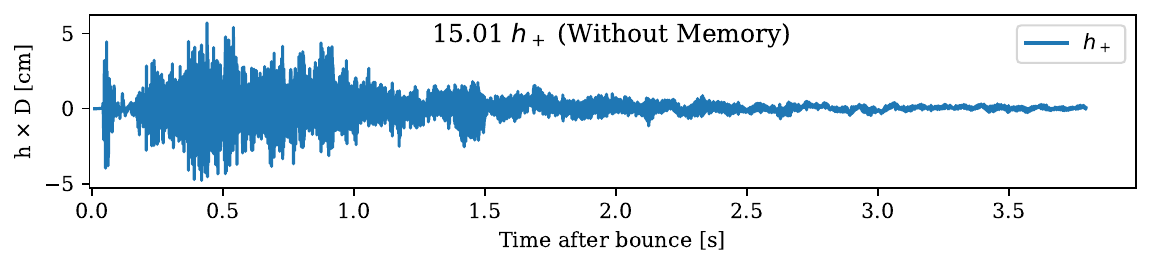}
    \includegraphics[width=0.9\linewidth]{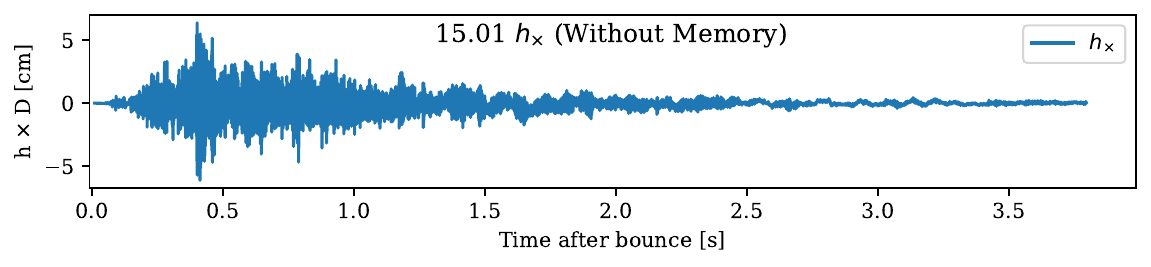}
    \caption{The $h_+$ and $h_\times$ matter strain of the 15.01 progenitor model in the x-direction with the low-frequency memory subtracted from the original signal. By isolating the high-frequency component of the signal due to turbulence, we observe that the memory-subtracted strain indeed returns to zero by the end of the simulation. \label{butter_subtract}}
    \end{figure*}

In Figure \ref{butter_examples}, we plot the filtered, memory component of the signal (orange) against the original strain (blue) for the 15.01 model in the x-direction, where we can more clearly observe how the low-frequency component does not return to zero, even after the main burst of the explosion. In contrast, we plot the GW signal with the low-frequency component removed for the 15.01 progenitor in Figure \ref{butter_subtract} to understand the appearance of the signal if there were no matter gravitational memory. In this plot, the strain returns to zero by the end of the simulation, as there is no permanent alteration to the surrounding spacetime. 
    
We note that previous works such as \citet{gill2024}, \citet{richardson2024}, and \citet{Richardson_2022} employed a tapering function to extend the memory component of the signal until it reaches zero, as the sudden drop from a nonzero strain at the end of the simulation introduces artifacts that increase the calculated SNR value. However, artificially extending a signal also artificially increases its SNR. Instead of a tapering function, we employ an exponential windowing function convolved with the Fourier transform in eq. \eqref{fourier} to avoid the abrupt discontinuity in the finite signal. 

In Table \ref{SNR_mem_table}, we calculate the SNR averaged over the solid angle for the matter memory component of the signal. We observe that while the matter memory SNR is close to zero, and thus cannot be detected for all models with the Advanced LIGO detector at 10 kiloparsecs, the ET and DECIGO detectors should be able to detect this memory component for some higher-mass progenitors, with DECIGO being the detector most capable of detecting matter memory. This stands in contrast to the results of Table \ref{SNR_reg_table}, where ET and CE were the detectors which most effectively captured the matter strain. We note that the nonexploding 12.25 and 14 models have near-negligible matter memory SNR values across all detectors, a result consistent with the lack of nonzero displacement observed at late times in their strains in Figure \ref{all_strains}. Our results concerning the detectability of memory with future Earth-based detectors are in agreement with the results of \citet{richardson2024} and \citet{powell2024}, but contrasts with \citet{gill2024}'s claim that matter memory can be detected only with lunar detectors. 

\begin{deluxetable}{ccccc}
\tablecolumns{5}
\tablewidth{0pt}
\label{SNR_mem_table}

\begin{minipage}{\textwidth}
  \centering
  \textbf{Matter Memory SNR} \\  
\end{minipage}

\tablehead{\colhead{Progenitor} & \colhead{aLIGO} & \colhead{ET} & \colhead{CE} & \colhead{DECIGO}}
\startdata
$9a $ & 0.00 & 1.14 & 0.17 & 2.53 \\
$9b $& 0.00 & 0.91 & 0.20 & 1.77 \\
$9.25$  & 0.01 & 1.86 & 0.33 &  2.82\\
$9.5$ & 0.02 & 6.38 & 1.19 & 8.10 \\
$11$ & 0.03 & 9.97 & 1.78 & 16.59 \\
$12.25$ & 0.00 & 0.10 & 0.02 & 0.05 \\
$14$ & 0.00 & 0.19 & 0.03 & 0.15 \\
$15.01$  & 0.02 & 7.38 & 1.26 & 10.50 \\
$16$ & 0.02 & 6.15 & 1.20 & 9.73 \\
$17$ &  0.10 & 30.41 & 5.61 & 48.38\\
$18$ & 0.04 & 11.53 & 1.98 & 18.67 \\
$18.5$ & 0.09 & 26.27 & 4.88 & 38.99 \\
$19$ & 0.04 & 11.15 & 1.91 & 16.09\\
$19.56$ & 0.23 &66.53& 11.53 & 95.51 \\
$20$ & 0.03 & 7.53 & 1.31 & 10.36 \\
$21.68$ & 0.07 & 17.33 & 3.39 & 22.82\\
$23$ & 0.03 & 7.51 & 1.43 & 11.08 \\
$24$ & 0.12 & 34.83 & 6.23 & 50.25\\
$25$ & 0.06 & 17.98 & 3.20 & 30.18\\
$40$  & 0.16 & 45.93 & 7.83 & 64.52 \\
$60$ & 0.04 & 12.80 & 2.34 & 22.56 \\
\enddata
\caption{SNR calculated at D=10 kpc for the matter memory component of all progenitor models with the Advanced LIGO, ET, CE, and DECIGO detectors averaged over all viewing angles. While the near-zero SNR values in the second column indicate that Advanced LIGO will not be able to detect any of the matter memory at 10 kiloparsecs (kpc), some future detectors such as ET and DECIGO will be able to detect the matter memory component of the higher-mass progenitor models. The nonexploding models (12.25 and 14) contain negligible memory effects across all detectors. }
\end{deluxetable}

\bigskip\bigskip\bigskip\bigskip

\section{Polarization Angle}
\label{polar}
Another significant feature of the GW strain from CCSNe events is the \textit{polarization angle}, which can serve as another way to detect gravitational memory. In this section, we adopt the polarization angle convention of \citet{Hayama_2018}, who define the polarization angle $\theta$ to be $\theta = \arctan{(h_+ / h_{\times})}$ where $\theta \in [-\pi, \pi]$. Note that this is not the same definition of the ``polarization angle" $\psi$ of the antenna pattern functions in \eqref{antenna_pattern}, which is defined independently of any properties of the signal itself \citep{Isi_2023}.

\begin{figure*}[htpb!]
    \centering
    \includegraphics[width=0.9\linewidth]{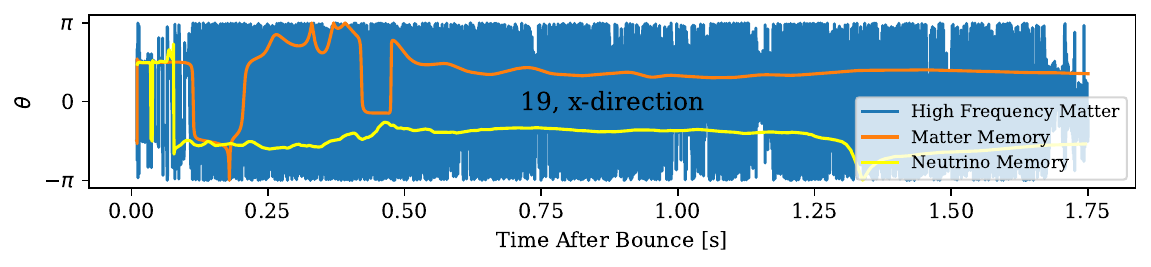}
    \includegraphics[width=0.9\linewidth]{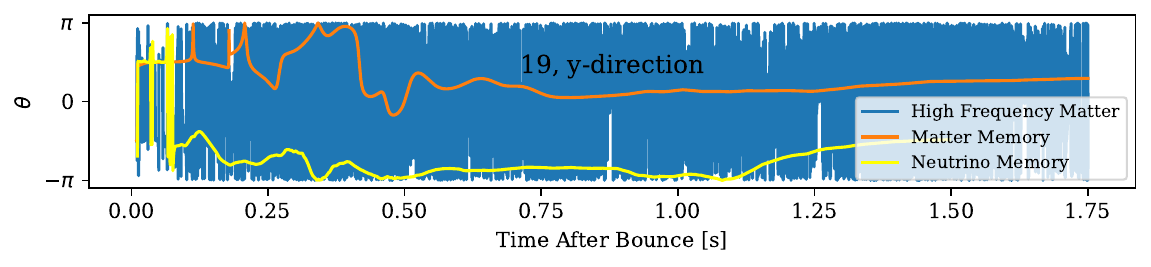}
    \includegraphics[width=0.9\linewidth]{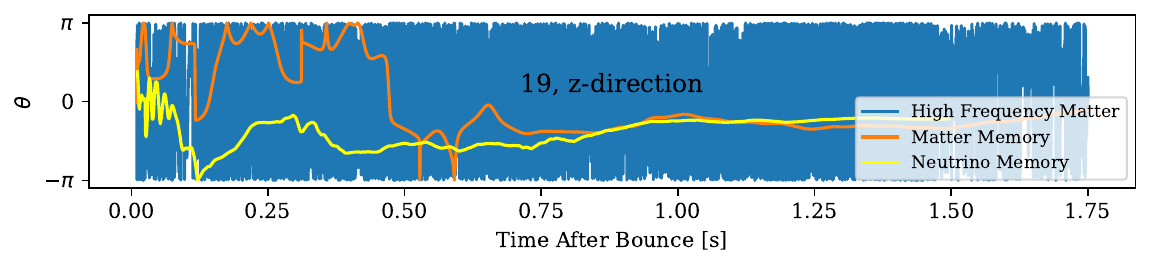}
    \caption{Polarization angle $\theta = \arctan{(h_+ / h_{\times})}$ as a function of time calculated from the high frequency (blue) and memory component (orange) for the 19 progenitor model viewed from the x-, y-, and z-directions. Here, $\theta \in [-\pi, \pi]$. The polarization angle oscillates wildly for the high-frequency component of the signal and effectively ``averages" itself out. On the other hand, the polarization angle calculated from the matter memory and neutrino memory components oscillate on much longer timescales within the first 0.7 seconds and then settle down to a more constant value at later times. This stability and lack of averaging-out of the polarization angle in the low-frequency regime thus serves as a unique signature of the memory effect. 
    \label{polar_examples_19}}
    \end{figure*}

\begin{figure*}[htb!]
    \centering
    \includegraphics[width=0.9\linewidth]{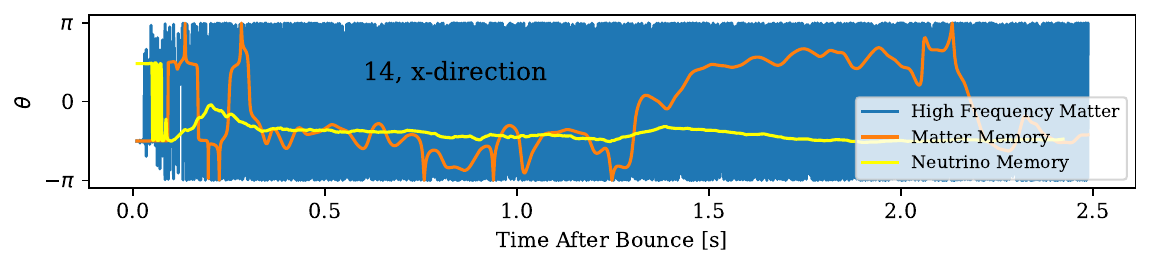}
    \includegraphics[width=0.9\linewidth]{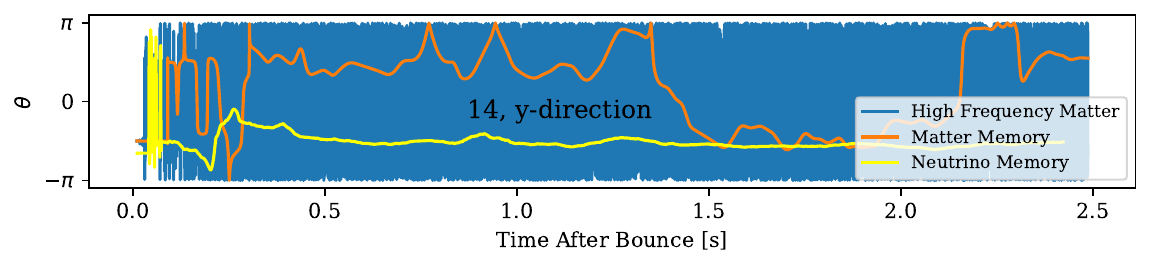}
    \includegraphics[width=0.9\linewidth]{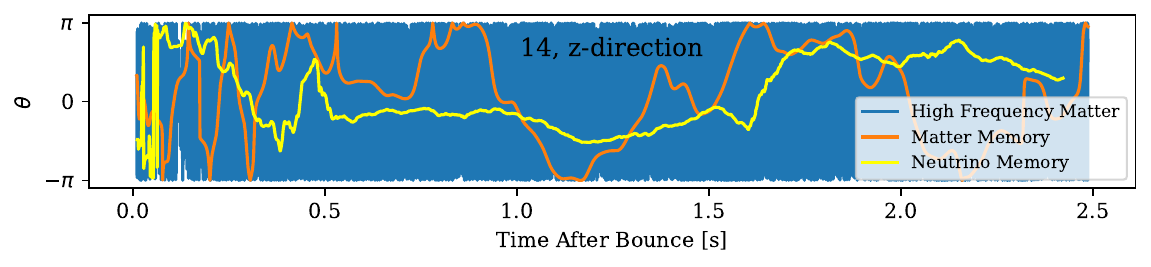}
    \caption{Polarization angle $\theta = \arctan{(h_+ / h_{\times})}$ as a function of time calculated from the high frequency (blue) and low-frequency matter memory component (orange) for the 14 progenitor model viewed from the x-, y-, and z-directions where $\theta \in [-\pi, \pi]$. Once again, the matter memory component of the polarization angle oscillates on a much longer timescale compared to the high-frequency component of the polarization angle, which oscillates at a frequency of around 10$^3$ Hz and effectively ``averages" itself out. Therefore, the polarization angle once again serves as a detectable, distinguishing factor of matter memory. However, it is important to note that unlike with the exploding 19 model in Figure \ref{polar_examples_19}, the magnitude of the memory strain from this nonexploding model is very low.
    \label{polar_examples_14}}
    \end{figure*}

In Figures \ref{polar_examples_19} and \ref{polar_examples_14}, we plot the polarization angle $\theta$ as a function of time for both the high-frequency (blue) and the low-frequency matter memory component of the signal (orange) as viewed from the x-, y-, and z-directions for the exploding 19 and non-exploding 14 progenitor models. The polarization angle calculated from the neutrino strain is also plotted (yellow) and will be discussed in Section \ref{nu_polarization_angle}.

In Figure \ref{polar_examples_19}, while the polarization angle oscillates wildly for the high-frequency component of the signal, the polarization angle calculated from the memory strain is more stable and settles down to a constant value less than 0.75 seconds after bounce. This stability of the polarization angle in the low-frequency regime is a unique signature of both the matter and the neutrino memories and can be used to identify memory when observing GWs from CCSNe. More quantitatively, we can also compare the approximate variation timescales of the polarization angles from the high-frequency and matter memory components. For the 19 progenitor model, the timescale of the high frequency component of the polarization angle variation is less than one millisecond for the first two seconds, oscillating so rapidly that there is an ``averaging" out effect, making the polarization signal almost unmeasurable. This ``averaging" out is a result of the matter accreting onto the proto-neutron star in all directions just after explosion. On the other hand, the timescale for the low frequency polarization angle variation is much longer (at around 0.04 seconds) in the z-direction and even longer in the x- and y- directions, before settling down at later times. Therefore, in addition to the stabilization of the polarization angle of the memory component at around 0.7 seconds after bounce, the significantly lower-frequency variation of this polarization angle for the first 0.7 seconds immediately after bounce is another characteristic signature of memory. Similarly, in Figure \ref{polar_examples_14}, the matter memory component of the polarization angle for the 14 model fluctuates only of order twenty times for the entire duration of the signal, compared to the approximately one millisecond timescale variation of the high-frequency component. While the matter memory does not settle to a constant value at late times as with the 19 model, the high-frequency component of the 14 model's polarization angle still fluctuates so rapidly that it effectively ``averages" itself out, once again serving as a distinguishing feature between the high-frequency and the memory components of the matter signal. 

Overall, the lower-frequency matter memory and, as we will demonstrate in the following section, neutrino memory strains enable us to understand the systematic anisotropy of the exploding matter morphology and neutrino emission of the CCSN, respectively. 

\bigskip
\bigskip

\section{Neutrino Memory}
\label{neutrino_memory}
\subsection{Neutrino Memory Strain}
\label{direction_memory}
In addition to GW strain caused by matter motions, the neutrinos released during the CCSN also produce a low-frequency GW memory strain of the form \citep{muller2012, vartanyan2020}
\begin{equation}
\label{neu_strain}
h_{+,\times}(t,\alpha, \beta)=\frac{2G}{c^4D}\int_0^tdt'\Lambda(t')\alpha_{+/\times}(t',\alpha,\beta) \,\, ,
\end{equation}
where $\Lambda$ is the neutrino luminosity, the anisotropy parameter $\alpha$ is given by
\begin{equation}
\alpha_{+/\times}(t,\alpha,\beta)=\frac{1}{\Lambda(t)}\int_{4\pi}d\Omega' W_{+/\times}(\Omega', \alpha, \beta)\frac{d\Lambda}{d\Omega'}(\Omega', t) \,\, ,
\end{equation}
and the geometric weight for the anisotropy parameter is 
\begin{equation}
W_{+,\times} = \frac{D_{+,\times}(\theta', \phi', \alpha, \beta)}{N(\theta', \phi', \alpha, \beta)} \,\, .
\end{equation}
The quantities $D_+$, $D_{\times}$, and $N$ are given by
\begin{equation}
\begin{aligned}
    D_+ &= [1+(\cos(\phi')\cos(\alpha)+\sin(\phi')\sin(\alpha))\sin(\theta')\sin(\beta) +\cos(\theta')\cos(\beta)][[(\cos(\phi')\cos(\alpha) \\
    &+\sin(\phi')\sin(\alpha))\sin(\theta')\cos(\beta)-\cos(\theta')\sin(\beta)]^2 - \sin^2(\theta')(\sin(\phi')\cos(\alpha -\cos(\phi')\sin(\alpha))^2] \,\, ,
\end{aligned}
\end{equation}

\begin{equation}
\begin{aligned}
D_{\times} &= [1+(\cos(\phi')\cos(\alpha)+\sin(\phi')\sin(\alpha))\sin(\theta')\sin(\beta) +\cos(\theta')\cos(\beta)]2[(\cos(\phi')\cos(\alpha) \\
    &+\sin(\phi')\sin(\alpha))\sin(\theta')\cos(\beta) - \cos(\theta')\sin(\beta)]\sin(\theta')(\sin(\phi')\cos(\alpha) - \cos(\phi')\sin(\alpha))^2 \,\, ,
\end{aligned}
\end{equation}
and 
\begin{equation}
\begin{aligned}
N &= [(\cos(\phi')\cos(\alpha)+\sin(\phi')\sin(\alpha))\sin(\theta')\cos(\beta) - \cos(\theta')\sin(\beta)]^2 + \sin^2(\theta')(\sin(\phi')\cos(\alpha) -\cos(\phi')\sin(\alpha))^2 \,\, .
\end{aligned}
\end{equation}

In the above equations, $\alpha \in [-\pi,\pi]$, $\beta \in [0, \pi]$ are viewing angles which are related to the source (primed) and observer (unprimed) coordinates by
\begin{equation}
\begin{aligned}
\label{viewing_angles}
\sin\theta\cos\phi &= (\cos\phi'\cos\alpha+\sin\phi'\sin\alpha)\sin\theta'\cos\beta - \cos\theta'\sin\beta \\
\sin\theta\sin\phi &= (\sin\phi'\cos\alpha-\cos\phi'\sin\alpha)\sin\theta' \\
\cos\theta &= (\cos\phi'\cos\alpha + \sin\phi'\sin\alpha)\sin\theta'\sin\beta +\cos\theta'\cos\beta \,\, .
\end{aligned}
\end{equation}
In Figure \ref{neu_strain_plot}, we plotted the $h_{+, \times}$ polarizations of the neutrino strain in the x-direction for all neutrino species combined. As with the strains due to matter motion, the magnitude of the strain roughly increases with progenitor mass. However, the neutrino strain times its distance can be over 1000 cm for certain progenitor models, making it several orders of magnitude larger than the matter strain, a result consistent with \citet{vartanyan2023}. Additionally, the strain due to neutrino emission is entirely characterized by low-frequency memory and lacks the burst of high-frequency emission characteristic of the matter strain. We note that the neutrino strain shown here is an order of magnitude larger than the neutrino strains shown in \citet{powell2024}. This discrepancy is due to the fact that our simulations are carried out up to six seconds post-bounce, whereas the simulations in \citet{powell2024} last for less than a second, thus demonstrating the need for longer simulations to capture the full GW signal. 

In Figure \ref{sensitivities}, we plotted the spectra for the neutrino strain for all progenitor models assuming a distance of 10 kiloparsecs (kpc) (bottom right), as well as the total matter and neutrino memory strain spectra (top left). The spectral densities due to neutrino emission peak at lower frequencies than the corresponding curves for the matter strains, and also reach larger values. Note that, due to the Nyquist sampling of the data, the spectra for the neutrino component of the signal are not reliable in the $> 500$ Hz range . 

\begin{figure*}[htbp!]
    \centering
    \includegraphics[width=0.49\linewidth]{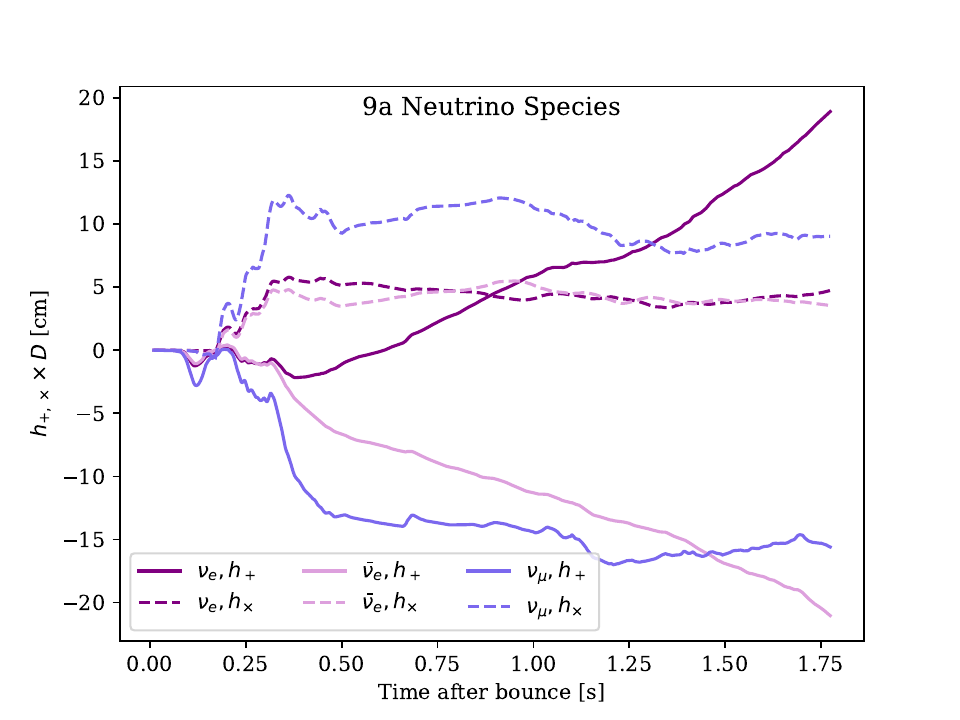}
    \includegraphics[width=0.49\linewidth]{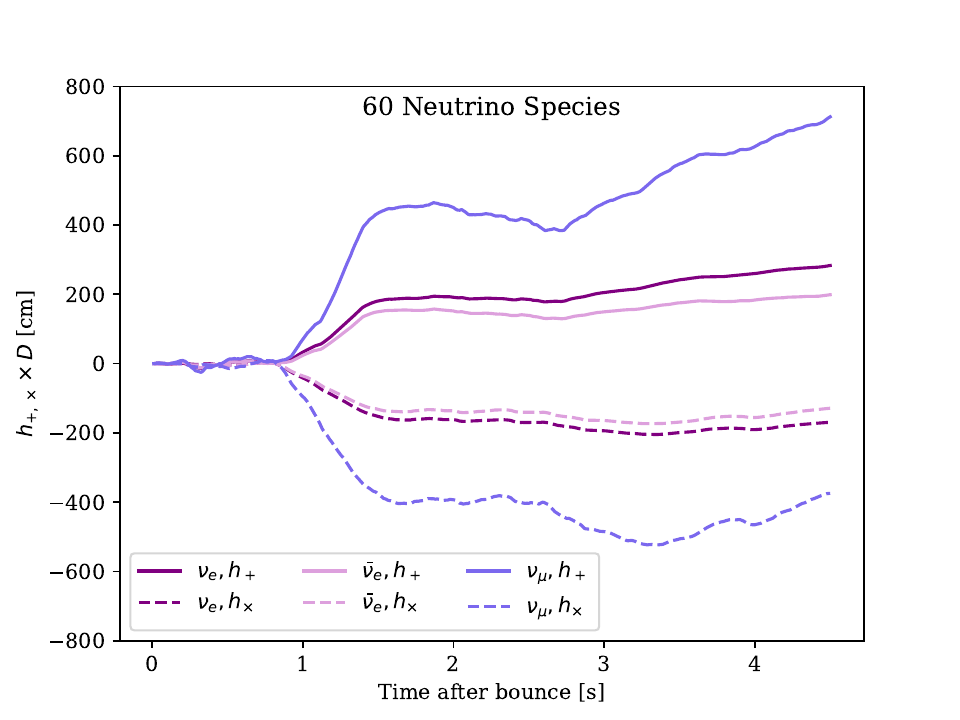}
    \caption{GW strains due each neutrino species $\nu_e, \bar{\nu}_e, and ``\nu_{\mu}$" (a bundling of $\nu_{\mu}$, $\bar{\nu}_{\mu}$, $\nu_{\tau}$, and $\bar{\nu}_{\tau}$ neutrinos) as a function of time after bounce for the 9a (left) and 60 (right) progenitor models. The plus polarization is plotted with solid lines and the cross polarization is plotted with dashed lines. Between both progenitor models, the heavy neutrino models species ``$\nu_{\mu}$" tends to have the larger strain across both polarizations, reaching close to 800 cm divided by distance for the 60 model, over an order of magnitude larger than the memory strain due to large-scale matter motions. The antineutrino species $\bar{\nu}_e$ has the weakest strain across both progenitors and polarizations. \label{neutrino_species}}
    \end{figure*}

To understand the contributions of the individual neutrino species, in Figure \ref{neutrino_species} we plot the $h_{+, \times}$ strains due to the $\nu_e$, $\bar{\nu}_e$, and ``$\nu_{\mu}$" neutrinos, the latter a bundling of the $\nu_{\mu}$, $\bar{\nu}_{\mu}$, $\nu_{\tau}$, and $\bar{\nu}_{\tau}$ neutrinos, as a function of time for the 9a and 60 progenitor models. We observe that the ``$\nu_{\mu}$" neutrino species tends to produce the greatest strains, whereas the antineutrino strain has the weakest strains. Each species, however, has a greater strain than the strain due to matter motions.

\subsection{Radiated Gravitational-Wave Energy Associated with Neutrino Memory}
\label{nu_memory_energy}
On the right side of Figure \ref{EGW_tot}, we plotted the total energy radiated in GWs due to neutrino emission for all neutrino species combined. As with the energy from matter motions, the 40 progenitor model radiates the greatest amount of energy while the 9a and 9b models radiate the least. However, for all models, the total radiated energy due to neutrino emission is roughly two orders of magnitude \textit{lower} than the energy radiated due to matter emission, due to their much lower frequencies. Additionally, for all models, the total radiated energy continues to increase over the course of the simulation, instead of plateauing to a constant value, demonstrating the importance of carrying out CCSN simulations to late times. 

\subsection{Neutrino Memory Polarization Angle}
\label{nu_polarization_angle}
In Figures \ref{polar_examples_19} and \ref{polar_examples_14}, we also plotted the polarization angle $\theta = \arctan{(h_+ / h_{\times})}$ as a function of time after bounce for the neutrino memory component of the strain for the 19 and 14 models as viewed from the x-, y-, and z-directions. In both models, the polarization angle of the neutrino memory varies even less than the matter memory, most clearly demonstrated for the 14 progenitor model. Therefore, the polarization angle serves as a important distinguishing feature of GW memory of both types. Moreover,  the relative lack of temporal variation in the neutrino memory may serve as a way (some day) to distinguish neutrino memory from matter memory. 
    
\subsection{Neutrino Memory SNR}
\label{nu_memory_snr}
As with the GW strain due to matter motion, we can calculate the SNR of the strain due to neutrino emission using eq. \eqref{SNR}. In Table \ref{neu_SNR_table}, we calculate the SNR of the GW strain due to neutrino emission for each progenitor model, averaged over the solid viewing angle and assuming a distance of 10 kiloparsecs for the Advanced LIGO, ET, CE, and DECIGO detectors. Due to the large neutrino strains observed in Figure \ref{neu_strain_plot}, the SNRs reported in this table are larger than those of the matter strain for all detectors in Table \ref{SNR_reg_table}, and at least an entire order of magnitude larger than those of the matter memory strain in Table \ref{SNR_mem_table}. Therefore, we find that neutrino memory is the most detectable component of a GW CCSN signal. While all of the future detectors have high neutrino memory SNR values for all progenitor models indicating confident and high-resolution signals, even Advanced LIGO is sensitive enough to detect neutrino memory signatures from higher-mass ($> 11 \: M_{\odot}$) progenitors at 10 kiloparsecs. However, DECIGO has the highest SNR values for all progenitors and is thus the most capable of detecting neutrino memory. 
\begin{deluxetable}{ccccc}
\tablecolumns{5}
\tablewidth{0pt}
\label{neu_SNR_table}

\begin{minipage}{\textwidth}
  \centering
  \textbf{Neutrino Memory SNR} \\  
\end{minipage}

\tablehead{\colhead{Progenitor} & \colhead{aLIGO} & \colhead{ET} & \colhead{CE} & \colhead{DECIGO}}
\startdata
$9a $ & 1.61 & 36.60 & 63.16 & 95.33 \\
$9b $ & 1.61 & 30.83 & 55.12 & 76.74 \\
$9.25$ & 3.20 & 58.83 & 104.73 & 158.75 \\
$9.5$ & 3.48 & 66.21& 116.15 & 218.59 \\
$11$ & 49.36 & 1224.23 & 2124.67 & 2855.68\\
$12.25$ & 12.59 & 267.80 & 464.63 & 572.55 \\
$14$ & 13.20 & 228.96 & 393.76 & 515.95 \\
$15.01$ & 17.20 & 404.96 &705.68 & 1036.50\\
$16$ & 28.75 & 707.89 &1228.56 & 1589.35\\
$17$ & 29.44 & 706.11 &1233.91 & 1899.58\\
$18$ & 17.64 & 422.50 & 732.06& 1061.24 \\
$18.5$ & 24.07 & 549.16 & 952.85& 1455.74\\
$19$ & 40.32 & 998.68 & 1734.43 & 2137.96 \\
$19.56$ & 42.53 & 1022.95 &1772.87 & 2279.80\\
$20$ & 26.79 & 613.95 & 1067.47 & 1385.11 \\
$21.68$ & 33.83 & 809.15 & 1410.45 & 2137.03\\
$23$ & 36.62 & 891.39 & 1550.95 &  2004.62 \\
$24$ & 29.45 & 699.32 & 1218.07 & 1573.65 \\
$25$ & 25.57 & 594.30 & 1032.49 & 1472.91\\
$40$ & 68.07 & 1674.30 & 2895.07 & 4038.80\\
$60$ & 41.65 & 1024.45 & 1780.21 &2507.53 \\
\enddata
\caption{SNR calculated for D = 10 kpc due to neutrino memory for all models with the aLIGO, ET, CE, and DECIGO detectors averaged over all viewing angles. While neutrino memory is detectable for all progenitors with future detectors, even current detectors such as aLIGO (despite the steep rise in its noise curve at low frequencies) should be able to detect the neutrino memory effect for CCSNe for sufficiently massive progenitors. The SNRs reported here are significantly larger than those of the matter or matter memory SNR, indicating that neutrino memory is the most detectable component of a CCSN GW signal. } 
\end{deluxetable}

\subsection{Neutrino Memory Detection Range}
\label{nu_memory_range}
\begin{figure*}[htbp]
    \centering
    \includegraphics[width=0.48\linewidth]{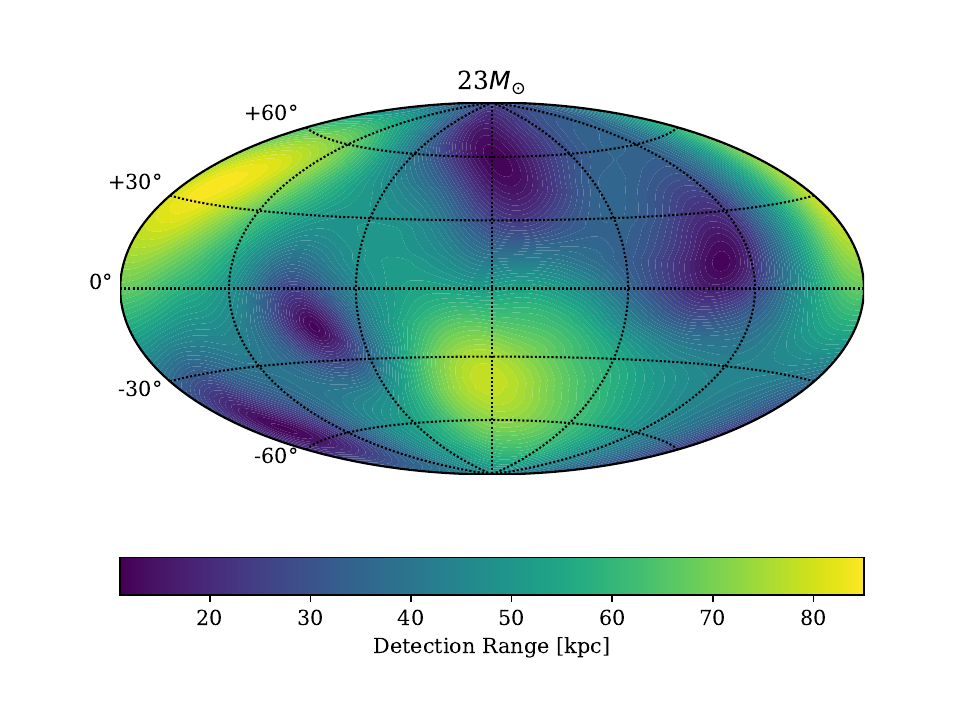}
    \includegraphics[width=0.48\linewidth]{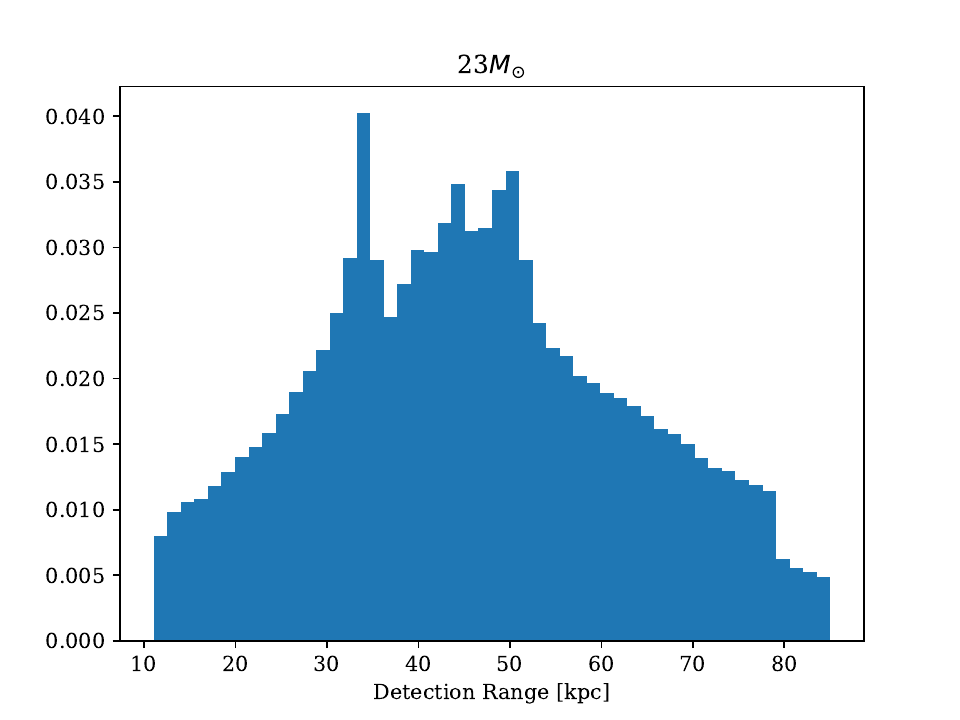}
    \caption{(Left) Detection range in kiloparsecs due to GWs from neutrino emission as a function of viewing angle ($\theta, \phi$) for the 23 $M_{\odot}$ progenitor model using an aLIGO detector. (Right) Distribution of detection ranges across all viewing angles. While the histogram demonstrates a preference for detection ranges around 30-40 kpc, it is not peaked at one particular value. Here, we observe a large variation in possible detection ranges from just over 10 kpc to 80 kpc, a range larger than the diameter of the Milky Way. Therefore, for the neutrino memory contribution to the GW signal, the angle at which the event can be observed from Earth has a significant impact on detectability $-$ several degrees may determine whether or not we observe the GW strain from neutrino emission. \label{neutrino_dopt}}
    \end{figure*}

To conclude our discussion of neutrino memory, we use eq. \eqref{d_opt} to calculate the distance out to which we can detect the GW strain from neutrino emission using the strains given by eq. \eqref{neu_strain}. In Figure \ref{neutrino_dopt}, we plot the detection range due to GW strain from neutrino emission as a function of viewing angle for the 23 progenitor model using the Advanced LIGO detector and then decompose the distribution into a histogram. Looking at the left figure, we can see that, as with the matter strain, the detection range due to neutrino memory has regions of observation that correspond to significantly greater and smaller distances. The spread in attainable detection range values is significantly larger than those due to matter motion alone, with the nearest and furthest possible detection ranges differing by over 60 kiloparsecs. While the histogram indicates a higher probability of the detection ranges around 30-40 kiloparsecs, it is not peaked at one particular value. Due to the large spread in possible detection ranges due to strain from neutrino emission for the 23 model, the angle at which the event can be observed from Earth has a significant impact on detectability, as several degrees may determine whether or not we observe the GW strain from the neutrino emission if the event occurs sufficiently far away.  

In Tables \ref{d_opt_table}, \ref{d_opt_table_ET}, \ref{d_opt_table_CE}, and \ref{d_opt_table_decigo}, we calculate the maximum, minimum, and solid angle-averaged detection ranges for neutrino memory for all progenitors for the Advanced LIGO, ET, CE, and DECIGO detectors, respectively. Once again, if considering a network of detectors, the results in these tables will increase by a factor related to the sensitivities of the detectors in the network and the orientation of the source relative to the detectors. In the last column, we calculated the combined average detection range due to the sum of the matter and neutrino memory strains. For all detectors, the detection range for both the matter and neutrino memory strains roughly increases with progenitor mass. The difference between the minimum and maximum matter and neutrino memory detection ranges also increases with progenitor mass, with the neutrino detection ranges varying by up to an order of magnitude between the minimum and maximum, indicating a strong degree of angular anisotropy. 

For the Advanced LIGO ranges in Table \ref{d_opt_table}, we observe that for low-mass progenitors, the detection ranges for both the matter and neutrino memory components are roughly equal and even combined do not span the entire Milky Way, which has a diameter of around 30 kiloparsecs. For heavier progenitors, the neutrino memory detection ranges are around three to four times larger than the matter detection ranges, and combined both easily cover the entire galaxy. 

While the ET and CE detectors in Tables \ref{d_opt_table_ET} and \ref{d_opt_table_CE}, respectively, have similar detection ranges, the CE telescope can see neutrino memory out to further distances. However, both can see the entire Milky Way with both the matter and neutrino memory detection ranges individually, but with the neutrino memory detection ranges being multiple times larger. For higher-mass progenitors, their combined detection ranges can reach even into our neighboring galaxy Andromeda, located around 765 kiloparsecs away \citep{Riess_2012, srivastava2019}. 

For DECIGO, the matter detection ranges are almost negligible compared to the neutrino memory ranges. The neutrino memory detection ranges are larger than those of both ET and CE for all progenitors and, thus, extend well into Andromeda and beyond, opening the possibility for distant CCSN GW detections. 

\begin{deluxetable*}{cccccccc}
\tablecolumns{8}
\tablewidth{0pt}
\label{d_opt_table}

\begin{minipage}{\textwidth}
  \centering
  \textbf{aLIGO Detection Ranges} \\  
\end{minipage}

\tablehead{\colhead{Progenitor} & \colhead{Matter} & \colhead{Matter} & \colhead{Matter} & \colhead{Neutrino} & \colhead{Neutrino} & \colhead{Neutrino} & \colhead{Combined} \\ 
\colhead{} & \colhead{Min} & \colhead{Max} & \colhead{Avg} & \colhead{Memory Min} & \colhead{Memory Max} & \colhead{Memory Avg} & \colhead{Avg}
}
\startdata
$9a$ &1.51 & 2.06 & 1.78 & 0.91 & 4.44 & 2.10 & 3.88\\
$9b$ & 1.69& 2.37& 2.07 & 1.36 & 3.75& 2.07 & 4.14 \\
$9.25$ & 3.33 & 3.71& 3.52& 3.13 &6.64 & 4.10 & 7.62\\
$9.5$ & 3.96 & 5.27& 4.66& 3.27& 6.47& 4.47& 9.13\\
$11$ & 9.37 & 11.03& 10.08 & 4.94 & 111.34& 65.44 & 75.52\\
$12.25$ & 5.45& 6.07& 5.74 & 8.85 &24.10& 16.35& 22.09\\
$14$ & 6.16 & 8.02& 7.08& 12.91& 21.60& 16.79& 23.87\\
$15.01$ & 11.10 & 12.88& 12.06 & 8.28 & 37.75 &22.64 & 34.70 \\
$16$ & 8.90 & 12.58 & 10.77& 6.09& 66.95&38.08 & 48.85\\
$17$ & 16.78 & 21.15 & 18.86& 10.47& 68.23& 38.90& 57.76\\
$18$ & 12.61 & 16.80 & 14.80 & 7.40 &43.71& 23.25& 38.05\\
$18.5$ & 11.64 & 16.09 &14.11 & 14.41&56.86&31.54 & 45.65\\
$19$ &14.00 & 17.63 & 15.81 & 8.66&87.86& 53.44& 69.25\\
$19.56$ &20.66& 26.90 & 23.97 & 15.80& 106.85&56.12 & 80.09\\
$20$ & 11.43 & 16.08 & 13.79 & 14.18&59.49& 35.14& 48.93\\
$21.68$ & 17.03 & 21.06 & 19.23 & 12.48&82.33 & 44.64& 63.87\\
$23$ & 9.29 & 12.53 &10.94 & 11.18& 90.17&48.44& 59.38\\
$24$ & 14.35 &18.09  &16.09 &11.99 &71.37& 38.83& 54.92\\
$25$ &11.70 & 17.28&14.69 & 14.50&54.33&33.58 & 48.27\\
$40$ & 23.19 & 27.00 &25.26&15.51 &159.53& 90.05& 115.31\\
$60$ &14.04 & 17.26 &15.60& 9.47& 92.17& 55.16& 70.76\\
\enddata
\caption{Maximum, minimum, and mean detection ranges in kpc, assuming a SNR of 8, for each progenitor mass for an Advanced LIGO detector (where the mean is taken over solid angle) for GWs from matter, neutrinos, and for both strain components combined. We define the detection range as the distance in kiloparsecs out to which an optimally-oriented source will have an SNR of eight. While the matter strain and neutrino memory contribute equally to the detection range for lower-mass models, for higher-mass models the detection range is dominated by the neutrino memory signal, meaning that most of the signal we observe out to large distances with aLIGO would be due to the neutrino memory rather than the matter strain. Comparing the maximum and minimum detection ranges across all viewing angles, the neutrino memory strain shows a wider spread in possible detection ranges compared to the total matter strain, demonstrating the greater angular anisotropy of the neutrino memory strain signal compared to the corresponding quantity for matter. } 
\end{deluxetable*}

\begin{deluxetable*}{cccccccc}
\tablecolumns{8}
\tablewidth{0pt}
\label{d_opt_table_ET}

\begin{minipage}{\textwidth}
  \centering
  \textbf{ET Detection Ranges} \\  
\end{minipage}

\tablehead{\colhead{Progenitor} & \colhead{Matter} & \colhead{Matter} & \colhead{Matter} &\colhead{Neutrino} & \colhead{Neutrino} & \colhead{Neutrino} & \colhead{Combined} \\ 
\colhead{} & \colhead{Min} & \colhead{Max} & \colhead{Avg} &\colhead{Memory Min} & \colhead{Memory Max} & \colhead{Memory Avg} & \colhead{Avg}
}
\startdata
$9a$ & 15.96 & 21.67 & 18.72& 16.11 & 103.83 &45.75 & 64.47\\
$9b$ & 17.86 & 24.75 & 21.63& 22.12 & 82.22 & 38.53& 60.16\\
$9.25$ & 33.71& 37.19& 35.30& 47.92&145.01 & 73.54& 108.84\\
$9.5$ & 40.15 & 52.43&46.54 & 46.90& 139.05& 82.77& 129.31 \\
$11$ & 86.20 & 99.73& 91.80&77.30 &2608.46& 1530.29& 1662.09\\
$12.25$ & 52.94& 58.43& 55.21 &118.10 &534.44&334.74 & 389.95\\
$14$ &59.00 & 77.59& 68.26& 169.46& 436.35& 286.19& 354.45\\
$15.01$ & 104.80& 121.88& 113.60 & 134.71&864.72& 506.20& 619.80 \\
$16$ & 85.43& 119.83& 102.99&98.69& 1569.17 & 884.86& 987.85\\
$17$ &157.96 & 198.90 &176.66 & 167.43 & 1575.26& 882.65& 1059.31\\
$18$ & 120.29& 158.64&139.99 & 120.97& 1014.90&528.13 & 668.12\\
$18.5$ &108.96 & 153.99& 133.38&232.81 &1308.77& 686.45& 819.83\\
$19$ & 132.07& 166.42& 149.27&137.92 &2070.83& 1248.35& 1397.62\\
$19.56$ &195.89 & 265.83&231.85& 241.62& 2502.41& 1278.68& 1510.53\\
$20$ & 105.51 & 150.35& 128.39 & 211.01& 1371.91& 767.43& 895.82\\
$21.68$ &149.93 &190.84&171.76& 176.60& 1905.45& 1011.43& 1183.19 \\
$23$ & 85.30&116.51& 101.27& 178.57& 2107.92& 1114.24& 1215.51\\
$24$ &132.86&172.41&152.48 & 196.13&1668.61& 874.15& 1026.63\\
$25$ & 109.74&161.45& 137.83& 225.88& 1260.74& 742.87& 880.70\\
$40$ & 215.27&258.24& 237.30&247.51 &3731.75&2092.88 & 2330.18\\
$60$ & 133.47&161.82& 147.17& 143.61& 2160.75& 1280.57 & 1427.74\\
\enddata
\caption{Maximum, minimum, and mean detection ranges in kpc for each progenitor mass for the future ET detector (where the mean is taken over the solid angle) for GWs from matter, neutrinos, and both strains combined. We define the detection range as the distance in kiloparsecs out to which an optimally-oriented source will have an SNR of eight. Given that the diameter of the Milky Way is less than 30 kpc and the distance to the nearest galaxy Andromeda is around 765 kpc, with both the matter and the neutrino contribution to the GW strain, with ET we can expect to observe GWs from CCSNe in our entire galaxy and potentially in our neighboring galaxy as well. } 
\end{deluxetable*}

\begin{deluxetable*}{cccccccc}
\tablecolumns{8}
\tablewidth{0pt}
\label{d_opt_table_CE}

\begin{minipage}{\textwidth}
  \centering
  \textbf{CE Detection Ranges} \\  
\end{minipage}

\tablehead{\colhead{Progenitor} & \colhead{Matter} & \colhead{Matter} & \colhead{Matter} &\colhead{Neutrino} & \colhead{Neutrino} & \colhead{Neutrino} & \colhead{Combined} \\ 
\colhead{} & \colhead{Min} & \colhead{Max} & \colhead{Avg} &\colhead{Memory Min} & \colhead{Memory Max} & \colhead{Memory Avg} & \colhead{Avg}
}
\startdata
$9a$ & 15.54&21.26 & 18.28 & 27.35 & 179.70 & 78.95 & 97.23\\
$9b$ & 17.20& 24.01& 20.95& 42.07& 143.88 & 68.90 & 89.85\\
$9.25$ & 34.26 & 38.40& 36.41& 83.93& 253.95& 130.91& 167.32\\
$9.5$ & 41.17 &55.02 & 48.77& 82.14 & 236.65 & 145.19& 193.96 \\
$11$ & 94.01 & 110.12 &101.34 &134.12 & 4527.47 & 2655.83& 2757.17\\
$12.25$ & 55.72& 61.78& 58.59 &201.01 &927.55&580.78 & 639.37\\
$14$ &62.61 & 82.80& 72.74& 287.93& 743.7& 492.20& 564.94\\
$15.01$ & 116.49& 136.36& 127.33 &231.74 &1516.47&882.10 & 1009.43\\
$16$ & 95.50&135.24 & 115.62&174.75 & 2723.40& 1535.71& 1651.33\\
$17$ & 181.96& 232.17& 204.44& 308.07& 2757.67&1542.38 & 1746.82\\
$18$ & 134.43& 181.95& 159.15 & 212.04& 1762.26& 915.07& 1074.22\\
$18.5$ &126.65 & 183.34& 156.81&408.82 &2278.85& 1191.07& 1347.88\\
$19$ & 147.91& 188.67 & 168.53&246.73& 3586.70& 2168.03& 2336.56\\
$19.56$ & 228.20 & 333.05 & 282.13& 416.88& 4326.75&2216.09 &2498.22\\
$20$ & 118.87 & 170.49& 145.17& 366.98& 2373.76& 1334.33 & 1479.50\\
$21.68$ & 173.30 & 217.37&197.36&319.64 & 3334.43& 1763.07& 1960.43\\
$23$ &98.34 &131.73& 115.27&306.57&3659.85&1938.69& 2039.56\\
$24$ & 152.05&207.27& 180.54 & 353.28& 2893.33& 1522.58& 1703.12\\
$25$ &125.83&185.55& 158.45& 411.83&2182.18& 1290.61& 1449.06\\
$40$ & 250.74&311.22& 281.80& 410.64& 6467.74& 3618.84& 3900.64\\
$60$ &148.41&183.65& 165.89& 245.61&3744.89 & 2225.27& 2391.16\\
\enddata
\caption{Maximum, minimum, and mean detection ranges in kpc for each progenitor mass for the future CE detector (where the mean is taken over the solid angle) for GWs from matter, neutrinos, and both strains combined. We define the detection range as the distance in kiloparsecs out to which an optimally-oriented source will have an SNR of eight. As with ET, given the diameter of the Milky Way and the distance to Andromeda, with the future CE telescope we should expect to observe GWs from CCSNe thruoghout the galaxy and potentially from Andromeda as well. } 
\end{deluxetable*}

\begin{deluxetable*}{cccccccc}
\tablecolumns{8}
\tablewidth{0pt}
\label{d_opt_table_decigo}

\begin{minipage}{\textwidth}
  \centering
  \textbf{DECIGO Detection Ranges} \\  
\end{minipage}

\tablehead{\colhead{Progenitor} & \colhead{Matter} & \colhead{Matter} & \colhead{Matter} &\colhead{Neutrino} & \colhead{Neutrino} & \colhead{Neutrino} & \colhead{Combined} \\ 
\colhead{} & \colhead{Min} & \colhead{Max} & \colhead{Avg} &\colhead{Memory Min} & \colhead{Memory Max} & \colhead{Memory Avg} & \colhead{Avg}
}
\startdata
$9a$ & 1.13& 4.73& 3.05& 24.41 & 302.67 & 119.17 & 122.22\\
$9b$ & 0.37& 3.63& 2.18& 29.26 & 217.97 & 95.93& 98.11\\
$9.25$ & 1.42 & 5.20& 3.45&84.78 & 375.68& 198.44& 201.89\\
$9.5$ & 1.35 &14.97 & 9.94&126.08 &499.34&273.23 & 283.17\\
$11$ & 4.66 & 33.02 & 20.62 & 283.23&6102.94& 3569.60&389.86\\
$12.25$ & 0.05& 0.08& 0.07 & 109.89&1206.24& 715.68& 715.75\\
$14$ & 0.16& 0.23& 0.20& 268.99& 985.73& 644.94& 645.14\\
$15.01$ &4.62 & 19.17& 13.01 &557.05 &1987.34& 1295.62& 1308.63\\
$16$ &6.10 & 16.97 & 12.13&276.54 & 3525.41&1986.69 & 1998.82\\
$17$ & 10.90&90.62  & 58.60& 482.22& 4012.28& 2374.48& 2433.08\\
$18$ &8.59 & 35.69& 23.16&474.79 &2467.43&1326.55 & 1349.71\\
$18.5$ &14.82 &73.13& 48.11& 901.18& 3158.85& 1819.68& 1867.79\\
$19$ &8.24&29.82&19.90& 392.32&4316.66&2672.45& 2692.35\\
$19.56$ &25.10&190.66& 117.63& 729.56& 5292.25& 2849.75& 2967.38\\
$20$ & 8.42&16.12& 12.80& 762.93&2693.84& 1731.38& 1744.18\\
$21.68$ & 11.68&43.06&27.76& 695.92& 4754.41& 2671.29& 2699.05\\
$23$ & 5.73&20.67&13.78&571.68&4676.78&2505.78& 2519.56\\
$24$ &12.93 &93.75& 62.13& 695.67&3597.12& 1967.06& 2029.19\\
$25$ &22.17&49.79&37.27&647.42 &2850.88&1841.14 & 1878.41\\
$40$ &18.69 &124.65&77.60& 722.16&8750.53& 5048.50& 5126.10\\
$60$ &9.06&43.81&27.97& 571.36&  4907.58&3134.42 & 3162.39 \\
\enddata
\caption{Maximum, minimum, and mean detection ranges in kpc for each progenitor mass for the proposed DECIGO detector (where the mean is taken over the solid angle) for GWs from matter, neutrinos, and both strains combined. We define the detection range as the distance in kiloparsecs out to which an optimally-oriented source will have an SNR of eight. For the DECIGO detector more than the aLIGO, ET, and CE detectors, the neutrino memory dominates the detection range. DECIGO will easily be able to detect GWs from the neutrino emission component from over a thousand kiloparsecs (out to the local group of galaxies),  but only out to a few kiloparsecs for the GWs from matter motion. Especially for the non-exploding BH-forming models 12.25 and 14, the detection range is almost entirely due to the neutrino memory component, with the matter strain contributing negligibly. } 
\end{deluxetable*}

\section{Mean Detection Rate}
\label{rate}
While previous studies have estimated the rate of CCSN events in the Milky Way to be anywhere ranging from 1.63 \citep{Rozwadowska_2021} to 3.2 \citep{Adams_2013} events per century, as demonstrated in the preceding section, depending on the mass of the progenitor and the distance at which the event occurs, there is no guarantee of a GW detection even when a CCSN occurs in the galaxy. In this section, we use the above detection ranges in Table \ref{d_opt_table} to develop a galactic density- and stellar population-weighted detection rate of GWs from CCSNe using the Advanced LIGO detector. 

Our procedure for calculating the rate is as follows. We begin with the galactic density model developed by \citet{McMillan2016}. For each progenitor, we integrate the density distribution in a sphere centered at the Earth out to the corresponding aLIGO detection range in Table \ref{d_opt_table}. We normalize each integral to the total mass of the galaxy to obtain the \textit{fraction} of the galaxy that aLIGO can see for each progenitor. We then combine each fraction by weighting them with the relative distribution of stellar masses modeled by \citet{salpeter1955}. Finally, we use this galactic density- and stellar population-weighted fraction to scale the galactic CCSN rate to a rate of CCSN GW events detectable by aLIGO. We repeat this procedure for the ``Matter Avg" detection ranges in the fourth column of Table \ref{d_opt_table}, as well as for the combined matter and neutrino memory detection ranges. 

While there are several galactic density models in the literature \citep{lian2022, Juric2008}, in this paper, we use the galactic density model given by \citet{McMillan2016} which conducts a Bayesian analysis using survey-fitted priors from \cite{Juric2008} and is constrained with velocity curves and galactic maser source data \citep{McMillan2016}. The model incorporates the galactic bulge, dark matter halo, think and thick stellar disks, and molecular gas disks. For our purposes, we are only concerned with the bulge and thin and thick disk components
    \begin{equation}
        \rho(R,z) = \rho_b(R, z) + \rho_{d,thin}(R,z) + f\rho_{d,thick}(R,z) \,\, ,
    \end{equation}
    where $\rho_b$ is the density of the bulge and $\rho_d$ is the density of the thick and thin stellar disks. The density of the bulge is given by
    \begin{equation}
    \begin{aligned}
        \rho_b(R, z) &= \frac{\rho_{0, b}}{(1 + r/r_0)^{\alpha}}e^{-(\frac{r}{r_{cut}})^2} \\
        r &= \sqrt{R^2 + (z / q)^2} \,\, ,
    \end{aligned}
    \end{equation}
    where the galactic survey-fitted parameters are $\rho_{0, b} = 9.84\times10^{10} \:\: M_{\odot}/kpc^3$, $\alpha = 1.8$, $r_0 = 0.075$ kpc, $r_{cut} = 2.1$ kpc, and $q=0.5$. The density of the disks is given by
    \begin{equation}
        \rho_d(R, z) = \frac{\Sigma_0}{2z_d}exp(-\frac{|z|}{z_d}-\frac{R}{R_d}) \,\, ,
    \end{equation}
    where the galactic survey-fitted parameters are $\Sigma_{0, thin} = 8.96\times10^{8} \:\: M_{\odot}/kpc^2$, $R_{d, thin} = 2.5$ kpc, $z_{d,thin} = 0.3$ kpc, $\Sigma_{0, thick} = 1.83\times10^{8} \:\: M_{\odot}/kpc^2$, $R_{d, thick} = 3.02$ kpc, and $z_{d, thick} = 0.9$ kpc. 

We assume that within this density distribution which we call $\rho(R, z)$, the Earth is located at galactic coordinates $(R_{\odot}, z_{\odot}) = (8.2, 0.027)$ kiloparsecs \citep{lian2022}. We then integrate this galactic density distribution over a sphere of radius $d$ corresponding to the ``Matter Avg" detection ranges in Table \ref{d_opt_table}. The integral, which we call $X(d)$, is performed in Earth-centered spherical coordinates over the galactocenteric cylindrical density profile $\rho(R, z)$ and is defined as follows:

\begin{equation}
\label{density_int}
X(d) = \int_{0}^d\int_0^{\pi}\int_0^{2\pi}\rho(R, z)r^2\sin\theta d\phi d\theta dr \,\, .
\end{equation}

\begin{figure}[ht]
\centering
\includegraphics[width=0.5\linewidth]
{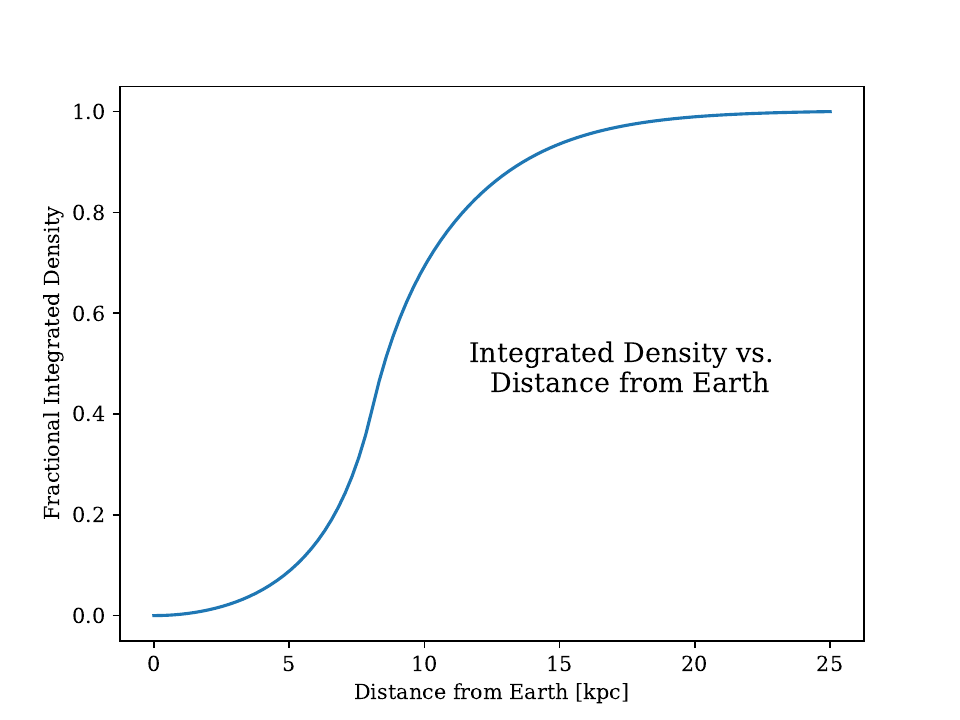}
\caption{Integrated galactic density as a function of distance from Earth using the \citet{McMillan2016} galactic density model (from eq. \eqref{density_int}), normalized to the total integrated density of the galaxy. Assuming the Earth is located around 8.2 kpc from the galactic center, there is a sharp increase in integrated density starting around 8 kpc when nearing the galactic bulge and the density profile starts to plateau around 13 kpc once the integral passes to the other side of the galactic bulge. \label{McMillian_int_density}}
\end{figure}

The details of this calculation are explained in the Appendix. We plot the results of $X$ as a function of distance from Earth using the \citet{McMillan2016} density profile normalized to the total integrated density of the galaxy in Figure \ref{McMillian_int_density}. The slope of the curve begins to steepen at around eight kiloparsecs due to the galactic bulge, and plateaus to one as the integral reaches the opposite edge of the galaxy.
    
\begin{figure}[ht]
\centering
\includegraphics[width=0.49\linewidth]{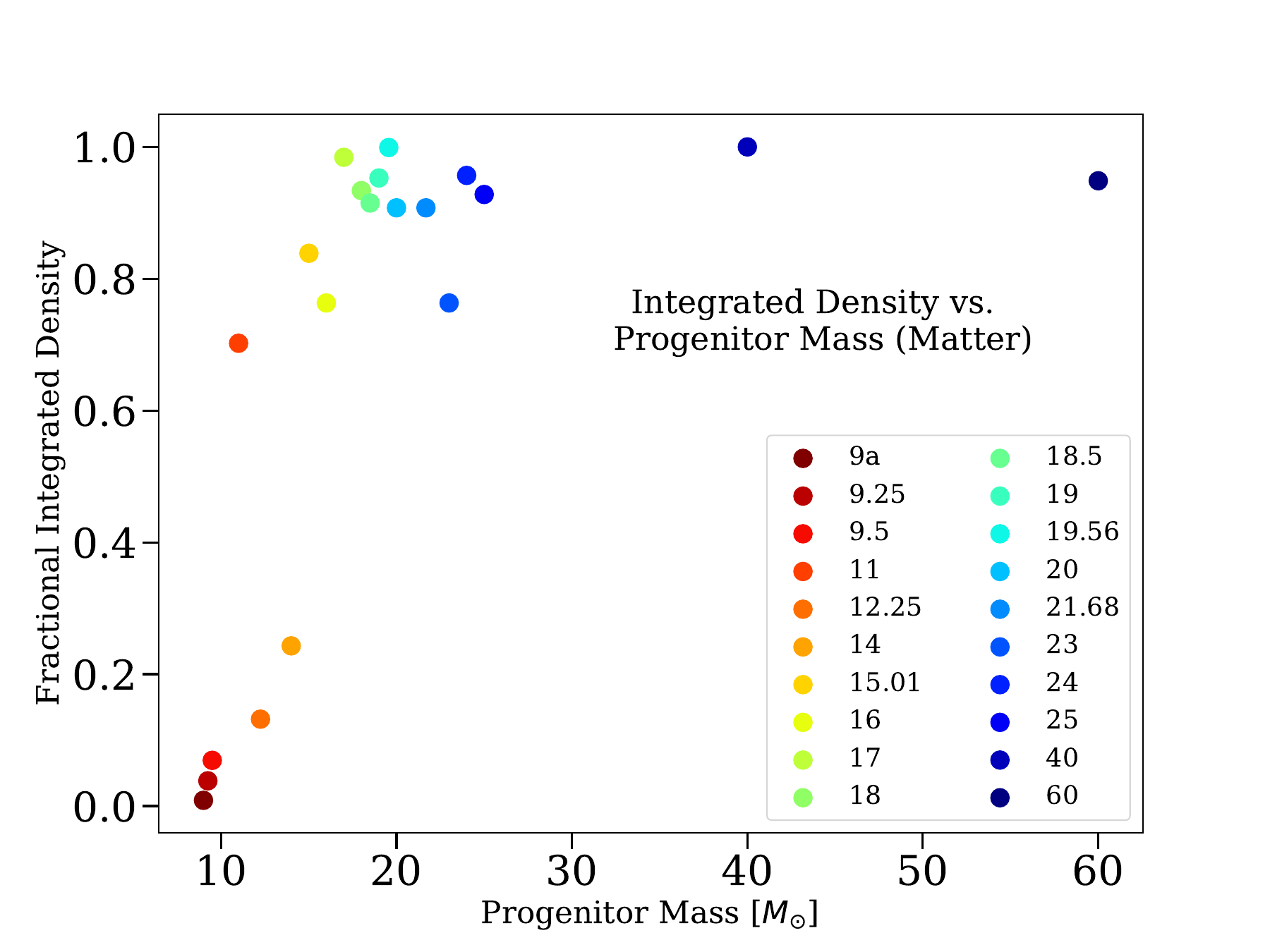}
\includegraphics[width=0.49\linewidth]{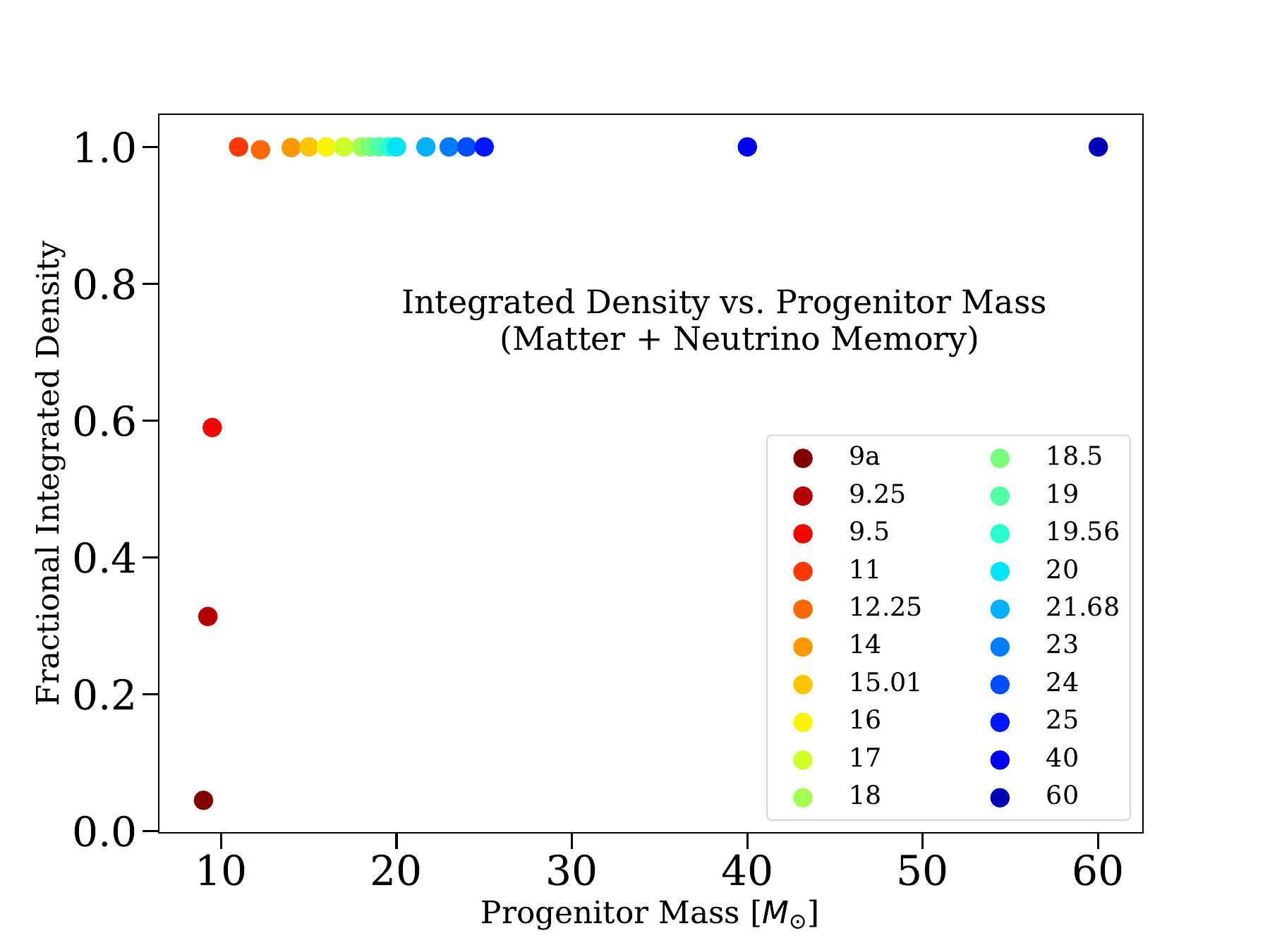}
\caption{(Left) Integrated galactic density as a function of progenitor mass using the ``Matter Avg" detection ranges. (Right) Integrated galactic density as a function of progenitor mass using the ``Combined Avg" detection ranges. Both plots are normalized to the total integrated density of the galaxy. The fraction of the galaxy in which aLIGO can detect the CCSN event roughly increases with progenitor mass. The left plot saturates to one at higher masses ($> 30 \: M_{\odot}$) whereas the right plot saturates to one at around 12.25 $M_{\odot}$. \label{int_densities}}
\end{figure}

On the left side of Figure \ref{int_densities}, we plot the normalized integrated densities $X$ as a function of progenitor mass for each progenitor model in our suite of simulations using the ``Matter Avg" detection ranges. While the results are not monotonic with progenitor mass, the integrated density general increases with progenitor mass. 

We then combine these fractional integrated densities into a single fraction by weighing them by the relative distribution of massive stars in the galaxy given by the Salpeter mass function \citep{salpeter1955}

\begin{equation}
    \begin{aligned}
    \label{salpeter}
    N(M_1, M_2) &= \xi_0\int_{M_1}^{M_2}M^{-2.35}dM \\
    &= \frac{\xi_0}{1.35}[M_1^{-1.35}-M_2^{-1.35}]\, , 
    \end{aligned}
    \end{equation}
which approximates the number of stars contained within the mass interval $[M_1, M_2]$. To make our calculation continuous, we use the best-fit curve $0.236 \log(M - 8.74) - 0.262$ to obtain an approximate analytic relationship between progenitor mass and fractional integrated density. Since this curve reaches one at around $M=32 \: M_{\odot}$, we carry out the final integral in two parts, using the best-fit curve until $M=32 \: M_{\odot}$ and then assuming all higher-mass progenitors identically correspond to an integrated density of one. All together, the total galactic CCSN rate is modified by a factor of:
\begin{equation}
    \begin{aligned}
    \label{rate_integrated1}
    \textrm{fractional rate} &= \int_{9M_{\odot}}^{\infty}\frac{X(M)}{X_{tot}}\frac{N(M)}{N_{tot}}dM \\
    &= \frac{1}{N_{tot}}\left[\int_{9M_{\odot}}^{32M_{\odot}}M^{-2.35}(0.236 \log(M - 8.72) - 0.262) dM + \int_{32M_{\odot}}^{\infty}M^{-2.35} (1) dM\right]\\
    &= 0.237 \,\, ,
    \end{aligned}
    \end{equation}
where $N_{tot} = \int_{9M_{\odot}}^{\infty}M^{-2.35}dM$ and $X_{tot}$ is the total integrated density normalization factor. For the matter strain seen by the Advanced LIGO detector and assuming the galactic CCSN rate from \citet{Adams_2013}, the expected galactic CCSN GW rate is $\frac{3.2}{\textrm{century}} * 0.237 = \frac{0.758}{\textrm{century}}$, which corresponds to a detection once every 132 years. For the galactic CCSN rate from \citet{Rozwadowska_2021}, this GW detection rate is $\frac{1.63}{\textrm{century}} * 0.237 = \frac{0.386}{\textrm{century}}$, or a detection once every 259 years.

We now repeat this calculation using the \textit{combined} matter and neutrino memory detection ranges in the last column of Table \ref{d_opt_table}. We plot the corresponding fractional integrated density versus progenitor mass on the right side of Figure \ref{int_densities}. In this plot, the integrated density saturates to one after around 12.25 $M_{\odot}$. Therefore, for this combined matter and neutrino memory calculation, we once again calculate eq. \eqref{rate_integrated1} in two parts, using the best-fit curve $0.249\log(M-8.93)+0.684$ until 12.25 $M_{\odot}$ and then assuming all higher-mass progenitors identically correspond to an integrated density of one. The final fraction modifying the CCSN GW rate using the combined matter and neutrino memory detection ranges is therefore:
    \begin{equation}
    \begin{aligned}
    \label{rate_integrated}
    \textrm{fractional rate} &= \frac{3.2}{\textrm{century}}\int_{9M_{\odot}}^{\infty}\frac{K(M)}{K_{tot}}\frac{N(M)}{N_{tot}}dM \\
    &= \frac{1}{N_{tot}}\left[\int_{9M_{\odot}}^{12.25M_{\odot}}M^{-2.35}(0.249\log(M-8.93) + 0.689)dM + \int_{12.25M_{\odot}}^{\infty}M^{-2.35} (1) dM\right] \\
    &= 0.921 \,\, .
    \end{aligned}
    \end{equation}
For the \citet{Adams_2013} estimate, this calculation corresponds to a rate of $\frac{3.2}{\textrm{century}} * 0.921 = \frac{2.95}{\textrm{century}}$, or a detection around once every 34 years. For the \citet{Rozwadowska_2021} galactic CCSN rate, this corresponds to a rate of $\frac{1.63}{\textrm{century}} * 0.921 = \frac{1.47}{\textrm{century}}$, or a detection once every 68 years. 

We emphasize that both detection rates calculated in this section rely on the SNR threshold $\rho = 8$. If we use a higher threshold or demand an SNR value sufficiently high enough to resolve features of the strain instead of merely detecting it, both rates calculated in this section will decrease significantly.

For next-generation detectors (ET, CE, DECIGO), the detection ranges for the combined matter and neutrino memory strain for every progenitor model extend to or beyond the diameter of the Milky Way, which is around 30 kiloparsecs. As noted in the preceeding section, some detection ranges may even reach into our neighboring galaxy Andromeda, which is 765 kpc away \citep{Riess_2012, srivastava2019}. Therefore, for next-generation detectors whose detection ranges cover the entire Milky Way for all progenitor models, while the strength of the observed signal will still depend on progenitor mass and distance, the observable CCSN GW rate in the Milky Way will be the same as the standard galactic CCSN rate, as the fraction in \eqref{rate_integrated} is one. If we consider the possibility of detections of CCSN events in Andromeda and other nearby galaxies, the rate will be even greater. 

\section{Conclusions}
\label{conclusions}
In this paper, we have examined the energy, signal-to-noise ratio, detection range, and angular anisotropy of the total matter and the neutrino memory gravitational-wave signatures of 21 3D CCSN progenitor models carried to late times. This model set captures just about the entire gravitational-wave signal of initially non-rotating supernova explosions. When examining the angular anisotropy of the CCSN GW strain, we find that the total inferred energy radiated in GWs by the supernova is a highly angle-dependent quantity, and that the range of possible inferrred energy values increases with progenitor mass. Even though the true total radiated energy is not degenerate with progenitor mass, there is significant overlap between various progenitors' inferred energy values depending on the viewing angle. Additionally, we find that, due to significant viewing-angle variations, signal-to-noise ratios and detection ranges are also viewing-angle-dependent quantities. Again, heavier progenitors have a wider spread in possible signal-to-noise ratios and detection ranges due to the greater anisotropy of their emissions. These results indicate that the angle of observation of the event can affect whether a CCSN GW signal is detectable.

When using a Butterworth filter to separate out the low-frequency matter memory component of the signal, we find that DECIGO, among all the planned new GW detectors, is the most capable of detecting matter memory, with matter memory being undetectable with Advanced LIGO and barely detectable with ET and CE for the higher-mass progenitors. Moreover, we find that the polarization angle $\theta = \arctan(h_+ / h_{\times})$ can serve as a powerful method of detecting both matter and neutrino memory, since the high-frequency component of the polarization angle oscillates so rapidly that it ``averages" itself out, leaving only the low-frequency memory components. 
When comparing SNR values, we find that the neutrino memory signals across all detectors are significantly larger than both the high-frequency matter and the matter memory components, indicating that neutrino memory is the most detectable component of the CCSN GW signal. When comparing detection ranges across all detectors, we find that while Advanced LIGO can detect combined matter and neutrino memory CCSN GW signals throughout the entire galaxy for higher-mass progenitors, all future detectors (ET, CE, and DECIGO) will be able to see the entire galaxy for all progenitors and might even be able to detect signals thousands of kiloparsecs away in the local group of galaxies. These large detection ranges are primarily due to the neutrino memory component of the signal, with the matter component contributing sub-dominantly to the detection ranges. This once again demonstrates that the neutrino memory should be the most detectable component of a CCSN GW signal.

Finally, we develop a galactic density- and stellar mass-weighted formalism to calculate the rate at which we can expect to detect CCSN GW signals with Advanced LIGO. With only the matter component of the signal, we find that the aLIGO detection rate is $\sim$24\% of the total galactic supernova rate. If we include the neutrino memory contribution of the signal, the aLIGO detection rate increases to $\sim$92\% of the total galactic rate. Since all future detectors are expected to detect events throughout the entire galaxy and beyond, we expect the detection rate for those detectors to track or even exceed the overall galactic supernova rate. 

It is important to emphasize that the detection range and rate results in this paper rely on the standard GW detectability threshold of $\rho = 8$. To characterize and resolve features of the CCSN GW signal, a much larger SNR threshold is required, thereby shortening the detection ranges and lengthening the detection timescales. However, the possibility of combining current and future detectors in a network serves as a way to boost the SNRs of the signals and can provide more optimistic detection prospects. 

While we have highlighted the significant anticipated angular anisotropies, the low-frequency memory, the polarization of the latter, and the detection prospects of CCSN GW signals overall, there still remains much work to be done. This includes ascertaining the dependence of the above quantities on, for example, the nuclear equation-of-state and the role of initial rotation, among other inputs to supernova modeling. In the future, we will also seek to examine methods of extracting various physical characteristics of the dynamics of the CCSN mechanism and phenomenon, using GW signals with sufficiently high signal-to-noise ratios. 

\section*{Data Availability}

The data presented in this paper is available upon reasonable request to the authors.  A subset of these data, including representative strains and quadrupole tensors as a function of time, can be found at \url{http://www.astro.princeton.edu/~burrows/gw.3d.2024.update/}.

\section*{Acknowledgments}
We acknowledge our ongoing and fruitful collaborations with Tianshu Wang and Christopher White. We also acknowledge support from the U.~S. Department of Energy Office of Science and the Office of Advanced Scientific Computing Research via the Scientific Discovery through Advanced Computing (SciDAC4) program and Grant DE-SC0018297 (subaward 00009650), support from the U.~S.\ National Science Foundation (NSF) under Grants AST-1714267 and PHY-1804048 (the latter via the Max-Planck/Princeton Center (MPPC) for Plasma Physics), and support from NASA under award JWST-GO-01947.011-A.  A generous award of computer time was provided by the INCITE program, using resources of the Argonne Leadership Computing Facility, a DOE Office of Science User Facility supported under Contract DE-AC02-06CH11357. We also acknowledge access to the Frontera cluster (under awards AST20020 and AST21003); this research is part of the Frontera computing project at the Texas Advanced Computing Center \citep{stanzione2020} under NSF award OAC-1818253. Finally, the
authors acknowledge computational resources provided by the
high-performance computer center at Princeton University, which is jointly supported by the Princeton Institute for Computational Science and Engineering (PICSciE) and the Princeton University Office of Information Technology, and our continuing allocation at the National Energy Research Scientific Computing Center (NERSC), which is supported by the Office of Science of the U.~S.\ Department of Energy under contract
DE-AC03-76SF00098.

\bibliography{JPrefs_new}

\begin{thebibliography}{}
\expandafter\ifx\csname natexlab\endcsname\relax\def\natexlab#1{#1}\fi
\providecommand{\url}[1]{\href{#1}{#1}}
\providecommand{\dodoi}[1]{doi:~\href{http://doi.org/#1}{\nolinkurl{#1}}}
\providecommand{\doeprint}[1]{\href{http://ascl.net/#1}{\nolinkurl{http://ascl.net/#1}}}
\providecommand{\doarXiv}[1]{\href{https://arxiv.org/abs/#1}{\nolinkurl{https://arxiv.org/abs/#1}}}

\bibitem[{{Aartsen} {et~al.}(2014){Aartsen}, {Ackermann}, {Adams}, {Aguilar}, {Ahlers}, {Ahrens}, {Altmann}, {Anderson}, {Arguelles}, {Arlen}, {Auffenberg}, {Bai}, {Barwick}, {Baum}, {Beatty}, {Becker Tjus}, {Becker}, {BenZvi}, {Berghaus}, {Berley}, {Bernardini}, {Bernhard}, {Besson}, {Binder}, {Bindig}, {Bissok}, {Blaufuss}, {Blumenthal}, {Boersma}, {Bohm}, {Bos}, {Bose}, {B{\"o}ser}, {Botner}, {Brayeur}, {Bretz}, {Brown}, {Casey}, {Casier}, {Chirkin}, {Christov}, {Christy}, {Clark}, {Classen}, {Clevermann}, {Coenders}, {Cowen}, {Cruz Silva}, {Danninger}, {Daughhetee}, {Davis}, {Day}, {de Andr{\'e}}, {De Clercq}, {De Ridder}, {Desiati}, {de Vries}, {de With}, {DeYoung}, {D{\'\i}az-V{\'e}lez}, {Dunkman}, {Eagan}, {Eberhardt}, {Eichmann}, {Eisch}, {Euler}, {Evenson}, {Fadiran}, {Fazely}, {Fedynitch}, {Feintzeig}, {Felde}, {Feusels}, {Filimonov}, {Finley}, {Fischer-Wasels}, {Flis}, {Franckowiak}, {Frantzen}, {Fuchs}, {Gaisser}, {Gallagher}, {Gerhardt}, {Gier}, {Gladstone}, {Gl{\"u}senkamp}, {Goldschmidt},
  {Golup}, {Gonzalez}, {Goodman}, {G{\'o}ra}, {Grandmont}, {Grant}, {Gretskov}, {Groh}, {Gro{\ss}}, {Ha}, {Haack}, {Haj Ismail}, {Hallen}, {Hallgren}, {Halzen}, {Hanson}, {Hebecker}, {Heereman}, {Heinen}, {Helbing}, {Hellauer}, {Hellwig}, {Hickford}, {Hill}, {Hoffman}, {Hoffmann}, {Homeier}, {Hoshina}, {Huang}, {Huelsnitz}, {Hulth}, {Hultqvist}, {Hussain}, {Ishihara}, {Jacobi}, {Jacobsen}, {Jagielski}, {Japaridze}, {Jero}, {Jlelati}, {Jurkovic}, {Kaminsky}, {Kappes}, {Karg}, {Karle}, {Kauer}, {Kelley}, {Kheirandish}, {Kiryluk}, {Kl{\"a}s}, {Klein}, {K{\"o}hne}, {Kohnen}, {Kolanoski}, {Koob}, {K{\"o}pke}, {Kopper}, {Kopper}, {Koskinen}, {Kowalski}, {Kriesten}, {Krings}, {Kroll}, {Kroll}, {Kunnen}, {Kurahashi}, {Kuwabara}, {Labare}, {Larsen}, {Larson}, {Lesiak-Bzdak}, {Leuermann}, {Leute}, {L{\"u}nemann}, {Mac{\'\i}as}, {Madsen}, {Maggi}, {Maruyama}, {Mase}, {Matis}, {McNally}, {Meagher}, {Medici}, {Meli}, {Meures}, {Miarecki}, {Middell}, {Middlemas}, {Milke}, {Miller}, {Mohrmann}, {Montaruli}, {Morse},
  {Nahnhauer}, {Naumann}, {Niederhausen}, {Nowicki}, {Nygren}, {Obertacke}, {Odrowski}, {Olivas}, {Omairat}, {O'Murchadha}, {Palczewski}, {Paul}, {Penek}, {Pepper}, {P{\'e}rez de los Heros}, {Pfendner}, {Pieloth}, {Pinat}, {Posselt}, {Price}, {Przybylski}, {P{\"u}tz}, {Quinnan}, {R{\"a}del}, {Rameez}, {Rawlins}, {Redl}, {Rees}, {Reimann}, {Resconi}, {Rhode}, {Richman}, {Riedel}, {Robertson}, {Rodrigues}, {Rongen}, {Rott}, {Ruhe}, {Ruzybayev}, {Ryckbosch}, {Saba}, {Sander}, {Sandroos}, {Santander}, {Sarkar}, {Schatto}, {Scheriau}, {Schmidt}, {Schmitz}, {Schoenen}, {Sch{\"o}neberg}, {Sch{\"o}nwald}, {Schukraft}, {Schulte}, {Schulz}, {Seckel}, {Sestayo}, {Seunarine}, {Shanidze}, {Sheremata}, {Smith}, {Soldin}, {Spiczak}, {Spiering}, {Stamatikos}, {Stanev}, {Stanisha}, {Stasik}, {Stezelberger}, {Stokstad}, {St{\"o}{\ss}l}, {Strahler}, {Str{\"o}m}, {Strotjohann}, {Sullivan}, {Taavola}, {Taboada}, {Tamburro}, {Tepe}, {Ter-Antonyan}, {Terliuk}, {Te{\v{s}}i{\'c}}, {Tilav}, {Toale}, {Tobin}, {Tosi}, {Tselengidou},
  {Unger}, {Usner}, {Vallecorsa}, {van Eijndhoven}, {Vandenbroucke}, {van Santen}, {Vehring}, {Voge}, {Vraeghe}, {Walck}, {Wallraff}, {Weaver}, {Wellons}, {Wendt}, {Westerhoff}, {Whelan}, {Whitehorn}, {Wichary}, {Wiebe}, {Wiebusch}, {Williams}, {Wissing}, {Wolf}, {Wood}, {Woschnagg}, {Xu}, {Xu}, {Yanez}, {Yodh}, {Yoshida}, {Zarzhitsky}, {Ziemann}, {Zierke}, {Zoll}, {Aasi}, {Abbott}, {Abbott}, {Abbott}, {Abernathy}, {Acernese}, {Ackley}, {Adams}, {Adams}, {Addesso}, {Adhikari}, {Affeldt}, {Agathos}, {Aggarwal}, {Aguiar}, {Ajith}, {Alemic}, {Allen}, {Allocca}, {Amariutei}, {Andersen}, {Anderson}, {Anderson}, {Anderson}, {Arai}, {Araya}, {Arceneaux}, {Areeda}, {Ast}, {Aston}, {Astone}, {Aufmuth}, {Augustus}, {Aulbert}, {Aylott}, {Babak}, {Baker}, {Ballardin}, {Ballmer}, {Barayoga}, {Barbet}, {Barish}, {Barker}, {Barone}, {Barr}, {Barsotti}, {Barsuglia}, {Barton}, {Bartos}, {Bassiri}, {Basti}, {Batch}, {Bauchrowitz}, {Bauer}, {Baune}, {Bavigadda}, {Behnke}, {Bejger}, {Beker}, {Belczynski}, {Bell}, {Bell},
  {Bergmann}, {Bersanetti}, {Bertolini}, {Betzwieser}, {Bilenko}, {Billingsley}, {Birch}, {Biscans}, {Bitossi}, {Biwer}, {Bizouard}, {Black}, {Blackburn}, {Blackburn}, {Blair}, {Bloemen}, {Bock}, {Bodiya}, {Boer}, {Bogaert}, {Bogan}, {Bojtos}, {Bond}, {Bondu}, {Bonelli}, {Bonnand}, {Bork}, {Born}, {Boschi}, {Bose}, {Bosi}, {Bradaschia}, {Brady}, {Braginsky}, {Branchesi}, {Brau}, {Briant}, {Bridges}, {Brillet}, {Brinkmann}, {Brisson}, {Brooks}, {Brown}, {Brown}, {Br{\"u}ckner}, {Buchman}, {Buikema}, {Bulik}, {Bulten}, {Buonanno}, {Burman}, {Buskulic}, {Buy}, {Cadonati}, {Cagnoli}, {Calder{\'o}n Bustillo}, {Calloni}, {Camp}, {Campsie}, {Cannon}, {Canuel}, {Cao}, {Capano}, {Carbognani}, {Carbone}, {Caride}, {Castaldi}, {Caudill}, {Cavagli{\`a}}, {Cavalier}, {Cavalieri}, {Celerier}, {Cella}, {Cepeda}, {Cesarini}, {Chakraborty}, {Chalermsongsak}, {Chamberlin}, {Chao}, {Charlton}, {Chassande-Mottin}, {Chen}, {Chen}, {Chincarini}, {Chiummo}, {Cho}, {Cho}, {Chow}, {Christensen}, {Chu}, {Chua}, {Chung}, {Ciani},
  {Clara}, {Clark}, {Clark}, {Clayton}, {Cleva}, {Coccia}, {Cohadon}, {Colla}, {Collette}, {Colombini}, {Cominsky}, {Constancio}, {Conte}, {Cook}, {Corbitt}, {Cornish}, {Corsi}, {Costa}, {Coughlin}, {Coulon}, {Countryman}, {Couvares}, {Coward}, {Cowart}, {Coyne}, {Coyne}, {Craig}, {Creighton}, {Croce}, {Crowder}, {Cumming}, {Cunningham}, {Cuoco}, {Cutler}, {Dahl}, {Dal Canton}, {Damjanic}, {Danilishin}, {D'Antonio}, {Danzmann}, {Dattilo}, {Daveloza}, {Davier}, {Davies}, {Daw}, {Day}, {Dayanga}, {DeBra}, {Debreczeni}, {Degallaix}, {Del{\'e}glise}, {Del Pozzo}, {Del Pozzo}, {Denker}, {Dent}, {Dereli}, {Dergachev}, {De Rosa}, {DeRosa}, {DeSalvo}, {Dhurandhar}, {D{\'\i}az}, {Dickson}, {Di Fiore}, {Di Lieto}, {Di Palma}, {Di Virgilio}, {Dolique}, {Dominguez}, {Donovan}, {Dooley}, {Doravari}, {Douglas}, {Downes}, {Drago}, {Drever}, {Driggers}, {Du}, {Ducrot}, {Dwyer}, {Eberle}, {Edo}, {Edwards}, {Effler}, {Eggenstein}, {Ehrens}, {Eichholz}, {Eikenberry}, {Endr{\H{o}}czi}, {Essick}, {Etzel}, {Evans}, {Evans},
  {Factourovich}, {Fafone}, {Fairhurst}, {Fan}, {Fang}, {Farinon}, {Farr}, {Farr}, {Favata}, {Fazi}, {Fehrmann}, {Fejer}, {Feldbaum}, {Feroz}, {Ferrante}, {Ferreira}, {Ferrini}, {Fidecaro}, {Finn}, {Fiori}, {Fisher}, {Flaminio}, {Fournier}, {Franco}, {Frasca}, {Frasconi}, {Frede}, {Frei}, {Freise}, {Frey}, {Fricke}, {Fritschel}, {Frolov}, {Fulda}, {Fyffe}, {Gair}, {Gammaitoni}, {Gaonkar}, {Garufi}, {Gehrels}, {Gemme}, {Gendre}, {Genin}, {Gennai}, {Ghosh}, {Giaime}, {Giardina}, {Giazotto}, {Gleason}, {Goetz}, {Goetz}, {Gondan}, {Gonz{\'a}lez}, {Gordon}, {Gorodetsky}, {Gossan}, {Go{\ss}ler}, {Gouaty}, {Gr{\"a}f}, {Graff}, {Granata}, {Grant}, {Gras}, {Gray}, {Greenhalgh}, {Gretarsson}, {Groot}, {Grote}, {Grover}, {Grunewald}, {Guidi}, {Guido}, {Gushwa}, {Gustafson}, {Gustafson}, {Ha}, {Hall}, {Hamilton}, {Hammer}, {Hammond}, {Hanke}, {Hanks}, {Hanna}, {Hannam}, {Hanson}, {Harms}, {Harry}, {Harry}, {Harstad}, {Hart}, {Hartman}, {Haster}, {Haughian}, {Heidmann}, {Heintze}, {Heitmann}, {Hello}, {Hemming}, {Hendry},
  {Heng}, {Heptonstall}, {Heurs}, {Hewitson}, {Hild}, {Hoak}, {Hodge}, {Hofman}, {Holt}, {Hopkins}, {Horrom}, {Hoske}, {Hosken}, {Hough}, {Howell}, {Hu}, {Huerta}, {Hughey}, {Husa}, {Huttner}, {Huynh}, {Huynh-Dinh}, {Idrisy}, {Ingram}, {Inta}, {Islas}, {Isogai}, {Ivanov}, {Iyer}, {Izumi}, {Jacobson}, {Jang}, {Jaranowski}, {Ji}, {Jim{\'e}nez-Forteza}, {Johnson}, {Jones}, {Jones}, {Jonker}, {Ju}, {K}, {Kalmus}, {Kalogera}, {Kandhasamy}, {Kang}, {Kanner}, {Karlen}, {Kasprzack}, {Katsavounidis}, {Katzman}, {Kaufer}, {Kaufer}, {Kaur}, {Kawabe}, {Kawazoe}, {K{\'e}f{\'e}lian}, {Keiser}, {Keitel}, {Kelley}, {Kells}, {Keppel}, {Khalaidovski}, {Khalili}, {Khazanov}, {Kim}, {Kim}, {Kim}, {Kim}, {Kim}, {Kim}, {King}, {King}, {Kinzel}, {Kissel}, {Klimenko}, {Kline}, {Koehlenbeck}, {Kokeyama}, {Kondrashov}, {Koranda}, {Korth}, {Kowalska}, {Kozak}, {Kringel}, {Kr{\'o}lak}, {Kuehn}, {Kumar}, {Kumar}, {Kumar}, {Kumar}, {Kuo}, {Kutynia}, {Lam}, {Landry}, {Lantz}, {Larson}, {Lasky}, {Lazzarini}, {Lazzaro}, {Leaci}, {Leavey},
  {Lebigot}, {Lee}, {Lee}, {Lee}, {Lee}, {Lee}, {Leonardi}, {Leong}, {Le Roux}, {Leroy}, {Letendre}, {Levin}, {Levine}, {Lewis}, {Li}, {Libbrecht}, {Libson}, {Lin}, {Littenberg}, {Lockerbie}, {Lockett}, {Lodhia}, {Loew}, {Logue}, {Lombardi}, {Lopez}, {Lorenzini}, {Loriette}, {Lormand}, {Losurdo}, {Lough}, {Lubinski}, {L{\"u}ck}, {Lundgren}, {Ma}, {Macdonald}, {MacDonald}, {Machenschalk}, {MacInnis}, {Macleod}, {Maga{\~n}a-Sandoval}, {Magee}, {Mageswaran}, {Maglione}, {Mailand}, {Majorana}, {Maksimovic}, {Malvezzi}, {Man}, {Manca}, {Mandel}, {Mandic}, {Mangano}, {Mangini}, {Mansell}, {Mantovani}, {Marchesoni}, {Marion}, {M{\'a}rka}, {M{\'a}rka}, {Markosyan}, {Maros}, {Marque}, {Martelli}, {Martin}, {Martin}, {Martinelli}, {Martynov}, {Marx}, {Mason}, {Masserot}, {Massinger}, {Matichard}, {Matone}, {Mavalvala}, {May}, {Mazumder}, {Mazzolo}, {McCarthy}, {McClelland}, {McGuire}, {McIntyre}, {McIver}, {McLin}, {Meacher}, {Meadors}, {Mehmet}, {Meidam}, {Meinders}, {Melatos}, {Mendell}, {Mercer}, {Meshkov},
  {Messenger}, {Meyer}, {Meyers}, {Mezzani}, {Miao}, {Michel}, {Mikhailov}, {Milano}, {Miller}, {Minenkov}, {Mingarelli}, {Mishra}, {Mitra}, {Mitrofanov}, {Mitselmakher}, {Mittleman}, {Moe}, {Moggi}, {Mohan}, {Mohapatra}, {Moraru}, {Moreno}, {Morgado}, {Morriss}, {Mossavi}, {Mours}, {Mow-Lowry}, {Mueller}, {Mueller}, {Mukherjee}, {Mullavey}, {Munch}, {Murphy}, {Murray}, {Mytidis}, {Nagy}, {Nardecchia}, {Naticchioni}, {Nayak}, {Necula}, {Nelemans}, {Neri}, {Neri}, {Newton}, {Nguyen}, {Nielsen}, {Nissanke}, {Nitz}, {Nocera}, {Nolting}, {Normandin}, {Nuttall}, {Ochsner}, {O'Dell}, {Oelker}, {Oh}, {Oh}, {Ohme}, {Omar}, {Oppermann}, {Oram}, {O'Reilly}, {Ortega}, {O'Shaughnessy}, {Osthelder}, {Ottaway}, {Ottens}, {Overmier}, {Owen}, {Padilla}, {Pai}, {Palashov}, {Palomba}, {Pan}, {Pan}, {Pankow}, {Paoletti}, {Papa}, {Paris}, {Pasqualetti}, {Passaquieti}, {Passuello}, {Pedraza}, {Pele}, {Penn}, {Perreca}, {Phelps}, {Pichot}, {Pickenpack}, {Piergiovanni}, {Pierro}, {Pinard}, {Pinto}, {Pitkin}, {Poeld}, {Poggiani},
  {Poteomkin}, {Powell}, {Prasad}, {Predoi}, {Premachandra}, {Prestegard}, {Price}, {Prijatelj}, {Privitera}, {Prodi}, {Prokhorov}, {Puncken}, {Punturo}, {Puppo}, {P{\"u}rrer}, {Qin}, {Quetschke}, {Quintero}, {Quitzow-James}, {Raab}, {Rabeling}, {R{\'a}cz}, {Radkins}, {Raffai}, {Raja}, {Rajalakshmi}, {Rakhmanov}, {Ramet}, {Ramirez}, {Rapagnani}, {Raymond}, {Razzano}, {Re}, {Recchia}, {Reed}, {Regimbau}, {Reid}, {Reitze}, {Reula}, {Rhoades}, {Ricci}, {Riesen}, {Riles}, {Robertson}, {Robinet}, {Rocchi}, {Roddy}, {Rolland}, {Rollins}, {Romano}, {Romanov}, {Romie}, {Rosi{\'n}ska}, {Rowan}, {R{\"u}diger}, {Ruggi}, {Ryan}, {Salemi}, {Sammut}, {Sandberg}, {Sanders}, {Sankar}, {Sannibale}, {Santiago-Prieto}, {Saracco}, {Sassolas}, {Sathyaprakash}, {Saulson}, {Savage}, {Scheuer}, {Schilling}, {Schilman}, {Schmidt}, {Schnabel}, {Schofield}, {Schreiber}, {Schuette}, {Schutz}, {Scott}, {Scott}, {Sellers}, {Sengupta}, {Sentenac}, {Sequino}, {Sergeev}, {Shaddock}, {Shah}, {Shahriar}, {Shaltev}, {Shao}, {Shapiro},
  {Shawhan}, {Shoemaker}, {Sidery}, {Siellez}, {Siemens}, {Sigg}, {Simakov}, {Singer}, {Singer}, {Singh}, {Sintes}, {Slagmolen}, {Slutsky}, {Smith}, {Smith}, {Smith}, {Smith-Lefebvre}, {Son}, {Sorazu}, {Souradeep}, {Staley}, {Stebbins}, {Steinke}, {Steinlechner}, {Steinlechner}, {Stephens}, {Steplewski}, {Stevenson}, {Stone}, {Stops}, {Strain}, {Straniero}, {Strigin}, {Sturani}, {Stuver}, {Summerscales}, {Susmithan}, {Sutton}, {Swinkels}, {Tacca}, {Talukder}, {Tanner}, {Tao}, {Tarabrin}, {Taylor}, {Tellez}, {Thirugnanasambandam}, {Thomas}, {Thomas}, {Thorne}, {Thorne}, {Thrane}, {Tiwari}, {Tokmakov}, {Tomlinson}, {Tonelli}, {Torres}, {Torrie}, {Travasso}, {Traylor}, {Tse}, {Tshilumba}, {Tuennermann}, {Ugolini}, {Unnikrishnan}, {Urban}, {Usman}, {Vahlbruch}, {Vajente}, {Valdes}, {Vallisneri}, {van Beuzekom}, {van den Brand}, {Van Den Broeck}, {van der Sluys}, {van Heijningen}, {van Veggel}, {Vass}, {Vas{\'u}th}, {Vaulin}, {Vecchio}, {Vedovato}, {Veitch}, {Veitch}, {Venkateswara}, {Verkindt}, {Vetrano},
  {Vicer{\'e}}, {Vincent-Finley}, {Vinet}, {Vitale}, {Vo}, {Vocca}, {Vorvick}, {Vousden}, {Vyachanin}, {Wade}, {Wade}, {Wade}, {Walker}, {Wallace}, {Walsh}, {Wang}, {Wang}, {Ward}, {Was}, {Weaver}, {Wei}, {Weinert}, {Weinstein}, {Weiss}, {Welborn}, {Wen}, {Wessels}, {West}, {Westphal}, {Wette}, {Whelan}, {White}, {Whiting}, {Wiesner}, {Wilkinson}, {Williams}, {Williams}, {Williams}, {Williams}, {Williamson}, {Willis}, {Willke}, {Wimmer}, {Winkler}, {Wipf}, {Wiseman}, {Wittel}, {Woan}, {Wolovick}, {Worden}, {Wu}, {Yablon}, {Yakushin}, {Yam}, {Yamamoto}, {Yancey}, {Yang}, {Yoshida}, {Yvert}, {Zadro{\.Z}ny}, {Zanolin}, {Zendri}, {Zhang}, {Zhang}, {Zhao}, {Zhu}, {Zhu}, {Zucker}, {Zuraw}, {Zweizig}, \& {IceCube Collaboration}}]{Aartsen2014}
{Aartsen}, M.~G., {Ackermann}, M., {Adams}, J., {et~al.} 2014, \prd, 90, 102002, \dodoi{10.1103/PhysRevD.90.102002}

\bibitem[{Aasi {et~al.}(2015)Aasi, Abbott, Abbott, Abbott, Abernathy, Ackley, Adams, Adams, Addesso, Adhikari, Adya, Affeldt, Aggarwal, Aguiar, Ain, Ajith, Alemic, Allen, Amariutei, Anderson, Anderson, Arai, Araya, Arceneaux, Areeda, Ashton, Ast, Aston, Aufmuth, Aulbert, Aylott, Babak, Baker, Ballmer, Barayoga, Barbet, Barclay, Barish, Barker, Barr, Barsotti, Bartlett, Barton, Bartos, Bassiri, Batch, Baune, Behnke, Bell, Bell, Benacquista, Bergman, Bergmann, Berry, Betzwieser, Bhagwat, Bhandare, Bilenko, Billingsley, Birch, Biscans, Biwer, Blackburn, Blackburn, Blair, Blair, Bock, Bodiya, Bojtos, Bond, Bork, Born, Bose, Brady, Braginsky, Brau, Bridges, Brinkmann, Brooks, Brown, Brown, Brown, Buchman, Buikema, Buonanno, Cadonati, Calderón~Bustillo, Camp, Cannon, Cao, Capano, Caride, Caudill, Cavaglià, Cepeda, Chakraborty, Chalermsongsak, Chamberlin, Chao, Charlton, Chen, Cho, Cho, Chow, Christensen, Chu, Chung, Ciani, Clara, Clark, Collette, Cominsky, Constancio, Cook, Corbitt, Cornish, Corsi, Costa,
  Coughlin, Countryman, Couvares, Coward, Cowart, Coyne, Coyne, Craig, Creighton, Creighton, Cripe, Crowder, Cumming, Cunningham, Cutler, Dahl, Dal~Canton, Damjanic, Danilishin, Danzmann, Dartez, Dave, Daveloza, Davies, Daw, DeBra, Del~Pozzo, Denker, Dent, Dergachev, DeRosa, DeSalvo, Dhurandhar, D´ıaz, Di~Palma, Dojcinoski, Dominguez, Donovan, Dooley, Doravari, Douglas, Downes, Driggers, Du, Dwyer, Eberle, Edo, Edwards, Edwards, Effler, Eggenstein, Ehrens, Eichholz, Eikenberry, Essick, Etzel, Evans, Evans, Factourovich, Fairhurst, Fan, Fang, Farr, Farr, Favata, Fays, Fehrmann, Fejer, Feldbaum, Ferreira, Fisher, Frei, Freise, Frey, Fricke, Fritschel, Frolov, Fuentes-Tapia, Fulda, Fyffe, Gair, Gaonkar, Gehrels, Gergely´, Giaime, Giardina, Gleason, Goetz, Goetz, Gondan, González, Gordon, Gorodetsky, Gossan, Goßler, Gräf, Graff, Grant, Gras, Gray, Greenhalgh, Gretarsson, Grote, Grunewald, Guido, Guo, Gushwa, Gustafson, Gustafson, Hacker, Hall, Hammond, Hanke, Hanks, Hanna, Hannam, Hanson, Hardwick, Harry,
  Harry, Hart, Hartman, Haster, Haughian, Hee, Heintze, Heinzel, Hendry, Heng, Heptonstall, Heurs, Hewitson, Hild, Hoak, Hodge, Hollitt, Holt, Hopkins, Hosken, Hough, Houston, Howell, Hu, Huerta, Hughey, Husa, Huttner, Huynh, Huynh-Dinh, Idrisy, Indik, Ingram, Inta, Islas, Isler, Isogai, Iyer, Izumi, Jacobson, Jang, Jawahar, Ji, Jiménez-Forteza, Johnson, Jones, Jones, Ju, Haris, Kalogera, Kandhasamy, Kang, Kanner, Katsavounidis, Katzman, Kaufer, Kaufer, Kaur, Kawabe, Kawazoe, Keiser, Keitel, Kelley, Kells, Keppel, Key, Khalaidovski, Khalili, Khazanov, Kim, Kim, Kim, Kim, Kim, King, King, Kinzel, Kissel, Klimenko, Kline, Koehlenbeck, Kokeyama, Kondrashov, Korobko, Korth, Kozak, Kringel, Krishnan, Krueger, Kuehn, Kumar, Kumar, Kuo, Landry, Lantz, Larson, Lasky, Lazzarini, Lazzaro, Le, Leaci, Leavey, Lebigot, Lee, Lee, Lee, Leong, Levin, Levine, Lewis, Li, Libbrecht, Libson, Lin, Littenberg, Lockerbie, Lockett, Logue, Lombardi, Lormand, Lough, Lubinski, Lück, Lundgren, Lynch, Ma, Macarthur, MacDonald,
  Machenschalk, MacInnis, Macleod, Magaña-Sandoval, Magee, Mageswaran, Maglione, Mailand, Mandel, Mandic, Mangano, Mansell, Márka, Márka, Markosyan, Maros, Martin, Martin, Martynov, Marx, Mason, Massinger, Matichard, Matone, Mavalvala, Mazumder, Mazzolo, McCarthy, McClelland, McCormick, McGuire, McIntyre, McIver, McLin, McWilliams, Meadors, Meinders, Melatos, Mendell, Mercer, Meshkov, Messenger, Meyers, Miao, Middleton, Mikhailov, Miller, Miller, Millhouse, Ming, Mirshekari, Mishra, Mitra, Mitrofanov, Mitselmakher, Mittleman, Moe, Mohanty, Mohapatra, Moore, Moraru, Moreno, Morriss, Mossavi, Mow-Lowry, Mueller, Mueller, Mukherjee, Mullavey, Munch, Murphy, Murray, Mytidis, Nash, Nayak, Necula, Nedkova, Newton, Nguyen, Nielsen, Nissanke, Nitz, Nolting, Normandin, Nuttall, Ochsner, O’Dell, Oelker, Ogin, Oh, Oh, Ohme, Oppermann, Oram, O’Reilly, Ortega, O’Shaughnessy, Osthelder, Ott, Ottaway, Ottens, Overmier, Owen, Padilla, Pai, Pai, Palashov, Pal-Singh, Pan, Pankow, Pannarale, Pant, Papa, Paris, Patrick,
  Pedraza, Pekowsky, Pele, Penn, Perreca, Phelps, Pierro, Pinto, Pitkin, Poeld, Post, Poteomkin, Powell, Prasad, Predoi, Premachandra, Prestegard, Price, Principe, Privitera, Prix, Prokhorov, Puncken, Pürrer, Qin, Quetschke, Quintero, Quiroga, Quitzow-James, Raab, Rabeling, Radkins, Raffai, Raja, Rajalakshmi, Rakhmanov, Ramirez, Raymond, Reed, Reid, Reitze, Reula, Riles, Robertson, Robie, Rollins, Roma, Romano, Romanov, Romie, Rowan, Rüdiger, Ryan, Sachdev, Sadecki, Sadeghian, Saleem, Salemi, Sammut, Sandberg, Sanders, Sannibale, Santiago-Prieto, Sathyaprakash, Saulson, Savage, Sawadsky, Scheuer, Schilling, Schmidt, Schnabel, Schofield, Schreiber, Schuette, Schutz, Scott, Scott, Sellers, Sengupta, Sergeev, Serna, Sevigny, Shaddock, Shahriar, Shaltev, Shao, Shapiro, Shawhan, Shoemaker, Sidery, Siemens, Sigg, Silva, Simakov, Singer, Singer, Singh, Sintes, Slagmolen, Smith, Smith, Smith, Smith-Lefebvre, Son, Sorazu, Souradeep, Staley, Stebbins, Steinke, Steinlechner, Steinlechner, Steinmeyer, Stephens,
  Steplewski, Stevenson, Stone, Strain, Strigin, Sturani, Stuver, Summerscales, Sutton, Szczepanczyk, Szeifert, Talukder, Tanner, Tápai, Tarabrin, Taracchini, Taylor, Tellez, Theeg, Thirugnanasambandam, Thomas, Thomas, Thorne, Thorne, Thrane, Tiwari, Tomlinson, Torres, Torrie, Traylor, Tse, Tshilumba, Ugolini, Unnikrishnan, Urban, Usman, Vahlbruch, Vajente, Valdes, Vallisneri, van Veggel, Vass, Vaulin, Vecchio, Veitch, Veitch, Venkateswara, Vincent-Finley, Vitale, Vo, Vorvick, Vousden, Vyatchanin, Wade, Wade, Wade, Walker, Wallace, Walsh, Wang, Wang, Wang, Ward, Warner, Was, Weaver, Weinert, Weinstein, Weiss, Welborn, Wen, Wessels, Westphal, Wette, Whelan, Whitcomb, White, Whiting, Wilkinson, Williams, Williams, Williamson, Willis, Willke, Wimmer, Winkler, Wipf, Wittel, Woan, Worden, Xie, Yablon, Yakushin, Yam, Yamamoto, Yancey, Yang, Zanolin, Zhang, Zhang, Zhang, Zhang, Zhao, Zhou, Zhu, Zucker, Zuraw, \& Zweizig}]{AdvLIGO}
Aasi, J., Abbott, B.~P., Abbott, R., {et~al.} 2015, Classical and Quantum Gravity, 32, 074001, \dodoi{10.1088/0264-9381/32/7/074001}

\bibitem[{Abadie {et~al.}(2012)Abadie, Abbott, Abbott, Abbott, Abernathy, Accadia, \& Acernese}]{ligo_sensitivity2012}
Abadie, J., Abbott, J., Abbott, R., {et~al.} 2012, Sensitivity Achieved by the LIGO and Virgo Gravitational Wave Detectors during LIGO's Sixth and Virgo's Second and Third Science Runs.
\newblock \doarXiv{1203.2674}

\bibitem[{{Abadie} {et~al.}(2010){Abadie}, {Abbott}, {Abbott}, {Abernathy}, {Accadia}, {Acernese}, {Adams}, {Adhikari}, {Ajith}, {Allen}, {Allen}, {Amador Ceron}, {Amin}, {Anderson}, {Anderson}, {Antonucci}, {Aoudia}, {Arain}, {Araya}, {Aronsson}, {Arun}, {Aso}, {Aston}, {Astone}, {Atkinson}, {Aufmuth}, {Aulbert}, {Babak}, {Baker}, {Ballardin}, {Ballmer}, {Barker}, {Barnum}, {Barone}, {Barr}, {Barriga}, {Barsotti}, {Barsuglia}, {Barton}, {Bartos}, {Bassiri}, {Bastarrika}, {Bauchrowitz}, {Bauer}, {Behnke}, {Beker}, {Belczynski}, {Benacquista}, {Bertolini}, {Betzwieser}, {Beveridge}, {Beyersdorf}, {Bigotta}, {Bilenko}, {Billingsley}, {Birch}, {Birindelli}, {Biswas}, {Bitossi}, {Bizouard}, {Black}, {Blackburn}, {Blackburn}, {Blair}, {Bland}, {Blom}, {Blomberg}, {Boccara}, {Bock}, {Bodiya}, {Bondarescu}, {Bondu}, {Bonelli}, {Bork}, {Born}, {Bose}, {Bosi}, {Boyle}, {Braccini}, {Bradaschia}, {Brady}, {Braginsky}, {Brau}, {Breyer}, {Bridges}, {Brillet}, {Brinkmann}, {Brisson}, {Britzger}, {Brooks}, {Brown},
  {Budzy{\'n}ski}, {Bulik}, {Bulten}, {Buonanno}, {Burguet-Castell}, {Burmeister}, {Buskulic}, {Byer}, {Cadonati}, {Cagnoli}, {Calloni}, {Camp}, {Campagna}, {Campsie}, {Cannizzo}, {Cannon}, {Canuel}, {Cao}, {Capano}, {Carbognani}, {Caride}, {Caudill}, {Cavagli{\`a}}, {Cavalier}, {Cavalieri}, {Cella}, {Cepeda}, {Cesarini}, {Chalermsongsak}, {Chalkley}, {Charlton}, {Chassande Mottin}, {Chelkowski}, {Chen}, {Chincarini}, {Christensen}, {Chua}, {Chung}, {Clark}, {Clark}, {Clayton}, {Cleva}, {Coccia}, {Colacino}, {Colas}, {Colla}, {Colombini}, {Conte}, {Cook}, {Corbitt}, {Corda}, {Cornish}, {Corsi}, {Costa}, {Coulon}, {Coward}, {Coyne}, {Creighton}, {Creighton}, {Cruise}, {Culter}, {Cumming}, {Cunningham}, {Cuoco}, {Dahl}, {Danilishin}, {Dannenberg}, {D'Antonio}, {Danzmann}, {Dari}, {Das}, {Dattilo}, {Daudert}, {Davier}, {Davies}, {Davis}, {Daw}, {Day}, {Dayanga}, {De Rosa}, {DeBra}, {Degallaix}, {del Prete}, {Dergachev}, {DeRosa}, {DeSalvo}, {Devanka}, {Dhurandhar}, {Di Fiore}, {Di Lieto}, {Di Palma}, {Emilio},
  {Di Virgilio}, {D{\'\i}az}, {Dietz}, {Donovan}, {Dooley}, {Doomes}, {Dorsher}, {Douglas}, {Drago}, {Drever}, {Driggers}, {Dueck}, {Dumas}, {Eberle}, {Edgar}, {Edwards}, {Effler}, {Ehrens}, {Engel}, {Etzel}, {Evans}, {Evans}, {Fafone}, {Fairhurst}, {Fan}, {Farr}, {Fazi}, {Fehrmann}, {Feldbaum}, {Ferrante}, {Fidecaro}, {Finn}, {Fiori}, {Flaminio}, {Flanigan}, {Flasch}, {Foley}, {Forrest}, {Forsi}, {Fotopoulos}, {Fournier}, {Franc}, {Frasca}, {Frasconi}, {Frede}, {Frei}, {Frei}, {Freise}, {Frey}, {Fricke}, {Friedrich}, {Fritschel}, {Frolov}, {Fulda}, {Fyffe}, {Gammaitoni}, {Garofoli}, {Garufi}, {Gemme}, {Genin}, {Gennai}, {Gholami}, {Ghosh}, {Giaime}, {Giampanis}, {Giardina}, {Giazotto}, {Gill}, {Goetz}, {Goggin}, {Gonz{\'a}lez}, {Gorodetsky}, {Go{\ss}ler}, {Gouaty}, {Graef}, {Granata}, {Grant}, {Gras}, {Gray}, {Greenhalgh}, {Gretarsson}, {Greverie}, {Grosso}, {Grote}, {Grunewald}, {Guidi}, {Gustafson}, {Gustafson}, {Hage}, {Hall}, {Hallam}, {Hammer}, {Hammond}, {Hanks}, {Hanna}, {Hanson}, {Harms}, {Harry},
  {Harry}, {Harstad}, {Haughian}, {Hayama}, {Heefner}, {Heitmann}, {Hello}, {Heng}, {Heptonstall}, {Hewitson}, {Hild}, {Hirose}, {Hoak}, {Hodge}, {Holt}, {Hosken}, {Hough}, {Howell}, {Hoyland}, {Huet}, {Hughey}, {Husa}, {Huttner}, {Huynh-Dinh}, {Ingram}, {Inta}, {Isogai}, {Ivanov}, {Jaranowski}, {Johnson}, {Jones}, {Jones}, {Jones}, {Ju}, {Kalmus}, {Kalogera}, {Kandhasamy}, {Kanner}, {Katsavounidis}, {Kawabe}, {Kawamura}, {Kawazoe}, {Kells}, {Keppel}, {Khalaidovski}, {Khalili}, {Khazanov}, {Kim}, {Kim}, {King}, {Kinzel}, {Kissel}, {Klimenko}, {Kondrashov}, {Kopparapu}, {Koranda}, {Kowalska}, {Kozak}, {Krause}, {Kringel}, {Krishnamurthy}, {Krishnan}, {Kr{\'o}lak}, {Kuehn}, {Kullman}, {Kumar}, {Kwee}, {Landry}, {Lang}, {Lantz}, {Lastzka}, {Lazzarini}, {Leaci}, {Leong}, {Leonor}, {Leroy}, {Letendre}, {Li}, {Li}, {Lin}, {Lindquist}, {Lockerbie}, {Lodhia}, {Lorenzini}, {Loriette}, {Lormand}, {Losurdo}, {Lu}, {Luan}, {Lubi{\'n}ski}, {Lucianetti}, {L{\"u}ck}, {Lundgren}, {Machenschalk}, {MacInnis}, {Mackowski},
  {Mageswaran}, {Mailand}, {Majorana}, {Mak}, {Man}, {Mandel}, {Mandic}, {Mantovani}, {Marchesoni}, {Marion}, {M{\'a}rka}, {M{\'a}rka}, {Maros}, {Marque}, {Martelli}, {Martin}, {Martin}, {Marx}, {Mason}, {Masserot}, {Matichard}, {Matone}, {Matzner}, {Mavalvala}, {McCarthy}, {McClelland}, {McGuire}, {McIntyre}, {McIvor}, {McKechan}, {Meadors}, {Mehmet}, {Meier}, {Melatos}, {Melissinos}, {Mendell}, {Men{\'e}ndez}, {Mercer}, {Merill}, {Meshkov}, {Messenger}, {Meyer}, {Miao}, {Michel}, {Milano}, {Miller}, {Minenkov}, {Mino}, {Mitra}, {Mitrofanov}, {Mitselmakher}, {Mittleman}, {Moe}, {Mohan}, {Mohanty}, {Mohapatra}, {Moraru}, {Moreau}, {Moreno}, {Morgado}, {Morgia}, {Morioka}, {Mors}, {Mosca}, {Moscatelli}, {Mossavi}, {Mours}, {MowLowry}, {Mueller}, {Mukherjee}, {Mullavey}, {M{\"u}ller-Ebhardt}, {Munch}, {Murray}, {Nash}, {Nawrodt}, {Nelson}, {Neri}, {Newton}, {Nishizawa}, {Nocera}, {Nolting}, {Ochsner}, {O'Dell}, {Ogin}, {Oldenburg}, {O'Reilly}, {O'Shaughnessy}, {Osthelder}, {Ottaway}, {Ottens}, {Overmier},
  {Owen}, {Page}, {Pagliaroli}, {Palladino}, {Palomba}, {Pan}, {Pankow}, {Paoletti}, {Papa}, {Pardi}, {Pareja}, {Parisi}, {Pasqualetti}, {Passaquieti}, {Passuello}, {Patel}, {Pedraza}, {Pekowsky}, {Penn}, {Peralta}, {Perreca}, {Persichetti}, {Pichot}, {Pickenpack}, {Piergiovanni}, {Pietka}, {Pinard}, {Pinto}, {Pitkin}, {Pletsch}, {Plissi}, {Poggiani}, {Postiglione}, {Prato}, {Predoi}, {Price}, {Prijatelj}, {Principe}, {Privitera}, {Prix}, {Prodi}, {Prokhorov}, {Puncken}, {Punturo}, {Puppo}, {Quetschke}, {Raab}, {Rabaste}, {Rabeling}, {Radke}, {Radkins}, {Raffai}, {Rakhmanov}, {Rankins}, {Rapagnani}, {Raymond}, {Re}, {Reed}, {Reed}, {Regimbau}, {Reid}, {Reitze}, {Ricci}, {Riesen}, {Riles}, {Roberts}, {Robertson}, {Robinet}, {Robinson}, {Robinson}, {Rocchi}, {Roddy}, {R{\"o}ver}, {Rogstad}, {Rolland}, {Rollins}, {Romano}, {Romano}, {Romie}, {Rosi{\'n}ska}, {Rowan}, {R{\"u}diger}, {Ruggi}, {Ryan}, {Sakata}, {Sakosky}, {Salemi}, {Sammut}, {Sancho de la Jordana}, {Sandberg}, {Sannibale}, {Santamar{\'\i}a},
  {Santostasi}, {Saraf}, {Sassolas}, {Sathyaprakash}, {Sato}, {Satterthwaite}, {Saulson}, {Savage}, {Schilling}, {Schnabel}, {Schofield}, {Schulz}, {Schutz}, {Schwinberg}, {Scott}, {Scott}, {Searle}, {Seifert}, {Sellers}, {Sengupta}, {Sentenac}, {Sergeev}, {Shaddock}, {Shapiro}, {Shawhan}, {Shoemaker}, {Sibley}, {Siemens}, {Sigg}, {Singer}, {Sintes}, {Skelton}, {Slagmolen}, {Slutsky}, {Smith}, {Smith}, {Smith}, {Somiya}, {Sorazu}, {Speirits}, {Stein}, {Stein}, {Steinlechner}, {Steplewski}, {Stochino}, {Stone}, {Strain}, {Strigin}, {Stroeer}, {Sturani}, {Stuver}, {Summerscales}, {Sung}, {Susmithan}, {Sutton}, {Swinkels}, {Talukder}, {Tanner}, {Tarabrin}, {Taylor}, {Taylor}, {Thomas}, {Thorne}, {Thorne}, {Thrane}, {Th{\"u}ring}, {Titsler}, {Tokmakov}, {Toncelli}, {Tonelli}, {Torres}, {Torrie}, {Tournefier}, {Travasso}, {Traylor}, {Trias}, {Trummer}, {Tseng}, {Ugolini}, {Urbanek}, {Vahlbruch}, {Vaishnav}, {Vajente}, {Vallisneri}, {van den Brand}, {Van Den Broeck}, {van der Putten}, {van der Sluys}, {van Veggel},
  {Vass}, {Vaulin}, {Vavoulidis}, {Vecchio}, {Vedovato}, {Veitch}, {Veitch}, {Veltkamp}, {Verkindt}, {Vetrano}, {Vicer{\'e}}, {Villar}, {Vinet}, {Vocca}, {Vorvick}, {Vyachanin}, {Waldman}, {Wallace}, {Wanner}, {Ward}, {Was}, {Wei}, {Weinert}, {Weinstein}, {Weiss}, {Wen}, {Wen}, {Wessels}, {West}, {Westphal}, {Wette}, {Whelan}, {Whitcomb}, {White}, {Whiting}, {Wilkinson}, {Willems}, {Williams}, {Willke}, {Winkelmann}, {Winkler}, {Wipf}, {Wiseman}, {Woan}, {Wooley}, {Worden}, {Yakushin}, {Yamamoto}, {Yamamoto}, {Yeaton-Massey}, {Yoshida}, {Yu}, {Yvert}, {Zanolin}, {Zhang}, {Zhang}, {Zhao}, {Zotov}, {Zucker}, {Zweizig}, {LIGO Scientific Collaboration}, \& {Virgo Collaboration}}]{abadie2010}
{Abadie}, J., {Abbott}, B.~P., {Abbott}, R., {et~al.} 2010, Classical and Quantum Gravity, 27, 173001, \dodoi{10.1088/0264-9381/27/17/173001}

\bibitem[{{Abbott} {et~al.}(2016{\natexlab{a}}){Abbott}, {Abbott}, {Abbott}, {Abernathy}, {Acernese}, {Ackley}, {Adams}, {Adams}, {Addesso}, \& {Adhikari}}]{abbott2016}
{Abbott}, B.~P., {Abbott}, R., {Abbott}, T.~D., {et~al.} 2016{\natexlab{a}}, \prl, 116, 061102, \dodoi{10.1103/PhysRevLett.116.061102}

\bibitem[{{Abbott} {et~al.}(2016{\natexlab{b}}){Abbott}, {Abbott}, {Abbott}, {Abernathy}, {Acernese}, {Ackley}, {Adams}, {Adams}, {Addesso}, {Adhikari}, {Adya}, {Affeldt}, {Agathos}, {Agatsuma}, {Aggarwal}, {Aguiar}, {Aiello}, {Ain}, {Ajith}, {Allen}, {Allocca}, {Altin}, {Anderson}, {Anderson}, {Arai}, {Araya}, {Arceneaux}, {Areeda}, {Arnaud}, {Arun}, {Ascenzi}, {Ashton}, {Ast}, {Aston}, {Astone}, {Aufmuth}, {Aulbert}, {Babak}, {Bacon}, {Bader}, {Baker}, {Baldaccini}, {Ballardin}, {Ballmer}, {Barayoga}, {Barclay}, {Barish}, {Barker}, {Barone}, {Barr}, {Barsotti}, {Barsuglia}, {Barta}, {Bartlett}, {Bartos}, {Bassiri}, {Basti}, {Batch}, {Baune}, {Bavigadda}, {Bazzan}, {Behnke}, {Bejger}, {Bell}, {Bell}, {Berger}, {Bergman}, {Bergmann}, {Berry}, {Bersanetti}, {Bertolini}, {Betzwieser}, {Bhagwat}, {Bhandare}, {Bilenko}, {Billingsley}, {Birch}, {Birney}, {Biscans}, {Bisht}, {Bitossi}, {Biwer}, {Bizouard}, {Blackburn}, {Blair}, {Blair}, {Blair}, {Bloemen}, {Bock}, {Bodiya}, {Boer}, {Bogaert}, {Bogan}, {Bohe},
  {Bojtos}, {Bond}, {Bondu}, {Bonnand}, {Boom}, {Bork}, {Boschi}, {Bose}, {Bouffanais}, {Bozzi}, {Bradaschia}, {Brady}, {Braginsky}, {Branchesi}, {Brau}, {Briant}, {Brillet}, {Brinkmann}, {Brisson}, {Brockill}, {Brooks}, {Brown}, {Brown}, {Brown}, {Buchanan}, {Buikema}, {Bulik}, {Bulten}, {Buonanno}, {Buskulic}, {Buy}, {Byer}, {Cadonati}, {Cagnoli}, {Cahillane}, {Calder{\'o}n Bustillo}, {Callister}, {Calloni}, {Camp}, {Cannon}, {Cao}, {Capano}, {Capocasa}, {Carbognani}, {Caride}, {Casanueva Diaz}, {Casentini}, {Caudill}, {Cavagli{\`a}}, {Cavalier}, {Cavalieri}, {Cella}, {Cepeda}, {Cerboni Baiardi}, {Cerretani}, {Cesarini}, {Chakraborty}, {Chalermsongsak}, {Chamberlin}, {Chan}, {Chao}, {Charlton}, {Chassande-Mottin}, {Chen}, {Chen}, {Cheng}, {Chincarini}, {Chiummo}, {Cho}, {Cho}, {Chow}, {Christensen}, {Chu}, {Chua}, {Chung}, {Ciani}, {Clara}, {Clark}, {Cleva}, {Coccia}, {Cohadon}, {Colla}, {Collette}, {Cominsky}, {Constancio}, {Conte}, {Conti}, {Cook}, {Corbitt}, {Cornish}, {Corpuz}, {Corsi}, {Cortese},
  {Costa}, {Coughlin}, {Coughlin}, {Coulon}, {Countryman}, {Couvares}, {Coward}, {Cowart}, {Coyne}, {Coyne}, {Craig}, {Creighton}, {Cripe}, {Crowder}, {Cumming}, {Cunningham}, {Cuoco}, {Dal Canton}, {Danilishin}, {D'Antonio}, {Danzmann}, {Darman}, {Dattilo}, {Dave}, {Daveloza}, {Davier}, {Davies}, {Daw}, {Day}, {DeBra}, {Debreczeni}, {Degallaix}, {De Laurentis}, {Del{\'e}glise}, {Del Pozzo}, {Denker}, {Dent}, {Dergachev}, {De Rosa}, {DeRosa}, {DeSalvo}, {Dhurandhar}, {D{\'\i}az}, {Di Fiore}, {Di Giovanni}, {Di Girolamo}, {Di Lieto}, {Di Pace}, {Di Palma}, {Di Virgilio}, {Dojcinoski}, {Dolique}, {Donovan}, {Dooley}, {Doravari}, {Douglas}, {Downes}, {Drago}, {Drever}, {Driggers}, {Du}, {Ducrot}, {Dwyer}, {Edo}, {Edwards}, {Effler}, {Eggenstein}, {Ehrens}, {Eichholz}, {Eikenberry}, {Engels}, {Essick}, {Etzel}, {Evans}, {Evans}, {Everett}, {Factourovich}, {Fafone}, {Fair}, {Fairhurst}, {Fan}, {Fang}, {Farinon}, {Farr}, {Farr}, {Favata}, {Fays}, {Fehrmann}, {Fejer}, {Ferrante}, {Ferreira}, {Ferrini}, {Fidecaro},
  {Fiori}, {Fiorucci}, {Fisher}, {Flaminio}, {Fletcher}, {Fournier}, {Frasca}, {Frasconi}, {Frei}, {Freise}, {Frey}, {Frey}, {Fricke}, {Fritschel}, {Frolov}, {Fulda}, {Fyffe}, {Gabbard}, {Gair}, {Gammaitoni}, {Gaonkar}, {Garufi}, {Gaur}, {Gehrels}, {Gemme}, {Genin}, {Gennai}, {George}, {Gergely}, {Germain}, {Ghosh}, {Ghosh}, {Giaime}, {Giardina}, {Giazotto}, {Gill}, {Glaefke}, {Goetz}, {Goetz}, {Gondan}, {Gonz{\'a}lez}, {Gonzalez Castro}, {Gopakumar}, {Gordon}, {Gorodetsky}, {Gossan}, {Gosselin}, {Gouaty}, {Grado}, {Graef}, {Graff}, {Granata}, {Grant}, {Gras}, {Gray}, {Greco}, {Green}, {Groot}, {Grote}, {Grunewald}, {Guidi}, {Guo}, {Gupta}, {Gupta}, {Gushwa}, {Gustafson}, {Gustafson}, {Hacker}, {Hall}, {Hall}, {Hammond}, {Haney}, {Hanke}, {Hanks}, {Hanna}, {Hannam}, {Hanson}, {Hardwick}, {Harms}, {Harry}, {Harry}, {Hart}, {Hartman}, {Haster}, {Haughian}, {Heidmann}, {Heintze}, {Heitmann}, {Hello}, {Hemming}, {Hendry}, {Heng}, {Hennig}, {Heptonstall}, {Heurs}, {Hild}, {Hoak}, {Hodge}, {Hofman}, {Hollitt},
  {Holt}, {Holz}, {Hopkins}, {Hosken}, {Hough}, {Houston}, {Howell}, {Hu}, {Huang}, {Huerta}, {Huet}, {Hughey}, {Husa}, {Huttner}, {Huynh-Dinh}, {Idrisy}, {Indik}, {Ingram}, {Inta}, {Isa}, {Isac}, {Isi}, {Islas}, {Isogai}, {Iyer}, {Izumi}, {Jacqmin}, {Jang}, {Jani}, {Jaranowski}, {Jawahar}, {Jim{\'e}nez-Forteza}, {Johnson}, {Jones}, {Jones}, {Jonker}, {Ju}, {Haris}, {Kalaghatgi}, {Kalmus}, {Kalogera}, {Kamaretsos}, {Kandhasamy}, {Kang}, {Kanner}, {Karki}, {Kasprzack}, {Katsavounidis}, {Katzman}, {Kaufer}, {Kaur}, {Kawabe}, {Kawazoe}, {K{\'e}f{\'e}lian}, {Kehl}, {Keitel}, {Kelley}, {Kells}, {Kennedy}, {Key}, {Khalaidovski}, {Khalili}, {Khan}, {Khan}, {Khan}, {Khazanov}, {Kijbunchoo}, {Kim}, {Kim}, {Kim}, {Kim}, {Kim}, {Kim}, {King}, {King}, {Kinzel}, {Kissel}, {Kleybolte}, {Klimenko}, {Koehlenbeck}, {Kokeyama}, {Koley}, {Kondrashov}, {Kontos}, {Korobko}, {Korth}, {Kowalska}, {Kozak}, {Kringel}, {Krishnan}, {Kr{\'o}lak}, {Krueger}, {Kuehn}, {Kumar}, {Kuo}, {Kutynia}, {Lackey}, {Landry}, {Lange}, {Lantz},
  {Lasky}, {Lazzarini}, {Lazzaro}, {Leaci}, {Leavey}, {Lebigot}, {Lee}, {Lee}, {Lee}, {Lee}, {Lenon}, {Leonardi}, {Leong}, {Leroy}, {Letendre}, {Levin}, {Levine}, {Li}, {Libson}, {Littenberg}, {Lockerbie}, {Loew}, {Logue}, {Lombardi}, {Lord}, {Lorenzini}, {Loriette}, {Lormand}, {Losurdo}, {Lough}, {L{\"u}ck}, {Lundgren}, {Luo}, {Lynch}, {Ma}, {MacDonald}, {Machenschalk}, {MacInnis}, {Macleod}, {Maga{\~n}a-Sandoval}, {Magee}, {Mageswaran}, {Majorana}, {Maksimovic}, {Malvezzi}, {Man}, {Mandel}, {Mandic}, {Mangano}, {Mansell}, {Manske}, {Mantovani}, {Marchesoni}, {Marion}, {M{\'a}rka}, {M{\'a}rka}, {Markosyan}, {Maros}, {Martelli}, {Martellini}, {Martin}, {Martin}, {Martynov}, {Marx}, {Mason}, {Masserot}, {Massinger}, {Masso-Reid}, {Mastrogiovanni}, {Matichard}, {Matone}, {Mavalvala}, {Mazumder}, {Mazzolo}, {McCarthy}, {McClelland}, {McCormick}, {McGuire}, {McIntyre}, {McIver}, {McManus}, {McWilliams}, {Meacher}, {Meadors}, {Meidam}, {Melatos}, {Mendell}, {Mendoza-Gandara}, {Mercer}, {Merilh}, {Merzougui},
  {Meshkov}, {Messenger}, {Messick}, {Metzdorff}, {Meyers}, {Mezzani}, {Miao}, {Michel}, {Middleton}, {Mikhailov}, {Milano}, {Miller}, {Miller}, {Millhouse}, {Minenkov}, {Ming}, {Mirshekari}, {Mishra}, {Mitra}, {Mitrofanov}, {Mitselmakher}, {Mittleman}, {Moggi}, {Mohan}, {Mohapatra}, {Montani}, {Moore}, {Moore}, {Moraru}, {Moreno}, {Morriss}, {Mossavi}, {Mours}, {Mow-Lowry}, {Mueller}, {Mueller}, {Muir}, {Mukherjee}, {Mukherjee}, {Mukherjee}, {Mukund}, {Mullavey}, {Munch}, {Murphy}, {Murray}, {Mytidis}, {Nardecchia}, {Naticchioni}, {Nayak}, {Necula}, {Nedkova}, {Nelemans}, {Neri}, {Neunzert}, {Newton}, {Nguyen}, {Nielsen}, {Nissanke}, {Nitz}, {Nocera}, {Nolting}, {Normandin}, {Nuttall}, {Oberling}, {Ochsner}, {O'Dell}, {Oelker}, {Ogin}, {Oh}, {Oh}, {Ohme}, {Oliver}, {Oppermann}, {Oram}, {O'Reilly}, {O'Shaughnessy}, {Ott}, {Ottaway}, {Ottens}, {Overmier}, {Owen}, {Pai}, {Pai}, {Palamos}, {Palashov}, {Palomba}, {Pal-Singh}, {Pan}, {Pankow}, {Pannarale}, {Pant}, {Paoletti}, {Paoli}, {Papa}, {Paris}, {Parker},
  {Pascucci}, {Pasqualetti}, {Passaquieti}, {Passuello}, {Patricelli}, {Patrick}, {Pearlstone}, {Pedraza}, {Pedurand}, {Pekowsky}, {Pele}, {Penn}, {Pereira}, {Perreca}, {Phelps}, {Piccinni}, {Pichot}, {Piergiovanni}, {Pierro}, {Pillant}, {Pinard}, {Pinto}, {Pitkin}, {Poggiani}, {Popolizio}, {Post}, {Powell}, {Prasad}, {Predoi}, {Premachandra}, {Prestegard}, {Price}, {Prijatelj}, {Principe}, {Privitera}, {Prix}, {Prodi}, {Prokhorov}, {Puncken}, {Punturo}, {Puppo}, {P{\"u}rrer}, {Qi}, {Qin}, {Quetschke}, {Quintero}, {Quitzow-James}, {Raab}, {Rabeling}, {Radkins}, {Raffai}, {Raja}, {Rakhmanov}, {Rapagnani}, {Raymond}, {Razzano}, {Re}, {Read}, {Reed}, {Regimbau}, {Rei}, {Reid}, {Reitze}, {Rew}, {Ricci}, {Riles}, {Robertson}, {Robie}, {Robinet}, {Rocchi}, {Rolland}, {Rollins}, {Roma}, {Romano}, {Romano}, {Romanov}, {Romie}, {Rosi{\'n}ska}, {Rowan}, {R{\"u}diger}, {Ruggi}, {Ryan}, {Sachdev}, {Sadecki}, {Sadeghian}, {Salconi}, {Saleem}, {Salemi}, {Samajdar}, {Sammut}, {Sanchez}, {Sandberg}, {Sandeen}, {Sanders},
  {Santamaria}, {Sassolas}, {Sathyaprakash}, {Saulson}, {Sauter}, {Savage}, {Sawadsky}, {Schale}, {Schilling}, {Schmidt}, {Schmidt}, {Schnabel}, {Schofield}, {Sch{\"o}nbeck}, {Schreiber}, {Schuette}, {Schutz}, {Scott}, {Scott}, {Sellers}, {Sentenac}, {Sequino}, {Sergeev}, {Serna}, {Setyawati}, {Sevigny}, {Shaddock}, {Shahriar}, {Shaltev}, {Shao}, {Shapiro}, {Shawhan}, {Sheperd}, {Shoemaker}, {Shoemaker}, {Siellez}, {Siemens}, {Sieniawska}, {Sigg}, {Silva}, {Simakov}, {Singer}, {Singer}, {Singh}, {Singh}, {Singhal}, {Sintes}, {Slagmolen}, {Smith}, {Smith}, {Smith}, {Son}, {Sorazu}, {Sorrentino}, {Souradeep}, {Srivastava}, {Staley}, {Steinke}, {Steinlechner}, {Steinlechner}, {Steinmeyer}, {Stephens}, {Stone}, {Strain}, {Straniero}, {Stratta}, {Strauss}, {Strigin}, {Sturani}, {Stuver}, {Summerscales}, {Sun}, {Sutton}, {Swinkels}, {Szczepa{\'n}czyk}, {Tacca}, {Talukder}, {Tanner}, {T{\'a}pai}, {Tarabrin}, {Taracchini}, {Taylor}, {Theeg}, {Thirugnanasambandam}, {Thomas}, {Thomas}, {Thomas}, {Thorne}, {Thorne},
  {Thrane}, {Tiwari}, {Tiwari}, {Tokmakov}, {Tomlinson}, {Tonelli}, {Torres}, {Torrie}, {T{\"o}yr{\"a}}, {Travasso}, {Traylor}, {Trifir{\`o}}, {Tringali}, {Trozzo}, {Tse}, {Turconi}, {Tuyenbayev}, {Ugolini}, {Unnikrishnan}, {Urban}, {Usman}, {Vahlbruch}, {Vajente}, {Valdes}, {van Bakel}, {van Beuzekom}, {van den Brand}, {Van Den Broeck}, {Vander-Hyde}, {van der Schaaf}, {van Heijningen}, {van Veggel}, {Vardaro}, {Vass}, {Vas{\'u}th}, {Vaulin}, {Vecchio}, {Vedovato}, {Veitch}, {Veitch}, {Venkateswara}, {Verkindt}, {Vetrano}, {Vicer{\'e}}, {Vinciguerra}, {Vine}, {Vinet}, {Vitale}, {Vo}, {Vocca}, {Vorvick}, {Voss}, {Vousden}, {Vyatchanin}, {Wade}, {Wade}, {Wade}, {Walker}, {Wallace}, {Walsh}, {Wang}, {Wang}, {Wang}, {Wang}, {Wang}, {Ward}, {Warner}, {Was}, {Weaver}, {Wei}, {Weinert}, {Weinstein}, {Weiss}, {Welborn}, {Wen}, {We{\ss}els}, {Westphal}, {Wette}, {Whelan}, {Whitcomb}, {White}, {Whiting}, {Williams}, {Williamson}, {Willis}, {Willke}, {Wimmer}, {Winkler}, {Wipf}, {Wittel}, {Woan}, {Worden}, {Wright},
  {Wu}, {Yablon}, {Yam}, {Yamamoto}, {Yancey}, {Yap}, {Yu}, {Yvert}, {Zadro{\.Z}ny}, {Zangrando}, {Zanolin}, {Zendri}, {Zevin}, {Zhang}, {Zhang}, {Zhang}, {Zhang}, {Zhao}, {Zhou}, {Zhou}, {Zhu}, {Zucker}, {Zuraw}, {Zweizig}, {LIGO Scientific Collaboration}, \& {Virgo Collaboration}}]{abbott2016_ccsn}
---. 2016{\natexlab{b}}, \prd, 94, 102001, \dodoi{10.1103/PhysRevD.94.102001}

\bibitem[{Abbott {et~al.}(2020)Abbott, Abbott, Abbott, Abraham, Acernese, Ackley, Adams, Adya, Affeldt, Agathos, Agatsuma, Aggarwal, Aguiar, Aiello, Ain, \& Ajith}]{Abbott_2020}
Abbott, B.~P., Abbott, R., Abbott, T.~D., {et~al.} 2020, Physical Review D, 101, \dodoi{10.1103/physrevd.101.084002}

\bibitem[{{Abbott} {et~al.}(2021){Abbott}, {Abbott}, {Acernese}, {Ackley}, {Adams}, {Adhikari}, {Adhikari}, {Adya}, {Affeldt}, {Agarwal}, {Agathos}, {Agatsuma}, {Aggarwal}, {Aguiar}, {Aiello}, {Ain}, {Ajith}, {Akutsu}, {Albanesi}, {Allocca}, {Altin}, {Amato}, {Anand}, {Anand}, {Ananyeva}, {Anderson}, {Anderson}, {Ando}, {Andrade}, {Andres}, {Andri{\'c}}, {Angelova}, {Ansoldi}, {Antelis}, {Antier}, {Appert}, {Arai}, {Arai}, {Arai}, {Araki}, {Araya}, {Araya}, {Areeda}, {Ar{\`e}ne}, {Aritomi}, {Arnaud}, {Aronson}, {Arun}, {Asada}, {Asali}, {Ashton}, {Aso}, {Assiduo}, {Aston}, {Astone}, {Aubin}, {Austin}, {Babak}, {Badaracco}, {Bader}, {Badger}, {Bae}, {Bae}, {Baer}, {Bagnasco}, {Bai}, {Baiotti}, {Baird}, {Bajpai}, {Ball}, {Ballardin}, {Ballmer}, {Balsamo}, {Baltus}, {Banagiri}, {Bankar}, {Barayoga}, {Barbieri}, {Barish}, {Barker}, {Barneo}, {Barone}, {Barr}, {Barsotti}, {Barsuglia}, {Barta}, {Bartlett}, {Barton}, {Bartos}, {Bassiri}, {Basti}, {Bawaj}, {Bayley}, {Baylor}, {Bazzan}, {B{\'e}csy}, {Bedakihale},
  {Bejger}, {Belahcene}, {Benedetto}, {Beniwal}, {Bennett}, {Bentley}, {Benyaala}, {Bergamin}, {Berger}, {Bernuzzi}, {Berry}, {Bersanetti}, {Bertolini}, {Betzwieser}, {Beveridge}, {Bhandare}, {Bhardwaj}, {Bhattacharjee}, {Bhaumik}, {Bilenko}, {Billingsley}, {Bini}, {Birney}, {Birnholtz}, {Biscans}, {Bischi}, {Biscoveanu}, {Bisht}, {Biswas}, {Bitossi}, {Bizouard}, {Blackburn}, {Blair}, {Blair}, {Blair}, {Bobba}, {Bode}, {Boer}, {Bogaert}, {Boldrini}, {Bonavena}, {Bondu}, {Bonilla}, {Bonnand}, {Booker}, {Boom}, {Bork}, {Boschi}, {Bose}, {Bose}, {Bossilkov}, {Boudart}, {Bouffanais}, {Bozzi}, {Bradaschia}, {Brady}, {Bramley}, {Branch}, {Branchesi}, {Brau}, {Breschi}, {Briant}, {Briggs}, {Brillet}, {Brinkmann}, {Brockill}, {Brooks}, {Brooks}, {Brown}, {Brunett}, {Bruno}, {Bruntz}, {Bryant}, {Bulik}, {Bulten}, {Buonanno}, {Buscicchio}, {Buskulic}, {Buy}, {Byer}, {Cadonati}, {Cagnoli}, {Cahillane}, {Bustillo}, {Callaghan}, {Callister}, {Calloni}, {Cameron}, {Camp}, {Canepa}, {Canevarolo}, {Cannavacciuolo}, {Cannon},
  {Cao}, {Cao}, {Capocasa}, {Capote}, {Carapella}, {Carbognani}, {Carlin}, {Carney}, {Carpinelli}, {Carrillo}, {Carullo}, {Carver}, {Diaz}, {Casentini}, {Castaldi}, {Caudill}, {Cavagli{\`a}}, {Cavalier}, {Cavalieri}, {Ceasar}, {Cella}, {Cerd{\'a}-Dur{\'a}n}, {Cesarini}, {Chaibi}, {Chakravarti}, {Subrahmanya}, {Champion}, {Chan}, {Chan}, {Chan}, {Chan}, {Chan}, {Chandra}, {Chanial}, {Chao}, {Charlton}, {Chase}, {Chassande-Mottin}, {Chatterjee}, {Chatterjee}, {Chatterjee}, {Chaturvedi}, {Chaty}, {Chatziioannou}, {Chen}, {Chen}, {Chen}, {Chen}, {Chen}, {Chen}, {Chen}, {Chen}, {Cheng}, {Cheong}, {Cheung}, {Chia}, {Chiadini}, {Chiang}, {Chiarini}, {Chierici}, {Chincarini}, {Chiofalo}, {Chiummo}, {Cho}, {Cho}, {Choudhary}, {Choudhary}, {Christensen}, {Chu}, {Chu}, {Chu}, {Chua}, {Chung}, {Ciani}, {Ciecielag}, {Cie{\'s}lar}, {Cifaldi}, {Ciobanu}, {Ciolfi}, {Cipriano}, {Cirone}, {Clara}, {Clark}, {Clark}, {Clarke}, {Clearwater}, {Clesse}, {Cleva}, {Coccia}, {Codazzo}, {Cohadon}, {Cohen}, {Cohen}, {Colleoni},
  {Collette}, {Colombo}, {Colpi}, {Compton}, {Constancio}, {Conti}, {Cooper}, {Corban}, {Corbitt}, {Cordero-Carri{\'o}n}, {Corezzi}, {Corley}, {Cornish}, {Corre}, {Corsi}, {Cortese}, {Costa}, {Cotesta}, {Coughlin}, {Coulon}, {Countryman}, {Cousins}, {Couvares}, {Coward}, {Cowart}, {Coyne}, {Coyne}, {Creighton}, {Creighton}, {Criswell}, {Croquette}, {Crowder}, {Cudell}, {Cullen}, {Cumming}, {Cummings}, {Cunningham}, {Cuoco}, {Cury{\l}o}, {Dabadie}, {Canton}, {Dall'Osso}, {D{\'a}lya}, {Dana}, {Daneshgaranbajastani}, {D'Angelo}, {Danilishin}, {D'Antonio}, {Danzmann}, {Darsow-Fromm}, {Dasgupta}, {Datrier}, {Datta}, {Dattilo}, {Dave}, {Davier}, {Davies}, {Davis}, {Davis}, {Daw}, {Dean}, {Debra}, {Deenadayalan}, {Degallaix}, {de Laurentis}, {Del{\'e}glise}, {Del Favero}, {de Lillo}, {de Lillo}, {Del Pozzo}, {Demarchi}, {de Matteis}, {D'Emilio}, {Demos}, {Dent}, {Depasse}, {de Pietri}, {De Rosa}, {de Rossi}, {Desalvo}, {de Simone}, {Dhurandhar}, {D{\'\i}az}, {Diaz-Ortiz}, {Didio}, {Dietrich}, {di Fiore}, {di
  Fronzo}, {di Giorgio}, {di Giovanni}, {di Giovanni}, {di Girolamo}, {di Lieto}, {Ding}, {di Pace}, {di Palma}, {di Renzo}, {Divakarla}, {Dmitriev}, {Doctor}, {D'Onofrio}, {Donovan}, {Dooley}, {Doravari}, {Dorrington}, {Drago}, {Driggers}, {Drori}, {Ducoin}, {Dupej}, {Durante}, {D'Urso}, {Duverne}, {Dwyer}, {Eassa}, {Easter}, {Ebersold}, {Eckhardt}, {Eddolls}, {Edelman}, {Edo}, {Edy}, {Effler}, {Eguchi}, {Eichholz}, {Eikenberry}, {Eisenmann}, {Eisenstein}, {Ejlli}, {Engelby}, {Enomoto}, {Errico}, {Essick}, {Estell{\'e}s}, {Estevez}, {Etienne}, {Etzel}, {Evans}, {Evans}, {Ewing}, {Fafone}, {Fair}, {Fairhurst}, {Farah}, {Farinon}, {Farr}, {Farr}, {Farrow}, {Fauchon-Jones}, {Favaro}, {Favata}, {Fays}, {Fazio}, {Feicht}, {Fejer}, {Fenyvesi}, {Ferguson}, {Fernandez-Galiana}, {Ferrante}, {Ferreira}, {Fidecaro}, {Figura}, {Fiori}, {Fishbach}, {Fisher}, {Fittipaldi}, {Fiumara}, {Flaminio}, {Floden}, {Fong}, {Font}, {Fornal}, {Forsyth}, {Franke}, {Frasca}, {Frasconi}, {Frederick}, {Freed}, {Frei}, {Freise}, {Frey},
  {Fritschel}, {Frolov}, {Fronz{\'e}}, {Fujii}, {Fujikawa}, {Fukunaga}, {Fukushima}, {Fulda}, {Fyffe}, {Gabbard}, {Gadre}, {Gair}, {Gais}, {Galaudage}, {Gamba}, {Ganapathy}, {Ganguly}, {Gao}, {Gaonkar}, {Garaventa}, {Garc{\'\i}a-N{\'u}{\~n}ez}, {Garc{\'\i}a-Quir{\'o}s}, {Garufi}, {Gateley}, {Gaudio}, {Gayathri}, {Ge}, {Gemme}, {Gennai}, {George}, {Gerberding}, {Gergely}, {Gewecke}, {Ghonge}, {Ghosh}, {Ghosh}, {Ghosh}, {Ghosh}, {Giacomazzo}, {Giacoppo}, {Giaime}, {Giardina}, {Gibson}, {Gier}, {Giesler}, {Giri}, {Gissi}, {Glanzer}, {Gleckl}, {Godwin}, {Goetz}, {Goetz}, {Gohlke}, {Goncharov}, {Gonz{\'a}lez}, {Gopakumar}, {Gosselin}, {Gouaty}, {Gould}, {Grace}, {Grado}, {Granata}, {Granata}, {Grant}, {Gras}, {Grassia}, {Gray}, {Gray}, {Greco}, {Green}, {Green}, {Gretarsson}, {Gretarsson}, {Griffith}, {Griffiths}, {Griggs}, {Grignani}, {Grimaldi}, {Grimm}, {Grote}, {Grunewald}, {Gruning}, {Guerra}, {Guidi}, {Guimaraes}, {Guix{\'e}}, {Gulati}, {Guo}, {Guo}, {Gupta}, {Gupta}, {Gupta}, {Gustafson}, {Gustafson},
  {Guzman}, {Ha}, {Haegel}, {Hagiwara}, {Haino}, {Halim}, {Hall}, {Hamilton}, {Hammond}, {Han}, {Haney}, {Hanks}, {Hanna}, {Hannam}, {Hannuksela}, {Hansen}, {Hansen}, {Hanson}, {Harder}, {Hardwick}, {Haris}, {Harms}, {Harry}, {Harry}, {Hartwig}, {Hasegawa}, {Haskell}, {Hasskew}, {Haster}, {Hattori}, {Haughian}, {Hayakawa}, {Hayama}, {Hayes}, {Healy}, {Heidmann}, {Heidt}, {Heintze}, {Heinze}, {Heinzel}, {Heitmann}, {Hellman}, {Hello}, {Helmling-Cornell}, {Hemming}, {Hendry}, {Heng}, {Hennes}, {Hennig}, {Hennig}, {Hernandez}, {Vivanco}, {Heurs}, {Hild}, {Hill}, {Himemoto}, {Hines}, {Hiranuma}, {Hirata}, {Hirose}, {Hochheim}, {Hofman}, {Hohmann}, {Holcomb}, {Holland}, {Hollows}, {Holmes}, {Holt}, {Holz}, {Hong}, {Hopkins}, {Hough}, {Hourihane}, {Howell}, {Hoy}, {Hoyland}, {Hreibi}, {Hsieh}, {Hsu}, {Huang}, {Huang}, {Huang}, {Huang}, {Huang}, {Huang}, {H{\"u}bner}, {Huddart}, {Hughey}, {Hui}, {Hui}, {Husa}, {Huttner}, {Huxford}, {Huynh-Dinh}, {Ide}, {Idzkowski}, {Iess}, {Ikenoue}, {Imam}, {Inayoshi}, {Ingram},
  {Inoue}, {Ioka}, {Isi}, {Isleif}, {Ito}, {Itoh}, {Iyer}, {Izumi}, {Jaberianhamedan}, {Jacqmin}, {Jadhav}, {Jadhav}, {James}, {Jan}, {Jani}, {Janquart}, {Janssens}, {Janthalur}, {Jaranowski}, {Jariwala}, {Jaume}, {Jenkins}, {Jenner}, {Jeon}, {Jeunon}, {Jia}, {Jin}, {Johns}, {Jones}, {Jones}, {Jones}, {Jones}, {Jones}, {Jonker}, {Ju}, {Jung}, {Jung}, {Junker}, {Juste}, {Kaihotsu}, {Kajita}, {Kakizaki}, {Kalaghatgi}, {Kalogera}, {Kamai}, {Kamiizumi}, {Kanda}, {Kandhasamy}, {Kang}, {Kanner}, {Kao}, {Kapadia}, {Kapasi}, {Karat}, {Karathanasis}, {Karki}, {Kashyap}, {Kasprzack}, {Kastaun}, {Katsanevas}, {Katsavounidis}, {Katzman}, {Kaur}, {Kawabe}, {Kawaguchi}, {Kawai}, {Kawasaki}, {K{\'e}f{\'e}lian}, {Keitel}, {Key}, {Khadka}, {Khalili}, {Khan}, {Khazanov}, {Khetan}, {Khursheed}, {Kijbunchoo}, {Kim}, {Kim}, {Kim}, {Kim}, {Kim}, {Kim}, {Kimball}, {Kimura}, {Kinley-Hanlon}, {Kirchhoff}, {Kissel}, {Kita}, {Kitazawa}, {Kleybolte}, {Klimenko}, {Knee}, {Knowles}, {Knyazev}, {Koch}, {Koekoek}, {Kojima}, {Kokeyama},
  {Koley}, {Kolitsidou}, {Kolstein}, {Komori}, {Kondrashov}, {Kong}, {Kontos}, {Koper}, {Korobko}, {Kotake}, {Kovalam}, {Kozak}, {Kozakai}, {Kozu}, {Kringel}, {Krishnendu}, {Kr{\'o}lak}, {Kuehn}, {Kuei}, {Kuijer}, {Kumar}, {Kumar}, {Kumar}, {Kumar}, {Kume}, {Kuns}, {Kuo}, {Kuo}, {Kuromiya}, {Kuroyanagi}, {Kusayanagi}, {Kuwahara}, {Kwak}, {Lagabbe}, {Laghi}, {Lalande}, {Lam}, {Lamberts}, {Landry}, {Lane}, {Lang}, {Lange}, {Lantz}, {La Rosa}, {Lartaux-Vollard}, {Lasky}, {Laxen}, {Lazzarini}, {Lazzaro}, {Leaci}, {Leavey}, {Lecoeuche}, {Lee}, {Lee}, {Lee}, {Lee}, {Lee}, {Lee}, {Lehmann}, {Lema{\^\i}tre}, {Leonardi}, {Leroy}, {Letendre}, {Levesque}, {Levin}, {Leviton}, {Leyde}, {Li}, {Li}, {Li}, {Li}, {Li}, {Li}, {Lin}, {Lin}, {Lin}, {Lin}, {Lin}, {Linde}, {Linker}, {Linley}, {Littenberg}, {Liu}, {Liu}, {Liu}, {Liu}, {Llamas}, {Llorens-Monteagudo}, {Lo}, {Lockwood}, {London}, {Longo}, {Lopez}, {Lopez Portilla}, {Lorenzini}, {Loriette}, {Lormand}, {Losurdo}, {Lott}, {Lough}, {Lousto}, {Lovelace}, {Lucaccioni},
  {L{\"u}ck}, {Lumaca}, {Lundgren}, {Luo}, {Lynam}, {Macas}, {Macinnis}, {MacLeod}, {MacMillan}, {Macquet}, {Hernandez}, {Magazz{\`u}}, {Magee}, {Maggiore}, {Magnozzi}, {Mahesh}, {Majorana}, {Makarem}, {Maksimovic}, {Maliakal}, {Malik}, {Man}, {Mandic}, {Mangano}, {Mango}, {Mansell}, {Manske}, {Mantovani}, {Mapelli}, {Marchesoni}, {Marchio}, {Marion}, {Mark}, {M{\'a}rka}, {M{\'a}rka}, {Markakis}, {Markosyan}, {Markowitz}, {Maros}, {Marquina}, {Marsat}, {Martelli}, {Martin}, {Martin}, {Martinez}, {Martinez}, {Martinez}, {Martinovic}, {Martynov}, {Marx}, {Masalehdan}, {Mason}, {Massera}, {Masserot}, {Massinger}, {Masso-Reid}, {Mastrogiovanni}, {Matas}, {Mateu-Lucena}, {Matichard}, {Matiushechkina}, {Mavalvala}, {McCann}, {McCarthy}, {McClelland}, {McClincy}, {McCormick}, {McCuller}, {McGhee}, {McGuire}, {McIsaac}, {McIver}, {McRae}, {McWilliams}, {Meacher}, {Mehmet}, {Mehta}, {Meijer}, {Melatos}, {Melchor}, {Mendell}, {Menendez-Vazquez}, {Menoni}, {Mercer}, {Mereni}, {Merfeld}, {Merilh}, {Merritt}, {Merzougui},
  {Meshkov}, {Messenger}, {Messick}, {Meyers}, {Meylahn}, {Mhaske}, {Miani}, {Miao}, {Michaloliakos}, {Michel}, {Michimura}, {Middleton}, {Milano}, {Miller}, {Miller}, {Miller}, {Millhouse}, {Mills}, {Milotti}, {Minazzoli}, {Minenkov}, {Mio}, {Mir}, {Miravet-Ten{\'e}s}, {Mishra}, {Mishra}, {Mistry}, {Mitra}, {Mitrofanov}, {Mitselmakher}, {Mittleman}, {Miyakawa}, {Miyamoto}, {Miyazaki}, {Miyo}, {Miyoki}, {Mo}, {Moguel}, {Mogushi}, {Mohapatra}, {Mohite}, {Molina}, {Molina-Ruiz}, {Mondin}, {Montani}, {Moore}, {Moraru}, {Morawski}, {More}, {Moreno}, {Moreno}, {Mori}, {Morisaki}, {Moriwaki}, {Mours}, {Mow-Lowry}, {Mozzon}, {Muciaccia}, {Mukherjee}, {Mukherjee}, {Mukherjee}, {Mukherjee}, {Mukherjee}, {Mukund}, {Mullavey}, {Munch}, {Mu{\~n}iz}, {Murray}, {Musenich}, {Muth}, {Muusse}, {Nadji}, {Nagano}, {Nagano}, {Nagar}, {Nakamura}, {Nakano}, {Nakano}, {Nakashima}, {Nakayama}, {Napolano}, {Nardecchia}, {Narikawa}, {Naticchioni}, {Nayak}, {Nayak}, {Negishi}, {Neil}, {Neilson}, {Nelemans}, {Nelson}, {Nery},
  {Neubauer}, {Neunzert}, {Ng}, {Ng}, {Nguyen}, {Nguyen}, {Nguyen}, {Quynh}, {Ni}, {Nichols}, {Nishizawa}, {Nissanke}, {Nitoglia}, {Nocera}, {Norman}, {North}, {Nozaki}, {Nuttall}, {Oberling}, {O'Brien}, {Obuchi}, {O'Dell}, {Oelker}, {Ogaki}, {Oganesyan}, {Oh}, {Oh}, {Oh}, {Ohashi}, {Ohishi}, {Ohkawa}, {Ohme}, {Ohta}, {Okada}, {Okutani}, {Okutomi}, {Olivetto}, {Oohara}, {Ooi}, {Oram}, {O'Reilly}, {Ormiston}, {Ormsby}, {Ortega}, {O'Shaughnessy}, {O'Shea}, {Oshino}, {Ossokine}, {Osthelder}, {Otabe}, {Ottaway}, {Overmier}, {Pace}, {Pagano}, {Page}, {Pagliaroli}, {Pai}, {Pai}, {Palamos}, {Palashov}, {Palomba}, {Pan}, {Pan}, {Panda}, {Pang}, {Pang}, {Pankow}, {Pannarale}, {Pant}, {Panther}, {Paoletti}, {Paoli}, {Paolone}, {Parisi}, {Park}, {Park}, {Parker}, {Pascucci}, {Pasqualetti}, {Passaquieti}, {Passuello}, {Patel}, {Pathak}, {Patricelli}, {Patron}, {Patrone}, {Paul}, {Payne}, {Pedraza}, {Pegoraro}, {Pele}, {Arellano}, {Penn}, {Perego}, {Pereira}, {Pereira}, {Perez}, {P{\'e}rigois}, {Perkins}, {Perreca},
  {Perri{\`e}s}, {Petermann}, {Petterson}, {Pfeiffer}, {Pham}, {Phukon}, {Piccinni}, {Pichot}, {Piendibene}, {Piergiovanni}, {Pierini}, {Pierro}, {Pillant}, {Pillas}, {Pilo}, {Pinard}, {Pinto}, {Pinto}, {Piotrzkowski}, {Pirello}, {Pitkin}, {Placidi}, {Planas}, {Plastino}, {Pluchar}, {Poggiani}, {Polini}, {Pong}, {Ponrathnam}, {Popolizio}, {Porter}, {Poulton}, {Powell}, {Pracchia}, {Pradier}, {Prajapati}, {Prasai}, {Prasanna}, {Pratten}, {Principe}, {Prodi}, {Prokhorov}, {Prosposito}, {Prudenzi}, {Puecher}, {Punturo}, {Puosi}, {Puppo}, {P{\"u}rrer}, {Qi}, {Quetschke}, {Quitzow-James}, {Raab}, {Raaijmakers}, {Radkins}, {Radulesco}, {Raffai}, {Rail}, {Raja}, {Rajan}, {Ramirez}, {Ramirez}, {Ramos-Buades}, {Rana}, {Rapagnani}, {Rapol}, {Ray}, {Raymond}, {Raza}, {Razzano}, {Read}, {Rees}, {Regimbau}, {Rei}, {Reid}, {Reid}, {Reitze}, {Relton}, {Renzini}, {Rettegno}, {Rezac}, {Ricci}, {Richards}, {Richardson}, {Richardson}, {Riemenschneider}, {Riles}, {Rinaldi}, {Rink}, {Rizzo}, {Robertson}, {Robie}, {Robinet},
  {Rocchi}, {Rodriguez}, {Rolland}, {Rollins}, {Romanelli}, {Romano}, {Romel}, {Romero-Rodr{\'\i}guez}, {Romero-Shaw}, {Romie}, {Ronchini}, {Rosa}, {Rose}, {Rosi{\'n}ska}, {Ross}, {Rowan}, {Rowlinson}, {Roy}, {Roy}, {Roy}, {Rozza}, {Ruggi}, {Ryan}, {Sachdev}, {Sadecki}, {Sadiq}, {Sago}, {Saito}, {Saito}, {Sakai}, {Sakai}, {Sakellariadou}, {Sakuno}, {Salafia}, {Salconi}, {Saleem}, {Salemi}, {Samajdar}, {Sanchez}, {Sanchez}, {Sanchez}, {Sanchis-Gual}, {Sanders}, {Sanuy}, {Saravanan}, {Sarin}, {Sassolas}, {Satari}, {Sato}, {Sato}, {Sauter}, {Savage}, {Sawada}, {Sawant}, {Sawant}, {Sayah}, {Schaetzl}, {Scheel}, {Scheuer}, {Schiworski}, {Schmidt}, {Schmidt}, {Schnabel}, {Schneewind}, {Schofield}, {Sch{\"o}nbeck}, {Schulte}, {Schutz}, {Schwartz}, {Scott}, {Scott}, {Seglar-Arroyo}, {Sekiguchi}, {Sekiguchi}, {Sellers}, {Sengupta}, {Sentenac}, {Seo}, {Sequino}, {Sergeev}, {Setyawati}, {Shaffer}, {Shahriar}, {Shams}, {Shao}, {Sharma}, {Sharma}, {Shawhan}, {Shcheblanov}, {Shibagaki}, {Shikauchi}, {Shimizu}, {Shimoda},
  {Shimode}, {Shinkai}, {Shishido}, {Shoda}, {Shoemaker}, {Shoemaker}, {Shyamsundar}, {Sieniawska}, {Sigg}, {Singer}, {Singh}, {Singh}, {Singha}, {Sintes}, {Sipala}, {Skliris}, {Slagmolen}, {Slaven-Blair}, {Smetana}, {Smith}, {Smith}, {Soldateschi}, {Somala}, {Somiya}, {Son}, {Soni}, {Soni}, {Sordini}, {Sorrentino}, {Sorrentino}, {Sotani}, {Soulard}, {Souradeep}, {Sowell}, {Spagnuolo}, {Spencer}, {Spera}, {Srinivasan}, {Srivastava}, {Srivastava}, {Staats}, {Stachie}, {Steer}, {Steinlechner}, {Steinlechner}, {Stops}, {Stover}, {Strain}, {Strang}, {Stratta}, {Strunk}, {Sturani}, {Stuver}, {Sudhagar}, {Sudhir}, {Sugimoto}, {Suh}, {Summerscales}, {Sun}, {Sun}, {Sunil}, {Sur}, {Suresh}, {Sutton}, {Suzuki}, {Suzuki}, {Swinkels}, {Szczepa{\'n}czyk}, {Szewczyk}, {Tacca}, {Tagoshi}, {Tait}, {Takahashi}, {Takahashi}, {Takamori}, {Takano}, {Takeda}, {Takeda}, {Talbot}, {Talbot}, {Tanaka}, {Tanaka}, {Tanaka}, {Tanaka}, {Tanaka}, {Tanasijczuk}, {Tanioka}, {Tanner}, {Tao}, {Tao}, {Mart{\'\i}n}, {Taranto}, {Tasson},
  {Telada}, {Tenorio}, {Terhune}, {Terkowski}, {Thirugnanasambandam}, {Thomas}, {Thomas}, {Thompson}, {Thondapu}, {Thorne}, {Thrane}, {Tiwari}, {Tiwari}, {Tiwari}, {Toivonen}, {Toland}, {Tolley}, {Tomaru}, {Tomigami}, {Tomura}, {Tonelli}, {Torres-Forn{\'e}}, {Torrie}, {E Melo}, {T{\"o}yr{\"a}}, {Trapananti}, {Travasso}, {Traylor}, {Trevor}, {Tringali}, {Tripathee}, {Troiano}, {Trovato}, {Trozzo}, {Trudeau}, {Tsai}, {Tsai}, {Tsang}, {Tsang}, {Tsao}, {Tse}, {Tso}, {Tsubono}, {Tsuchida}, {Tsukada}, {Tsuna}, {Tsutsui}, {Tsuzuki}, {Turbang}, {Turconi}, {Tuyenbayev}, {Ubhi}, {Uchikata}, {Uchiyama}, {Udall}, {Ueda}, {Uehara}, {Ueno}, {Ueshima}, {Unnikrishnan}, {Uraguchi}, {Urban}, {Ushiba}, {Utina}, {Vahlbruch}, {Vajente}, {Vajpeyi}, {Valdes}, {Valentini}, {Valsan}, {van Bakel}, {van Beuzekom}, {van den Brand}, {van den Broeck}, {Vander-Hyde}, {van der Schaaf}, {van Heijningen}, {Vanosky}, {van Putten}, {van Remortel}, {Vardaro}, {Vargas}, {Varma}, {Vas{\'u}th}, {Vecchio}, {Vedovato}, {Veitch}, {Veitch},
  {Venneberg}, {Venugopalan}, {Verkindt}, {Verma}, {Verma}, {Veske}, {Vetrano}, {Vicer{\'e}}, {Vidyant}, {Viets}, {Vijaykumar}, {Villa-Ortega}, {Vinet}, {Virtuoso}, {Vitale}, {Vo}, {Vocca}, {von Reis}, {von Wrangel}, {Vorvick}, {Vyatchanin}, {Wade}, {Wade}, {Wagner}, {Walet}, {Walker}, {Wallace}, {Wallace}, {Walsh}, {Wang}, {Wang}, {Wang}, {Ward}, {Warner}, {Was}, {Washimi}, {Washington}, {Watchi}, {Weaver}, {Webster}, {Weinert}, {Weinstein}, {Weiss}, {Weller}, {Wellmann}, {Wen}, {We{\ss}els}, {Wette}, {Whelan}, {White}, {Whiting}, {Whittle}, {Wilken}, {Williams}, {Williams}, {Williamson}, {Willis}, {Willke}, {Wilson}, {Winkler}, {Wipf}, {Wlodarczyk}, {Woan}, {Woehler}, {Wofford}, {Wong}, {Wu}, {Wu}, {Wu}, {Wu}, {Wysocki}, {Xiao}, {Xu}, {Yamada}, {Yamamoto}, {Yamamoto}, {Yamamoto}, {Yamamoto}, {Yamashita}, {Yamazaki}, {Yang}, {Yang}, {Yang}, {Yang}, {Yang}, {Yap}, {Yeeles}, {Yelikar}, {Ying}, {Yokogawa}, {Yokoyama}, {Yokozawa}, {Yoo}, {Yoshioka}, {Yu}, {Yu}, {Yuzurihara}, {Zadro{\.z}ny}, {Zanolin}, {Zeidler},
  {Zelenova}, {Zendri}, {Zevin}, {Zhan}, {Zhang}, {Zhang}, {Zhang}, {Zhang}, {Zhang}, {Zhao}, {Zhao}, {Zhao}, {Zhao}, {Zhou}, {Zhou}, {Zhu}, {Zhu}, {Zimmerman}, {Zucker}, {Zweizig}, {Ligo Scientific Collaboration}, {VIRGO Collaboration}, \& {Kagra Collaboration}}]{Abbott2021}
{Abbott}, R., {Abbott}, T.~D., {Acernese}, F., {et~al.} 2021, \prd, 104, 122004, \dodoi{10.1103/PhysRevD.104.122004}

\bibitem[{{Abbott} {et~al.}(2023){Abbott}, {Abbott}, {Acernese}, {Ackley}, {Adams}, {Adhikari}, {Adhikari}, {Adya}, {Affeldt}, {Agarwal}, {Agathos}, {Agatsuma}, {Aggarwal}, {Aguiar}, {Aiello}, {Ain}, {Ajith}, {Akcay}, {Akutsu}, {Albanesi}, {Allocca}, {Altin}, {Amato}, {Anand}, {Anand}, {Ananyeva}, {Anderson}, {Anderson}, {Ando}, {Andrade}, {Andres}, {Andri{\'c}}, {Angelova}, {Ansoldi}, {Antelis}, {Antier}, {Appert}, {Arai}, {Arai}, {Arai}, {Araki}, {Araya}, {Araya}, {Areeda}, {Ar{\`e}ne}, {Aritomi}, {Arnaud}, {Arogeti}, {Aronson}, {Arun}, {Asada}, {Asali}, {Ashton}, {Aso}, {Assiduo}, {Aston}, {Astone}, {Aubin}, {Austin}, {Babak}, {Badaracco}, {Bader}, {Badger}, {Bae}, {Bae}, {Baer}, {Bagnasco}, {Bai}, {Baiotti}, {Baird}, {Bajpai}, {Ball}, {Ballardin}, {Ballmer}, {Balsamo}, {Baltus}, {Banagiri}, {Bankar}, {Barayoga}, {Barbieri}, {Barish}, {Barker}, {Barneo}, {Barone}, {Barr}, {Barsotti}, {Barsuglia}, {Barta}, {Bartlett}, {Barton}, {Bartos}, {Bassiri}, {Basti}, {Bawaj}, {Bayley}, {Baylor}, {Bazzan},
  {B{\'e}csy}, {Bedakihale}, {Bejger}, {Belahcene}, {Benedetto}, {Beniwal}, {Bennett}, {Bentley}, {Benyaala}, {Bergamin}, {Berger}, {Bernuzzi}, {Berry}, {Bersanetti}, {Bertolini}, {Betzwieser}, {Beveridge}, {Bhandare}, {Bhardwaj}, {Bhattacharjee}, {Bhaumik}, {Bilenko}, {Billingsley}, {Bini}, {Birney}, {Birnholtz}, {Biscans}, {Bischi}, {Biscoveanu}, {Bisht}, {Biswas}, {Bitossi}, {Bizouard}, {Blackburn}, {Blair}, {Blair}, {Blair}, {Bobba}, {Bode}, {Boer}, {Bogaert}, {Boldrini}, {Bonavena}, {Bondu}, {Bonilla}, {Bonnand}, {Booker}, {Boom}, {Bork}, {Boschi}, {Bose}, {Bose}, {Bossilkov}, {Boudart}, {Bouffanais}, {Bozzi}, {Bradaschia}, {Brady}, {Bramley}, {Branch}, {Branchesi}, {Brandt}, {Brau}, {Breschi}, {Briant}, {Briggs}, {Brillet}, {Brinkmann}, {Brockill}, {Brooks}, {Brooks}, {Brown}, {Brunett}, {Bruno}, {Bruntz}, {Bryant}, {Bulik}, {Bulten}, {Buonanno}, {Buscicchio}, {Buskulic}, {Buy}, {Byer}, {Davies}, {Cadonati}, {Cagnoli}, {Cahillane}, {Bustillo}, {Callaghan}, {Callister}, {Calloni}, {Cameron}, {Camp},
  {Canepa}, {Canevarolo}, {Cannavacciuolo}, {Cannon}, {Cao}, {Cao}, {Capocasa}, {Capote}, {Carapella}, {Carbognani}, {Carlin}, {Carney}, {Carpinelli}, {Carrillo}, {Carullo}, {Carver}, {Diaz}, {Casentini}, {Castaldi}, {Caudill}, {Cavagli{\`a}}, {Cavalier}, {Cavalieri}, {Ceasar}, {Cella}, {Cerd{\'a}-Dur{\'a}n}, {Cesarini}, {Chaibi}, {Chakravarti}, {Subrahmanya}, {Champion}, {Chan}, {Chan}, {Chan}, {Chan}, {Chan}, {Chandra}, {Chanial}, {Chao}, {Chapman-Bird}, {Charlton}, {Chase}, {Chassande-Mottin}, {Chatterjee}, {Chatterjee}, {Chatterjee}, {Chaturvedi}, {Chaty}, {Chatziioannou}, {Chen}, {Chen}, {Chen}, {Chen}, {Chen}, {Chen}, {Chen}, {Chen}, {Cheng}, {Cheong}, {Cheung}, {Chia}, {Chiadini}, {Chiang}, {Chiarini}, {Chierici}, {Chincarini}, {Chiofalo}, {Chiummo}, {Cho}, {Cho}, {Choudhary}, {Choudhary}, {Christensen}, {Chu}, {Chu}, {Chu}, {Chua}, {Chung}, {Ciani}, {Ciecielag}, {Cie{\'s}lar}, {Cifaldi}, {Ciobanu}, {Ciolfi}, {Cipriano}, {Cirone}, {Clara}, {Clark}, {Clark}, {Clarke}, {Clearwater}, {Clesse}, {Cleva},
  {Coccia}, {Codazzo}, {Cohadon}, {Cohen}, {Cohen}, {Colleoni}, {Collette}, {Colombo}, {Colpi}, {Compton}, {Constancio}, {Conti}, {Cooper}, {Corban}, {Corbitt}, {Cordero-Carri{\'o}n}, {Corezzi}, {Corley}, {Cornish}, {Corre}, {Corsi}, {Cortese}, {Costa}, {Cotesta}, {Coughlin}, {Coulon}, {Countryman}, {Cousins}, {Couvares}, {Coward}, {Cowart}, {Coyne}, {Coyne}, {Creighton}, {Creighton}, {Criswell}, {Croquette}, {Crowder}, {Cudell}, {Cullen}, {Cumming}, {Cummings}, {Cunningham}, {Cuoco}, {Cury{\l}o}, {Dabadie}, {Canton}, {Dall'Osso}, {D{\'a}lya}, {Dana}, {Daneshgaranbajastani}, {D'Angelo}, {Danila}, {Danilishin}, {D'Antonio}, {Danzmann}, {Darsow-Fromm}, {Dasgupta}, {Datrier}, {Dattilo}, {Dave}, {Davier}, {Davis}, {Davis}, {Daw}, {de Alarc{\'o}n}, {Dean}, {Debra}, {Deenadayalan}, {Degallaix}, {de Laurentis}, {Del{\'e}glise}, {Del Favero}, {de Lillo}, {de Lillo}, {Del Pozzo}, {Demarchi}, {de Matteis}, {D'Emilio}, {Demos}, {Dent}, {Depasse}, {de Pietri}, {De Rosa}, {de Rossi}, {Desalvo}, {de Simone}, {Dhurandhar},
  {D{\'\i}az}, {Diaz-Ortiz}, {Didio}, {Dietrich}, {di Fiore}, {di Fronzo}, {di Giorgio}, {di Giovanni}, {di Giovanni}, {di Girolamo}, {di Lieto}, {Ding}, {di Pace}, {di Palma}, {di Renzo}, {Divakarla}, {Dmitriev}, {Doctor}, {D'Onofrio}, {Donovan}, {Dooley}, {Doravari}, {Dorrington}, {Drago}, {Driggers}, {Drori}, {Ducoin}, {Dupej}, {Durante}, {D'Urso}, {Duverne}, {Dwyer}, {Eassa}, {Easter}, {Ebersold}, {Eckhardt}, {Eddolls}, {Edelman}, {Edo}, {Edy}, {Effler}, {Eguchi}, {Eichholz}, {Eikenberry}, {Eisenmann}, {Eisenstein}, {Ejlli}, {Engelby}, {Enomoto}, {Errico}, {Essick}, {Estell{\'e}s}, {Estevez}, {Etienne}, {Etzel}, {Evans}, {Evans}, {Ewing}, {Fafone}, {Fair}, {Fairhurst}, {Farah}, {Farinon}, {Farr}, {Farr}, {Farrow}, {Fauchon-Jones}, {Favaro}, {Favata}, {Fays}, {Fazio}, {Feicht}, {Fejer}, {Fenyvesi}, {Ferguson}, {Fernandez-Galiana}, {Ferrante}, {Ferreira}, {Fidecaro}, {Figura}, {Fiori}, {Fishbach}, {Fisher}, {Fittipaldi}, {Fiumara}, {Flaminio}, {Floden}, {Fong}, {Font}, {Fornal}, {Forsyth}, {Franke},
  {Frasca}, {Frasconi}, {Frederick}, {Freed}, {Frei}, {Freise}, {Frey}, {Fritschel}, {Frolov}, {Fronz{\'e}}, {Fujii}, {Fujikawa}, {Fukunaga}, {Fukushima}, {Fulda}, {Fyffe}, {Gabbard}, {Gabella}, {Gadre}, {Gair}, {Gais}, {Galaudage}, {Gamba}, {Ganapathy}, {Ganguly}, {Gao}, {Gaonkar}, {Garaventa}, {Garc{\'\i}a}, {Garc{\'\i}a-N{\'u}{\~n}ez}, {Garc{\'\i}a-Quir{\'o}s}, {Garufi}, {Gateley}, {Gaudio}, {Gayathri}, {Ge}, {Gemme}, {Gennai}, {George}, {George}, {Gerberding}, {Gergely}, {Gewecke}, {Ghonge}, {Ghosh}, {Ghosh}, {Ghosh}, {Ghosh}, {Giacomazzo}, {Giacoppo}, {Giaime}, {Giardina}, {Gibson}, {Gier}, {Giesler}, {Giri}, {Gissi}, {Glanzer}, {Gleckl}, {Godwin}, {Goetz}, {Goetz}, {Gohlke}, {Golomb}, {Goncharov}, {Gonz{\'a}lez}, {Gopakumar}, {Gosselin}, {Gouaty}, {Gould}, {Grace}, {Grado}, {Granata}, {Granata}, {Grant}, {Gras}, {Grassia}, {Gray}, {Gray}, {Greco}, {Green}, {Green}, {Gretarsson}, {Gretarsson}, {Griffith}, {Griffiths}, {Griggs}, {Grignani}, {Grimaldi}, {Grimm}, {Grote}, {Grunewald}, {Gruning}, {Guerra},
  {Guidi}, {Guimaraes}, {Guix{\'e}}, {Gulati}, {Guo}, {Guo}, {Gupta}, {Gupta}, {Gupta}, {Gustafson}, {Gustafson}, {Guzman}, {Ha}, {Haegel}, {Hagiwara}, {Haino}, {Halim}, {Hall}, {Hamilton}, {Hammond}, {Han}, {Haney}, {Hanks}, {Hanna}, {Hannam}, {Hannuksela}, {Hansen}, {Hansen}, {Hanson}, {Harder}, {Hardwick}, {Haris}, {Harms}, {Harry}, {Harry}, {Hartwig}, {Hasegawa}, {Haskell}, {Hasskew}, {Haster}, {Hattori}, {Haughian}, {Hayakawa}, {Hayama}, {Hayes}, {Healy}, {Heidmann}, {Heidt}, {Heintze}, {Heinze}, {Heinzel}, {Heitmann}, {Hellman}, {Hello}, {Helmling-Cornell}, {Hemming}, {Hendry}, {Heng}, {Hennes}, {Hennig}, {Hennig}, {Hernandez}, {Hernandez Vivanco}, {Heurs}, {Hild}, {Hill}, {Himemoto}, {Hines}, {Hiranuma}, {Hirata}, {Hirose}, {Hochheim}, {Hofman}, {Hohmann}, {Holcomb}, {Holland}, {Holley-Bockelmann}, {Hollows}, {Holmes}, {Holt}, {Holz}, {Hong}, {Hopkins}, {Hough}, {Hourihane}, {Howell}, {Hoy}, {Hoyland}, {Hreibi}, {Hsieh}, {Hsu}, {Huang}, {Huang}, {Huang}, {Huang}, {Huang}, {Huang}, {H{\"u}bner},
  {Huddart}, {Hughey}, {Hui}, {Hui}, {Husa}, {Huttner}, {Huxford}, {Huynh-Dinh}, {Ide}, {Idzkowski}, {Iess}, {Ikenoue}, {Imam}, {Inayoshi}, {Ingram}, {Inoue}, {Ioka}, {Isi}, {Isleif}, {Ito}, {Itoh}, {Iyer}, {Izumi}, {Jaberianhamedan}, {Jacqmin}, {Jadhav}, {Jadhav}, {James}, {Jan}, {Jani}, {Janquart}, {Janssens}, {Janthalur}, {Jaranowski}, {Jariwala}, {Jaume}, {Jenkins}, {Jenner}, {Jeon}, {Jeunon}, {Jia}, {Jin}, {Johns}, {Johnson-McDaniel}, {Jones}, {Jones}, {Jones}, {Jones}, {Jones}, {Jonker}, {Ju}, {Jung}, {Jung}, {Junker}, {Juste}, {Kaihotsu}, {Kajita}, {Kakizaki}, {Kalaghatgi}, {Kalogera}, {Kamai}, {Kamiizumi}, {Kanda}, {Kandhasamy}, {Kang}, {Kanner}, {Kao}, {Kapadia}, {Kapasi}, {Karat}, {Karathanasis}, {Karki}, {Kashyap}, {Kasprzack}, {Kastaun}, {Katsanevas}, {Katsavounidis}, {Katzman}, {Kaur}, {Kawabe}, {Kawaguchi}, {Kawai}, {Kawasaki}, {K{\'e}f{\'e}lian}, {Keitel}, {Key}, {Khadka}, {Khalili}, {Khan}, {Khazanov}, {Khetan}, {Khursheed}, {Kijbunchoo}, {Kim}, {Kim}, {Kim}, {Kim}, {Kim}, {Kim}, {Kimball},
  {Kimura}, {Kinley-Hanlon}, {Kirchhoff}, {Kissel}, {Kita}, {Kitazawa}, {Kleybolte}, {Klimenko}, {Knee}, {Knowles}, {Knyazev}, {Koch}, {Koekoek}, {Kojima}, {Kokeyama}, {Koley}, {Kolitsidou}, {Kolstein}, {Komori}, {Kondrashov}, {Kong}, {Kontos}, {Koper}, {Korobko}, {Kotake}, {Kovalam}, {Kozak}, {Kozakai}, {Kozu}, {Kringel}, {Krishnendu}, {Kr{\'o}lak}, {Kuehn}, {Kuei}, {Kuijer}, {Kulkarni}, {Kumar}, {Kumar}, {Kumar}, {Kumar}, {Kume}, {Kuns}, {Kuo}, {Kuo}, {Kuromiya}, {Kuroyanagi}, {Kusayanagi}, {Kuwahara}, {Kwak}, {Lagabbe}, {Laghi}, {Lalande}, {Lam}, {Lamberts}, {Landry}, {Lane}, {Lang}, {Lange}, {Lantz}, {La Rosa}, {Lartaux-Vollard}, {Lasky}, {Laxen}, {Lazzarini}, {Lazzaro}, {Leaci}, {Leavey}, {Lecoeuche}, {Lee}, {Lee}, {Lee}, {Lee}, {Lee}, {Lee}, {Lehmann}, {Lema{\^\i}tre}, {Leonardi}, {Leroy}, {Letendre}, {Levesque}, {Levin}, {Leviton}, {Leyde}, {Li}, {Li}, {Li}, {Li}, {Li}, {Li}, {Lin}, {Lin}, {Lin}, {Lin}, {Lin}, {Linde}, {Linker}, {Linley}, {Littenberg}, {Liu}, {Liu}, {Liu}, {Liu}, {Llamas},
  {Llorens-Monteagudo}, {Lo}, {Lockwood}, {Loh}, {London}, {Longo}, {Lopez}, {Portilla}, {Lorenzini}, {Loriette}, {Lormand}, {Losurdo}, {Lott}, {Lough}, {Lousto}, {Lovelace}, {Lucaccioni}, {L{\"u}ck}, {Lumaca}, {Lundgren}, {Luo}, {Lynam}, {Macas}, {Macinnis}, {MacLeod}, {MacMillan}, {Macquet}, {Hernandez}, {Magazz{\`u}}, {Magee}, {Maggiore}, {Magnozzi}, {Mahesh}, {Majorana}, {Makarem}, {Maksimovic}, {Maliakal}, {Malik}, {Man}, {Mandic}, {Mangano}, {Mango}, {Mansell}, {Manske}, {Mantovani}, {Mapelli}, {Marchesoni}, {Marchio}, {Marion}, {Mark}, {M{\'a}rka}, {M{\'a}rka}, {Markakis}, {Markosyan}, {Markowitz}, {Maros}, {Marquina}, {Marsat}, {Martelli}, {Martin}, {Martin}, {Martinez}, {Martinez}, {Martinez}, {Martinovic}, {Martynov}, {Marx}, {Masalehdan}, {Mason}, {Massera}, {Masserot}, {Massinger}, {Masso-Reid}, {Mastrogiovanni}, {Matas}, {Mateu-Lucena}, {Matichard}, {Matiushechkina}, {Mavalvala}, {McCann}, {McCarthy}, {McClelland}, {McClincy}, {McCormick}, {McCuller}, {McGhee}, {McGuire}, {McIsaac}, {McIver},
  {McRae}, {McWilliams}, {Meacher}, {Mehmet}, {Mehta}, {Meijer}, {Melatos}, {Melchor}, {Mendell}, {Menendez-Vazquez}, {Menoni}, {Mercer}, {Mereni}, {Merfeld}, {Merilh}, {Merritt}, {Merzougui}, {Meshkov}, {Messenger}, {Messick}, {Meyers}, {Meylahn}, {Mhaske}, {Miani}, {Miao}, {Michaloliakos}, {Michel}, {Michimura}, {Middleton}, {Milano}, {Miller}, {Miller}, {Miller}, {Millhouse}, {Mills}, {Milotti}, {Minazzoli}, {Minenkov}, {Mio}, {Mir}, {Miravet-Ten{\'e}s}, {Mishra}, {Mishra}, {Mistry}, {Mitra}, {Mitrofanov}, {Mitselmakher}, {Mittleman}, {Miyakawa}, {Miyamoto}, {Miyazaki}, {Miyo}, {Miyoki}, {Mo}, {Modafferi}, {Moguel}, {Mogushi}, {Mohapatra}, {Mohite}, {Molina}, {Molina-Ruiz}, {Mondin}, {Montani}, {Moore}, {Moraru}, {Morawski}, {More}, {Moreno}, {Moreno}, {Mori}, {Morisaki}, {Moriwaki}, {Morr{\'a}s}, {Mours}, {Mow-Lowry}, {Mozzon}, {Muciaccia}, {Mukherjee}, {Mukherjee}, {Mukherjee}, {Mukherjee}, {Mukherjee}, {Mukund}, {Mullavey}, {Munch}, {Mu{\~n}iz}, {Murray}, {Musenich}, {Muusse}, {Nadji}, {Nagano},
  {Nagano}, {Nagar}, {Nakamura}, {Nakano}, {Nakano}, {Nakashima}, {Nakayama}, {Napolano}, {Nardecchia}, {Narikawa}, {Naticchioni}, {Nayak}, {Nayak}, {Negishi}, {Neil}, {Neilson}, {Nelemans}, {Nelson}, {Nery}, {Neubauer}, {Neunzert}, {Ng}, {Ng}, {Nguyen}, {Nguyen}, {Nguyen}, {Quynh}, {Ni}, {Nichols}, {Nishizawa}, {Nissanke}, {Nitoglia}, {Nocera}, {Norman}, {North}, {Nozaki}, {Siles}, {Nuttall}, {Oberling}, {O'Brien}, {Obuchi}, {O'Dell}, {Oelker}, {Ogaki}, {Oganesyan}, {Oh}, {Oh}, {Oh}, {Ohashi}, {Ohishi}, {Ohkawa}, {Ohme}, {Ohta}, {Okada}, {Okutani}, {Okutomi}, {Olivetto}, {Oohara}, {Ooi}, {Oram}, {O'Reilly}, {Ormiston}, {Ormsby}, {Ortega}, {O'Shaughnessy}, {O'Shea}, {Oshino}, {Ossokine}, {Osthelder}, {Otabe}, {Ottaway}, {Overmier}, {Pace}, {Pagano}, {Page}, {Pagliaroli}, {Pai}, {Pai}, {Palamos}, {Palashov}, {Palomba}, {Pan}, {Pan}, {Panda}, {Pang}, {Pang}, {Pankow}, {Pannarale}, {Pant}, {Panther}, {Paoletti}, {Paoli}, {Paolone}, {Parisi}, {Park}, {Park}, {Parker}, {Pascucci}, {Pasqualetti}, {Passaquieti},
  {Passuello}, {Patel}, {Pathak}, {Patricelli}, {Patron}, {Paul}, {Payne}, {Pedraza}, {Pegoraro}, {Pele}, {Arellano}, {Penn}, {Perego}, {Pereira}, {Pereira}, {Perez}, {P{\'e}rigois}, {Perkins}, {Perreca}, {Perri{\`e}s}, {Petermann}, {Petterson}, {Pfeiffer}, {Pham}, {Phukon}, {Piccinni}, {Pichot}, {Piendibene}, {Piergiovanni}, {Pierini}, {Pierro}, {Pillant}, {Pillas}, {Pilo}, {Pinard}, {Pinto}, {Pinto}, {Piotrzkowski}, {Piotrzkowski}, {Pirello}, {Pitkin}, {Placidi}, {Planas}, {Plastino}, {Pluchar}, {Poggiani}, {Polini}, {Pong}, {Ponrathnam}, {Popolizio}, {Porter}, {Poulton}, {Powell}, {Pracchia}, {Pradier}, {Prajapati}, {Prasai}, {Prasanna}, {Pratten}, {Principe}, {Prodi}, {Prokhorov}, {Prosposito}, {Prudenzi}, {Puecher}, {Punturo}, {Puosi}, {Puppo}, {P{\"u}rrer}, {Qi}, {Quetschke}, {Quitzow-James}, {Qutob}, {Raab}, {Raaijmakers}, {Radkins}, {Radulesco}, {Raffai}, {Rail}, {Raja}, {Rajan}, {Ramirez}, {Ramirez}, {Ramos-Buades}, {Rana}, {Rapagnani}, {Rapol}, {Ray}, {Raymond}, {Raza}, {Razzano}, {Read}, {Rees},
  {Regimbau}, {Rei}, {Reid}, {Reid}, {Reitze}, {Relton}, {Renzini}, {Rettegno}, {Reza}, {Rezac}, {Ricci}, {Richards}, {Richardson}, {Richardson}, {Riemenschneider}, {Riles}, {Rinaldi}, {Rink}, {Rizzo}, {Robertson}, {Robie}, {Robinet}, {Rocchi}, {Rodriguez}, {Rolland}, {Rollins}, {Romanelli}, {Romano}, {Romel}, {Romero-Rodr{\'\i}guez}, {Romero-Shaw}, {Romie}, {Ronchini}, {Rosa}, {Rose}, {Rosi{\'n}ska}, {Ross}, {Rowan}, {Rowlinson}, {Roy}, {Roy}, {Roy}, {Rozza}, {Ruggi}, {Ruiz-Rocha}, {Ryan}, {Sachdev}, {Sadecki}, {Sadiq}, {Sago}, {Saito}, {Saito}, {Sakai}, {Sakai}, {Sakellariadou}, {Sakuno}, {Salafia}, {Salconi}, {Saleem}, {Salemi}, {Samajdar}, {Sanchez}, {Sanchez}, {Sanchez}, {Sanchis-Gual}, {Sanders}, {Sanuy}, {Saravanan}, {Sarin}, {Sassolas}, {Satari}, {Sathyaprakash}, {Sato}, {Sato}, {Sauter}, {Savage}, {Sawada}, {Sawant}, {Sawant}, {Sayah}, {Schaetzl}, {Scheel}, {Scheuer}, {Schiworski}, {Schmidt}, {Schmidt}, {Schnabel}, {Schneewind}, {Schofield}, {Sch{\"o}nbeck}, {Schulte}, {Schutz}, {Schwartz}, {Scott},
  {Scott}, {Seglar-Arroyo}, {Sekiguchi}, {Sekiguchi}, {Sellers}, {Sengupta}, {Sentenac}, {Seo}, {Sequino}, {Sergeev}, {Setyawati}, {Shaffer}, {Shahriar}, {Shams}, {Shao}, {Sharma}, {Sharma}, {Shawhan}, {Shcheblanov}, {Shibagaki}, {Shikauchi}, {Shimizu}, {Shimoda}, {Shimode}, {Shinkai}, {Shishido}, {Shoda}, {Shoemaker}, {Shoemaker}, {Shyamsundar}, {Sieniawska}, {Sigg}, {Singer}, {Singh}, {Singh}, {Singha}, {Sintes}, {Sipala}, {Skliris}, {Slagmolen}, {Slaven-Blair}, {Smetana}, {Smith}, {Smith}, {Soldateschi}, {Somala}, {Somiya}, {Son}, {Soni}, {Soni}, {Sordini}, {Sorrentino}, {Sorrentino}, {Sotani}, {Soulard}, {Souradeep}, {Sowell}, {Spagnuolo}, {Spencer}, {Spera}, {Srinivasan}, {Srivastava}, {Srivastava}, {Staats}, {Stachie}, {Steer}, {Steinhoff}, {Steinlechner}, {Steinlechner}, {Stevenson}, {Stops}, {Stover}, {Strain}, {Strang}, {Stratta}, {Strunk}, {Sturani}, {Stuver}, {Sudhagar}, {Sudhir}, {Sugimoto}, {Suh}, {Sullivan}, {Sullivan}, {Summerscales}, {Sun}, {Sun}, {Sunil}, {Sur}, {Suresh}, {Sutton}, {Suzuki},
  {Suzuki}, {Swinkels}, {Szczepa{\'n}czyk}, {Szewczyk}, {Tacca}, {Tagoshi}, {Tait}, {Takahashi}, {Takahashi}, {Takamori}, {Takano}, {Takeda}, {Takeda}, {Talbot}, {Talbot}, {Tanaka}, {Tanaka}, {Tanaka}, {Tanaka}, {Tanaka}, {Tanasijczuk}, {Tanioka}, {Tanner}, {Tao}, {Tao}, {Mart{\'\i}n}, {Taranto}, {Tasson}, {Telada}, {Tenorio}, {Terhune}, {Terkowski}, {Thirugnanasambandam}, {Thomas}, {Thomas}, {Thomas}, {Thompson}, {Thondapu}, {Thorne}, {Thrane}, {Tiwari}, {Tiwari}, {Tiwari}, {Toivonen}, {Toland}, {Tolley}, {Tomaru}, {Tomigami}, {Tomura}, {Tonelli}, {Torres-Forn{\'e}}, {Torrie}, {E Melo}, {T{\"o}yr{\"a}}, {Trapananti}, {Travasso}, {Traylor}, {Trevor}, {Tringali}, {Tripathee}, {Troiano}, {Trovato}, {Trozzo}, {Trudeau}, {Tsai}, {Tsai}, {Tsang}, {Tsang}, {Tsao}, {Tse}, {Tso}, {Tsubono}, {Tsuchida}, {Tsukada}, {Tsuna}, {Tsutsui}, {Tsuzuki}, {Turbang}, {Turconi}, {Tuyenbayev}, {Ubhi}, {Uchikata}, {Uchiyama}, {Udall}, {Ueda}, {Uehara}, {Ueno}, {Ueshima}, {Unnikrishnan}, {Uraguchi}, {Urban}, {Ushiba}, {Utina},
  {Vahlbruch}, {Vajente}, {Vajpeyi}, {Valdes}, {Valentini}, {Valsan}, {van Bakel}, {van Beuzekom}, {van den Brand}, {van den Broeck}, {Vander-Hyde}, {van der Schaaf}, {van Heijningen}, {Vanosky}, {van Putten}, {van Remortel}, {Vardaro}, {Vargas}, {Varma}, {Vas{\'u}th}, {Vecchio}, {Vedovato}, {Veitch}, {Veitch}, {Venneberg}, {Venugopalan}, {Verkindt}, {Verma}, {Verma}, {Veske}, {Vetrano}, {Vicer{\'e}}, {Vidyant}, {Viets}, {Vijaykumar}, {Villa-Ortega}, {Vinet}, {Virtuoso}, {Vitale}, {Vo}, {Vocca}, {von Reis}, {von Wrangel}, {Vorvick}, {Vyatchanin}, {Wade}, {Wade}, {Wagner}, {Walet}, {Walker}, {Wallace}, {Wallace}, {Walsh}, {Wang}, {Wang}, {Wang}, {Ward}, {Warner}, {Was}, {Washimi}, {Washington}, {Watchi}, {Weaver}, {Webster}, {Weinert}, {Weinstein}, {Weiss}, {Weller}, {Weller}, {Wellmann}, {Wen}, {We{\ss}els}, {Wette}, {Whelan}, {White}, {Whiting}, {Whittle}, {Wilken}, {Williams}, {Williams}, {Williams}, {Williamson}, {Willis}, {Willke}, {Wilson}, {Winkler}, {Wipf}, {Wlodarczyk}, {Woan}, {Woehler}, {Wofford},
  {Wong}, {Wu}, {Wu}, {Wu}, {Wu}, {Wysocki}, {Xiao}, {Xu}, {Yamada}, {Yamamoto}, {Yamamoto}, {Yamamoto}, {Yamamoto}, {Yamashita}, {Yamazaki}, {Yang}, {Yang}, {Yang}, {Yang}, {Yang}, {Yap}, {Yeeles}, {Yelikar}, {Ying}, {Yokogawa}, {Yokoyama}, {Yokozawa}, {Yoo}, {Yoshioka}, {Yu}, {Yu}, {Yuzurihara}, {Zadro{\.z}ny}, {Zanolin}, {Zeidler}, {Zelenova}, {Zendri}, {Zevin}, {Zhan}, {Zhang}, {Zhang}, {Zhang}, {Zhang}, {Zhang}, {Zhao}, {Zhao}, {Zhao}, {Zhao}, {Zheng}, {Zhou}, {Zhou}, {Zhu}, {Zhu}, {Zimmerman}, {Zlochower}, {Zucker}, {Zweizig}, {Ligo Scientific Collaboration}, Collaboration, \& {Kagra Collaboration}}]{LVK_O3}
---. 2023, Physical Review X, 13, 041039, \dodoi{10.1103/PhysRevX.13.041039}

\bibitem[{{Accadia} {et~al.}(2012){Accadia}, {Acernese}, {Alshourbagy}, {Amico}, {Antonucci}, {Aoudia}, {Arnaud}, {Arnault}, {Arun}, {Astone}, {Avino}, {Babusci}, {Ballardin}, {Barone}, {Barrand}, {Barsotti}, {Barsuglia}, {Basti}, {Bauer}, {Beauville}, {Bebronne}, {Bejger}, {Beker}, {Bellachia}, {Belletoile}, {Beney}, {Bernardini}, {Bigotta}, {Bilhaut}, {Birindelli}, {Bitossi}, {Bizouard}, {Blom}, {Boccara}, {Boget}, {Bondu}, {Bonelli}, {Bonnand}, {Boschi}, {Bosi}, {Bouedo}, {Bouhou}, {Bozzi}, {Bracci}, {Braccini}, {Bradaschia}, {Branchesi}, {Briant}, {Brillet}, {Brisson}, {Brocco}, {Bulik}, {Bulten}, {Buskulic}, {Buy}, {Cagnoli}, {Calamai}, {Calloni}, {Campagna}, {Canuel}, {Carbognani}, {Carbone}, {Cavalier}, {Cavalieri}, {Cecchi}, {Cella}, {Cesarini}, {Chassande-Mottin}, {Chatterji}, {Chiche}, {Chincarini}, {Chiummo}, {Christensen}, {Clapson}, {Cleva}, {Coccia}, {Cohadon}, {Colacino}, {Colas}, {Colla}, {Colombini}, {Conforto}, {Corsi}, {Cortese}, {Cottone}, {Coulon}, {Cuoco}, {D'Antonio}, {Daguin}, {Dari},
  {Dattilo}, {David}, {Davier}, {Day}, {Debreczeni}, {De Carolis}, {Dehamme}, {Del Fabbro}, {Del Pozzo}, {del Prete}, {Derome}, {De Rosa}, {DeSalvo}, {Dialinas}, {Di Fiore}, {Di Lieto}, {Emilio}, {Di Virgilio}, {Dietz}, {Doets}, {Dominici}, {Dominjon}, {Drago}, {Drezen}, {Dujardin}, {Dulach}, {Eder}, {Eleuteri}, {Enard}, {Evans}, {Fabbroni}, {Fafone}, {Fang}, {Ferrante}, {Fidecaro}, {Fiori}, {Flaminio}, {Forest}, {Forte}, {Fournier}, {Fournier}, {Franc}, {Francois}, {Frasca}, {Frasconi}, {Freise}, {Gaddi}, {Galimberti}, {Gammaitoni}, {Ganau}, {Garnier}, {Garufi}, {G{\'a}sp{\'a}r}, {Gemme}, {Genin}, {Gennai}, {Gennaro}, {Giacobone}, {Giazotto}, {Giordano}, {Giordano}, {Girard}, {Gouaty}, {Grado}, {Granata}, {Granata}, {Grave}, {Greverie}, {Groenstege}, {Guidi}, {Hamdani}, {Hayau}, {Hebri}, {Heidmann}, {Heitmann}, {Hello}, {Hemming}, {Hennes}, {Hermel}, {Heusse}, {Holloway}, {Huet}, {Iannarelli}, {Jaranowski}, {Jehanno}, {Journet}, {Karkar}, {Ketel}, {Voet}, {Kovalik}, {Kowalska}, {Kreckelbergh}, {Krolak},
  {Lacotte}, {Lagrange}, {La Penna}, {Laval}, {Le Marec}, {Leroy}, {Letendre}, {Li}, {Lieunard}, {Liguori}, {Lodygensky}, {Lopez}, {Lorenzini}, {Loriette}, {Losurdo}, {Loupias}, {Mackowski}, {Maiani}, {Majorana}, {Magazz{\`u}}, {Maksimovic}, {Malvezzi}, {Man}, {Mancini}, {Mansoux}, {Mantovani}, {Marchesoni}, {Marion}, {Marin}, {Marque}, {Martelli}, {Masserot}, {Massonnet}, {Matone}, {Matone}, {Mazzoni}, {Menzinger}, {Michel}, {Milano}, {Minenkov}, {Mitra}, {Mohan}, {Montorio}, {Morand}, {Moreau}, {Moreau}, {Morgado}, {Morgia}, {Mosca}, {Moscatelli}, {Mours}, {Mugnier}, {Mul}, {Naticchioni}, {Neri}, {Nocera}, {Pacaud}, {Pagliaroli}, {Pai}, {Palladino}, {Palomba}, {Paoletti}, {Paoletti}, {Paoli}, {Pardi}, {Parguez}, {Parisi}, {Pasqualetti}, {Passaquieti}, {Passuello}, {Perciballi}, {Perniola}, {Persichetti}, {Petit}, {Pichot}, {Piergiovanni}, {Pietka}, {Pignard}, {Pinard}, {Poggiani}, {Popolizio}, {Pradier}, {Prato}, {Prodi}, {Punturo}, {Puppo}, {Qipiani}, {Rabaste}, {Rabeling}, {R{\'a}cz}, {Raffaelli},
  {Rapagnani}, {Rapisarda}, {Re}, {Reboux}, {Regimbau}, {Reita}, {Remilleux}, {Ricci}, {Ricciardi}, {Richard}, {Ripepe}, {Robinet}, {Rocchi}, {Rolland}, {Romano}, {Rosi{\'n}ska}, {Roudier}, {Ruggi}, {Russo}, {Salconi}, {Sannibale}, {Sassolas}, {Sentenac}, {Solimeno}, {Sottile}, {Sperandio}, {Stanga}, {Sturani}, {Swinkels}, {Tacca}, {Taddei}, {Taffarello}, {Tarallo}, {Tissot}, {Toncelli}, {Tonelli}, {Torre}, {Tournefier}, {Travasso}, {Tremola}, {Turri}, {Vajente}, {van den Brand}, {Van Den Broeck}, {van der Putten}, {Vasuth}, {Vavoulidis}, {Vedovato}, {Verkindt}, {Vetrano}, {V{\'e}ziant}, {Vicer{\'e}}, {Vinet}, {Vilalte}, {Vitale}, {Vocca}, {Ward}, {Was}, {Yamamoto}, {Yvert}, {Zendri}, \& {Zhang}}]{Virgo_design}
{Accadia}, T., {Acernese}, F., {Alshourbagy}, M., {et~al.} 2012, Journal of Instrumentation, 7, 3012, \dodoi{10.1088/1748-0221/7/03/P03012}

\bibitem[{Adams {et~al.}(2013)Adams, Kochanek, Beacom, Vagins, \& Stanek}]{Adams_2013}
Adams, S.~M., Kochanek, C.~S., Beacom, J.~F., Vagins, M.~R., \& Stanek, K.~Z. 2013, The Astrophysical Journal, 778, 164, \dodoi{10.1088/0004-637x/778/2/164}

\bibitem[{{Afle} {et~al.}(2023){Afle}, {Kundu}, {Cammerino}, {Coughlin}, {Brown}, {Vartanyan}, \& {Burrows}}]{afle2023}
{Afle}, C., {Kundu}, S.~K., {Cammerino}, J., {et~al.} 2023, Physical Review D, 107, \dodoi{https://doi.org/10.1103/PhysRevD.107.123005}

\bibitem[{{Andresen} {et~al.}(2017){Andresen}, {M{\"u}ller}, {M{\"u}ller}, \& {Janka}}]{andresen2017}
{Andresen}, H., {M{\"u}ller}, B., {M{\"u}ller}, E., \& {Janka}, H.~T. 2017, \mnras, 468, 2032, \dodoi{10.1093/mnras/stx618}

\bibitem[{{Arca Sedda} {et~al.}(2020){Arca Sedda}, {Berry}, {Jani}, {Amaro-Seoane}, {Auclair}, {Baird}, {Baker}, {Berti}, {Breivik}, {Burrows}, {Caprini}, {Chen}, {Doneva}, {Ezquiaga}, {Saavik Ford}, {Katz}, {Kolkowitz}, {McKernan}, {Mueller}, {Nardini}, {Pikovski}, {Rajendran}, {Sesana}, {Shao}, {Tamanini}, {Vartanyan}, {Warburton}, {Witek}, {Wong}, \& {Zevin}}]{DECIGO_curves}
{Arca Sedda}, M., {Berry}, C. P.~L., {Jani}, K., {et~al.} 2020, Classical and Quantum Gravity, 37, 215011, \dodoi{10.1088/1361-6382/abb5c1}

\bibitem[{{Bizouard} {et~al.}(2021){Bizouard}, {Maturana-Russel}, {Torres-Forn{\'e}}, {Obergaulinger}, {Cerd{\'a}-Dur{\'a}n}, {Christensen}, {Font}, \& {Meyer}}]{bizouard2021}
{Bizouard}, M.-A., {Maturana-Russel}, P., {Torres-Forn{\'e}}, A., {et~al.} 2021, \prd, 103, 063006, \dodoi{10.1103/PhysRevD.103.063006}

\bibitem[{Bovy {et~al.}(2016)Bovy, Rix, Schlafly, Nidever, Holtzman, Shetrone, \& Beers}]{Bovy_2016}
Bovy, J., Rix, H.-W., Schlafly, E.~F., {et~al.} 2016, The Astrophysical Journal, 823, 30, \dodoi{10.3847/0004-637x/823/1/30}

\bibitem[{Branchesi {et~al.}(2023)Branchesi, Maggiore, Alonso, Badger, Banerjee, Beirnaert, Belgacem, Bhagwat, Boileau, Borhanian, Brown, Leong~Chan, Cusin, Danilishin, Degallaix, De~Luca, Dhani, Dietrich, Dupletsa, Foffa, Franciolini, Freise, Gemme, Goncharov, Ghosh, Gulminelli, Gupta, Kumar~Gupta, Harms, Hazra, Hild, Hinderer, Siong~Heng, Iacovelli, Janquart, Janssens, Jenkins, Kalaghatgi, Koroveshi, Li, Li, Loffredo, Maggio, Mancarella, Mapelli, Martinovic, Maselli, Meyers, Miller, Mondal, Muttoni, Narola, Oertel, Oganesyan, Pacilio, Palomba, Pani, Pasqualetti, Perego, Périgois, Pieroni, Piccinni, Puecher, Puppo, Ricciardone, Riotto, Ronchini, Sakellariadou, Samajdar, Santoliquido, Sathyaprakash, Steinlechner, Steinlechner, Utina, Van Den~Broeck, \& Zhang}]{EinsteinTelescope}
Branchesi, M., Maggiore, M., Alonso, D., {et~al.} 2023, Journal of Cosmology and Astroparticle Physics, 2023, 068, \dodoi{10.1088/1475-7516/2023/07/068}

\bibitem[{{Bruel} {et~al.}(2023){Bruel}, {Bizouard}, {Obergaulinger}, {Maturana-Russel}, {Torres-Forn{\'e}}, {Cerd{\'a}-Dur{\'a}n}, {Christensen}, {Font}, \& {Meyer}}]{bruel2023}
{Bruel}, T., {Bizouard}, M.-A., {Obergaulinger}, M., {et~al.} 2023, \prd, 107, 083029, \dodoi{10.1103/PhysRevD.107.083029}

\bibitem[{{Burrows} \& {Hayes}(1996)}]{burrows_hayes_1996}
{Burrows}, A., \& {Hayes}, J. 1996, \prl, 76, 352, \dodoi{10.1103/PhysRevLett.76.352}

\bibitem[{{Burrows} \& {Vartanyan}(2021)}]{burrows2021}
{Burrows}, A., \& {Vartanyan}, D. 2021, Nature

\bibitem[{{Burrows} {et~al.}(2023){Burrows}, {Vartanyan}, \& {Wang}}]{burrows2023_bh}
{Burrows}, A., {Vartanyan}, D., \& {Wang}, T. 2023, \apj, 957, 68, \dodoi{10.3847/1538-4357/acfc1c}

\bibitem[{{Burrows} {et~al.}(2024){Burrows}, {Wang}, \& {Vartanyan}}]{burrows2024}
{Burrows}, A., {Wang}, T., \& {Vartanyan}, D. 2024, Astrophys. J. Lett.

\bibitem[{Christodoulou(1991)}]{Christodoulou1991}
Christodoulou, D. 1991, Phys. Rev. Lett., 67, 1486, \dodoi{10.1103/PhysRevLett.67.1486}

\bibitem[{{Couch}(2013)}]{couch2013}
{Couch}, S.~M. 2013, \apj, 775, 35, \dodoi{10.1088/0004-637X/775/1/35}

\bibitem[{{Epstein}(1978)}]{Epstein1978}
{Epstein}, R. 1978, \apj, 223, 1037, \dodoi{10.1086/156337}

\bibitem[{{Evans} {et~al.}(2021){Evans}, {Adhikari}, {Afle}, {Ballmer}, {Biscoveanu}, {Borhanian}, {Brown}, {Chen}, {Eisenstein}, {Gruson}, {Gupta}, {Hall}, {Huxford}, {Kamai}, {Kashyap}, {Kissel}, {Kuns}, {Landry}, {Lenon}, {Lovelace}, {McCuller}, {Ng}, {Nitz}, {Read}, {Sathyaprakash}, {Shoemaker}, {Slagmolen}, {Smith}, {Srivastava}, {Sun}, {Vitale}, \& {Weiss}}]{CosmicExplorer}
{Evans}, M., {Adhikari}, R.~X., {Afle}, C., {et~al.} 2021, arXiv e-prints, arXiv:2109.09882, \dodoi{10.48550/arXiv.2109.09882}

\bibitem[{Fairhurst {et~al.}(2010)Fairhurst, Kalogera, Mandel, \& Weinstein}]{Fairhurst2010}
Fairhurst, S., Kalogera, V., Mandel, I., \& Weinstein, A. 2010, An Astrophysical Metric for LIGO Open Data Release, Tech. rep., LIGO

\bibitem[{{Finn} \& {Evans}(1990)}]{finn1990}
{Finn}, L., \& {Evans}, C. 1990, The Astrophysical Journal, 351, 588, \dodoi{10.1086/168497}

\bibitem[{{Flanagan} \& {Hughes}(1998)}]{flanagan_hughes_1998}
{Flanagan}, E.~E., \& {Hughes}, S.~A. 1998, Physical Review D, 57

\bibitem[{{Fuller} {et~al.}(2015){Fuller}, {Klion}, {Abdikamalov}, \& {Ott}}]{fuller2015}
{Fuller}, J., {Klion}, H., {Abdikamalov}, E., \& {Ott}, C.~D. 2015, \mnras, 450, 414, \dodoi{10.1093/mnras/stv698}

\bibitem[{{Gill}(2024)}]{gill2024}
{Gill}, K. 2024, arXiv e-prints, arXiv:2405.13211, \dodoi{10.48550/arXiv.2405.13211}

\bibitem[{Hayama {et~al.}(2018)Hayama, Kuroda, Kotake, \& Takiwaki}]{Hayama_2018}
Hayama, K., Kuroda, T., Kotake, K., \& Takiwaki, T. 2018, Monthly Notices of the Royal Astronomical Society: Letters, 477, L96–L100, \dodoi{10.1093/mnrasl/sly055}

\bibitem[{Isi(2023)}]{Isi_2023}
Isi, M. 2023, Classical and Quantum Gravity, 40, 203001, \dodoi{10.1088/1361-6382/acf28c}

\bibitem[{{Jakobus} {et~al.}(2023){Jakobus}, {M{\"u}ller}, {Heger}, {Zha}, {Powell}, {Motornenko}, {Steinheimer}, \& {Stoecker}}]{jakobus}
{Jakobus}, P., {M{\"u}ller}, B., {Heger}, A., {et~al.} 2023, arXiv e-prints, arXiv:2301.06515.
\newblock \doarXiv{2301.06515}

\bibitem[{Jurić {et~al.}(2008)Jurić, Ivezić, Brooks, Lupton, Schlegel, Finkbeiner, Padmanabhan, Bond, Sesar, Rockosi, Knapp, Gunn, Sumi, Schneider, Barentine, Brewington, Brinkmann, Fukugita, Harvanek, Kleinman, Krzesinski, Long, Neilsen, Nitta, Snedden, \& York}]{Juric2008}
Jurić, M., Ivezić, Z., Brooks, A., {et~al.} 2008, The Astrophysical Journal, 673, 864–914, \dodoi{10.1086/523619}

\bibitem[{Kawamura {et~al.}(2008)Kawamura, Ando, Nakamura, Tsubono, Tanaka, Funaki, Seto, Numata, Sato, Ioka, Kanda, Takashima, Agatsuma, Akutsu, Akutsu, s~Aoyanagi, Arai, Arase, Araya, Asada, Aso, Chiba, Ebisuzaki, Enoki, Eriguchi, Fujimoto, Fujita, Fukushima, Futamase, Ganzu, Harada, Hashimoto, Hayama, Hikida, Himemoto, Hirabayashi, Hiramatsu, Hong, Horisawa, Hosokawa, Ichiki, Ikegami, Inoue, Ishidoshiro, Ishihara, Ishikawa, Ishizaki, Ito, Itoh, Kamagasako, Kawashima, Kawazoe, Kirihara, Kishimoto, Kiuchi, Kobayashi, Kohri, Koizumi, Kojima, Kokeyama, Kokuyama, Kotake, Kozai, Kudoh, Kunimori, Kuninaka, Kuroda, i~Maeda, Matsuhara, Mino, Miyakawa, Miyoki, Morimoto, Morioka, Morisawa, Moriwaki, Mukohyama, Musha, Nagano, Naito, Nakagawa, Nakamura, Nakano, Nakao, Nakasuka, Nakayama, Nishida, Nishiyama, Nishizawa, Niwa, Ohashi, Ohishi, Ohkawa, Okutomi, Onozato, Oohara, Sago, Saijo, Sakagami, i~Sakai, Sakata, Sasaki, Sato, Shibata, Shinkai, Somiya, Sotani, Sugiyama, Suwa, Tagoshi, Takahashi, Takahashi, Takahashi,
  Takahashi, Takahashi, Takahashi, Takamori, Takano, Taniguchi, Taruya, Tashiro, Tokuda, Tokunari, Toyoshima, Tsujikawa, Tsunesada, i~Ueda, Utashima, Yamakawa, Yamamoto, Yamazaki, Yokoyama, Yoo, Yoshida, \& Yoshino}]{Decigo}
Kawamura, S., Ando, M., Nakamura, T., {et~al.} 2008, Journal of Physics: Conference Series, 120, 032004, \dodoi{10.1088/1742-6596/120/3/032004}

\bibitem[{Kotake(2013)}]{Kotake2013}
Kotake, K. 2013, Comptes Rendus. Physique, 14, 318, \dodoi{10.1016/j.crhy.2013.01.008}

\bibitem[{{Kuroda} {et~al.}(2014){Kuroda}, {Takiwaki}, \& {Kotake}}]{kuroda:14}
{Kuroda}, T., {Takiwaki}, T., \& {Kotake}, K. 2014, \prd, 89, 044011, \dodoi{10.1103/PhysRevD.89.044011}

\bibitem[{{Lian} {et~al.}(2022){Lian}, {Zasowsku}, {Mackereth}, {Imig}, {Holtzman}, {Beaton}, C., Katia, G., Danny, {Lane}, {Masters}, {Nitschelm}, \& {Roman-Lopes}}]{lian2022}
{Lian}, J., {Zasowsku}, G., {Mackereth}, T., {et~al.} 2022, Monthly Notices of the Royal Astronomical Society

\bibitem[{Marek {et~al.}(2009)Marek, Janka, \& Müller}]{Marek_2009}
Marek, A., Janka, H.-T., \& Müller, E. 2009, Astronomy \& Astrophysics, 496, 475–494, \dodoi{10.1051/0004-6361/200810883}

\bibitem[{Matthew {et~al.}(2020)Matthew, Stuani, Vitale, \& Hall}]{LIGO_curves}
Matthew, E., Stuani, M., Vitale, S., \& Hall, E. 2020, {Unofficial sensitivity curves (ASD) for aLIGO, Kagra, Virgo, Voyager, Cosmic Explorer, and Einstein Telescope}, Tech. rep., LIGO Scientific Collaboration

\bibitem[{{McMillan}(2017)}]{McMillan2016}
{McMillan}, P.~J. 2017, \mnras, 465, 76, \dodoi{10.1093/mnras/stw2759}

\bibitem[{{Mezzacappa} \& {Zanolin}(2024)}]{Mezzacappa2024}
{Mezzacappa}, A., \& {Zanolin}, M. 2024, arXiv e-prints, arXiv:2401.11635, \dodoi{10.48550/arXiv.2401.11635}

\bibitem[{Mezzacappa {et~al.}(2020)Mezzacappa, Marronetti, Landfield, Lentz, Yakunin, Bruenn, Hix, Messer, Endeve, Blondin, \& Harris}]{Mezzacappa2020}
Mezzacappa, A., Marronetti, P., Landfield, R.~E., {et~al.} 2020, Phys. Rev. D, 102, 023027, \dodoi{10.1103/PhysRevD.102.023027}

\bibitem[{{Moenchmeyer} {et~al.}(1991){Moenchmeyer}, {Schaefer}, {Mueller}, \& {Kates}}]{Moenchmeyer1991}
{Moenchmeyer}, R., {Schaefer}, G., {Mueller}, E., \& {Kates}, R.~E. 1991, \aap, 246, 417

\bibitem[{{Moore} {et~al.}(2014){Moore}, {Cole}, \& {Berry}}]{moore2014}
{Moore}, C.~J., {Cole}, R.~H., \& {Berry}, C. P.~L. 2014, Classical and Quantum Gravity, 32, \dodoi{10.1088/0264-9381/32/1/015014}

\bibitem[{{Morozova} {et~al.}(2018){Morozova}, {Radice}, {Burrows}, \& {Vartanyan}}]{morozova2018}
{Morozova}, V., {Radice}, D., {Burrows}, A., \& {Vartanyan}, D. 2018, The Astrophysical Journal, 861, 19, \dodoi{https://doi.org/10.48550/arXiv.1801.01914}

\bibitem[{Motizuki {et~al.}(2004)Motizuki, Madokoro, \& Shimizu}]{Motizuki_2004}
Motizuki, Y., Madokoro, H., \& Shimizu, T. 2004, EAS Publications Series, 11, 163–171, \dodoi{10.1051/eas:2004011}

\bibitem[{{Mueller}(1982)}]{muller1982}
{Mueller}, E. 1982, \aap, 114, 53

\bibitem[{{M{\"u}ller} {et~al.}(2013){M{\"u}ller}, {Janka}, \& {Marek}}]{muller2013}
{M{\"u}ller}, B., {Janka}, H.-T., \& {Marek}, A. 2013, The Astrophysical Journal, 766, 43, \dodoi{10.1088/0004-637x/766/1/43}

\bibitem[{{M{\"u}ller} \& {Janka}(1997)}]{muller1997}
{M{\"u}ller}, E., \& {Janka}, H.~T. 1997, \aap, 317

\bibitem[{{M{\"u}ller} {et~al.}(2012){M{\"u}ller}, {Janka}, \& {Wongwathanarat}}]{muller2012}
{M{\"u}ller}, E., {Janka}, H.~T., \& {Wongwathanarat}, A. 2012, \aap, 537, A63, \dodoi{10.1051/0004-6361/201117611}

\bibitem[{{M{\"u}ller} {et~al.}(2004){M{\"u}ller}, {Rampp}, {Buras}, {Janka}, \& {Shoemaker}}]{muller2004}
{M{\"u}ller}, E., {Rampp}, M., {Buras}, R., {Janka}, H.~T., \& {Shoemaker}, D.~H. 2004, \apj, 603, 221, \dodoi{10.1086/381360}

\bibitem[{{Murphy} {et~al.}(2009){Murphy}, {Ott}, \& {Burrows}}]{murphy2009}
{Murphy}, J.~W., {Ott}, C.~D., \& {Burrows}, A. 2009, \apj, 707, 1173, \dodoi{10.1088/0004-637X/707/2/1173}

\bibitem[{O'Connor \& Ott(2011)}]{O'Connor_2011}
O'Connor, E., \& Ott, C.~D. 2011, The Astrophysical Journal, 730, 70, \dodoi{10.1088/0004-637X/730/2/70}

\bibitem[{{Oohara} {et~al.}(1997){Oohara}, {Natamura}, \& {Shibata}}]{oohara1997}
{Oohara}, K.-i., {Natamura}, T., \& {Shibata}, M. 1997, Progress of Theoretical Physics Supplement, 128, 183–249, \dodoi{https://doi.org/10.1143/PTPS.128.183}

\bibitem[{{Ott}(2009)}]{ott2009}
{Ott}, C.~D. 2009, Classical and Quantum Gravity, \dodoi{0.1088/0264-9381/26/6/063001}

\bibitem[{Pajkos {et~al.}(2023)Pajkos, VanCamp, Pan, Vartanyan, Deppe, \& Couch}]{Pajkos2023}
Pajkos, M.~A., VanCamp, S.~J., Pan, K.-C., {et~al.} 2023, Astrophys. J.
\newblock \doarXiv{2306.01919}

\bibitem[{{Powell} \& {M{\"u}ller}(2024)}]{powell2024}
{Powell}, J., \& {M{\"u}ller}, B. 2024, arXiv e-prints, arXiv:2406.09691, \dodoi{10.48550/arXiv.2406.09691}

\bibitem[{Radice {et~al.}(2019)Radice, Morozova, Burrows, Vartanyan, \& Nagakura}]{Radice_2019}
Radice, D., Morozova, V., Burrows, A., Vartanyan, D., \& Nagakura, H. 2019, The Astrophysical Journal Letters, 876, L9, \dodoi{10.3847/2041-8213/ab191a}

\bibitem[{Rakhmanov \& Klimenko(2003)}]{Rakhmanov2003}
Rakhmanov, M., \& Klimenko, S. 2003, Angular Correlation of LIGO Hanford and Livingston Interferometers, Tech. rep., California Institute of Technology and Massachusetts Institute of Technology

\bibitem[{{Richardson} {et~al.}(2024){Richardson}, {Andresen}, {Mezzacappa}, {Zanolin}, {Benjamin}, {Marronetti}, {Lentz}, \& {Szczepanczyk}}]{richardson2024}
{Richardson}, C.~J., {Andresen}, H., {Mezzacappa}, A., {et~al.} 2024, arXiv e-prints, arXiv:2404.02131, \dodoi{10.48550/arXiv.2404.02131}

\bibitem[{Richardson {et~al.}(2022)Richardson, Zanolin, Andresen, Szczepańczyk, Gill, \& Wongwathanarat}]{Richardson_2022}
Richardson, C.~J., Zanolin, M., Andresen, H., {et~al.} 2022, Physical Review D, 105, \dodoi{10.1103/physrevd.105.103008}

\bibitem[{{Richers} {et~al.}(2017){Richers}, {Ott}, {Abdikamalov}, {O'Conner}, \& {Sullivan}}]{richers2017}
{Richers}, S., {Ott}, C.~D., {Abdikamalov}, E., {O'Conner}, E., \& {Sullivan}, C. 2017, Phys Review D

\bibitem[{Riess {et~al.}(2012)Riess, Fliri, \& Valls-Gabaud}]{Riess_2012}
Riess, A.~G., Fliri, J., \& Valls-Gabaud, D. 2012, The Astrophysical Journal, 745, 156, \dodoi{10.1088/0004-637x/745/2/156}

\bibitem[{Rozwadowska {et~al.}(2021)Rozwadowska, Vissani, \& Cappellaro}]{Rozwadowska_2021}
Rozwadowska, K., Vissani, F., \& Cappellaro, E. 2021, New Astronomy, 83, 101498, \dodoi{10.1016/j.newast.2020.101498}

\bibitem[{{Salpeter}(1955)}]{salpeter1955}
{Salpeter}, E.~E. 1955, The Astrophysical Journal, 121, \dodoi{10.1086/145971}

\bibitem[{{Schutz}(2011)}]{schutz2011}
{Schutz}, B.~F. 2011, Classical and Quantum Gravity, 1, 34, \dodoi{10.1088/0264-9381/28/12/125023}

\bibitem[{Shibagaki {et~al.}(2021)Shibagaki, Kuroda, Kotake, \& Takiwaki}]{Shibagaki_2021}
Shibagaki, S., Kuroda, T., Kotake, K., \& Takiwaki, T. 2021, Monthly Notices of the Royal Astronomical Society, 502, 3066–3084, \dodoi{10.1093/mnras/stab228}

\bibitem[{Skinner {et~al.}(2019)Skinner, Dolence, Burrows, Radice, \& Vartanyan}]{Skinner2019}
Skinner, M.~A., Dolence, J.~C., Burrows, A., Radice, D., \& Vartanyan, D. 2019, The Astrophysical Journal Supplement Series, 241, 7, \dodoi{10.3847/1538-4365/ab007f}

\bibitem[{{Somiya}(2012)}]{Kagra_design}
{Somiya}, K. 2012, Classical and Quantum Gravity, 29, 124007, \dodoi{10.1088/0264-9381/29/12/124007}

\bibitem[{{Srivastava} {et~al.}(2019){Srivastava}, {Ballmer}, {Brown}, {Afle}, {Burrows}, {Radice}, \& {Vartanyan}}]{srivastava2019}
{Srivastava}, V., {Ballmer}, S., {Brown}, D.~A., {et~al.} 2019, \prd, 100, 043026, \dodoi{10.1103/PhysRevD.100.043026}

\bibitem[{Stanzione {et~al.}(2020)Stanzione, West, Evans, Minyard, Ghattas, \& Panda}]{stanzione2020}
Stanzione, D., West, J., Evans, R.~T., {et~al.} 2020, in {PEARC '20}, Practice and Experience in Advanced Research Computing, Portland, OR, 106--111

\bibitem[{{Sukhbold} {et~al.}(2016){Sukhbold}, {Ertl}, {Woosley}, {Brown}, \& {Janka}}]{Sukhbold2016}
{Sukhbold}, T., {Ertl}, T., {Woosley}, S.~E., {Brown}, J.~M., \& {Janka}, H.~T. 2016, \apj, 821, 38, \dodoi{10.3847/0004-637X/821/1/38}

\bibitem[{{Sukhbold} {et~al.}(2018){Sukhbold}, {Woosley}, \& {Heger}}]{Sukhbold2018}
{Sukhbold}, T., {Woosley}, S.~E., \& {Heger}, A. 2018, \apj, 860, 93, \dodoi{10.3847/1538-4357/aac2da}

\bibitem[{{Szczepa{\'n}czyk} {et~al.}(2023){Szczepa{\'n}czyk}, {Zheng}, {Antelis}, {Benjamin}, {Bizouard}, {Casallas-Lagos}, {Cerd{\'a}-Dur{\'a}n}, {Davis}, {Gondek-Rosi{\'n}ska}, {Klimenko}, {Moreno}, {Obergaulinger}, {Powell}, {Ramirez}, {Ratto}, {Richarson}, {Rijal}, {Stuver}, {Szewczyk}, {Vedovato}, {Zanolin}, {Bartos}, {Bhaumik}, {Bulik}, {Drago}, {Font}, {De Colle}, {Garc{\'\i}a-Bellido}, {Gayathri}, {Hughey}, {Mitselmakher}, {Mishra}, {Mukherjee}, {Nguyen}, {Chan}, {Di Palma}, {Piotrzkowski}, \& {Singh}}]{szczepanczyk2023}
{Szczepa{\'n}czyk}, M.~J., {Zheng}, Y., {Antelis}, J.~M., {et~al.} 2023, arXiv e-prints, arXiv:2305.16146, \dodoi{10.48550/arXiv.2305.16146}

\bibitem[{Szczepańczyk {et~al.}(2021)Szczepańczyk, Antelis, Benjamin, Cavaglià, Gondek-Rosińska, Hansen, Klimenko, Morales, Moreno, Mukherjee, Nurbek, Powell, Singh, Sitmukhambetov, Szewczyk, Valdez, Vedovato, Westhouse, Zanolin, \& Zheng}]{szczepanczyk2021}
Szczepańczyk, M.~J., Antelis, J.~M., Benjamin, M., {et~al.} 2021, Physical Review D, 104, \dodoi{10.1103/physrevd.104.102002}

\bibitem[{Takami {et~al.}(2015)Takami, Rezzolla, \& Baiotti}]{Takami2014}
Takami, K., Rezzolla, L., \& Baiotti, L. 2015, Phys. Rev. D, 91, 064001, \dodoi{10.1103/PhysRevD.91.064001}

\bibitem[{Takiwaki \& Kotake(2018)}]{Takiwaki_2018}
Takiwaki, T., \& Kotake, K. 2018, Monthly Notices of the Royal Astronomical Society: Letters, 475, L91–L95, \dodoi{10.1093/mnrasl/sly008}

\bibitem[{{Takiwaki} \& {Kotake}(2018)}]{tk18}
{Takiwaki}, T., \& {Kotake}, K. 2018, \mnras, \dodoi{10.1093/mnrasl/sly008}

\bibitem[{Thorne(1992)}]{Thorne1992}
Thorne, K.~S. 1992, Phys. Rev. D, 45, 520, \dodoi{10.1103/PhysRevD.45.520}

\bibitem[{{Torres-Forn{\'e}} {et~al.}(2019){Torres-Forn{\'e}}, {Cerd{\'a}-Dur{\'a}n}, {Obergaulinger}, {M{\"u}ller}, \& {Font}}]{torres2019}
{Torres-Forn{\'e}}, A., {Cerd{\'a}-Dur{\'a}n}, P., {Obergaulinger}, M., {M{\"u}ller}, B., \& {Font}, J.~A. 2019, \prl, 123, 051102, \dodoi{10.1103/PhysRevLett.123.051102}

\bibitem[{{Turner}(1978)}]{Turner1978}
{Turner}, M.~S. 1978, \nat, 274, 565, \dodoi{10.1038/274565a0}

\bibitem[{{Vartanyan} \& {Burrows}(2020)}]{vartanyan2020}
{Vartanyan}, D., \& {Burrows}. 2020, The Astrophysical Journal

\bibitem[{{Vartanyan} {et~al.}(2023){Vartanyan}, {Burrows}, {Wang}, {Coleman}, \& {White}}]{vartanyan2023}
{Vartanyan}, D., {Burrows}, A., {Wang}, T., {Coleman}, M.~S., \& {White}, C.~J. 2023, Phys Review D, 107, 34, \dodoi{10.1103/PhysRevD.107.103015}

\bibitem[{{Wang} \& {Burrows}(2023)}]{wang2023}
{Wang}, T., \& {Burrows}, A. 2023, \apj, 954, 114, \dodoi{10.3847/1538-4357/ace7b2}

\bibitem[{{Wheeler}(1966)}]{wheeler1966}
{Wheeler}, J.~A. 1966, \araa, 4, 393, \dodoi{10.1146/annurev.aa.04.090166.002141}

\bibitem[{{Yakunin} {et~al.}(2010){Yakunin}, {Marronetti}, {Mezzacappa}, {Bruenn}, {Lee}, {Chertkow}, {Hix}, {Blondin}, {Lentz}, {Messer}, \& {Yoshida}}]{yakunin:10}
{Yakunin}, K.~N., {Marronetti}, P., {Mezzacappa}, A., {et~al.} 2010, Classical and Quantum Gravity, 27, 194005, \dodoi{10.1088/0264-9381/27/19/194005}

\end{thebibliography}
\bibliographystyle{aasjournal}

\section*{Appendix A: Galactic Density and Rate Calculation}
\subsection*{McMillan (2017) Model}
Unlike other galactic density models such as \citet{Juric2008}, \citet{lian2022}, and \citet{Bovy_2016}, instead of proposing a parametrization and directly fitting free parameters to survey data, \citet{McMillan2016} uses a parametric model developed by previous work and data from \citet{Juric2008} as priors in a Markov Chain Monte Carlo method to determine the probability density function of the parameters $\Sigma_{thin}$, $\Sigma_{thick}$, $R_{d, thin}$, and $R_{d, thick}$, while holding constant the other parameters at values derived in previous work. While the full model includes a dark matter halo and molecular gas, for the purposes of CCSN rates, we need to include only the components of the Milky Way containing stars, which are the bulge, thin disk, and thick disk:
    \begin{equation}
        \rho(R,z) = \rho_b(R, z) + \rho_{d,thin}(R,z) + f\rho_{d,thick}(R,z) \,\, .
    \end{equation}
The expression for the bulge is given by
    \begin{equation}
    \begin{aligned}
        \rho_b(R, z) &= \frac{\rho_{0, b}}{(1 + r/r_0)^{\alpha}}e^{-(\frac{r}{r_{cut}})^2} \\
        r &= \sqrt{R^2 + (z / q)^2}\, ,
    \end{aligned}
    \end{equation}
where $\rho_{0, b} = 9.84\times10^{10} M_{\odot}/kpc^3$, $\alpha = 1.8$, $r_0 = 0.075$ kpc, $r_{cut} = 2.1$kpc, and $q=0.5$. While \citet{McMillan2016} acknowledges the presence of a bar at the center of the galaxy, an axisymmetric model is used for simplicity, an assumption which we adopt as well. The expression for both disks is given by
\begin{equation}
    \rho_d(R, z) = \frac{\Sigma_0}{2z_d}exp(-\frac{|z|}{z_d}-\frac{R}{R_d})
\end{equation}
where $\Sigma_{0, thin} = 8.96\times10^{8} M_{\odot}/kpc^2$, $R_{d, thin} = 2.5$kpc, $z_{d,thin} = 0.3$kpc, $\Sigma_{0, thick} = 1.83\times10^{8} M_{\odot}/kpc^2$, $R_{d, thick} = 3.02$kpc, and $z_{d, thick} = 0.9$kpc.

\begin{figure}[ht]
    \centering
    \includegraphics[width=0.6\linewidth]{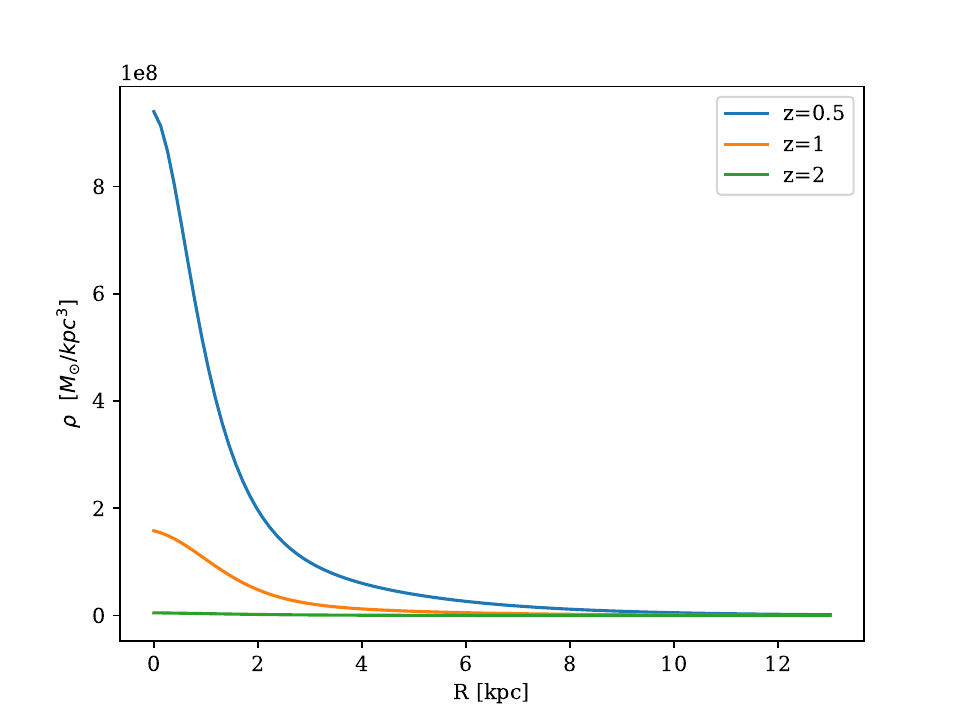}
    \caption{Milky Way density profile as described by \citet{McMillan2016} as a function of galactic radius $R$ (in kpc) for several values of vertical height $z$. The increase in density close to R=0 is due to the bulge at the center of the galaxy, and the effect of the bulge is less evident for larger values of z. \label{McMillan_density}}
    \end{figure}

In Figure \ref{McMillan_density}, we plot this density model as a function of galactic radius with the given parameters 

\subsection*{Integration Procedure for Galaxy Fraction}
In this section, we describe how we perform a spherical integral centered at a pointed located away from the axis of symmetry of a cylindrically-symmetric density profile.  

The \citet{McMillan2016} model uses galactic cylindrical coordinates $(R, \theta, z)$. In this coordinate system, Earth is located at $(R_{\odot}, z_{\odot}) = (x_{\odot}, 0, z_{\odot})$ where we have arbitrarily set $\theta = 0$ for Earth's location. We then convert to spherical coordinates $(r, \alpha, \beta)$ which are centered at $(x_{\odot}, 0, z_{\odot})$. The galactocentric Cartesian coordinates can be written in terms of these spherical coordinates using
\begin{equation}
\begin{aligned}
x &= r\sin\beta\cos\alpha + x_{\odot} \\
y &= r\sin\beta\sin\alpha \\
z &= r\cos\beta + z_{\odot} \,\, .
\end{aligned}
\end{equation}
We can now convert these Cartesian coordinates to the original cylindrical coordinates to obtain
\begin{equation}
\begin{aligned}
R(r, \alpha, \beta) &= \sqrt{(r\sin\beta\cos\alpha + x_{\odot})^2+(r\sin\beta\sin\alpha)^2} \\
z(r, \alpha, \beta) &= r\cos\beta + z_{\odot} \,\, .
\end{aligned}
\end{equation}
Finally, we can integrate the above cylindrical density profiles over an off-center sphere as
\begin{equation}
    \int_0^{d}\int_0^{\pi}\int_{0}^{2\pi}\rho(R(r, \alpha, \beta), z(r, \alpha, \beta))r^2\sin\beta d\alpha d\beta dr \,\, .
\end{equation}
This integral is plotted as a function of $d$ in Figure \ref{McMillian_int_density}.

\end{document}